\documentclass[CERN,manyauthors]{cernphprep}

\usepackage[comma,square,numbers,sort&compress]{natbib}
\usepackage[pdftitle={Study of hard and electromagnetic processes at SPS energy},
        pdfauthor={NA60+ Collaboration},
        pdfsubject={},
        bookmarksopen=false]{hyperref}
\usepackage{lineno}
\usepackage{bm}
\usepackage{xcolor}
\usepackage{textcomp}
\usepackage{parskip}
\usepackage[export]{adjustbox}
\usepackage{booktabs}

\usepackage{enumitem}
\setlist{nolistsep}

\usepackage{makecell}

\usepackage{siunitx}
\newcommand{\pp}{\ensuremath{\rm pp}\xspace}

\newcommand{\pA}{\ensuremath{\text{pA}}\xspace} 
\newcommand{\pBe}{\ensuremath{\text{p--Be}}\xspace}
\newcommand{\pPb}{\ensuremath{\text{p--Pb}}\xspace}

\newcommand{\AuAu}{\ensuremath{\text{Au--Au}}\xspace}

\newcommand{\PbPb}{\ensuremath{\text{Pb--Pb}}\xspace}
\newcommand{\InIn}{\ensuremath{\text{In--In}}\xspace}

\newcommand{\sqrtsNN}{\ensuremath{\sqrt{s_{\mathrm{\scriptscriptstyle NN}}}}\xspace}

\newcommand{\Elab}{\ensuremath{E_{\rm lab}}\xspace}

\newcommand{\aone}{\ensuremath{\mathrm{a}_1}\xspace}
\newcommand{\jpsi}{\ensuremath{\mathrm{J}/\psi}\xspace}
\newcommand{\chic}{\ensuremath{\chi_{\rm c}}\xspace}

\newcommand{\psiP}{\ensuremath{\psi\text{(2S)}}\xspace}

\newcommand{\ccbar}{\ensuremath{\mathrm{c\overline{c}}}\xspace}

\newcommand{\Dmeson}[1]{\ensuremath{\mathrm{D}^{#1}}\xspace}

\newcommand{\Dzero}{\Dmeson{0}}
\newcommand{\Dplus}{\Dmeson{+}}

\newcommand{\Ds}{\ensuremath{\mathrm{D}^{+}_{\rm s}}\xspace}

\newcommand{\lambdac}{\ensuremath{\Lambda_{\rm c}}\xspace}
\newcommand{\lambdacplus}{\ensuremath{\Lambda_{\rm c}^+}\xspace}

\newcommand{\mumu}{\ensuremath{\mu^+\mu^-}\xspace}

\newcommand{\raa}{\ensuremath{R_{\mathrm{AA}}}\xspace}

\DeclareSIUnit{\clight}{\ensuremath{\mathit{c}}}
\DeclareSIUnit\nucleon{nucleon}
\DeclareSIUnit{\AGeV}{\giga\electronvolt\per\nucleon}
\DeclareSIUnit\barn{b}

\newcommand{\MeV}{\ensuremath{~\si{\mega\electronvolt}}\xspace}
\newcommand{\GeV}{\ensuremath{~\si{\giga\electronvolt}}\xspace}
\newcommand{\TeV}{\ensuremath{~\si{\tera\electronvolt}}\xspace}

\newcommand{\GeVc}{\ensuremath{~\si{\giga\electronvolt\per\clight}}\xspace}

\newcommand{\MeVcc}{\ensuremath{~\si{\mega\electronvolt\per\clight\squared}}\xspace}
\newcommand{\GeVcc}{\ensuremath{~\si{\giga\electronvolt\per\clight\squared}}\xspace}
\newcommand{\mb}{\ensuremath{~\si{\milli\barn}}\xspace}
\newcommand{\mub}{\ensuremath{~\si{\micro\barn}}\xspace}
\newcommand{\mbinv}{\ensuremath{~\si[per-mode=reciprocal]{\per\milli\barn}}\xspace}

\newcommand{\eg}{e.\,g.\xspace}%
\newcommand{\ie}{i.\,e.\xspace}%

\newcommand{\dd}{\ensuremath{\mathrm{d}}}

\newcommand{\beq}{\begin{equation}}
\newcommand{\eeq}{\end{equation}}
\newcommand{\beqn}{\begin{eqnarray}}
\newcommand{\eeqn}{\end{eqnarray}}

\newcommand{\beqa}{\begin{eqnarray}}
\newcommand{\eeqa}{\end{eqnarray}}
\def\lsim{\raise0.3ex\hbox{$<$\kern-0.75em\raise-1.1ex\hbox{$\sim$}}}
\def\gsim{\raise0.3ex\hbox{$>$\kern-0.75em\raise-1.1ex\hbox{$\sim$}}}

\newcommand{\Npart}{\ensuremath{N_{\mathrm{part}}}\xspace}

\newcommand{\taa}{\ensuremath{T_{\mathrm{AA}}}\xspace}

\newcommand{\pt}{\ensuremath{p_{\mathrm{T}}}\xspace}

\newcommand{\T}[1]{\ensuremath{T_{\mathrm{#1}}}\xspace}
\newcommand{\Tc}{\T{c}}
\newcommand{\Tpc}{T_{\rm pc}}

\newcommand{\muB}{\ensuremath{\mu_{\rm B}}\xspace}
\newcommand{\pythia}{\textsc{Pythia}\xspace}
\newcommand{\powheg}{\textsc{Powheg}\xspace}
\newcommand{\fluka}{\textsc{Fluka}\xspace}

\begin{document}%


\begin{titlepage}
\title{\vspace{-2.0cm}Letter of Intent: the NA60+ experiment}
\ShortTitle{LoI: NA60+}
\bigskip
\Collaboration{NA60+ Collaboration\thanks{See \hyperref[app:collab]{Appendix} for the list of collaboration members. \\ \indent\indent E-mail contacts: {\tt enrico.scomparin@to.infn.it, gianluca.usai@ca.infn.it}}}
\ShortAuthor{NA60+ Collaboration}

\bigskip
\begin{figure}[h!]
    \centering
    \includegraphics[width=0.9\textwidth]{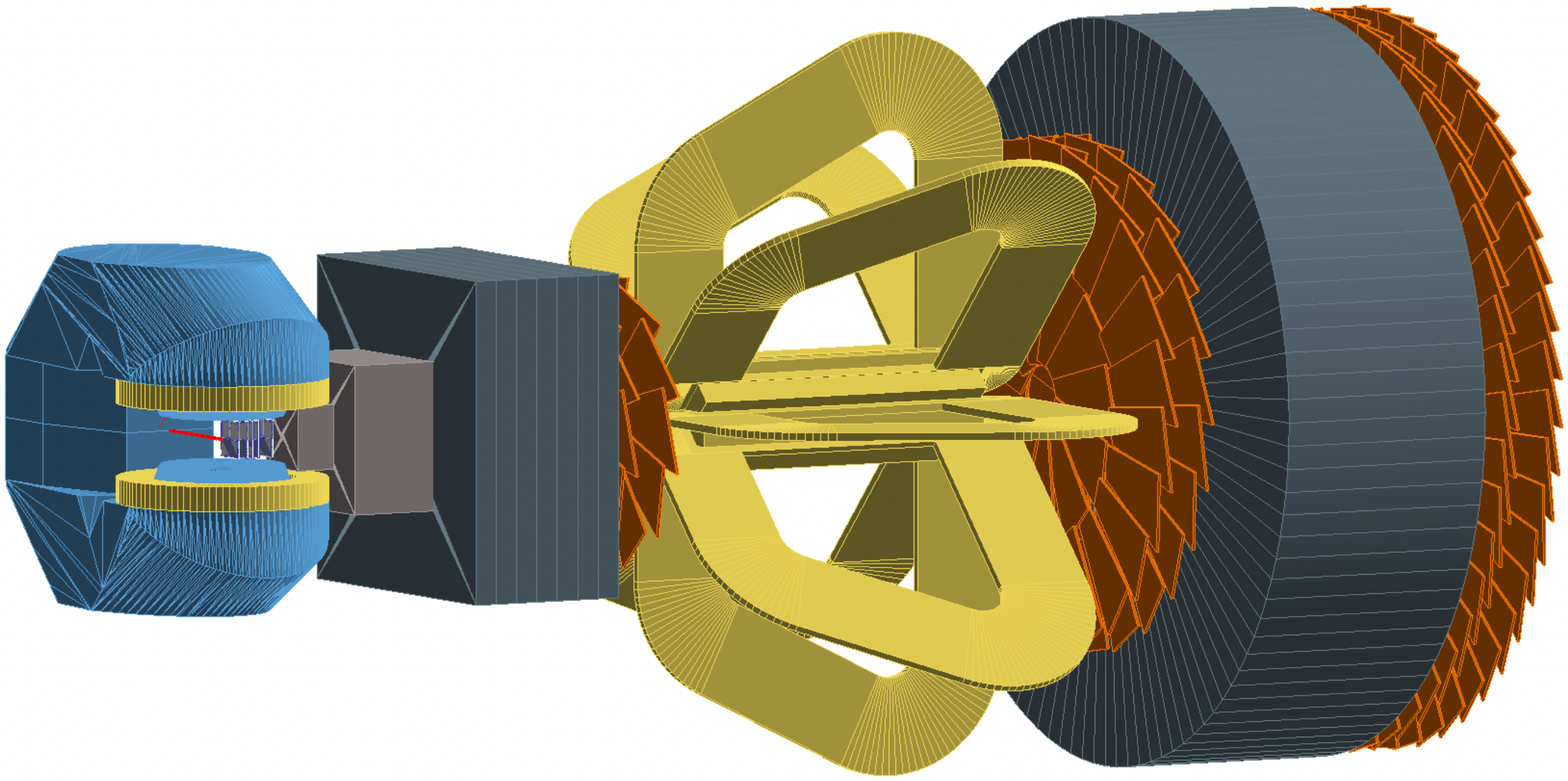}
\end{figure}

\begin{abstract}
\medskip
We propose a new fixed-target experiment for the study of electromagnetic and hard probes of the Quark-Gluon Plasma (QGP) in heavy-ion collisions at the CERN SPS. The experiment aims at performing measurements of the dimuon spectrum from threshold up to the charmonium region, and of hadronic decays of charm and strange hadrons. It is based on a muon spectrometer, which includes a toroidal magnet and six planes of tracking detectors, coupled to a vertex spectrometer, equipped with Si MAPS immersed in a dipole field. High luminosity is an essential requirement for the experiment, with the goal of taking data with 10$^6$ incident ions/s, at collision energies ranging from $\sqrt{s_{\rm NN}} = 6.3$ GeV ($E_{\rm lab}= 20$ A\,GeV) to top SPS energy ($\sqrt{s_{\rm NN}} = 17.3$ GeV, $E_{\rm lab}= 158$ A\,GeV). This document presents the physics motivation, the foreseen experimental set-up including integration and radioprotection studies, the current detector choices together with the status of the corresponding R\&D, and the outcome of physics performance studies. A preliminary cost evaluation is also carried out.
\end{abstract}
\vfill

{\centering\normalsize December 24, 2022\par}
\vfill
\clearpage
\end{titlepage}

\pagenumbering{roman}
\tableofcontents
\cleardoublepage

\pagenumbering{arabic}
\setcounter{page}{1}

\section*{Executive summary 
}
\label{sec:executivesummary}
\addcontentsline{toc}{section}{Executive summary 
}
\vskip 0.5cm
Experiments with ultrarelativistic heavy-ion beams allow the study of the Quark-Gluon Plasma (QGP), a state of matter where quarks and gluons in thermal equilibrium are deconfined over length scales much larger than the typical hadron size. The properties of this state and of the phase transition from hadronic matter to QGP represent an important field of research in the QCD domain and their investigation is actively pursued at both collider and fixed-target experiments. While the QGP studied by experiments at collider energies at RHIC and LHC is characterized by a large initial temperature (up to $T\sim 500$ MeV) and a zero net baryonic density (or equivalently zero 
baryo-chemical potential $\mu_{\rm B}$), collisions in the center-of-mass energy range per nucleon-nucleon collision $6 < \sqrt{s_{\rm{NN}}} < 17.3$ GeV, as available at the CERN SPS, may lead to the formation of a QGP characterized by a smaller initial $T$ and non-zero $\mu_{\rm B}$. When increasing $\mu_{\rm B}$, the transition from hadronic matter to QGP is expected to change from a rapid crossover to a first-order phase transition, with the presence of a critical point separating the two regimes. 

 Here we describe our intent to propose an experiment at the CERN SPS, presently denoted as NA60+, to study specific observables related to the high-$\mu_{\rm B}$ QGP formation and to the corresponding phase transition. More in detail, we would like to study electromagnetic probes of the QGP, via the measurement of the muon pair spectrum, as well as open and hidden charm production. The former give access to the temperature of the deconfined medium as well as to the modification of the hadronic spectrum due to the restoration of the chiral symmetry of QCD close to the phase transition. The latter give constraints on the transport properties of the QGP (open charm) and on the modification of the QCD binding in a deconfined medium (charmonium). None of these observables can be accurately measured in the SPS energy range by any other existing or presently foreseen experimental program. The above mentioned processes will be studied as a function of the collision energy with high-intensity beams, by means of an energy scan to be carried out with Pb--Pb collisions from $\sqrt{s_{\rm{NN}}}=6$ GeV (corresponding to $E_{\rm lab}\sim$20  A\,GeV) or even lower, if provided, up to top SPS energy ($\sqrt{s_{\rm{NN}}}=17.3$ GeV, $E_{\rm lab}=$158  A\,GeV). In addition, such an experiment can access strange hadron and, for the first time, hypernuclei production in this energy range.

The concept of the experimental set-up is inspired by the former NA60 experiment and will be based on a muon spectrometer, covering approximately one unit of rapidity, coupled to a vertex spectrometer. The muon spectrometer will include a warm toroidal magnet, and its tracking system will be based on six stations, two of them positioned upstream and two downstream of the toroid, and finally two stations downstream of a graphite absorber. Well-established detector techniques are foreseen for this system, with the choice between GEMs and MWPCs presently under discussion. A further thick absorber, made of graphite and BeO will be positioned upstream of the muon spectrometer, to filter out hadrons produced in the Pb--Pb interaction. Its thickness will be varied according to collision energy, allowing an efficient containment of the hadronic showers. At the same time, the muon spectrometer position will be shifted, in order to ensure a similar center-of-mass rapidity coverage at the various energies. Finally, the target region will be immersed in a dipole field, provided by the refurbished MEP48 magnet, presently stored at CERN. The vertex spectrometer, positioned immediately downstream of the targets, will consist of a series of stations (from 5 up to 10) of high-granularity and low material budget monolithic active pixel sensors (MAPS), that will allow an efficient tracking of the large number of produced charged particles, with $({\rm d}N_{\rm ch}/{\rm d}\eta)\sim$400 for central Pb--Pb collisions at top SPS energy. By matching, in coordinate and momentum space, tracks in the muon spectrometer with the corresponding tracks in the vertex spectrometer, it will be possible to perform an accurate measurement of the muon kinematics, reaching, as an example, resolutions $<10$ MeV at the $\omega$-meson mass and $\sim 30$ MeV at the J/$\psi$. Furthermore, the high-granularity of the vertex detectors will allow accurate measurements of hadronic decays of open charm and strange hadrons.

An R\&D program has already started for the definition of the detector aspects. For the MAPS, studies are advancing in the frame of a collaboration with ALICE, with the goal of producing, via a stitching technique, large surface and low-material budget detectors. For the muon tracker, where more traditional techniques as GEMs or MWPCs are being considered, first prototypes have been constructed and will be tested on SPS hadron beams. For the toroidal magnet, a working prototype with a scale 1:5 was already built and tested, to assess the feasibility of a device with the needed geometry, and provide the necessary information for the design of the full-scale object.  

The experiment can be located on the H8 beam line, in the PPE138 experimental zone of the EHN1 hall. Integration and radiation protection studies have shown the feasibility of such a solution, with the possibility of sustaining Pb beam intensities of the order of $10^7$ per 10 s spill, provided that an adequate shielding is built around the experimental set-up.

The experiment plans to perform measurements of Pb--Pb collisions at a single collision energy for each data taking period allocated for heavy-ion running in the CERN accelerator complex, typically one month per year. The experimental program will need a minimum of 6-7 years of data taking, to provide a fine enough energy scan for the 
characterization of the QGP at various baryo-chemical potential and the search of signals related to the first order phase transition. Measurements with a Pb beam need to be complemented by corresponding data taking periods with a proton beam incident on various nuclear targets, collecting an equivalent luminosity and providing reference data for the correct quantitative interpretation of nuclear collision results.

We foresee a time schedule of the experiment that will lead to data taking after the LHC Long Shutdown 3, once beams in the SPS will become available again. In this optics, the next step after this LoI would be the submission of a Technical Proposal to the SPSC, ideally by 2024.

The document is structured as follows: after a short overview on QCD in the region of large 
baryo-chemical potentials (Ch.~\ref{Overview}), we will elaborate in more detail the physics case for the unique measurements that will be accessible to NA60+ (Ch.~\ref{RareProbes}). In Ch.~\ref{NA60plus} the proposed layout of NA60+ will be introduced, together with a presentation of the foreseen data taking strategy and a discussion on the  role of the experiment in the frame of the various projects aiming at accessing the high-$\mu_{\rm B}$ region. Studies of physics performance for the measurements the experiment is aiming at will be the main subject of Ch.~\ref{PhysicsPerformance}. In Ch.~\ref{Detectors} the currently foreseen technical choices for the various detector systems, and in particular for the vertex and the muon spectrometer, will be discussed. Chapter~\ref{ExperimentalSite} contains a description of the integration of the experiment in the PPE138 zone, including the proposed beam set-up and a discussion of radiation protection issues. Finally, in Ch.~\ref{TimelineCost} a tentative timeline for the realization of the experiment is presented, together with a preliminary cost estimate.

\newpage
\section{Overview of the QCD phase diagram and general landscape for high-\texorpdfstring{$\mu_{\rm B}$}{muB} studies}
\label{Overview}
\vskip 0.4cm

\begin{figure}[!b]
\begin{center}
\includegraphics[height=0.31\linewidth,valign=t]{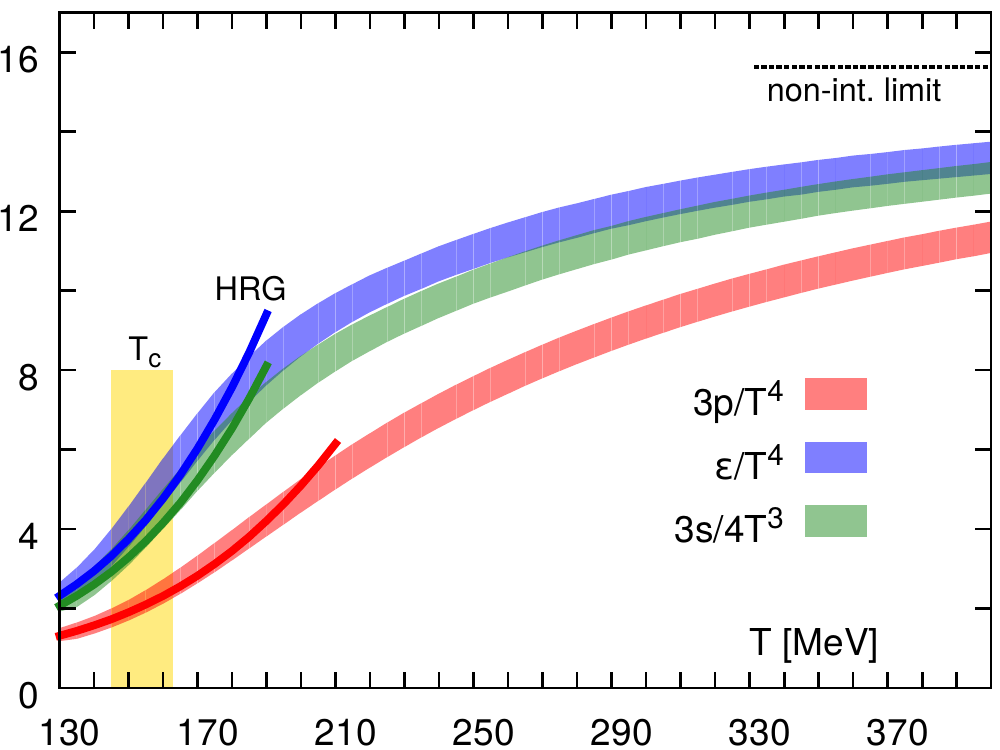}
\includegraphics[height=0.34\linewidth,valign=t]{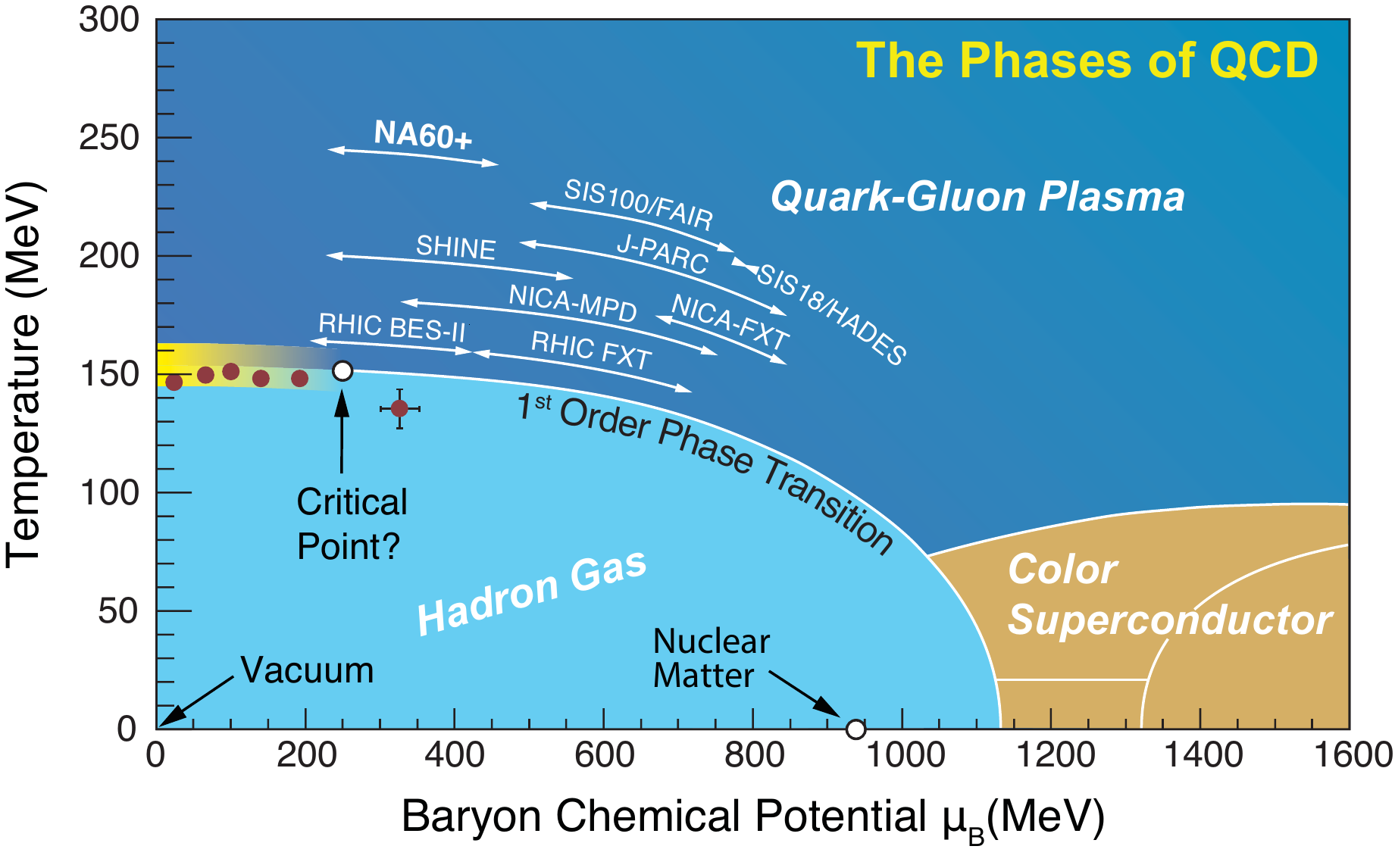}
\caption{(Left) The temperature dependence of the energy density (blue), scaled by the 4th power of temperature, at vanishing \muB computed in lattice-QCD~\cite{Bazavov:2014pvz}, is characterized by a rise in the effective number of active degrees of freedom, indicating a cross-over transition from hadronic matter to a QGP.
(Right) Sketch of the QCD phase diagram (courtesy of Thomas Ullrich). The arrows are suggestive of the \muB coverage of different experiments and do not provide information on the $T$ coverage.}
\label{fig:QCDEOSmu}
\end{center}
\end{figure}

In the Standard Model of particle physics, Quantum Chromodynamics (QCD) occupies a special role. Its running coupling constant implies that QCD systems are strongly coupled at typical scales of 1~fm or so, giving rise to the fundamental phenomena of confinement and the generation of about 98\% of the visible mass in the universe. In the early universe, at about 10 $\mu$s after the Big Bang, these phenomena emerged when a hot plasma of quarks and gluons converted into massive hadrons. This transition has by now been well established by numerical simulations of a lattice-discretized QCD partition function, as a cross-over transition at a pseudo-critical temperature of $T_{\rm pc}\simeq$~155\,MeV~\cite{HotQCD:2018pds}, see Fig.~\ref{fig:QCDEOSmu} (left panel). High-energy heavy-ion collisions at the SPS, RHIC, and the LHC have enabled detailed investigations of hot QCD matter. Among the main findings are that this medium is very strongly coupled, with transport coefficients that are close to universal lower bounds predicted by quantum mechanics. The microscopic understanding of these transport properties based on the in-medium forces of QCD, and how they relate to the phase transition, remains a central question that will be addressed with  upgrade programs at RHIC and the LHC.

While the high-energy frontier of heavy-ion collisions probes the QCD medium at high temperatures and small baryo-chemical potentials, $\mu_B\simeq 0$, it is of high interest to investigate the regime of large baryon densities. Highly compressed nuclear matter at relatively low temperatures still exists in the universe today inside neutron stars and their mergers. Rather little is known about QCD matter at high $\mu_B$, but theoretical calculations suggest a potentially rich phase structure including the emergence of a first-order transition along with a second-order critical endpoint~\cite{Fukushima:2010bq}. By lowering the collision energies, heavy-ion experiments provide unique opportunities for systematic studies of a substantial part of the QCD phase diagram at high $\mu_B$ (Fig.~\ref{fig:QCDEOSmu}, right panel), thereby also promising to unravel connections between astrophysical systems and the early universe.  

The standard way to characterize the phase structure of QCD is through its underlying condensate structure which in turn, manifests itself in the excitation spectrum. In the vacuum, the formation of gluon condensate(s) is believed to cause the linearly rising potential between the color charges of partons which ultimately leads to the appearance of colorless hadrons as the effective degrees of freedom, \ie, confinement. In addition, the formation of quark-antiquark condensate(s) induces the breaking of the approximate chiral symmetry of the QCD lagrangian for the near-massless light quarks. As a consequence, chiral multiplets in the hadronic spectrum are split by typically 0.5 GeV, e.g., the nucleon and $N^*(1535)$ or the $\rho$ and $a_1$. Lattice-QCD calculations for $\mu_B = 0$ show that the quark-antiquark condensate decreases rather steeply around the pseudo-critical temperature of $\Tpc\simeq 155$\,MeV, and that chiral symmetry is essentially restored for temperatures $T\gsim180$\,MeV~\cite{Borsanyi:2010bp,Bhattacharya:2014ara}. A long-standing question is how chiral symmetry restoration manifests itself in the hadron spectrum, i.e., what its observable consequences are. There is growing evidence now that, for small chemical potentials, the ground-state mass in the chiral multiplets is rather stable, and that chiral restoration is realized through ``burning off" the mass {\em splitting}~\cite{Hohler:2013eba,Aarts:2017rrl}. This would also imply that the visible mass in the universe is essentially generated by the gluon condensate and that the latter may persist to substantially higher temperatures than the quark condensate. Indirect evidence for such a scenario is that remnants of the confining force may persist well above $\Tpc$ and play a critical role in generating the remarkable transport properties of the strongly coupled QGP at vanishing $\mu_B$ as probed at the high-energy frontier of heavy-ion collisions~\cite{He:2022ywp}.
A fundamental question is how the QCD condensate structure evolves as the chemical potential is increased, providing exciting opportunities especially if a first order-transition arises, in particular, whether chiral restoration and deconfinement develop into separate transitions, and how the transport properties of the QCD medium respond to these changes.

Based on existing data on light-hadron production~\cite{Bzdak:2019pkr,An:2021wof}, the collision energy regime of $\sqrt{s_\mathrm{NN}}\simeq 5-20$\,GeV, broadly corresponding to the one investigated by NA60+, is well suited to address these questions: the chemical potentials are expected to reach $\sim$500\,MeV or more, while the produced energy densities are likely large enough to produce QCD matter in the parton-to-hadron transition regions and trigger chiral symmetry restoration. Extensive studies at the high-energy frontier have demonstrated the power of heavy-flavor (HF) particles to study key aspects of this medium: the diffusion of open HF particles enables direct access to a fundamental transport coefficient (the HF diffusion coefficient), while the in-medium kinetics of quarkonia encodes their melting and regeneration governed by their in-medium potential. In addition, the radiation of dileptons is the only known observable that provides direct information about a spectral function of the QCD medium, which, in particular, reveals the fate of the $\rho$-meson as chiral symmetry is restored. Dilepton radiation is also an excellent tool to determine the temperatures and lifetime of the fireball, which are key quantities to establish the conditions of the ambient medium, including anomalous behavior related to the onset of a first-order transition.

\newpage
\section{Rare probes of the QGP: concepts and observables}
\label{RareProbes}
\vskip 0.4cm
In this chapter we will elaborate the physics case for the unique measurements that NA60+ will be able to carry out. In Sec.~\ref{subsec:obsthermalradiation} we lay out how precision measurements of dimuon spectra can probe the realization of chiral restoration in hot and dense QCD matter. We also discuss how the radiation from the fireball can yield critical information on its early temperatures and lifetime, including the potential discovery of a first-order phase transition in the system. 
In Sec.~\ref{subsec:obsopencharm} we discuss how measurements of open charm hadrons in the energy range covered by NA60+ can illuminate the dynamics of heavy quarks in a baryon-rich medium, in a complementary way with respect to collider experiments. Further important studies include the hadronization mechanisms of charm quarks, the measurement of the total charm cross section and the investigation of partonic nuclear shadowing at large Bjorken-$x$ values (0.1--0.3).
In Sec.~\ref{subsec:obsquarkonium} we show how measurements of charmonium states in nuclear collisions, never attempted below top SPS energy, may allow to pin down the beam energy threshold for the onset of their suppression and relate this information to the temperature extracted from thermal dimuon studies. 
In Sec.~\ref{subsec:obsstrangeness} we discuss an essential role of high-statistics measurements of (multi-) strange hadrons in the energy range covered by NA60+ in confirming the physics picture emerging from studies at RHIC and LHC energies where strangeness production only depends on the associated hadronic multiplicity generated in the event and not on the specific collision system. Finally, in Sec.~\ref{hypernucleiintro} we motivate the interest in performing measurements of the production of hypernuclei, showing that NA60+ could play a significant role in extending our knowledge toward the region of high masses (A = 6--7).

\subsection{Thermal radiation}
\label{subsec:obsthermalradiation}
\vskip 0.2cm
\subsubsection[Chiral symmetry restoration: measurement of \texorpdfstring{$\rho\text{--}\mathrm{a}_1$}{rho--a1} chiral mixing]{Chiral symmetry restoration: measurement of \texorpdfstring{$\mathbf{\rho\text{--}a_1}$}{rho--a1} chiral mixing}
\label{subsec:chiralsymmetry}

The $\rho$ meson, whose strong coupling to the $\pi^+\pi^-$ channel results in a lifetime of only 1.3 fm/c in vacuum, is vigorously  regenerated in the much longer-lived fireball in nuclear collisions, and thus considered since long as the prime probe
for “in-medium modifications” of hadron properties through its direct decay into dileptons. 
It was suggested already 40 years ago~\cite{Pisarski:1981mq} that the in-medium modifications of the $\rho$ spectral function (the imaginary part of its propagator) are related to the restoration of the spontaneously broken chiral symmetry.
Precision measurements of the $\rho$ spectral properties were performed at the CERN SPS by CERES in Pb-Au collisions~\cite{CERESNA45:1997tgc,CERES:2005uih} and by NA60 in In-In collisions~\cite{Arnaldi:2006jq,Arnaldi:2008er,Specht:2010xu} at a centre-of-mass energy per nucleon-nucleon collision of $\sqrt{s_{NN}} = 17.3$ GeV.
Both the CERES and NA60 data on the fireball radiation are dominated by a strongly medium-modified $\rho$ contribution for $M < 1$ GeV and agree well with a microscopic many-body model predicting a very strong broadening with essentially no mass shift~\cite{Rapp:1999us,vanHees:2007th}. This suggests that the way chiral partners mix is ultimately realized by a degeneracy of the chiral partners at the mass of the ground-state, and accompanied by a ``complete melting” signaling a transition to partonic degrees of freedom~\cite{Rapp:1999us}.

\begin{figure}
    \centering
    \includegraphics[width=\textwidth]{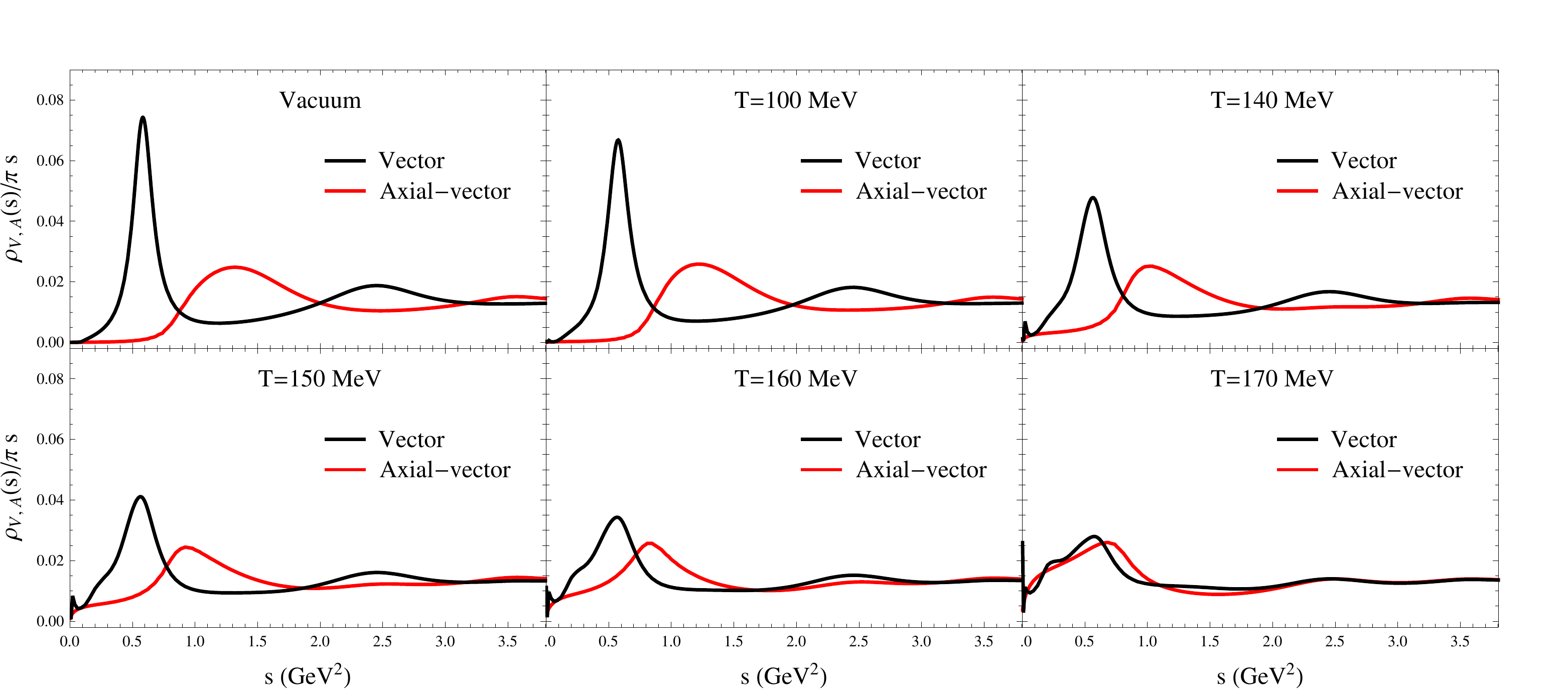}
    \caption[Temperature evolution of spectral functions]{Temperature evolution of vector and 
    axialvector spectral functions (non-linear realization) \cite{Hohler:2013eba}.}
    \label{fig:rho-a1}
\end{figure}

In recent years, further theoretical evaluations have been performed to understand if this scenario is rigorously compatible with chiral symmetry restoration in hadronic matter (HM). In Ref.~\cite{Hohler:2013eba} these calculations have been implemented into  first-principle Weinberg sum rules which relate (moments of) the difference between the vector and axialvector spectral functions to chiral order parameters. With the decrease of the chiral condensate as an input taken from finite-temperature lattice-QCD computations, it was found that the axialvector spectral function (red curves in Fig.~\ref{fig:rho-a1}) degenerates with the vector channel through a strong broadening accompanied by a mass drop of the $a_1$ meson toward the $\rho$ meson, see
Fig.~\ref{fig:rho-a1}. 
Independent corroboration of this mechanism has also been obtained from direct lattice-QCD computations of the nucleon and $\Delta$ correlation functions and their chiral partners~\cite{Aarts:2017rrl,Aarts:2020vyb}, and recently also in the vector-axialvector channel~\cite{Garcia-Mascaraque:2021sbt}.

The broadening of the $\rho$-meson spectral function observed at SPS~\cite{CERESNA45:1997tgc,CERESNA45:2002gnc,CERES:2006wcq,Arnaldi:2006jq,Arnaldi:2008er,Specht:2010xu} and later at RHIC energies~\cite{STAR:2013pwb,PHENIX:2015vek} is therefore consistent with chiral symmetry restoration. However, an unambiguous way to observe chiral symmetry restoration would be to measure not only the $\rho$-meson spectral function but also the one of the $a_{\rm 1}$. Unfortunately, the latter cannot be reconstructed exclusively in heavy-ion collisions.
However, the so called $\rho - a_{1}$ chiral mixing mechanism provides access to the properties of the $a_{\rm 1}$, albeit indirectly. For hot hadronic matter, it has been shown that to leading order in temperature~\cite{DEY1990620} (and also to leading order in nucleon density~\cite{Krippa:1997ss,Chanfray:1998ws}) the medium effects in the vector and axialvector channels are coupled through a mutual mixing of their correlation functions with each other. In essence, the presence of pions in the surrounding medium (both real and virtual) will “admix” the axialvector channel into the vector channel (and vice versa) and produce dileptons. In particular, these are $\pi + a_1$ annihilation processes which are most prominent in the “dip region” of the vector spectral function in vacuum,  for masses $M\simeq0.9-1.4$\,GeV, see the black line in the upper left panel of Fig.~\ref{fig:rho-a1}. 
As temperature increases, this dip gets filled and, close to the pseudo-critical temperature, the vector spectral function 
essentially flattens out signaling  the approach to chiral symmetry restoration, see the lower right panel of Fig.~\ref{fig:rho-a1}.
While this is a relatively small effect, the change in this region is sensitive to the mixing of the chiral partners $\rho$ and $a_1$ and therefore to chiral symmetry restoration.
The mixing effect can be identified by a precise experimental study of the mass region of the thermal dilepton spectrum for invariant masses between 0.9 and 1.4 GeV. However, the sensitivity can be limited by the fact that the experimentally measurable thermal dilepton spectrum arises by the convolution of the spectral function with a Boltzmann factor over the entire space-time evolution of the fireball. In addition, the sensitivity could be reduced by background sources.
Pb-Pb collisions at low energies are ideal to study this effect. Dilepton production at high energies, including top SPS energy, RHIC and LHC, receive substantial contributions from QGP radiation, which produces an exponential spectrum at any mass. This is an intrinsic background that cannot be subtracted. In order to understand in detail the effect related to chiral mixing, it is then necessary to perform a systematic study as a function of collision energy, in particular towards lower energies where the QGP radiation becomes suppressed and possibly negligible. From an experimental point of view, the ideal procedure would be to study the effect for a system produced close to the phase boundary. Here the effects of chiral restoration will be maximal. In presence of a first-order transition the system will spend an even larger amount of time in the mixed phase to burn latent heat, while at RHIC and LHC energies, the system is produced at initial temperatures well above the pseudo-critical temperature and it is then pushed fast across the phase boundary by the strong radial flow.

\begin{figure}
    \centering
    \includegraphics[width=0.6\textwidth]{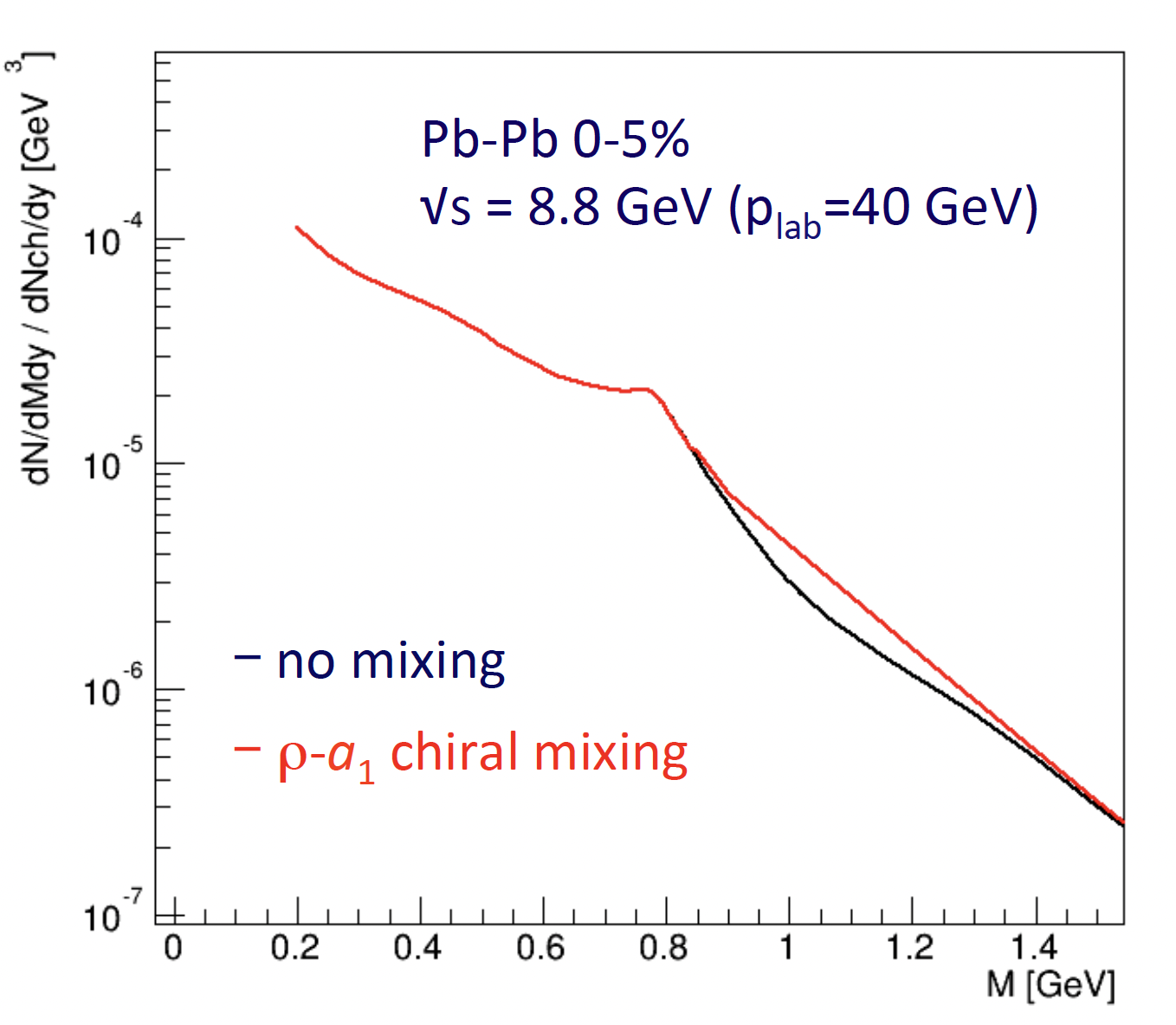}
    \caption[Thermal dilepton mass spectra]{Thermal dilepton mass spectra for two different scenarios: only $\rho$ broadening and $\rho$ broadening with chiral mixing of the $\rho$ and a$_1$.}
    \label{fig:physics:dileptons}
\end{figure}

The required experimental sensitivity to this effect is based on a calculation of the thermal dilepton yield with no chiral mixing and full chiral mixing at $\sqrt s_{NN}$ = 8.8 GeV within the framework of Ref.~\cite{vanHees:2007th,Rapp:2014hha}. Fig.~\ref{fig:physics:dileptons} displays the total thermal dilepton spectrum - experimentally measurable - for 
the aforementioned cases. 
Full mixing leads to an enhancement of the yield by 20-25\% 
with respect to calculations with an in-medium spectral functions without mixing. In order to discern it, an experimental accuracy of $\sim$10$\%$ for the yield measurement will be necessary.

Thus, the experimental strategy relies on a precision study of the mass spectrum for $0.85<M <1.4$ GeV, relating it to the caloric curve (see Sec.~\ref{subsec:caloriccurve}) with the temperature extracted from the same spectrum for $M >1.5$ GeV to tag the system temperature close to the phase boundary. 
An overall understanding of the $\rho$ spectral function as a function of collision energy is at the same time mandatory. Precision measurements of the $\rho$ for masses below the $\omega$ in combination with the Weinberg (and QCD) sum rules constrain the theoretical predictions at larger mass and are sensitive to additional mechanisms of chiral symmetry restoration. 

Measurements in elementary collision systems (pA) are also crucial to establish a vacuum reference and, in particular, to subtract the Drell-Yan contribution to the dilepton spectrum which is important at low energies.

 Finally, multi-differential measurements as a function of pair transverse momentum, sensitive to radial flow, azimuthal angle ($v_{\rm 2}$) and polarization variables~\cite{Speranza:2018osi}, will help establish a clear connection between $\rho$-$a_{\rm 1}$ chiral mixing and dilepton excess above the $\omega$ mass by mapping radial and elliptic flow effects. The emission profile will be constrained using the correlation of small (large) flow with early (late) production time. 

\subsubsection{Hadron-parton phase transition: measurement of the strongly interacting matter caloric curve at high \texorpdfstring{$\mu_{\mathbf{\textit B}}$}{muB}
 }
\label{subsec:caloriccurve}
The measurement of a caloric curve has been successfully used to establish evidence for a first-order phase transition from the liquid self-bound nuclear ground state to a gas of unbound nucleons~\cite{PhysRevLett.75.1040}.
We present here a method to perform the first measurement of a caloric curve for the phase transition between hadronic matter and the QGP.
The temperature measurement, performed as a function of collision energy, is based on a precise {\it dimuon} thermometer that is {\em independent} of the blue shift effect imparted on momentum spectra.

\begin{figure}[ht]
\begin{center}
\includegraphics[width=0.55\linewidth]{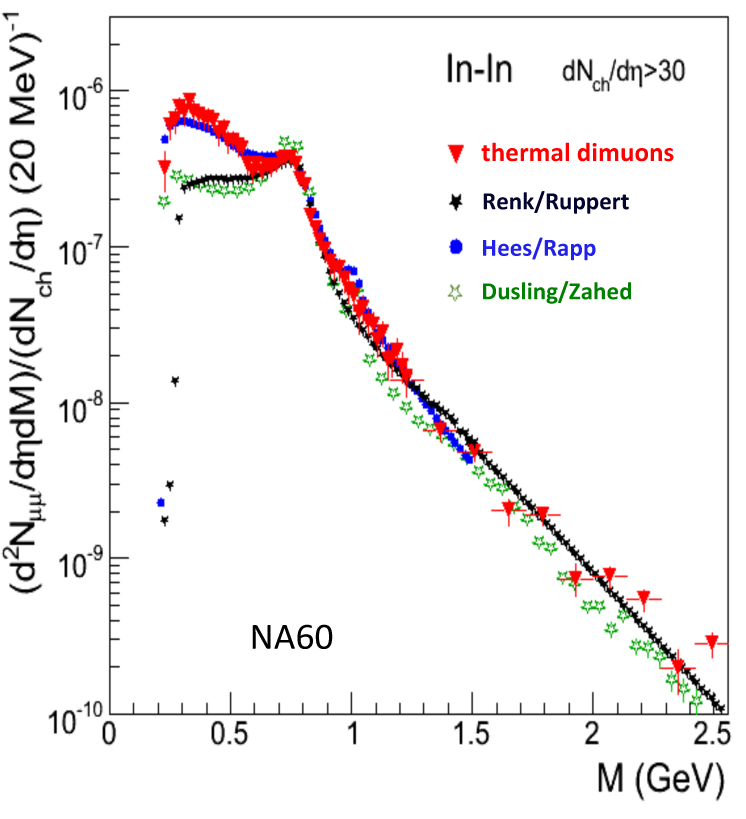}
\caption{Acceptance-corrected  mass  spectrum  of  thermal dimuons measured by NA60 in \InIn collisions at $\sqrtsNN = 17.3\GeV$~\cite{Specht:2010xu}.
Data are integrated over \pt.
The  theoretical  model  results  are  labelled  according  to the  authors  Hees/Rapp~\cite{Rapp:1999us,vanHees:2007th}, Renk/Ruppert~\cite{PhysRevC.77.024907,PhysRevLett.100.162301}, Dusling/Zahed~\cite{PhysRevC.75.024908,PhysRevC.80.014902}.}
\label{fig:NA60dilepton}
\end{center}
\end{figure}

For dilepton masses above 1.5\GeVcc, overlapping resonances lead to a continuum-like spectral density corresponding to a description in terms of quarks and gluons (hadron--parton duality).
Here, medium effects on the electromagnetic spectral function ($\Pi_{\rm EM}$) are parametrically small, being suppressed by powers of $(T/M)$, thus providing a stable thermometer of the Boltzmann factor.
With $\Pi_{\rm EM}\propto M^2$, and in non-relativistic approximation, one has $\dd N/\dd M \propto M^{3/2}\exp(-M/\T{slope})$~\cite{Rapp:2014hha}, which only depends on mass and is thus by construction a Lorentz-invariant, i.e., immune to any collective motion of the expanding source.
The parameter \T{slope} in the spectral shape of the mass spectrum is a space-time average of the time-dependent temperature $T$ during the fireball evolution.
The choice of the intermediate-mass region (IMR), \numrange{1.5}{2.5}\GeVcc, implies $T\ll M$ and thus strongly enhances the sensitivity to the early high-$T$ phases of the evolution.
This method has been exploited by NA60 to measure the medium temperature in \InIn collisions at $\sqrtsNN = 17.3\GeV$.
A fit of the mass spectrum of Fig.~\ref{fig:NA60dilepton} (the fit line is not displayed) has given  $\T{slope} = \SI{205\pm12}{\MeV}$~\cite{Arnaldi:2008er,Specht:2010xu}. To date, this remains the only explicit measurement of a temperature above $\Tpc$, thus showing that the QGP is produced at this collision energy.

The experimental programme of NA60+ proposes to perform an energy scan in the interval $\sqrtsNN=\SIrange{6}{17}{\GeV}$ ($\Elab = \SIrange{20}{160}{\AGeV}$), with particular focus on $\sqrtsNN<10{\GeV}$, which is believed  to be essential to map out the phase transition regime at high \muB, with the possible discovery of a plateau in the caloric curve built with dilepton slopes \T{slope}.
 
The evolution of the initial temperature and \T{slope} vs collision energy has been studied theoretically in the interval $\sqrtsNN = \SIrange{6}{200}{\GeV}$ in Ref.~\cite{Rapp:2014hha} utilizing a thermal fireball with cross-over transition and in the interval $\sqrtsNN = \SIrange{2}{6}{\GeV}$ with a coarse-graining method of a transport model~\cite{PhysRevC.92.014911}.
These calculations provide a baseline for the caloric curve expected in case of a cross-over transition. Equally important, they give a quantitative indication of the reliability of the \T{slope} measurement for the definition of a caloric curve. The average temperature \T{slope} from the mass fit is about 30\% below the corresponding initial one at $\sqrtsNN = 200\GeV$ but the two temperatures are rather close to the pseudo-critical temperature below $\sqrtsNN = 10\GeV$, with their difference reducing to less than 15\%~\cite{Rapp:2014hha}. This shows that \T{slope} from IMR dileptons (M$>$1.5 GeV) are indeed strongly “biased” toward the early stages, rather than an “average” temperature over the entire fireball evolution. It also shows that, since $ \T{slope} \sim\T{initial}\sim T_{\rm pc}$, the system is created close to the phase boundary for $\sqrtsNN < 10\GeV$,  which is a unique asset of this energy regime in exploring the QCD phase transition. It has further been checked that the implementation of a first-order transition changes the results for \T{slope} by a few MeV. This is so since the lattice-QCD based extrapolation already includes a strong cross-over transition characteristic for the change in degrees of freedom in the system. It is this feature that the measurement will be able to map out. With an experimental precision on $T$ of a few MeV, as targeted by NA60+, one will have excellent capability to identify the transition region in this critical part of the QCD phase diagram.

\subsubsection{Elliptic flow of thermal dileptons}
\label{sec:v2_thermal}
One of the most exciting observations in heavy-ion collisions at RHIC and the LHC is the large anisotropic flow of hadrons produced in non-central collisions. 
The spatial anisotropy of the initial nuclear overlap region is converted into momenta through azimuthally anisotropic pressure gradients, which are larger along the short compared to the long axis of the initial reaction zone. 
The second Fourier coefficient of the azimuthal distribution, which is called elliptic flow ($v_2$), is developed in the early stages of the fireball expansion, over the first $\sim$5\,fm/c. 
The large value of the $v_2$ parameter observed at RHIC and the LHC suggests the formation of a strongly coupled QGP that behaves like an almost perfect fluid, which is well described by relativistic viscous  hydrodynamics~\cite{Heinz:2013th,Gale:2013da}. 
However, the hydrodynamic generation of the ``elliptic flow" occurs in the early stages of the collision, while the hadronic observables are also affected by the late stages via hadronization and rescattering. 
As a consequence, key parameters of hydrodynamical simulations as extracted from hadronic spectra, like the shear and bulk viscosity coefficients or the initial conditions, remain subject to significant uncertainties. 
Thermal electromagnetic radiation (direct photons and virtual photons, i.e., dileptons), being penetrating, can be utilized to study the time dependence of the elliptic flow.

\begin{figure}[ht]
\begin{center}
\includegraphics[width=0.49\textwidth]{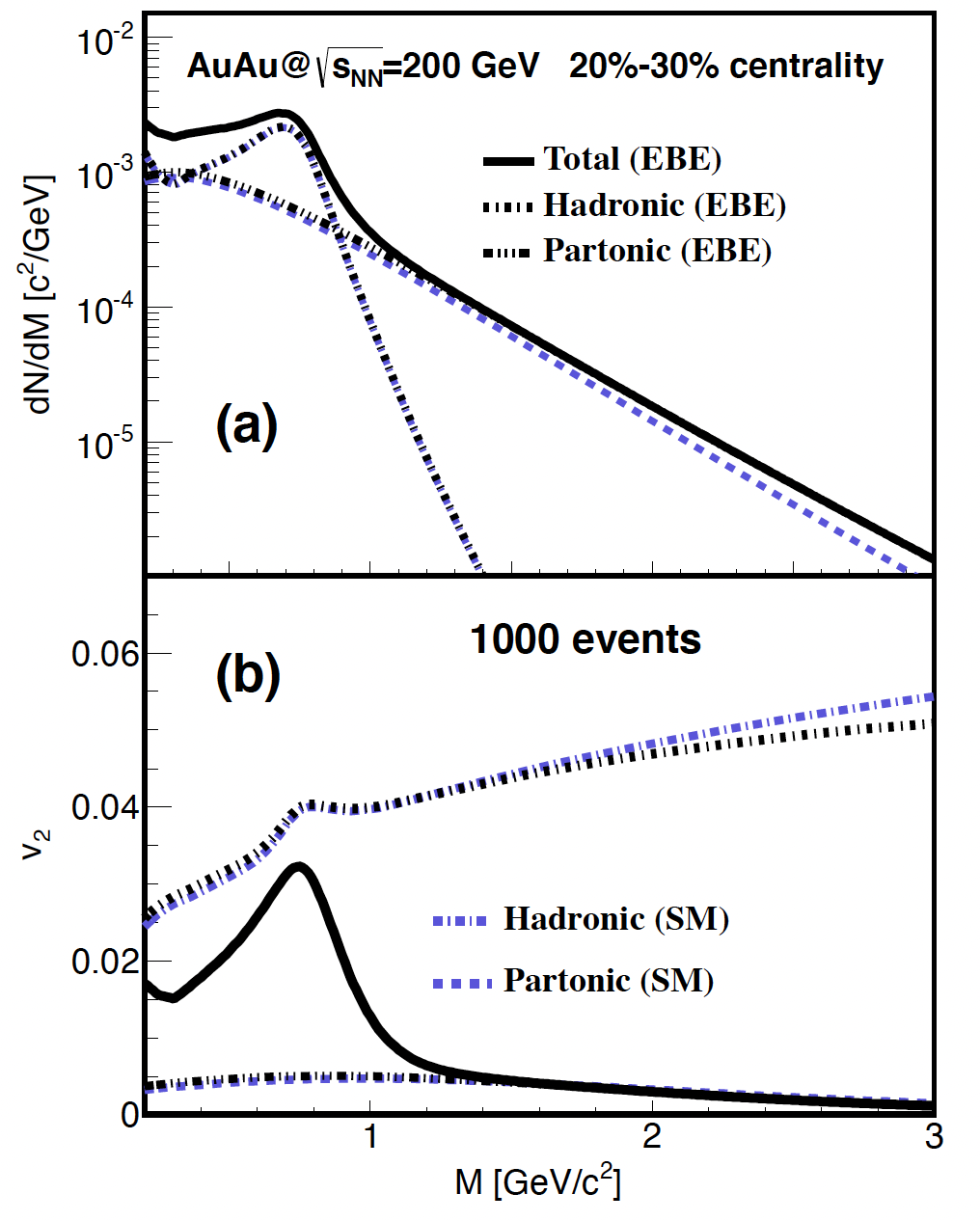}

\caption{Invariant mass spectrum (top) and ellipic flow (bottom) of the radiation from the hadronic and QGP phases from semi-central collisions at RHIC energies. Figure taken from Ref.~\cite{Xu:2014ada}.} 
\label{fig:v2_thermal_xu}
\end{center}
\end{figure}

Figure~\ref{fig:v2_thermal_xu} shows the invariant mass dependence of the elliptic flow from QGP and hadronic radiation at the RHIC energies, calculated according to the model presented in Ref.~\cite{Xu:2014ada}. 
By taking advantage of the emission profile of thermal dileptons as a function of their invariant mass, key information on the different stages of the fireball expansion can be extracted, with QGP radiation being the main component of the thermal spectrum in the IMR,  and hadronic radiation in the low-mass region. 
In particular, the elliptic flow parameter in the hadronic phase is expected to increase with mass as a consequence of the mass effect on the radial flow, while the $v_2$ of the QGP is expected to be much smaller, especially at lower collision energies, due to the limited lifetime of the QGP.
A measurement of the elliptic flow of thermal dileptons versus mass can therefore give direct information on the earliest stages of the collision, and map out the transition to hadronic matter. 
Due to lack of statistics, no measurement of the $v_2$ of thermal dileptons has been performed up to now. A high statistics, high precision measurement is necessary to extract this key parameter.

\subsubsection{Thermal dilepton excitation function and fireball lifetime}
\label{sec:fireball_lifetime}
A precise measurement of the excitation function of thermal-dilepton yields provides a unique opportunity to measure the true lifetime of the interacting fireball in heavy-ion collisions, as dileptons are emitted throughout its evolution~\cite{Heinz:1990jw,Hung:1994eq}. In particular, it was shown~\cite{Rapp:2014hha} that the integrated thermal ``excess” radiation (beyond final-state decays) in the mass region $0.3<M<0.7\GeVcc$ is sensitive to all emission stages and therefore tracks the total fireball lifetime remarkably well, within an accuracy of $\sim$10\%.
The NA60 measurement in \InIn collisions at $\sqrtsNN = 17.3\GeV$ allowed the fireball lifetime to be constrained with hitherto unprecedented precision: $\tau_{\rm fb} = \SI{7\pm1}{\femto\metre\per\clight}$~\cite{Rapp:2014hha}. In particular, the calculation shows that the low-mass thermal radiation yield and the fireball lifetime have the same smooth decreasing trend as a function of decreasing collision energy in case of a cross over transition   (see Fig~\ref{fig:fig3-thermal-performance} in Section~\ref{thermal_dimu_performances}).
Thus, a precise measurement of the total thermal dilepton yield in the window $0.3<M<0.7\GeVcc$ can be utilized as an additional tool to study the phase diagram. It has long been known that in the presence of a soft mixed phase in a first-order transition, the pressure gradients in the system are small and thus stall the fireball expansion~\cite{Hung:1994eq}. Again, this effect is especially pronounced if the system starts out near the soft region delaying the build-up of collective flow altogether (which is expected to be the case for the collision energies proposed here). The lifetime increase would be directly reflected in the low-mass dilepton yield, and would be signalled by an increased lifetime 
in the collision-energy regime where the mixed phase forms, which is different from the smooth decreasing trend if no mixed phase occurs. 
A further important observable to understand if such a (possibly non-monotonous, 
i.e. maximum) feature is indeed due to a mixed phase is the radial flow 
imprinted in the final-state spectra of hadrons.  Specifically, an increase of $\tau$ is not expected to lead to an increase of the total amount of radial flow at freeze-out, leaving hadron spectra unchanged or even softened.

\subsection{Transport properties of the QGP and hadronic phase: open charm 
}
\vskip 0.2cm
\label{subsec:obsopencharm}
Measurements of production of heavy-flavour (charm and beauty) hadrons at 
RHIC and the LHC are providing unprecedented insights into the properties of hot QCD matter at small baryo-chemical potential, and on the hadronisation of the QGP.
With a focus on the low- and intermediate-momentum region, where heavy quarks are expected to undergo a Brownian motion due to multiple soft scatterings with the medium constituents~\cite{Rapp:2008qc}, measurements of \pt distributions, compared to \pp collisions, and of azimuthal anisotropies of $\mathrm{D}$ mesons are being used to extract fundamental transport coefficients of the QGP, such as the heavy-quark diffusion coefficient (see, \eg, Ref.~\cite{Rapp:2018qla}).
Measurements of the relative abundances of different charm-hadron species, and in particular of strange-charm \Ds mesons~\cite{STAR:2021tte,ALICE:2021kfc} and \lambdac baryons~\cite{STAR:2019ank,ALICE:2021bib} in comparison with \Dzero and \Dplus mesons, are used to characterize the hadronisation mechanisms of charm quarks and the role of quark recombination~\cite{Plumari:2017ntm,He:2019vgs}.

\begin{figure}[ht]
\begin{center}
\includegraphics[width=0.55\textwidth]{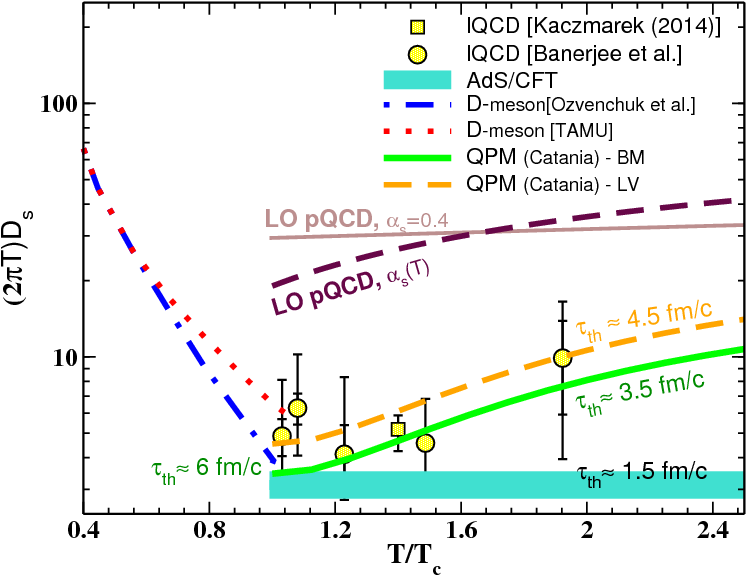}
\caption{Theoretical calculations of the charm-quark diffusion coefficient as a function of temperature in hot hadronic matter ($T/\Tc<1$)~\cite{He:2011yi} and in a QGP ($T/\Tc>1$) compared to lattice QCD calculations.
Figure taken from Ref.~\cite{Scardina:2017ipo}.
} 
\label{fig:charmmotiv}
\end{center}
\end{figure}

The investigation of the QCD matter properties with charm quarks would benefit substantially from measurements in heavy-ion collisions at lower centre-of-mass energies, which would allow one to i) probe the medium at lower temperatures as compared to the studies at colliders and ii) explore the region of finite \muB, in which the baryons of the colliding nuclei are ``stopped'' in the collision region.
As shown in Fig.~\ref{fig:charmmotiv}, the charm-quark spatial diffusion coefficient is predicted to depend on the temperature of the medium, with a minimum value in the vicinity of the pseudo-critical temperature, $\Tpc \simeq 155\MeV$~\cite{He:2011yi,Scardina:2017ipo}.
In particular, the spatial diffusion coefficient is expected to be larger in the hadronic phase than in the late QGP phases prior to hadronisation.
It should then be possible to investigate this feature in \PbPb collisions at SPS energies, where the lower initial temperature of the fireball enhances the sensitivity to the properties of the QGP at temperatures close to $\Tpc$ and where the hadronic phase with $T<\Tpc$ represents a larger part of the space-time evolution of the collisions as compared to measurements at collider energies.
This specific sensitivity to interactions of charm hadrons in the hadronic phase can provide important input also for precision estimates of heavy-quark diffusion coefficients at collider energies, where a rather extensive hadronic evolution from $\Tpc$ down to the kinetic freeze-out temperature of $T \sim 100\MeV$ occurs.
In addition, the question of charm-quark thermalisation in the shorter-lived medium that is formed in heavy-ion collisions at lower centre-of mass energies could be addressed by measuring the \pt distributions and azimuthal anisotropy of D mesons and by searching for the features induced by collective behaviours.
Concerning the hadronisation mechanisms, recombination effects can be studied by reconstructing different charm-hadron species to test the expectation of enhancements of the $\Ds/\mathrm{D}$ and $\lambdacplus/\mathrm{D}$ ratios relative to the ones observed in \pp collisions.
The enhancement could be larger at SPS than at RHIC and LHC energies, because of the larger net-quark content of the fireball.

In order to study the hadronisation mechanism of charm quarks in the QGP and to obtain an accurate determination of the total \ccbar production cross section, which also constitutes an important reference for the charmonium studies, it is crucial to measure the production yields of different ground-state meson and baryon states, namely \Dzero, \Dplus, \Ds, \lambdacplus, and possibly $\Xi_{\rm c}^{0,+}$.
The total production cross section of \ccbar pairs in hadronic collisions at centre-of-mass energies below 20\GeV has never been measured with high precision because the yields at these energies are very small.
The only measurements in nucleus--nucleus collisions at the SPS were obtained by the NA60 experiment in \InIn collisions (using intermediate-mass dimuons, with an uncertainty of about 20\%)~\cite{Arnaldi:2008er} and by the NA49 experiment in \PbPb collisions (an upper limit using reconstructed $\rm D^0$ decays)~\cite{Alt:2005zu}.

An interesting and unique possibility at center-of-mass energies that are not (too) far above the charm production threshold is that the diffusion of charm particles (quark and/or hadrons) can drive them to higher momenta and lead to an {\em increase} (even divergence) of the nuclear modification factor as the maximal $p_T$ set by the kinematic limit in \pp collisions is approached (or surpassed)~\cite{Inghirami:2018vqd}. This would not only be a qualitatively new signature but also provide very sensitive information on the diffusion properties of charm particles at higher momenta. Moreover, the rise of the nuclear modification factor with $p_T$  would be different for different charm hadron species, depending on their kinematic limits in \pp as well as their diffusion properties.
Along similar lines, the $J/\psi/\mathrm{D}$ ratio could be of high interest. As one lowers the collision energy toward the production threshold in \pp collisions, one would expect an increase in this ratio as the threshold for $J/\psi$ production is lower than that for $\mathrm{D\overline{D}}$ (this is still true, although less pronounced, for the $J/\psi/\mathrm{\overline{D}}$ ratio because $\mathrm{\overline{D}}$ mesons are preferentially produced in association with a $\lambdacplus$ and $\lambdacplus$ baryons are expected to be produced more abundantly than $\lambdac^-$ at SPS energies). It would be very interesting to find out whether, and if so at which energies, such an enhancement develops in \pp, and, most importantly, how it is affected by the medium produced in the heavy-ion environment. In particular, due to the sensitivity of the thresholds to mass variations one would be very sensitive to potential mass changes (or, more generally, modified spectral functions) of the charm particles in the produced medium.

\begin{figure}[ht]
\begin{center}
\includegraphics[width=0.45\textwidth]{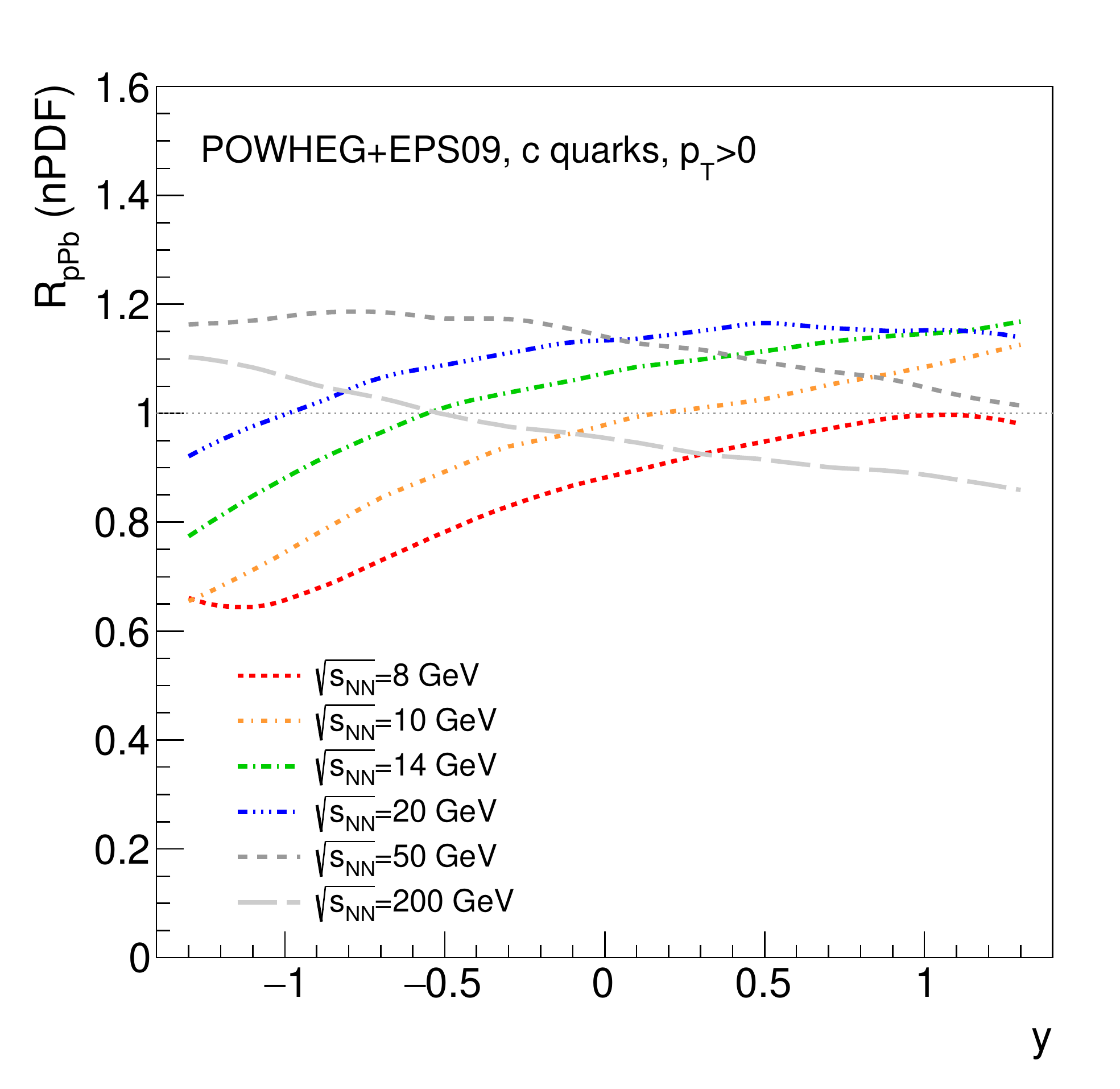}
\includegraphics[width=0.45\textwidth]{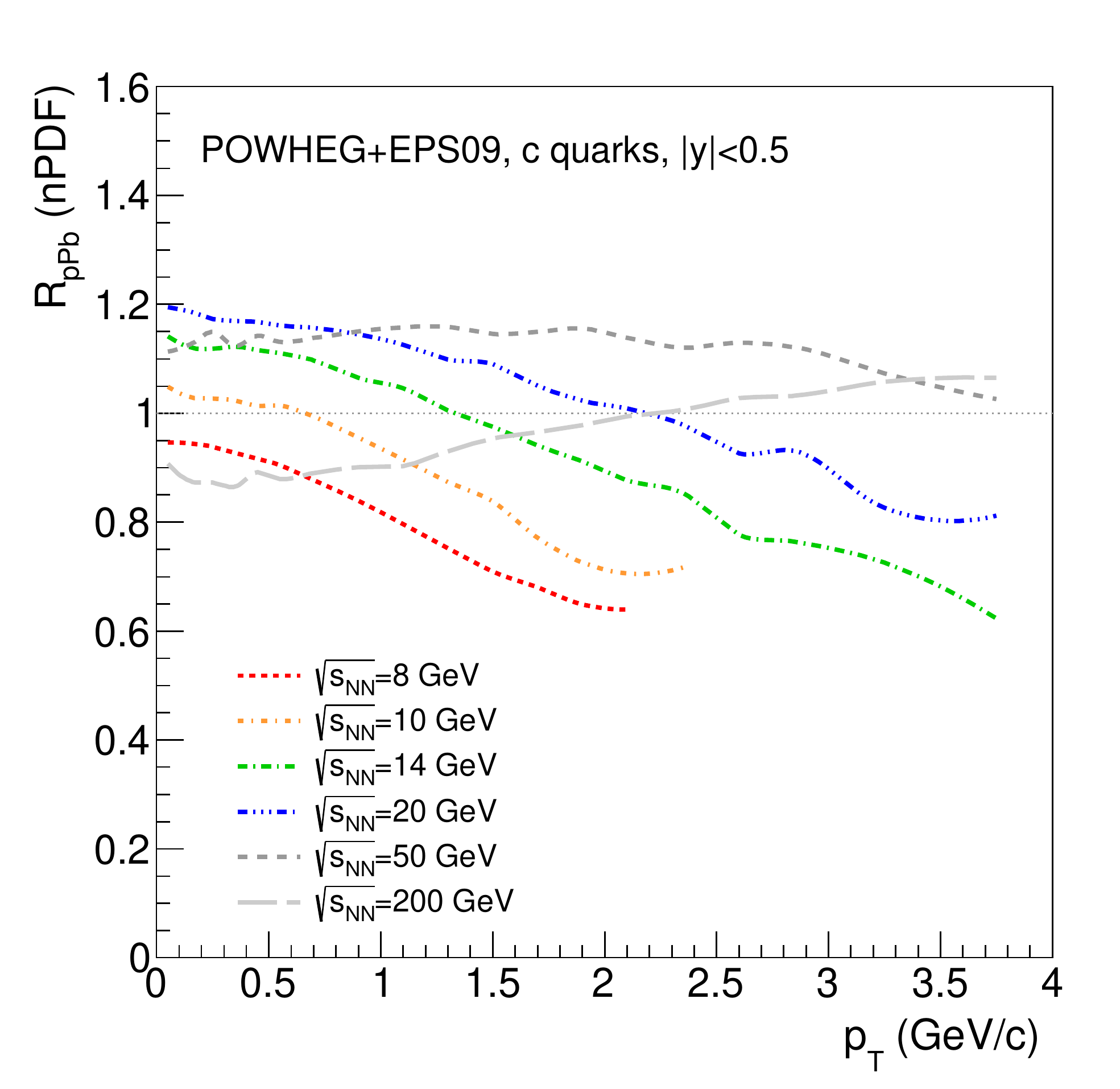}
\caption{Predictions for the effect of the nuclear PDFs (EPS09) on the charm-quark nuclear modification factor in p--Pb collisions at different collision enegies as a function of rapidity (left) and \pt\ (right).
} 
\label{fig:charmnPDF}
\end{center}
\end{figure}

Finally, measurements of D-meson production in proton--nucleus collisions at SPS energies can provide constraints on parameterisations of the nuclear modification of parton distribution functions (PDFs) at $Q^2 \sim \SIrange{10}{40}{\GeV\squared}$ and large Bjorken-$x$ of $x_{\rm Bj} \sim \numrange{0.1}{0.3}$, depending on $p_{\rm T,c} \sim \numrange{0}{3}\GeVc$.
In this kinematic region, which is poorly constrained by existing data, the PDFs in large nuclei are expected to change from enhancement (anti-shadowing) at $x_{\rm Bj} \sim 0.1$ to suppression (``EMC effect'') at $x_{\rm Bj} \sim 0.3$ (see \eg\ Ref.~\cite{Eskola:2016oht}).
NA60+ could provide precise input via ratios of the $\mathrm{D}$-meson production cross sections in \pPb collisions (maximal nuclear effects) and \pBe collisions (minimal nuclear effects).
The effect on charm production in p--Pb collisions predicted from POWHEG pQCD calculations with the EPS09 parameterisation of the nuclear PDFs (nPDFs), quantified by the nuclear modification factor $R_{\rm pPb}$, is illustrated in Fig.~\ref{fig:charmnPDF}. Measurements as a function of rapidity and \pt at different collision energies will allow to cover a large range of $x_{\rm Bj}$ and provide constraints in the domains of the EMC effect and anti-shadowing.
Furthermore, as discussed in detail in the next section about charmonium production, the measurements of $p_{\rm T}$ and rapidity distributions of charm hadrons in proton--nucleus collisions can provide a sensitive test for the predicted intrinsic-charm component in the nucleon wave function~\cite{Brodsky:1980pb}.

\subsection{Deconfinement threshold: charmonium suppression (\texorpdfstring{J/$\psi$, $\psi(2S)$, $\chi_{\rm c}$}{charmonia})
}
\label{subsec:obsquarkonium}
\begin{figure}[ht]
\begin{center}
\includegraphics[width=0.45\linewidth]{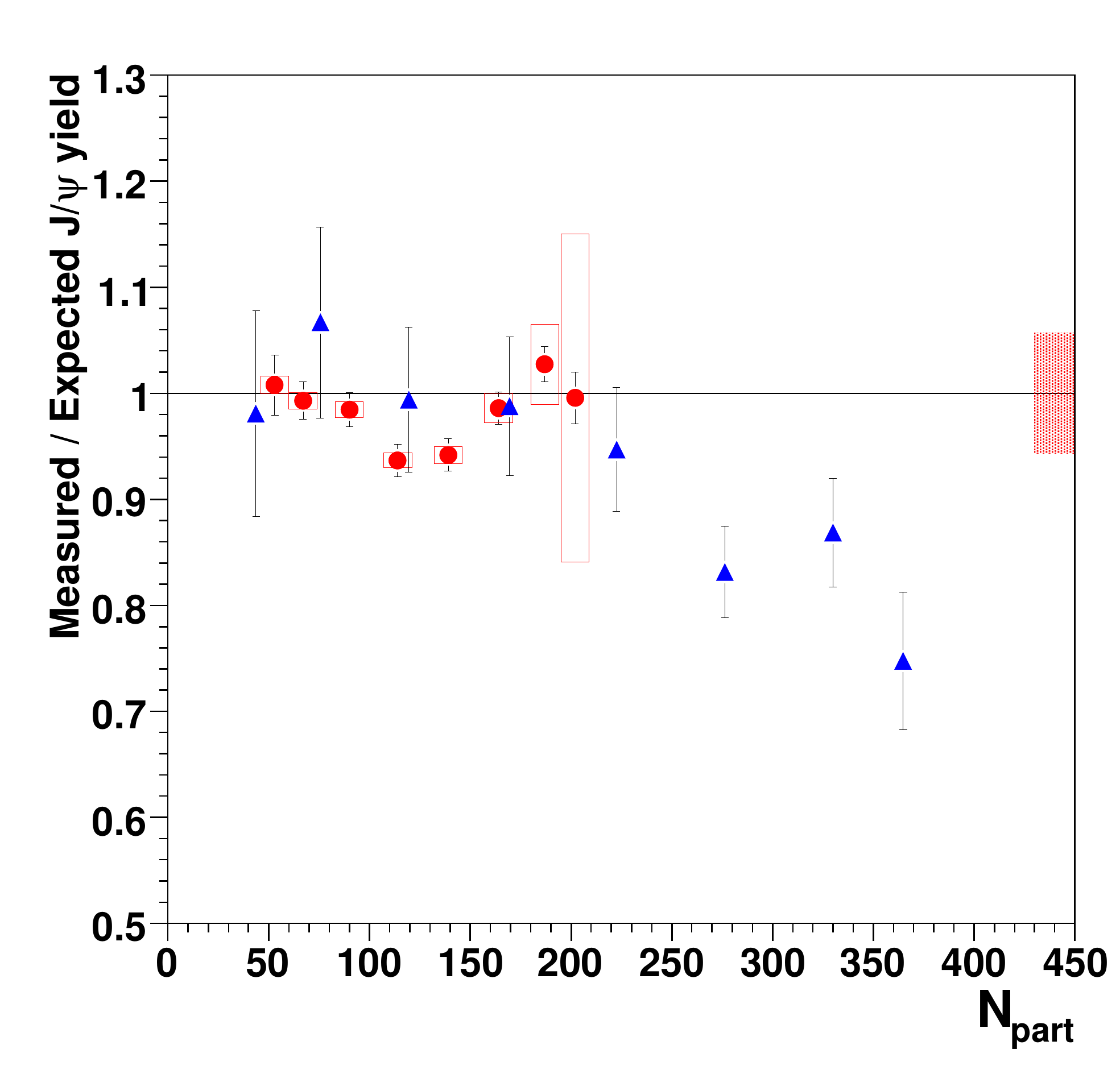}
\includegraphics[width=0.43\linewidth]{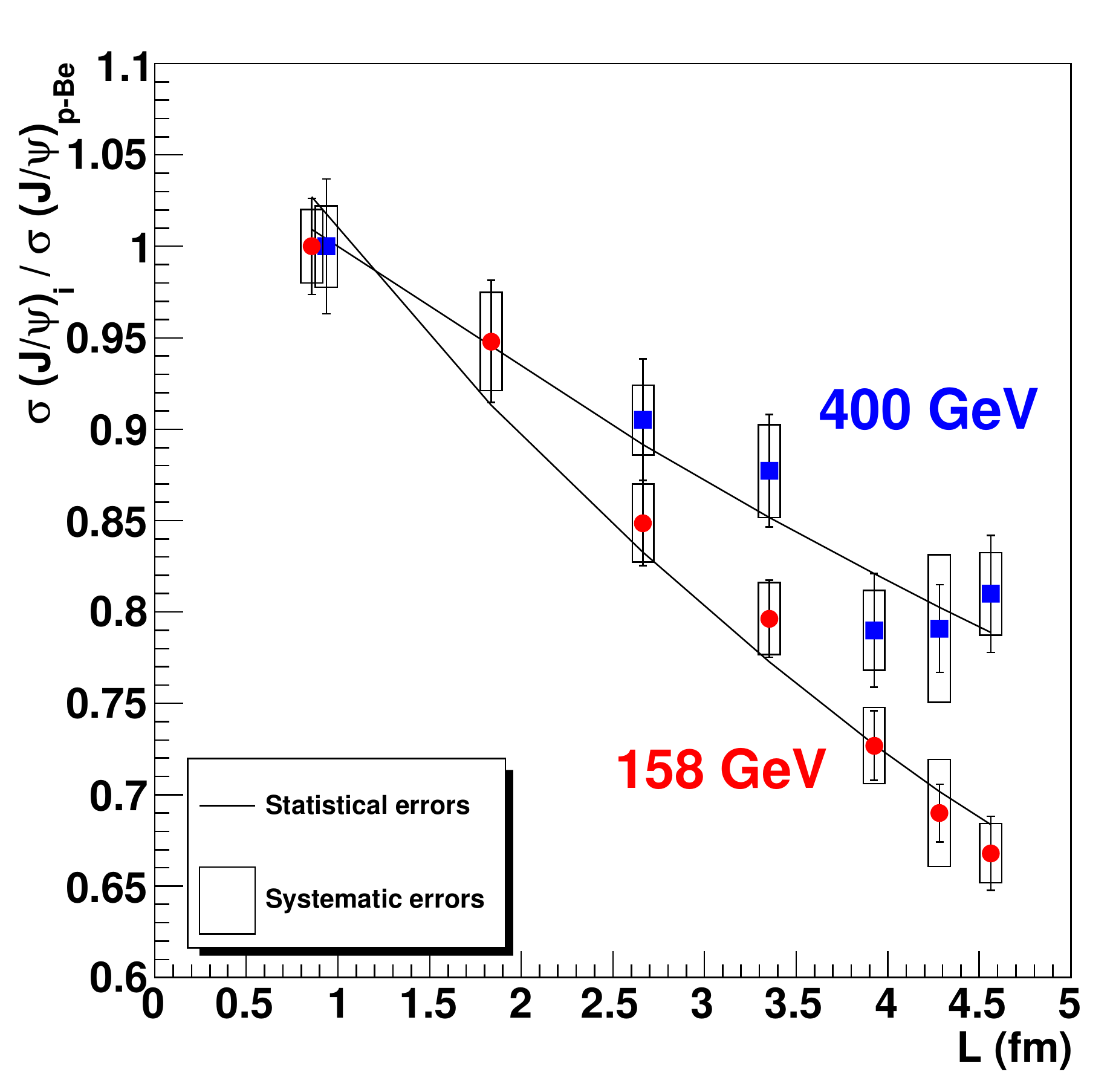}
\caption{(Left) Anomalous \jpsi suppression in \InIn (circles), measured by the NA60 experiment, and \PbPb collisions (triangles), measured by the NA50 experiment, as a function of the number of participant nucleons \Npart~\cite{Alessandro:2004ap,Arnaldi:2007zz}.
(Right) Cross sections measured by NA60 for \jpsi production in \pA collisions, normalized to the \pBe\ \jpsi cross section~\cite{Arnaldi:2010ky}.
Data are plotted as a function of $L$, the mean path of nuclear matter crossed by the \ccbar pair, which is calculated using the Glauber model of the collision geometry.}
\label{fig:NA60Jpsi}
\end{center}
\end{figure}

\vskip 0.2cm
A suppression of heavy-quarkonium states due to the screening of the colour interaction in a deconfined medium has been considered since early studies~\cite{Matsui:1986dk} as one of the key signatures for the formation of a QGP. This picture has then evolved towards the current modern understanding in terms of  dissociation reactions as the main suppression mechanism~\cite{Rothkopf:2019ipj}. The strength of the suppression effects affects differently the various quarkonia according to their binding energy.
Detailed experimental investigations, in particular for the \jpsi meson, were first performed at the top SPS energy of $\sqrtsNN = 17.3\GeV$~\cite{Alessandro:2004ap,Arnaldi:2007zz} by the NA50 (Pb--Pb) and NA60 (In--In) collaborations. A ${\sim}30$\% suppression of the \jpsi production that went beyond carefully assessed cold-nuclear-matter effects was observed in central \PbPb collisions, as shown in the left panel of Fig.~\ref{fig:NA60Jpsi}. The size of such an ``anomalous'' suppression of inclusive \jpsi  production is qualitatively consistent with the expected fraction of the \jpsi yield coming from the decay of the relatively weakly bound \chic and \psiP charmonium states. Therefore, this result implies that such states are melted in the QGP, while ``direct'' \jpsi, due to their stronger binding energy, can survive in the deconfined medium produced at SPS energy~\cite{Rapp:2017chc}.

More recently, extensive sets of measurements were also obtained, for both charmonium and bottomonium states, at RHIC (Au--Au and smaller collision systems at $\sqrtsNN$ up to 200\GeV) and the LHC (Pb--Pb at $\sqrtsNN$ up to 5.02\TeV)~\cite{Adare:2011yf,Adam:2016rdg}. At these energies, in spite of the higher initial QGP temperature, that would lead to a stronger suppression, the strong increase of the charm production cross section induces a recombination effect of the dissociated \ccbar pair~\cite{BraunMunzinger:2000px,Thews:2000rj}, that becomes dominant at LHC energy. As a consequence, the net suppression effects become smaller than those observed at top SPS energy.

As of today, heavy quarkonium production in nucleus--nucleus collisions has not been studied below top SPS energy. NA60+ proposes to carry out a measurement of charmonium production down to an energy of the incident heavy-ion beam of approximately $\Elab = \SI{50}{\AGeV}$ ($\sqrtsNN = 8.8\GeV$), or even lower, depending on the available integrated luminosity. While for \jpsi and \psiP the decay to muon pairs will be studied, with BRs of $\sim 6$\% and 0.8\% respectively, the \chic states can be accessed via their radiative decay $\chi_{\rm c}\rightarrow {\rm J/\psi}\gamma$, with a BR up to 34\% for the $\chi_{\rm c1}$, by detecting the lepton pair from the photon conversion in the vertex spectrometer.

In this low-energy domain, directly produced \jpsi, due to their strong binding energy, are not expected to be significantly affected by the QGP. However, the study of higher-mass charmonium states remains of pivotal interest, by means of their direct detection and/or observing a modified \jpsi yield due to the suppression of their feed-down decays.
Due to the progressive decrease of the initial temperature of the system when moving to lower collision energies, also dissociation effects on the \chic and \psiP states should eventually become small, and one of the main goals that we propose is the detection of the beam energy threshold for the onset of their suppression. By correlating this information with the corresponding measurement of the temperature via thermal dimuons, one could experimentally identify the ``threshold" temperature for the melting of those charmonium states. In this way, a crucial test of the corresponding theoretical predictions (constrained by 
first-principle studies from lattice QCD) can be carried out~\cite{Rapp:2008tf,Mocsy:2013syh}.
On the phenomenology side, quantitative calculations for charmonium resonances in the low SPS energy range are still in their infancy (see~\cite{ECTstar_2021} for a recent overview). Contrary to collider energies, the QGP formation time may become larger than the charmonium production time, so that a description of the pre-equilibrium phase of the system, including the influence of the large \muB environment (quark excess) in the dissociation processes, becomes necessary. Also, inelastic interactions with the hadronic medium, sometimes also referred to as comover interactions, are expected to become more important at low energies, as the QGP lifetime presumably becomes rather small compared to hadronic phase, and need to be taken into account. Especially, the studies of the $\psiP$, where a substantial suppression could still be expected, will be critical in clarifying the hierarchy of charmonium kinetics in QCD matter.

Another important aspect in the study of quarkonium production in the medium is represented by the so-called cold-nuclear-matter effects. They include various QCD-related phenomena, connected with the initial state, \eg, the nuclear modification of PDFs (shadowing), and also final-state effects, such as the break-up in nuclear matter of a colour singlet/octet pre-resonant state or of the final quarkonium resonance. Past fixed-target experiments for p--A collision at various facilities (SPS~\cite{NA50:2006rdp}, Tevatron~\cite{NuSea:1999mrl}, HERA~\cite{HERA-B:2008ymp}) collected extended sets of data, but the observations still lack a comprehensive interpretation.

In the SPS energy domain, there are already strong indications for an increase of the size of cold-nuclear-matter effects on the produced \ccbar pair, mostly related to final-state break-up and therefore inducing a suppression of charmonium, when the collision energy is decreased. In the right panel of Fig.~\ref{fig:NA60Jpsi}, NA60 results on the \jpsi cross section in \pA collisions at 158 and 400\GeV incident energy~\cite{Arnaldi:2010ky}, normalized to the corresponding value for \pBe, show a sizeable suppression at both collision energies, significantly stronger at 158 GeV. 

We propose to extend such measurements to collision energies below top SPS energy with a twofold interest. First, break-up effects in cold nuclear matter are not related to QGP formation, therefore they must be corrected for when evaluating any ``anomalous'' suppression in nucleus--nucleus collisions. Therefore, such data are mandatory for a correct interpretation of nucleus--nucleus results, and the possible decrease of QGP-related effects at low energy makes an accurate measurement of cold nuclear matter effects even more desirable.
Second, as detailed above, the overall interpretation of the \pA data in terms of various physics effects remains elusive until today, so that data at lower collision energy, accessing specific kinematic configurations can pose significant constraints on the size of those effects.

Finally, the study of \jpsi production at low SPS energy represents an ideal testing ground for the observation of an intrinsic charm component in the nucleon wavefunction. This effect, originally proposed by Brodsky et al.~\cite{Brodsky:1980pb}, and investigated in deep-inelastic scattering experiments (see EMC~\cite{EuropeanMuon:1981obg} results), may lead to an enhanced charm production at large $x_{\rm F}$. Very recently, evidence for the presence of intrinsic charm in the proton was reported~\cite{Ball:2022qks}. While at collider energy the region where the effects of intrinsic charm can be observed is pushed to very large rapidity, for fixed-target configurations at low energy an enhancement much closer to midrapidity is expected. Calculations~\cite{Vogt:2021vsc} of the \jpsi nuclear modification factor in p--A collisions at low SPS energy show that already with a probability of intrinsic charm contribution in the proton of 0.1\% the effect of intrinsic charm should become dominant.

\subsection{QGP chemistry: strangeness production
}
\label{subsec:obsstrangeness}
\vskip 0.2cm
The characterization of the QGP system in terms of yields and transverse momentum spectra of different hadron species is of primary interest in the determination of key parameters such as chemical and kinetic freeze-out temperatures, chemical potentials and viscosity.

Among all different hadrons, those containing more than one strange quark ($\Xi$, $\Omega$ and $\phi$) are particularly interesting, as strangeness is not present in the valence content of the colliding nuclei and is therefore produced in the hard scattering or in the hadronization process. At the same time, strange quarks are sufficiently light to be created copiously at center of mass energies higher than few GeV, thus allowing precise measurements of strange-hadron production rates, transverse momentum spectra and correlations to general properties of the event (such as charged particle multiplicity or event plane orientation).

An enhanced production of multi-strange hadrons in case of QGP formation was first proposed by Rafelski and M{\"u}ller \cite{Rafelski1986}, with the argument that the q-value for the formation of two strange hadrons is lower in case of a deconfined medium, rather than in a hadron gas. Moreover, considerations on the chemical equilibration time were also pointing to a higher production in presence of a deconfined medium.
Strangeness enhancement was first observed at the SPS \cite{NA57:2004nxc,WA971999}, confirmed at RHIC \cite{STAR:2007cqw} and further scrutinized at the LHC \cite{ALICE:2016fzo,ALICE:2019avo,ALICE:2015mpp,ALICE:2013xmt}.

\begin{figure}[ht]
\begin{center}
\includegraphics[width=0.31\linewidth]{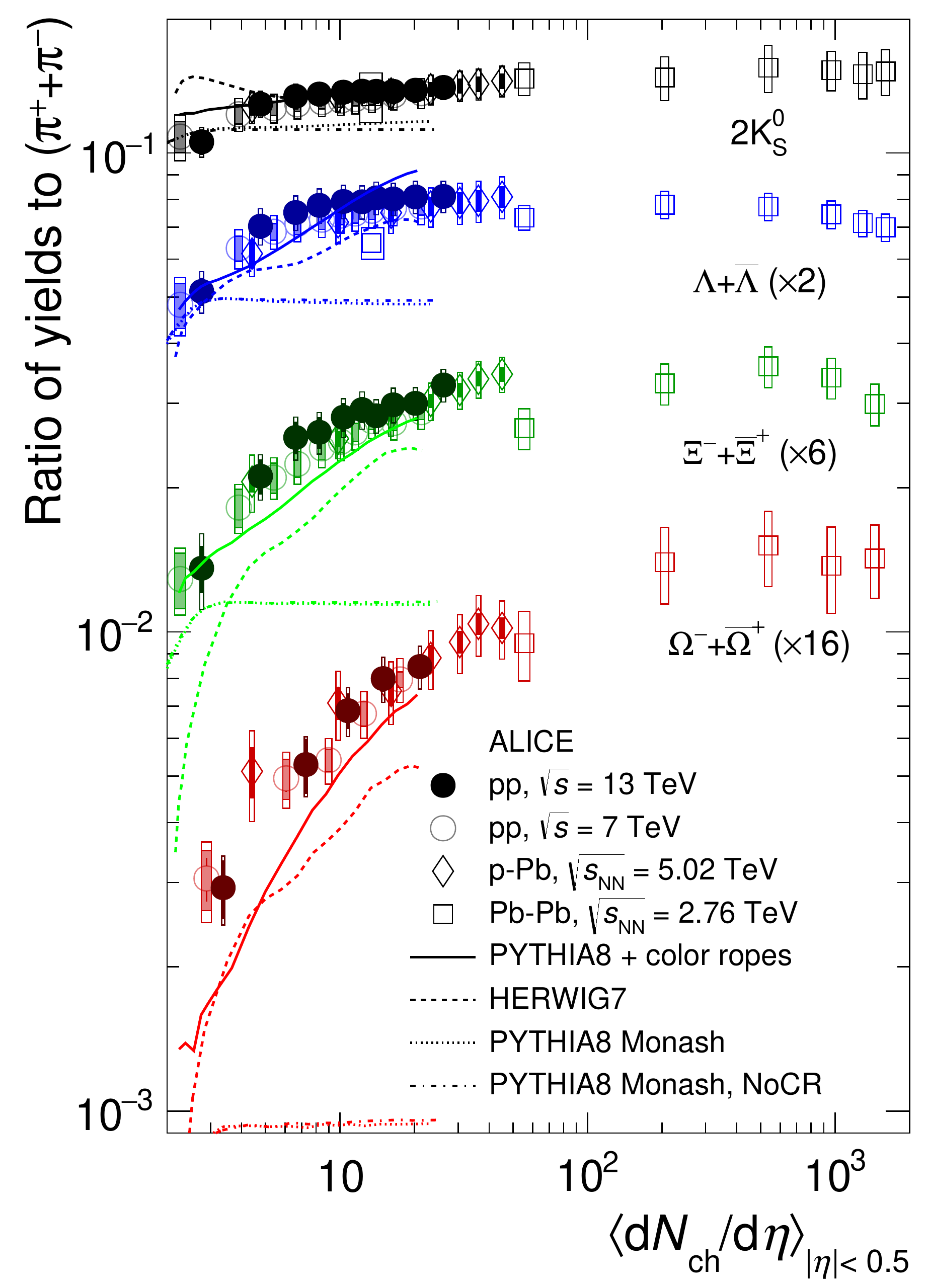}
\includegraphics[width=0.61\linewidth]{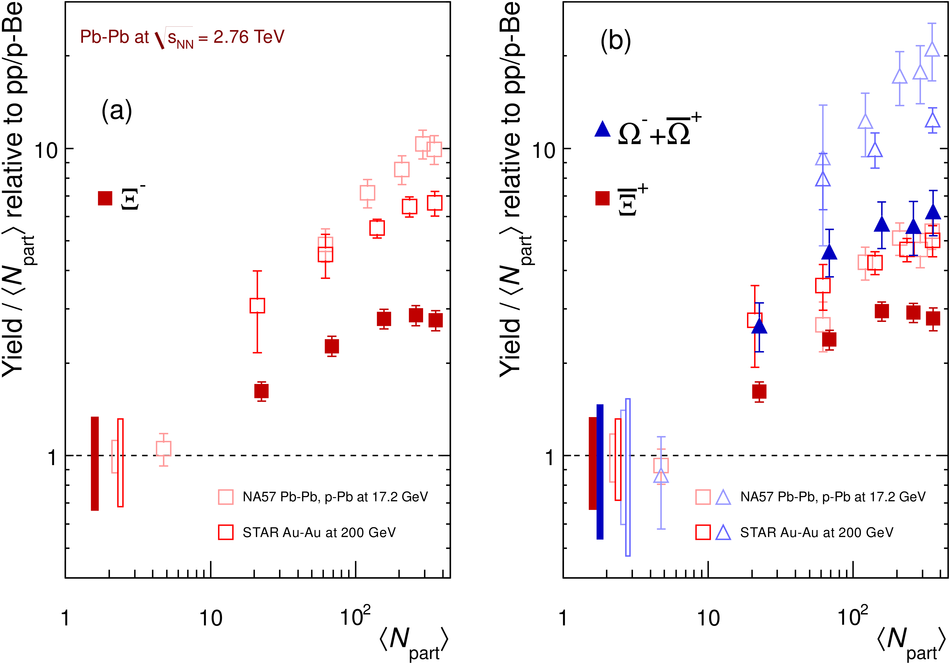}
\caption{(left) Integrated strange hadron-to-pion ratios as a function of $\langle d{\it N_{ch}}/d\eta \rangle$ measured by ALICE in pp, p--Pb, and Pb--Pb collisions~\cite{ALICE:2016fzo}. Different lines represent predictions from different MC generators for pp collisions at $\sqrt{s}$=13~TeV. (center and right) $\Xi$ and $\Omega$ yield ratios in heavy ion and p-A collisions as a function of $N_{\rm part}$ from STAR~\cite{STAR:2007cqw} and NA57~\cite{NA57:2004nxc} experiments.}
\label{fig:strenhalice}
\end{center}
\end{figure}

The most recent LHC and RHIC results have shown that the ratio of multi-strange hadron to pion yields is enhanced in heavy-ion collisions with respect to minimum bias hadronic interactions, more importantly for hadrons with higher strangeness content. The enhancement is proportional to the charged particle multiplicity created in the collision and does not depend on the colliding system nor on the collision energy, when analyzing pp, p-Pb, Au-Au, Xe-Xe and Pb-Pb interactions at center of mass energies ranging from hundreds of GeV to tens of TeV (see Fig.~\ref{fig:strenhalice})).
An opposite trend is observed for strongly decaying resonances, whose production yields decrease with multiplicity, with a slope which depends on the particle lifetime \cite{ALICE:2019xyr,STAR:2022gbl,ALICE:2018qdv}. This effect is connected to the re-scattering probability of the decay products in the dense hadronic medium produced after chemical freeze-out \cite{Torrieri:2001ue,Knospe:2015nva}.
Moreover, strange particles were shown to participate to the common expansion of the formed medium, exhibiting radial and anisotropic flow ($\it{v}_2$).

The $\phi$ meson plays a pivotal role in these studies as it contains hidden strangeness and it has the largest lifetime among all resonances. The $\phi$/$\pi$ ratio turns out to be enhanced with multiplicity with an intermediate slope between $\Lambda$/$\pi$ and $\Xi$/$\pi$, thus behaving effectively as a particle with S$\simeq$1.5. Additionally, having a mass similar to the one of the proton, the $\phi$ meson serves as a powerful tool in probing the features in the $\it{v}_2$ ordering of different identified hadrons: recent results \cite{ALICE:2018yph} have shown that it behaves as protons at low $\it{p}_T$ (approximate mass ordering due to flow) and switches to the meson behaviour at higher $\it{p}_T$ (baryon-meson splitting where re-combination dominates).

These findings can be interpreted by means of phenomenological models implementing different underlying physics, often connected to final state effects in the hadronization process \cite{Sjostrand:2014zea,Bierlich:2014xba,Pierog:2013ria,Kanakubo:2019ogh}. Interpretations based on pure scaling with the number of multi-parton interactions (MPI) are recently emerging as well \cite{Loizides:2021ima}. At present a complete quantitative description of strange particle production by means of microscopic phenomenological models is still missing. 
At the same time, statistical hadronization in the grand-canonical formalism can be applied to describe hadron yields in large hadronizing systems~\cite{Andronic:2017pug}, and extensions of the statistical treatment at lower multiplicities (corresponding to smaller hadronizing systems) are under development~\cite{Vovchenko:2019kes}.

The extension of this picture to lower center of mass energies is not straightforward. As previously mentioned, SPS experiments first observed strangeness enhancement when performing the ratio of strange baryon yields in Pb-Pb to those measured in smaller collision systems (e.g. p-Be). The enhancement was shown to depend on the number of participating nucleons ($N_{\rm part}$) and to be higher for hadrons with higher strangeness content. Nonetheless, the energy dependence turned out to be inverse: lower $\sqrt{s}$ collisions featured higher strangeness enhancement at equal $N_{\rm part}$ (see Fig.~\ref{fig:strenhalice}(right)). This puzzle was solved when noticing that the normalization to the yields in smaller collision systems, though useful to highlight the strangeness enhancement pattern at a fixed energy, makes the comparison of different $\sqrt{s}$ difficult, as the probability for strange hadron production dramatically increases with energy in p-p($\rm{\bar{p}}$) and p-A. For this reason, the ALICE experiment introduced the normalization to particles with smaller or no strangeness content (e.g. pions in \cite{ALICE:2016fzo} and $K^0_S$ in \cite{ALICE:2019avo}) as a new standard in this field. The relative particle production yield can be compared at different energies, and multiplicity is used as a scaling variable which can be easily measured in different collision systems.
Recently, results on strange hadron production at $\sqrt{s_{NN}}$ varying from 7.7 to 39 GeV were published by the STAR collaboration \cite{STAR:2019bjj} in the context of the Beam Energy Scan (BES) program. This extensive work was not focused on the determination of the multiplicity dependence of particle yield ratios, thus making the comparison with LHC results rather difficult. Moreover, statistics collected at low center of mass energies are limited by the interaction rate achieved at RHIC, thus leading to low precision in the determination of the $\Omega$ $p_{\rm{T}}$ spectrum and yield.

New high statistics studies of $K^0_S$, $\Lambda$, $\Xi$, $\Omega$ and resonances (such as $K^*$ and $\phi$) production in A-A collisions at the SPS are very important in solidifying or reverting the picture emerged at top RHIC and LHC energies. In particular, Pb-Pb and p-Pb(Be) interactions at center-of-mass energies of few GeV could probe the multiplicity region around 10-100 particles at mid-rapidity, which overlaps to the high-multiplicity pp and minimum-bias p-Pb collision regions at the LHC. 
Adopting the new standard in the normalization of particle yields and using multiplicity as a scaling variable, one would have the unique opportunity to probe the hadronization process of a large system at intermediate to low final state multiplicities and in an energy range which is one to three orders of magnitude lower that those probed at RHIC and LHC respectively.
Additionally, the study of $\it{v}_2$ for different strange particles (including $\phi$) would allow to test with unprecedented precision the hydrodinamic description of the produced medium at low center of mass energy.
Finally, rate imbalance between particles and anti-particles would allow a precise determination of the baryo-chemical potential and would probe production probabilities in an energy region where baryon number conservation depletes anti-particle yields.

\subsection{Studies on hyperon-nucleon interactions: production of hypernuclei}
\label{hypernucleiintro}
\vskip 0.2cm
Hypernuclei are bound states of nucleons and hyperons that are particularly interesting because they can be used as experimental probes of the hyperon-nucleon (Y–N) interaction.
The knowledge of the Y–N interaction is fundamental due to its connection to the modelling of dense astrophysical objects like neutron stars \cite{schaffner-bielich_2020, Lattimer:2004pg}. Indeed, in the inner core of neutron stars the creation of hyperons is energetically favoured compared to purely nucleonic matter \cite{Tolos:2020aln}. However, the presence of hyperons as additional degrees of freedom leads to a considerable softening of the equation of state (EOS) of hot matter, and consequently, the resulting EOS prohibits the formation of neutron stars with mass larger than two solar masses \cite{schaffner-bielich_2020, Tolos:2020aln}. This is usually referred to as the "hyperon puzzle" in neutron stars. 
The theoretical effort in solving this puzzle includes the introduction of repulsive three-body forces between hyperons and nucleons~\cite{Lonardoni:2014bwa,Logoteta:2019utx} to counter balance the large gravitational pressure and explain the existence of the observed supermassive neutron stars. 

Ultimately, all the theories on the Y-N interactions, including the multi-body forces, have to be benchmarked and tuned against the measurements of the properties of hypernuclei and, in the last two decades, heavy-ion collisions proved to be a suitable environment for this kind of studies \cite{STAR:2010gyg,Rappold:2013fic,ALICE:2015oer,STAR:2017gxa,ALICE:2019vlx,STAR:2019wjm,STAR:2021orx,STAR:2022zrf}.
Low energy heavy-ion collisions are particularly interesting in this respect, as the high baryon density environment favours the production of hypernuclear clusters \cite{Rappold:2013fic, STAR:2021orx,STAR:2022zrf}.
Furthermore,  a heavy ion collision experiments like NA60+ may be able to detect many different hypernuclear species thanks to its high-precision vertex tracker.

\begin{figure}[t!]
    \centering
    \includegraphics[width=0.8\textwidth]{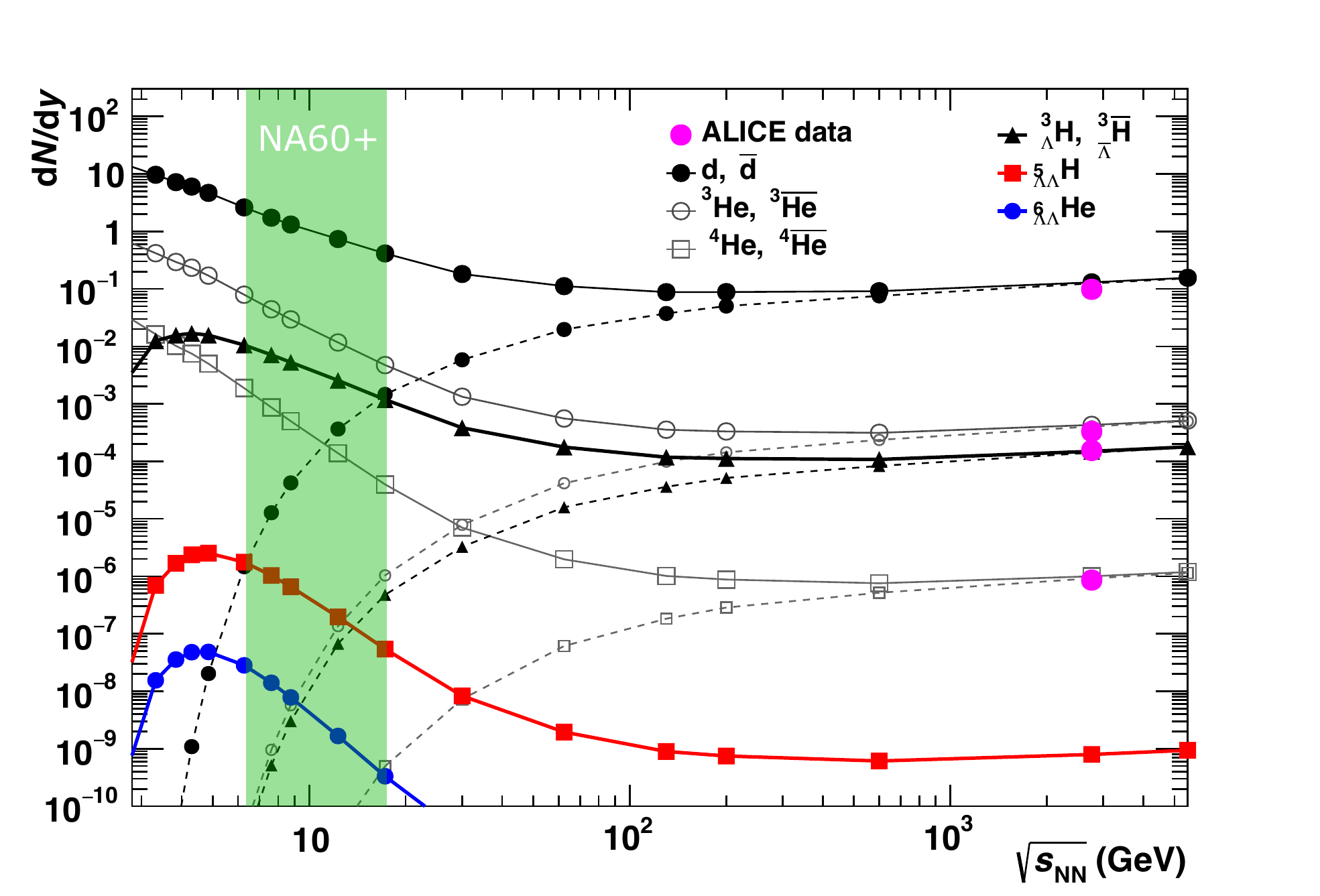}
    \caption{Statistical Hadronisation Model predictions \cite{Andronic:2010qu,Braun-Munzinger:2018hat} for the yield of nuclei, antinuclei and antinuclei in 10\% most central Pb--Pb collisions as a function of the collision energy. The green box highlights the energy region explored by NA60+.}
    \label{fig:SHMnuclei}
\end{figure}

Figure \ref{fig:SHMnuclei} derived from 
Refs.~\cite{Andronic:2010qu,Braun-Munzinger:2018hat} highlights the expected yield for light hypernuclei at the collision energies covered by NA60+, according to the Statistical Hadronisation Model (SHM).
Thanks to the large integrated luminosity, a copious amount of hypernuclei is expected to be detected in NA60+. For the lightest hypernucleus, the $\mathrm{^{3}_{\Lambda}H}$, more than $10^8$ could be inspected by the experiment at each collision energy. The yield of heavier hypernuclei is reduced by approximately a factor 100 for each additional nucleon in the hypernuclear cluster. As such, within the proposed integrated luminosity, NA60+ will be able to study in detail the properties of the hypernuclei with mass number up to A = 6 and it might be able to detect hypernuclei with mass number A = 7.
This wealth of produced hypernuclei enables a full hypernuclear physics programme that can be summarised in three parts:
\begin{itemize}
	\item Precise characterisation of known states: the properties of $\Lambda$ hypernuclei ($\mathrm{^{3}_{\Lambda}H}$, $\mathrm{^{4}_{\Lambda}H}$, $\mathrm{^{4}_{\Lambda}He}$ and $\mathrm{^{5}_{\Lambda}He}$) will be accessible with unprecedented precision, including the study of the $\Lambda$ separation energies of these states and the eventual charge symmetry breaking effect between sibling-hypernuclei \cite{Gal:2015bfa, STAR:2022zrf}.
	\item Properties and confirmation of poorly known/unknown hypernuclei: A = 6 hypernuclear states have very few measurements of their properties \cite{HypernuclearDataBase}. In the case of  $\mathrm{^{6}_{\Lambda}H}$ and $\mathrm{^{6}_{\Lambda\Lambda}H}$, a confirmation of their existence and the measurement of their properties is in reach of NA60+.
	\item Discovery of light $\Xi$ and $\Sigma$ hypernuclei, which are bound according to theory \cite{Hiyama:2019kpw, Le:2021gxa} (e.g. NNN$\Xi$): the excellent tracking capabilities of the NA60+ apparatus, combined with the high expected yield for such states will enable a wide search programme for all these yet-to-be-discovered hypernuclear states.

\end{itemize}

\newpage

\section{The NA60+ experiment: detector concept and general features 
}
\label{NA60plus}
\vskip 0.4cm

The physics topics described in the previous chapter can be studied by means of a new experimental set-up, which includes:
\begin{itemize}
    \item a vertex spectrometer, for a precise measurement of the momentum and production angle of the large amount of produced charged particles (${\rm d}N_{\rm ch}/{\rm d}\eta >400$ in central Pb--Pb collisions at top SPS energy);
    \item a muon spectrometer, which measures muon tracks which are filtered by a thick hadron absorber, positioned downstream of the vertex spectrometer.
\end{itemize}

Matching the candidate muon tracks in the muon spectrometer, in coordinate and momentum space, with the corresponding track in the vertex spectrometer, the muon kinematics can be precisely accessed, minimizing the effect of energy loss and multiple scattering in the hadron absorber. The precision tracking in the vertex spectrometer also allows the reconstruction of selected two- and three-body decay topologies, as those from strange and charmed hadrons. 
The detector concept resembles rather closely the previous NA60 experiment, which took data for top SPS energy In--In and p--A collisions in 2003-2004, and performed measurements of dilepton production of still unsurpassed accuracy~\cite{Arnaldi:2006jq,Arnaldi:2008er}. The experiment was installed in the ECN3 underground hall and was finally dismounted in 2010, to allow the preparation/installation of the NA62 experiment.

\begin{figure}[ht]
\begin{center}
\includegraphics[width=0.945\linewidth]{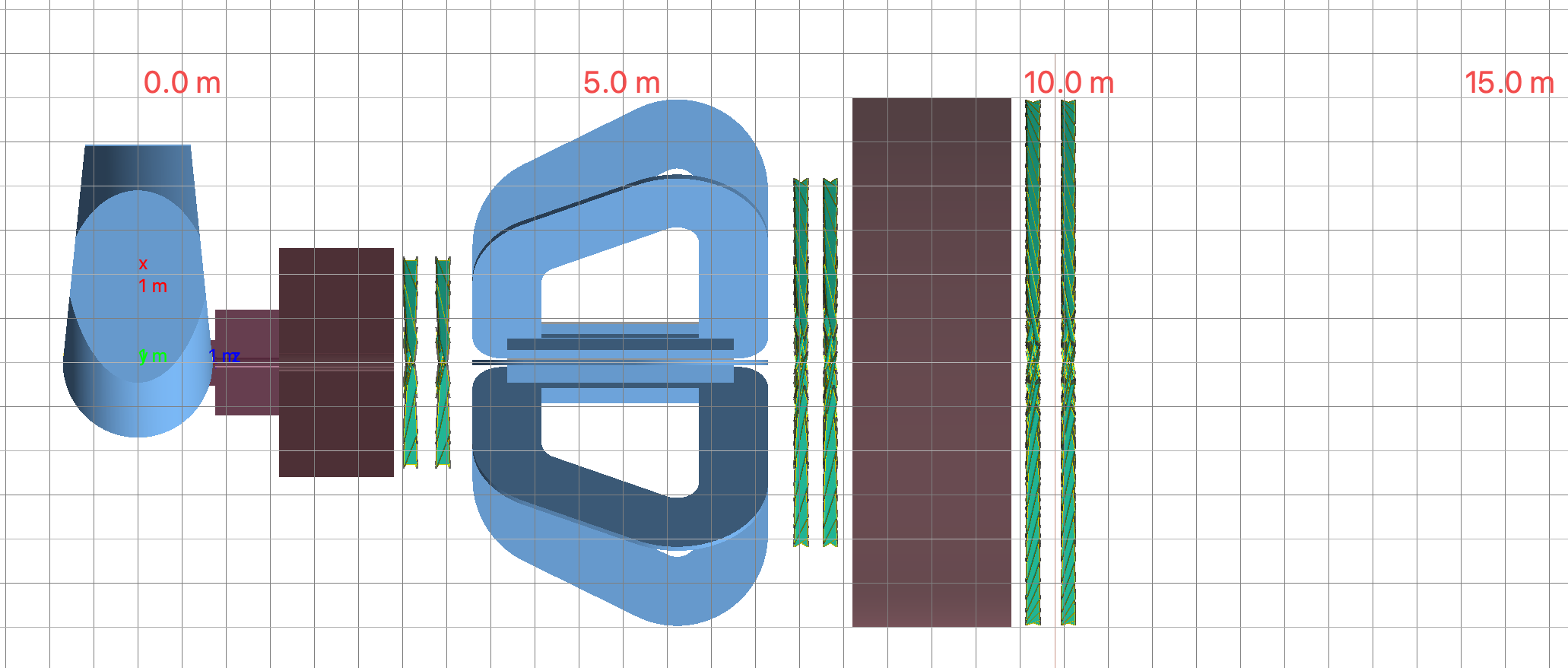}
\includegraphics[width=0.94\linewidth]{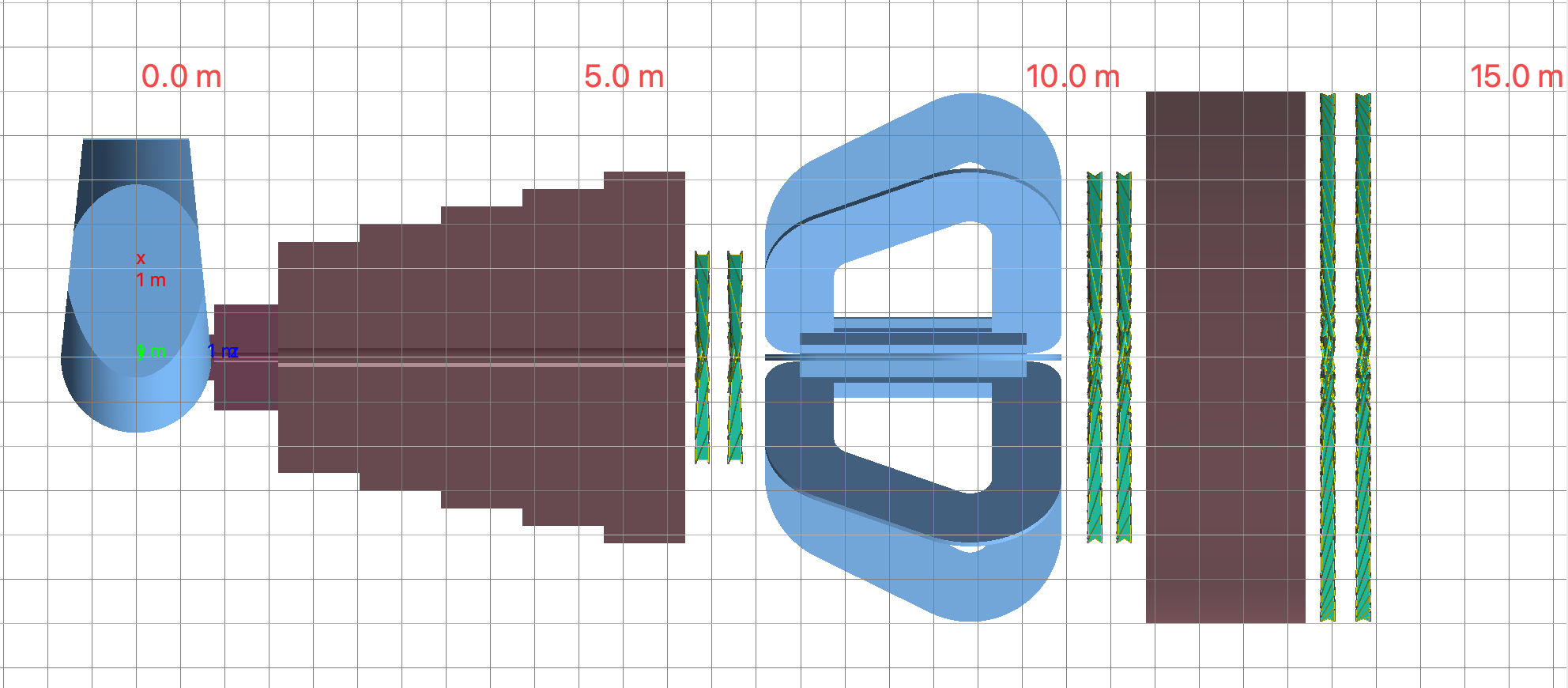}
\caption{GEANT4 rendering of the NA60+ experimental apparatus (top view). The top figure represents the  set-up adapted to low-energy collisions, with a thinner hadron absorber and the muon spectrometer relatively closer to the target and the bottom figure shows the set-up intended for high-energy collisions. The vertex spectrometer is visible in Fig.~\ref{fig:NA60concept_vertex}.}
\label{fig:NA60concept}
\end{center}
\end{figure}

The new experiment, currently denoted as NA60+, will extend and improve the physics program of NA60, by performing measurements of dileptons and heavy-quark production over all the available SPS energy range and increasing at all energies the precision of the top SPS-energy NA60 results. This will be possible thanks to state-of-the-art experimental technologies and to the use of a high-intensity beam (of the order of 10$^7$ Pb ions per spill). A conceptual drawing of the set-up is shown in Fig.~\ref{fig:NA60concept}. In order to keep a constant rapidity acceptance around $y\sim 0$  in the center-of-mass system of the collision, the muon spectrometer needs to be moved downstream when the beam energy is increased. At the same time, the thickness of the hadron absorber will also be increased, to cope with the larger hadron multiplicity. The position of the vertex spectrometer does not need to be modified, due to its intrinsically larger angular acceptance.  In the following Sec.~\ref{Explayout} we will briefly summarize the foreseen choices for the various elements of the experimental set-up (technical aspects and more details will then be extensively covered in Chapter~\ref{Detectors}). Then, in Sec.~\ref{Beamconditions}, we will discuss the foreseen beam requirements for the NA60+ data taking. Finally, in Sec.~\ref{Competition} we will analyze the role of NA60+ compared to other experiments either at the SPS (NA61) or at other facilities where the exploration of the high-$\mu_{\rm B}$ region of the QCD phase diagram can be performed.

\begin{figure}[ht]
\begin{center}
\includegraphics[width=0.95\linewidth]{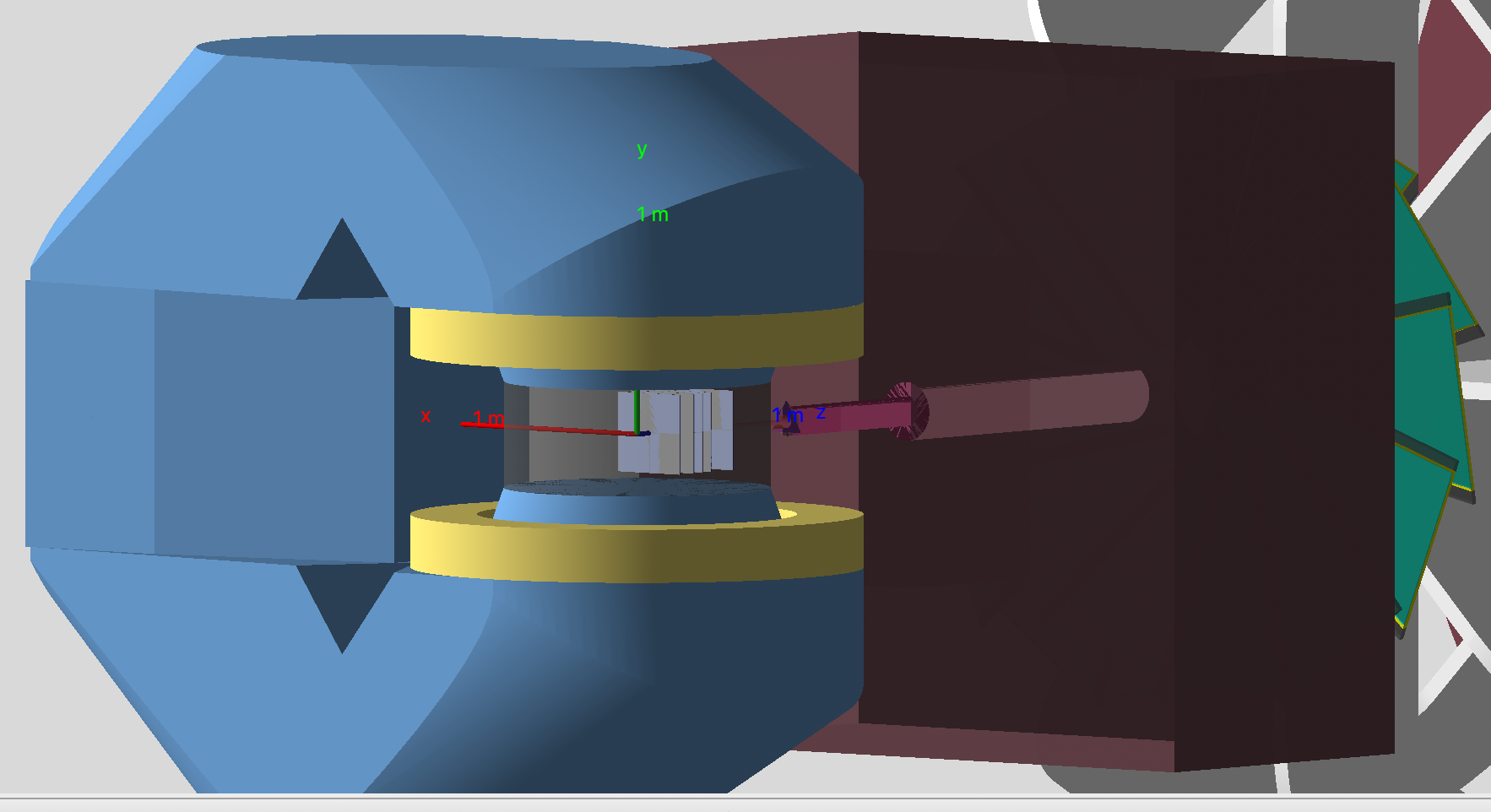}
\caption{GEANT4 drawing of the vertex region of the NA60+ setup. The vertex spectrometer immersed in  the dipole field of MEP48 is shown.}
\label{fig:NA60concept_vertex}
\end{center}
\end{figure}

\subsection{Experimental layout}
\label{Explayout}
\vskip 0.2cm

\subsubsection{Target system}
\label{TargetSystem}

For the heavy-ion runs, the target system will be composed of five 1.5 mm thick Pb disks spaced by 12 mm. The first one has a radius of 3 mm, 
while the other targets have 1 mm diameter. Due to their small transverse size, they should be aligned with respect to the beam axis with a precision of $\sim100$ $\mu$m.
The inelastic interaction probability of this system is ${\sim}15\%$. 

For proton--nucleus runs, the system will be composed of a number of sub-targets of different nuclear species like Be, Cu, In, W and  Pb, simultaneously exposed to an incident proton beam. The sub-targets will have a diameter of \SI{\sim1}{\mm} with a spacing of 12 mm.
The individual target thicknesses are chosen so as to collect event samples of similar sizes for each nuclear species, assuming a total interaction length up to ~${\sim}10\%$. The integrated  luminosity per nucleon-nucleon interaction should be similar to that of Pb--Pb collisions, implying a $\sim 2$ order of magnitude larger beam intensity. 

\subsubsection{Dipole magnet}

The magnetic field for the momentum measurement in the vertex spectrometer will be provided by a dipole magnet. The current choice is the MEP48 magnet, originally built for the PS170 experiment and now stored at CERN, which can deliver a 1.5 T field over a 400 mm gap. 

\subsubsection{Vertex spectrometer}
The vertex telescope consists of 5 identical silicon pixel planes positioned at $7<z<38~\si{\cm}$ starting from the most downstream target.
The absorber  starts at \SI{\sim 45}{cm} from the interaction point, providing a good rejection of background muons from pion and kaon decays.
The planes are immersed in the dipole field of MEP48, providing a field integral of about 1.2 Tm.
Each plane, featuring a material budget of 0.1\% X$_0$ and intrinsic spatial resolution of $\sim$5 $\mu$m,  is formed by 4 large area monolithic pixel sensors of 15x15 cm$^2$ each.
The total active area is $\sim$ 0.5 m$^2$.
A GEANT4 rendering of the silicon telescope immersed in the dipole field of MEP48 is shown in Fig.~\ref{fig:NA60concept_vertex}. Deatils on detector studies are reported in Sec.~\ref{sec:vertex_telescope}.

\subsubsection{Muon spectrometer}

The muon spectrometer should provide a precise measurement of the candidate muon tracks. Its dimensions match the angular acceptance of the vertex spectrometer and are also constrained by the size of the experimental area, discussed in Sec.~\ref{Proposed_zone_layout}, which limits the half transverse-size to $\sim 3$ m in the horizontal direction and to $\sim 2.8$ m in the vertical direction. The limitation in the vertical direction is due to height of the beam line above the floor of the PPE138 experimental area (285 cm), currently foreseen for the installation of the experiment. 

A thick hadron absorber is positioned upstream of the muon spectrometer. It has to fulfil the contrasting requirements of relatively high density and limited Z, the latter request being connected with the necessity of limiting the multiple scattering of the muons, that would decrease the matching efficiency between tracks in the muon and in the vertex spectrometers. The hadron absorber will include an upstream section composed of BeO, followed by graphite. The thickness of the graphite section will be increased when moving from low to high collision energies. At very forward rapidity a plug made of tungsten will dump the non-interacting beam particles as well as their fragmentation products. 

The set-up of the spectrometer includes six tracking stations. The first two stations (MS0, MS1) are located after the hadron absorber and upstream of the toroidal magnet, while the following ones (MS2, MS3) are installed downstream of it, providing in this way four space points. Following a design typical of this kind of spectrometers (NA50/60, ALICE), a thick graphite wall allows further filtering of hadrons that may have survived the hadron absorber, and is followed by two final tracking stations (MS4, MS5). 
Preliminary detector studies, with GEM and/or MWPC  as candidate technical solutions, are reported in Sec.~\ref{sec:muonspectrometer}. Preliminary

\subsubsection{Toroidal magnet}

The magnetic field for the measurement of the momenta of the candidate muons in the spectrometer will be provided by a magnet generating a toroidal field. The device used by NA60 (ACM) does not possess an angular aperture covering the desired acceptance at low SPS energy and therefore does not represent a viable choice. In the current design of the NA60+ experiment we foresee a warm magnet with an angular aperture of 0.29 rad, composed of eight radial sectors, each one consisting of a number of windings, in order to reach the desired current. The strength of the magnetic field is $\sim 0.37$ T at a radial distance of 1 m, with a $1/r$ dependence of the field. The total length of the magnet is 335 cm. The non-negligible technical challenges of such a project have led to the realization of a prototype in scale 1:5, to be considered as a testing bench for the possible solutions for the full-scale object. The technology choices, the prototype performance and the prospects for the final object will be described in Sec.~\ref{Toroid}.

\subsection{Beam energy scan and data taking conditions}
\label{Beamconditions}
\vskip 0.2cm

A fundamental aspect of the NA60+ experiment is the possibility of collecting data with a high beam intensity over all the energy range accessible to the SPS. Studies of the optics of the H8 beam line, that will be discussed in more detail in Sec.~\ref{beam_setup}, have shown that a primary Pb beam with an intensity of the order of 10$^7$/spill can be delivered, in the energy range $20<E<160$ GeV/nucleon. An extension to lower energy is currently under study. At the same time, the transverse dimensions of the beam need to be sub-millimetric, due to the constraints created by the geometry of the vertex spectrometer, dictated by the need of ensuring an angular  coverage corresponding to one unit of rapidity at least. For this reason, each one of the five MAPS station has a central hole with a 6--8 mm diameter. As it will be shown in Sec.~\ref{beam_setup}, values $\sigma_{\rm x}\sim\sigma_{\rm y}$ from 0.2 to 0.4 mm could be reached when moving from top SPS energy down to 20-30 GeV. Dedicated beam tests at the H8 beam line are foreseen, in order to validate the current calculations.

The physics performance studies discussed in Ch.~\ref{PhysicsPerformance} show that in order to get enough statistics for the foreseen physics program, $\sim10^{12}$ incident Pb ions on a 15\% interaction probability Pb target will be necessary for each energy.  This scenario can be reached, assuming a 9 s spill every $\sim$25 s, in about 30 days, with a beam intensity of $\sim 10^7$ Pb ions/spill. However, this duty cycle corresponds to having ions delivered only to the SPS in the supercycle. A more realistic situation is having a spill every $\sim 40$~s. In order to stay with a similar integrated beam intensity, one would therefore need a beam intensity of up to $\sim 2\times 10^7$ Pb ions/spill. 

In addition to the Pb-beam data taking, a corresponding period with a proton beam at each energy is mandatory, to collect reference data needed for the interpretation of the heavy-ion results. A high-purity (primary) beam is needed. Assuming 3000 spills/day to be delivered, the necessary integrated luminosity could be collected in $\sim 22$ days of beam at an intensity of $\sim 8\cdot 10^8$ per spill (this estimate is still preliminary).  

A tentative break-out for the first years of data taking, including energies and total number of particles on target is shown in Table~\ref{tab:NA60+beam}. The exact order and the precise value of the various beam energies is not to be considered as mandatory, except for starting at top SPS energy. This would allow obtaining a physics calibration point, by a comparison of the main results with those of the former NA60 experiment.

\begin{table} [htb]
   \caption{
Lead ion energies foreseen for the first phase of NA60+ data taking. It is assumed that Year 1 is the first year after the end of LS3 (likely in 2029). The exact order of the energies from Year 2 onward is flexible. We assume a two-year running time at the lowest energy. The number of days for the Pb beam assumes a 25 s SPS supercycle and 10$^7$ Pb ions per spill.}
\vspace{0.5cm}
\centering
\begin{tabular}{ c | c | c | c | c | c | c }
\hline
\hline
 &  Year 1 & Year 2 & Year 3 & Year 4-5 & Year 6 & Year 7    \\
\hline
Beam energy (A GeV) &  160 & 40 & 120 & 20 (30) & 80 & 60   \\
\hline
Momentum per charge (GeV/c/Z) &  406 & 101 & 304 & 50.7 (76.1) & 203 & 152   \\
\hline
Pb ions on target & \multicolumn{6}{c}{$\sim 10^{12}$ per energy ($\sim 30$ days)} \\
\hline 
protons on target & \multicolumn{6}{c}{$5-6\cdot 10^{13}$ per energy ($\sim 22$ days)} \\
    \hline
    \hline
  \end{tabular}
  \label{tab:NA60+beam}
\end{table}

\subsubsection{Trigger strategy}

The former NA60 experiment included a dimuon trigger, together  with a strongly pre-scaled beam trigger based on a minimum energy deposition in a zero-degree calorimeter. In NA60+, there are physics signals not related to dimuon production, and in particular open charm production via the measurement of hadronic decays in the vertex spectrometer. We plan 
tjournalctl --vacuum-time=10do define a minimum-bias interaction trigger, related to the charged-particle multiplicity. The current idea is using either a scintillation counter(s) located in the vertex spectrometer region, as done by the former NA57 experiment~\cite{WA97:2000kwf}, or a Cerenkov detector, as used in the CERES/NA45 experiment~\cite{etde_593433}. 

\subsection{Role of NA60+ in the experimental landscape of \texorpdfstring{high-$\mu_{\rm B}$}{highmuB} studies
}
\label{Competition}
\vskip 0.2cm

\begin{figure}[ht]
\begin{center}
\includegraphics[width=0.9\linewidth]{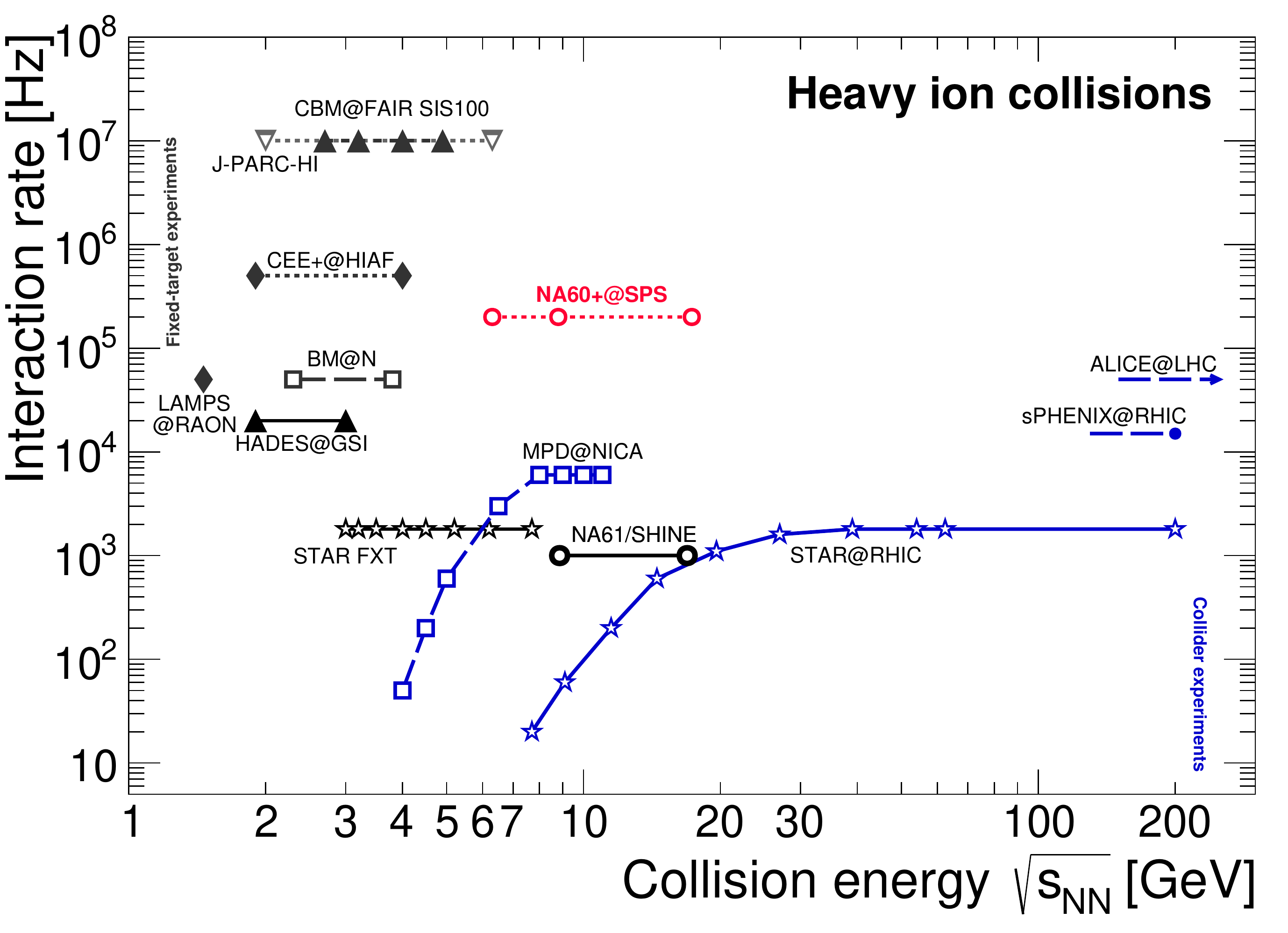}
\caption{Scheme of existing and foreseen experiments with high-energy nuclear beams. The horizontal axis shows the center-of-mass energy per N-N collision that is/will be covered, while on the vertical axis the available/expected interaction rate is shown (courtesy of T. Galatyuk~\cite{Galatyuk:2019lcf}).}
\label{fig:Tetyana}
\end{center}
\end{figure}

The NA60+ project will be part of a world-wide experimental program for the study of the properties of the Quark-Gluon Plasma in the region of relatively large baryochemical potential. As previously mentioned in this document (see e.g., Ch.~\ref{Overview}) these studies require much lower center-of-mass energy compared to the top energy that can be reached at the RHIC and LHC ion colliders. In Fig.~\ref{fig:Tetyana} the current list of existing and foreseen experiments is presented, in terms of energy coverage and interaction rate for nuclear collisions~\cite{Galatyuk:2019lcf}. The latter quantity is related to the detector capabilities and to the luminosity delivered by the corresponding facilities.

In the SPS energy range, which approximately covers the interval $6<\sqrt{s_{\rm NN}}<17$ GeV, the NA60+ experiment, with a foreseen interaction rate $>10^5$ s$^{-1}$ represents the better candidate for the study of rare processes. Other experiments that cover the same energy range, or at least part of it, may reach an interaction rate from about one to two orders of magnitude smaller (MPD~\cite{Golovatyuk:2016zps} at the forthcoming NICA collider~\cite{web:NICA} and NA61/SHINE~\cite{NA61:2014lfx} at the CERN SPS, respectively). Existing data from the RHIC beam energy scan in collider mode (BES-II), as well as in fixed-target configuration (STAR FXT) were collected at a maximum interaction rate of the order of 10$^3$ s$^{-1}$~\cite{Tlusty:2018rif}.

Below the SPS energy range, the CBM experiment~\cite{Friman:2011zz} at the forthcoming FAIR facility~\cite{Durante:2019hzd} will access a lower energy domain that nicely complements the one of NA60+, with an extremely large interaction rate (up to $10^7$ s$^{-1}$). Similar performances in the same energy range are also expected in a future physics program with acceleration of heavy-ion beams at J-PARC~\cite{Sako:2019hzh}. 

The experiments mentioned in this brief summary are expected to explore an overall interval of baryochemical potential $230<\mu_{\rm B}<800$ MeV, with NA60+ more specifically covering the range $230<\mu_{\rm B}<560$ MeV~\cite{Galatyuk:2019lcf}. In this situation it will be possible to explore a region of the QCD phase diagram where a first-order phase  transition between hadronic matter and QGP is foreseen. This zone is expected to terminate in a second-order critical point. Discovering signals of the first-order phase transition and the location of the critical point represent one of the hottest topics of relativistic heavy-ion physics. The NA60+ project, thanks to its rich and specific physics program that was described in Ch.~\ref{RareProbes}, will play a unique role in this endeavour by providing for the first time accurate data on dileptons, open charm and charmonia below top SPS energy. The study of strange particle production in heavy-ion collisions, pioneered by the NA57~\cite{NA57:2010tnk} and NA49~\cite{NA49:2008ysv} experiments and more recently investigated at SPS energies by NA61~\cite{NA61SHINE:2020czq}, will also be part of the NA60+ program. Finally, a program aiming at the measurement of various hypernuclear states is among the goals of NA60+.

\newpage
\section{Physics performance studies}
\label{PhysicsPerformance}
\vskip 0.4cm
In this Chapter, results on the physics performance of the NA60+ experiment will be described. We start by a brief description of the current framework used for these studies (Sec.~\ref{simframework}). We then show results for observables that are accessed via hadronic measurements, i.e. open charm, strangeness and hypernuclei production (Sec.~\ref{hadronicmeasurements}). In Sec.~\ref{dimuonmeasurements} we will discuss results on leptonic measurements, including dilepton studies and charmonium production.

\subsection{Simulation frameworks}
\label{simframework}
\vskip 0.2cm

\subsubsection{Fast simulation/reconstruction 
}

The detector performance was studied using a dedicated fast simulation and reconstruction tool (FSRT). It consists of: 
\begin{itemize}
  \setlength\itemsep{1em}
  \item A layout description module, which allows describing the experimental set-up as a combination of thin sensitive and extended passive material layers with normal orientation to the beam direction ($z$-axis) as well as the definition of regions with dipole and toroidal magnetic fields. The input card reader adapted from the NA60root package (simulation and reconstruction software framework of the NA60 experiment) allows on-the-fly initialization of different layouts stored in human-readable text files. The description of sensitive detector layers includes their acceptance coverage in $r$ and $\phi$, intrinsic resolutions along different axes (e.g. $x$, $y$ for the vertex telescope, $r$, $r\phi$ for the muon spectrometer) as well as randomly applied hit inefficiencies.
  
  \item A fast simulation engine which, starting from the initial kinematics and position of the probe particle performs its transport through the detector and registers hit positions at all sensitive layers. The transport accounts for the Coulomb multiple scattering in Gaussian approximation and deterministic ionization energy loss (no energy loss straggling is simulated). For the cross-checks and more reliable results accounting for all aspects of particle propagation in the materials, there is a possibility to perform \fluka ~\cite{Bohlen:2014buj,Ferrari:2005zk} or GEANT4 ~\cite{GEANT4:2002zbu} simulations through an throughequivalent layout and import the position and kinematics of the transported particle at each sensitive layer it crosses. The hits from the ``signal'' probe and optional background particles are smeared by the assigned intrinsic resolutions at every sensitive plane and stored for further tracking.
  
  In order to study the performance of the muon chamber read-out elements  made of two strip (U,V) and one wire (W) planes the simulated layout was extended with the explicit implementation of such triplets in the trapezoidal modules. The fired strips and wires are independently registered, and then all possible intersections of U, V and W fired channels are built in order to emulate the true space-points combinatorics seen in this kind of detectors.  

  \item A fast reconstruction module using the Kalman filtering both for track finding and fitting. It adapts the ALICE experiment Kalman barrel track model to a fixed target forward layout by swapping relevant axes. The reconstruction starts by creating the track seeds for every pair of points in the last two muon stations (after the toroidal magnet), validating them with the hits in the first station after the muon wall and determining the approximate momentum from the (bending) angle between the straight-line extrapolation of this seed to the nominal bending plane of the toroid and the vector connecting this extrapolation point to the target. Then every seed is propagated to the most downstream station of the spectrometer and followed towards the absorber with subsequent Kalman updates. The seeds which did not find a matching hit in the muon stations before the toroid are suppressed. In case of ambiguities, the hit with the best matching $\chi^2$ is selected. All tracks reconstructed in the muon spectrometer are propagated through the absorber towards the target in successive steps of extrapolation and Kalman updates by the hits found within the extrapolation tube. For every hit matching the seed on a given plane, a separate new seed is created with global $\chi^2$ of its parent seed incremented by track-hit $\chi^2$. In absence of the matching hits at a given plane, the seed $\chi^2$ is incremented by a penalty term. Thus, the reconstruction creates a tree of initial seeds propagation hypotheses, which grows as it approaches the target. The branches which do not acquire enough hits or with too large accumulated $\chi^2$ are eliminated. Finally, the ``winner'' track is selected at the target position as the branch with the smallest $\chi^2/NDF$. 
\end{itemize}

Only a single probe particle (optionally overlaid on the underlying background event)  is simulated and reconstructed at once, with all available hits participating in seeding and track following. Thus, in order to study the dimuon performance, each muon from the pair provided by the generator is simulated and tracked independently, then the two ``winner'' tracks (if any) are combined.

\subsubsection{GEANT4 simulation
}

A full and detailed MC simulation of the detector performance is critical for optimizing the detector layout and its elements, understanding the backgrounds, and building the basis for working out the corrections required for producing physics results. 
The geometry model of the NA60+ experiment was implemented in GEANT4~\cite{Allison:2006ve} version 10.07.p01 using the QGSP\_BERT\_HP physics list. Figure~\ref{fig:NA60concept} in Ch.~\ref{NA60plus} shows a conceptual design of the NA60+ simulation model for the low- and high-energy setups. It includes the following instrumentation components, detector systems, and infrastructure:

\begin{itemize}
  \setlength\itemsep{1em}

\item The vertex region (see Fig.~\ref{fig:NA60concept_vertex} in Ch.~\ref{NA60plus}) and detector support structures, implemented according to the design envisaged for NA60+ and described in the relevant sections of this document. 
Some simplification was made for
supporting structures in the areas not directly exposed to the signal or beam particles.

\item The dipole and toroidal magnet hardware models, based on technical drawings of the existing MEP48 magnet and of the current design  of the toroidal magnet, respectively. The fields that are implemented in the present version of the simulation fill the nominal volume of the magnets and are uniform in the case of the dipole and decrease with $1/r$ in the case of the toroid. Their strength is chosen based on the documentation for the magnets.

\item The absorbers, including the main geometric dimensions and choice of the materials, i.e. beryllium oxide, graphite, and tungsten.

\item The muon tracking chambers, based on the concept currently considered for the NA60+ muon spectrometer and on the materials that are used for the existing MWPC prototype (see section \ref{sec:muonspectrometer}). The supporting structure for the muon chamber is currently not implemented. 
\end{itemize}

\subsubsection{FLUKA simulations  
}
\label{flukastudies}

The \fluka particle transport code~\cite{Battistoni:2015epi} represents the state-of-the-art tool for 
calculations of particle transport and interactions with matter. It covers an extended range of energies, from the keV to the TeV range.

For an experiment like NA60+, operating with high intensity ion beams and with primary interactions that generate hundreds of charged particles per unit of rapidity, it is mandatory to precisely estimate the background sources that can affect the physics measurements. Those include the contribution from $\delta$-rays in the vertex spectrometer, as well as punch-through hadrons and background muons in the muon spectrometer.
These background sources directly affect the physics performance of the experiment and are also important in the definition of the needed resolution of the tracking detectors.

The NA60+ set-up, including both passive and active elements, was described using the Combinatorial Geometry package and the recent FLUKA2021.2.3 version of the code was used. 

The background effects on the physics performance studies described in the next sections are evaluated by injecting \fluka Pb-Pb events in the fast simulation framework on top of the generated signal sources. Furthermore, standalone simulations are used for the evaluation of the particle fluence in the tracking detectors, making use of the scoring options provided by \fluka. The results of these latter studies will be reported in Section~\ref{FLUKArate}.

\subsection{Hadronic measurements}
\label{hadronicmeasurements}
\vskip 0.2cm

\subsubsection{Open charm
}

Open-charm hadrons can be fully reconstructed with the NA60+ apparatus via their decays into two or three charged hadrons.
In particular, the following decays could enable the measurement of non-strange and strange D mesons as well as $\lambdac$ baryons: $\Dzero \to {\rm K}^-\pi^+$, $\Dplus\to {\rm K}^-\pi^+\pi^+$, ${\Ds \to \phi\pi^+\to {\rm K}^{+} {\rm K}^{-} \pi^{+}}$, $\lambdacplus \to {\rm p K^{-}}\pi^+$ and their charge conjugates.
The measurements of the production yields of different meson and baryon species open the possibility to provide specific insight into the hadronization mechanism of charm quarks in a QGP with large baryochemical potential.
In addition, these measurements are also relevant for an accurate determination of the total \ccbar production cross section, which would serve as a test of pQCD calculations in an unexplored region of \sqrtsNN and would also provide a natural reference for charmonium studies.

The charm-hadron decay particles (pions, kaons and protons) are detected by reconstructing their tracks in the silicon-pixel detectors of the vertex telescope (VT).
The $\mathrm{D}$-meson and \lambdac candidates are built by combining pairs or triplets of tracks with the proper charge signs.
The huge combinatorial background can be reduced via geometrical selections on the displaced decay-vertex topology, exploiting the fact that the mean proper decay lengths $\mathit{c}\tau$ of open-charm hadrons are of about \SIrange{60}{310}{\um}~\cite{PhysRevD.98.030001}, and therefore their decay vertices are typically displaced by a few hundred~\si{\um} from the primary interaction vertex.
Among the measurements proposed by NA60+, open-charm hadron studies are those that impose the strongest constraints on the design of the VT detectors, which should provide good resolution on the track parameters in order to allow us to separate the secondary tracks produced in open-charm hadron decays from the primary ones originating from the interaction point.

\paragraph{Two-prong decays: D$^0$}

\begin{figure}[tb]
\begin{center}
\includegraphics[width=0.9\linewidth]{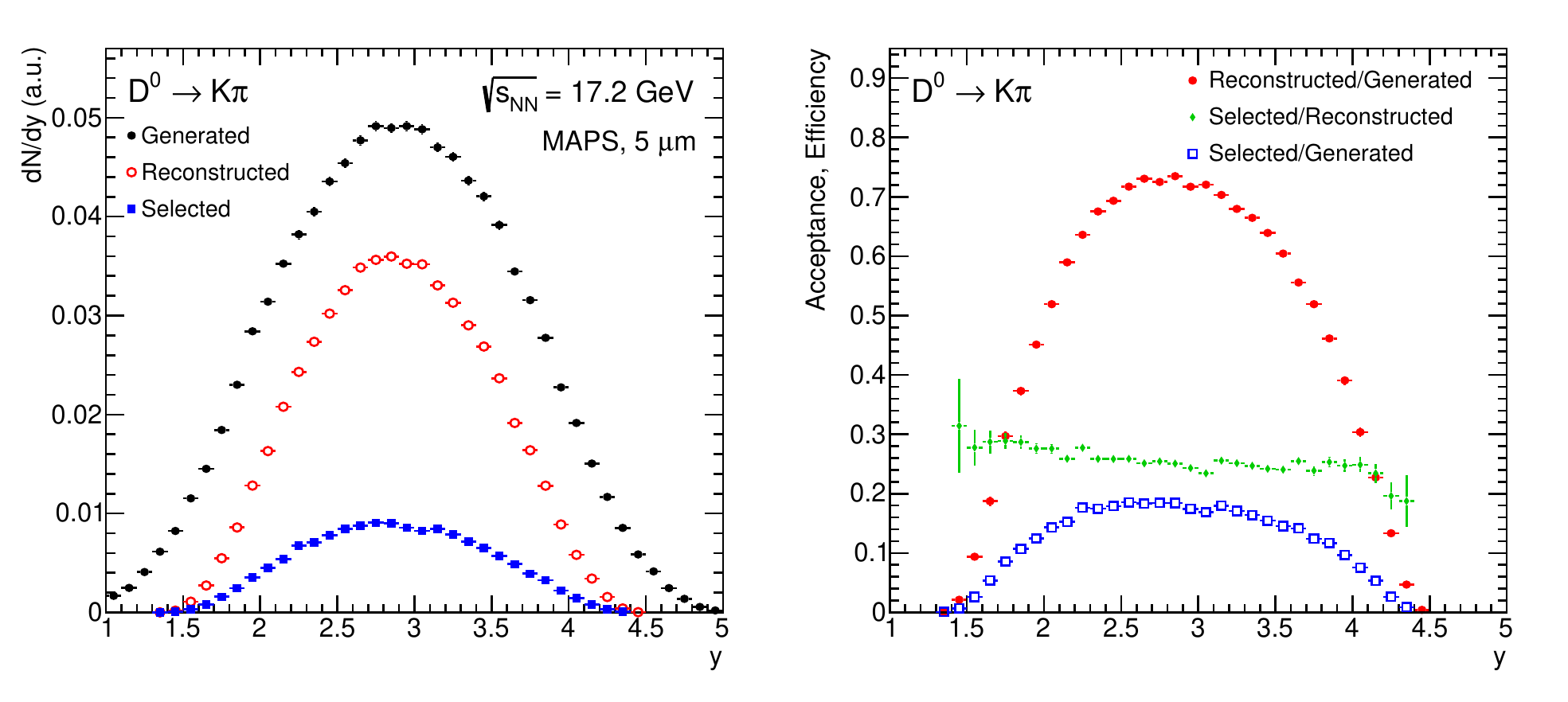}
\caption{Left: Rapidity distributions for $\Dzero \rightarrow \mathrm{K}^- \pi^+$ decays at $\sqrtsNN = 17.3$\GeV at generation level, after reconstruction of decay products in the VT, and after selection cuts based on the decay vertex topology for the case of the MAPS detectors in the VT. Right: acceptance, reconstruction and selection efficiency for $\Dzero \rightarrow \mathrm{K}^- \pi^+$ decays.}
\label{fig:D0acceff}
\end{center}
\end{figure}

Benchmark studies were carried out for the measurement of $\Dzero \to {\rm K}^-\pi^+$ in the 5\% most central \PbPb collisions at two different beam energies: \SIlist{158;60}{\AGeV}, corresponding to $\sqrtsNN = 17.3$ and 10.6\GeV, respectively.
The \Dzero and $\overline{\Dzero}$ mesons were simulated with \pt and rapidity distributions obtained with the {\sc Powheg-Box} event generator~\cite{Alioli:2010xd} for the hard-scattering and \pythia~6 for the parton shower and hadronization.
The \powheg simulations were performed with the \textsc{Cteq}6 PDF and a charm quark mass of 1.5\GeVcc.
The decay $\Dzero \to {\rm K}^-\pi^+$ was simulated with EvtGen~\cite{Lange:2001uf} and the kaon and pion were propagated through the VT utilizing the fast simulation of NA60+.
The rapidity distributions of \Dzero mesons at $\sqrtsNN = 17.3$\GeV at the generator level and after propagation and reconstruction of the kaon and pion decay products through the VT are reported in Fig.~\ref{fig:D0acceff}, together with their ratio, which shows that the acceptance times reconstruction efficiency for $\Dzero \to {\rm K}^-\pi^+$ is about 70\% at midrapidity.
The combinatorial background was estimated by simulating pions, kaons and protons with multiplicity, \pt and rapidity distributions taken from the parameterisations published by the NA49 collaboration~\cite{Afanasiev:2002mx,Alt:2006dk}.
The decay vertex position was computed by propagating the decay tracks to their point of closest approach and using the elements of the track covariance matrix as weights.
The combinatorial background was built by pairing tracks with opposite charge sign.
Particle identification is not available and therefore for each pair of tracks two candidates (\Dzero and $\overline{\Dzero}$) are produced depending on the mass assigned to the tracks (${\rm K}^-\pi^+$ and $\pi^-{\rm K}^+$).
The number of primary particles ($\pi$, $\mathrm{K}$, $\mathrm{p}$) per central \PbPb collision at $\sqrtsNN = 17.3\GeV$ is about 1200, which produces about $3.5\times 10^5$ background candidates per event, out of which about 8000 have an invariant mass within 60\MeVcc from the \Dzero-meson mass (1.865\GeVcc~\cite{PhysRevD.98.030001}).
The yield per event at $\sqrtsNN = 17.3\GeV$, estimated assuming a total charm cross section of $\sigma_{\ccbar}=5$\mub (based on~\cite{Lourenco:2006vw,Vogt:2001nh} and \powheg) and a fraction of charm quarks hadronizing into \Dzero mesons of  $f({\rm c}\to \Dzero)\sim0.55$, is about 0.006.
The \Dzero yield per event was calculated as $N_{\Dzero}= 2 \cdot \taa \cdot \sigma_{\ccbar} \cdot f({\rm c}\to \Dzero) \cdot {\rm BR}$, where $\taa=26.9\mbinv$ is the average nuclear overlap function for the 5\% most central \PbPb collisions, $\mathrm{BR} = 3.89\pm0.04\%$ is the branching ratio of the $\Dzero \rightarrow \mathrm{K}^- \pi^+$ decay~\cite{PhysRevD.98.030001}, and the factor of 2 accounts for the fact that the measurement includes both \Dzero and $\overline{\Dzero}$ mesons.
The signal-to-background ratio is therefore about \num{7e-7} and needs to be enhanced with the kinematical and geometrical selections.

\begin{figure}[tb]
\begin{center}
\includegraphics[width=0.9\linewidth]{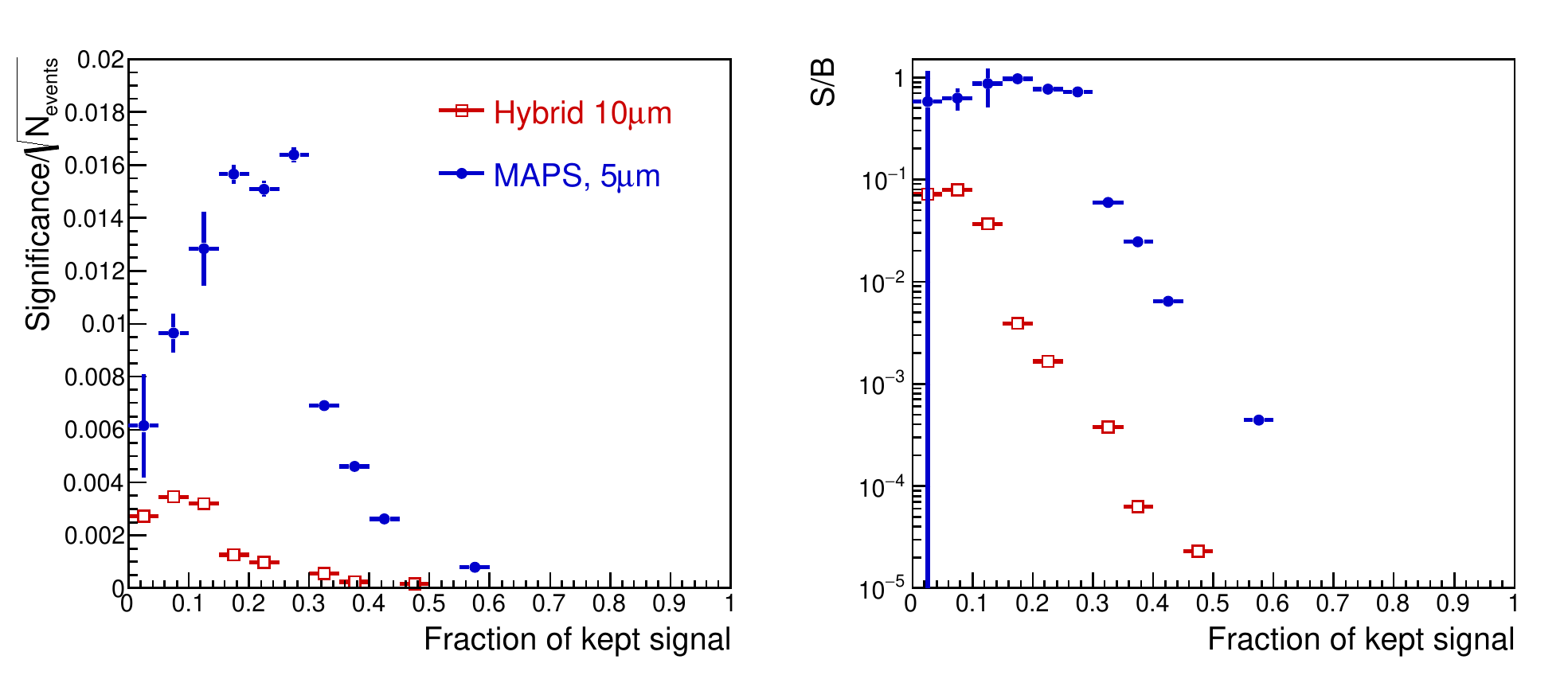}
\caption{\Dzero-meson statistical significance (left, normalized to the square root of the number of events) and signal-over-background ratio (right) as a function of the selection efficiency for the 5\% most central \PbPb collisions at $\sqrtsNN = 17.3\GeV$ for the two considered pixel designs.
The signal and the background yields are evaluated in a $2\sigma$ invariant-mass region around the \Dzero peak.}
\label{fig:D0signif}
\end{center}
\end{figure}

The following variables were utilized for the candidate selection: \pt of the decay tracks, the cosine of the $\theta^{\ast}$ angle (the angle between the kaon momentum in the \Dzero reference frame and the \Dzero flight line), the impact parameter of decay tracks and their product, the distance of closest approach between the decay tracks, the decay length (\ie\ the distance between primary and secondary vertices), the cosine of the pointing angle, and the \Dzero impact parameter.
Several selection sets were tested, and for each of them the signal-selection efficiency, the statistical significance of the signal ($S/\sqrt{S+B}$) and the signal-to-background ratio ($S/B$) were computed.
In Fig.~\ref{fig:D0signif}, the significance per event and the $S/B$ at $\sqrtsNN = 17.3\GeV$ are shown as a function of the efficiency for two different configurations of the VT detectors, one based on hybrid pixel sensors with \SI{10}{\um} spatial resolution and one based on Monolithic Active Pixels (MAPS, see Sec.~\ref{sec:vertex_telescope}), which feature a better spatial resolution (\SI{5}{\um}) and a reduced material budget (${\sim}0.1\%~X_0$ vs ${\sim}1\%~X_0$).
The performance is substantially better with the MAPS detector, which provides a better resolution on the decay track momentum, on the decay vertex position (\SIrange{10}{15}{\um} in the plane transverse to the beam line with MAPS vs \SIrange{30}{40}{\um} with hybrid pixels) and therefore on the \Dzero invariant mass (10\MeVcc vs 24\MeVcc).

\begin{figure}[tb]
\begin{center}
\includegraphics[width=0.45\linewidth]{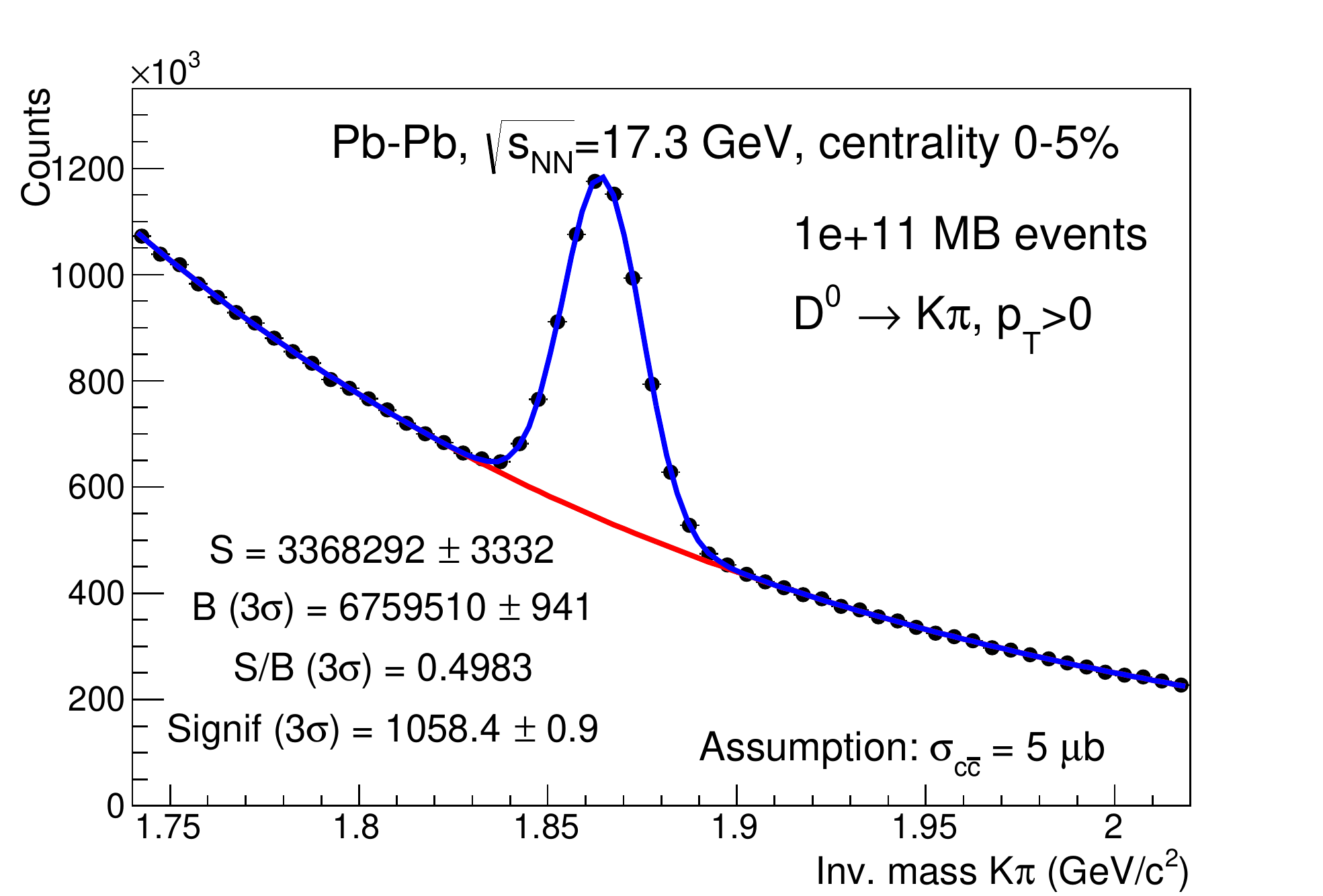}
\includegraphics[width=0.45\linewidth]{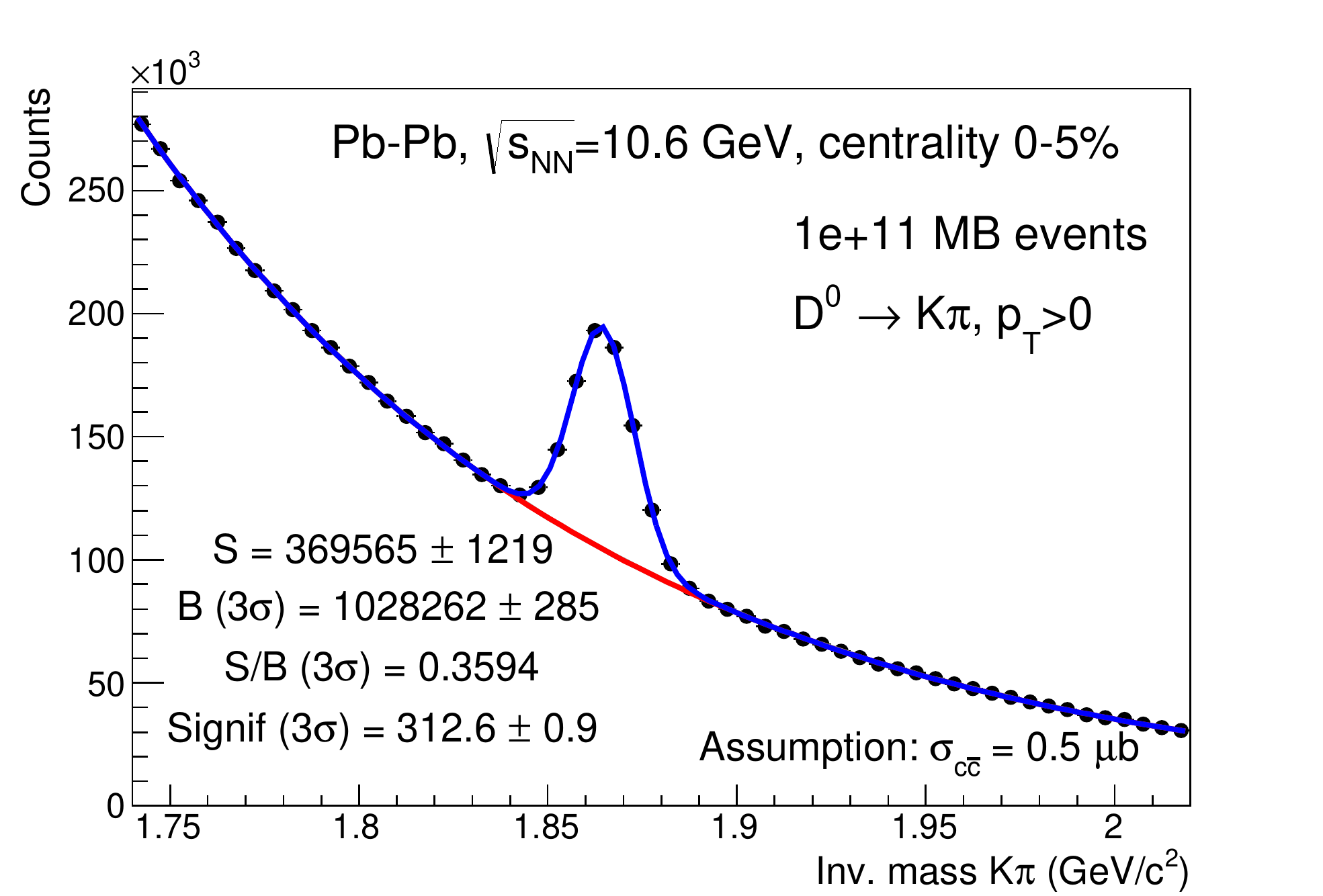}
\caption{Projection for the invariant-mass distribution of \Dzero candidates in \num{5e9} central \PbPb collisions at beam energies of 160 (left) and \SI{60}{\AGeV} (right) for the case of the VT detector based on MAPS with \SI{5}{\um} spatial resolution.}
\label{fig:D0invmass}
\end{center}
\end{figure}

The rapidity distribution of $\Dzero \rightarrow \mathrm{K}^- \pi^+$ decays passing the candidate selections for the set of cuts that provides the best significance in the case of the MAPS detector is shown in Fig.~\ref{fig:D0acceff}, together with the ratios to the rapidity distributions for generated and reconstructed decays. The fraction of candidates with reconstructed decay products, which pass the selections on the displaced decay vertex topology is about 25\% with a mild rapidity dependence for the considered selection criteria.
The left panel of Fig.~\ref{fig:D0invmass} shows a projection for the invariant-mass distribution of \Dzero candidates in \num{5e9} central \PbPb collisions at $\sqrtsNN = 17.3\GeV$, corresponding to a sample of \num{e11} minimum bias (MB) collisions, which can be collected in one month of data taking.
The MAPS detector would enable a measurement of the \Dzero-meson yield in central \PbPb collisions with a statistical precision much better than 1\%, which would allow also for studies in \pt and $y$ intervals and for the determination of the elliptic flow coefficient $v_2$ of D mesons with percent level statistical uncertainty.
In the right panel of Fig.~\ref{fig:D0invmass}, the projected performance for the 5\% most central \PbPb collisions at $\sqrtsNN = 10.6\GeV$ is shown for the case of MAPS detectors.
For this performance study we assumed $\sigma_{\ccbar}=0.5\mub$ and the combinatorial background was simulated based on the interpolation of NA49 measurements at \SIlist{40;80}{\AGeV} incident energy.
It demonstrates that the integrated \Dzero-meson production yield can be measured with a statistical precision better than 1\% at collision energies at which the charm cross section is poorly known experimentally.

\paragraph{Three-prong decays: D$^+$, D$_{\rm s}^+$, $\Lambda_{\rm c}^+$}

Performance studies were also carried out to study the feasibility of open-charm reconstruction via three-body decay channels of charmed hadrons, which would enable measurements of \Dplus and \Ds mesons and of \lambdacplus baryons.
A performance study was carried out for \Ds mesons, which have a shorter lifetime and are less abundantly produced as compared to \Dplus mesons, and thus they provide a more challenging benchmark for charm-meson reconstruction from three-body decays.
The simulations and the selections were performed utilizing the same approach described above for \Dzero mesons.
The main differences concern the variables used in the selection of the signal.
In particular, among the geometrical selections on the displaced decay vertex topology, the product of the impact parameters of the same-sign decay particles was used, which presents an asymmetric and predominantly negative distribution for the signal and a narrower and symmetric around zero distribution for the background.
Furthermore, a selection on the compatibility of the invariant mass of the K$^+$K$^-$ decay products with the mass of the $\phi$ meson was applied.

\begin{figure}[tb!]
\begin{center}
\includegraphics[width=0.9\linewidth]{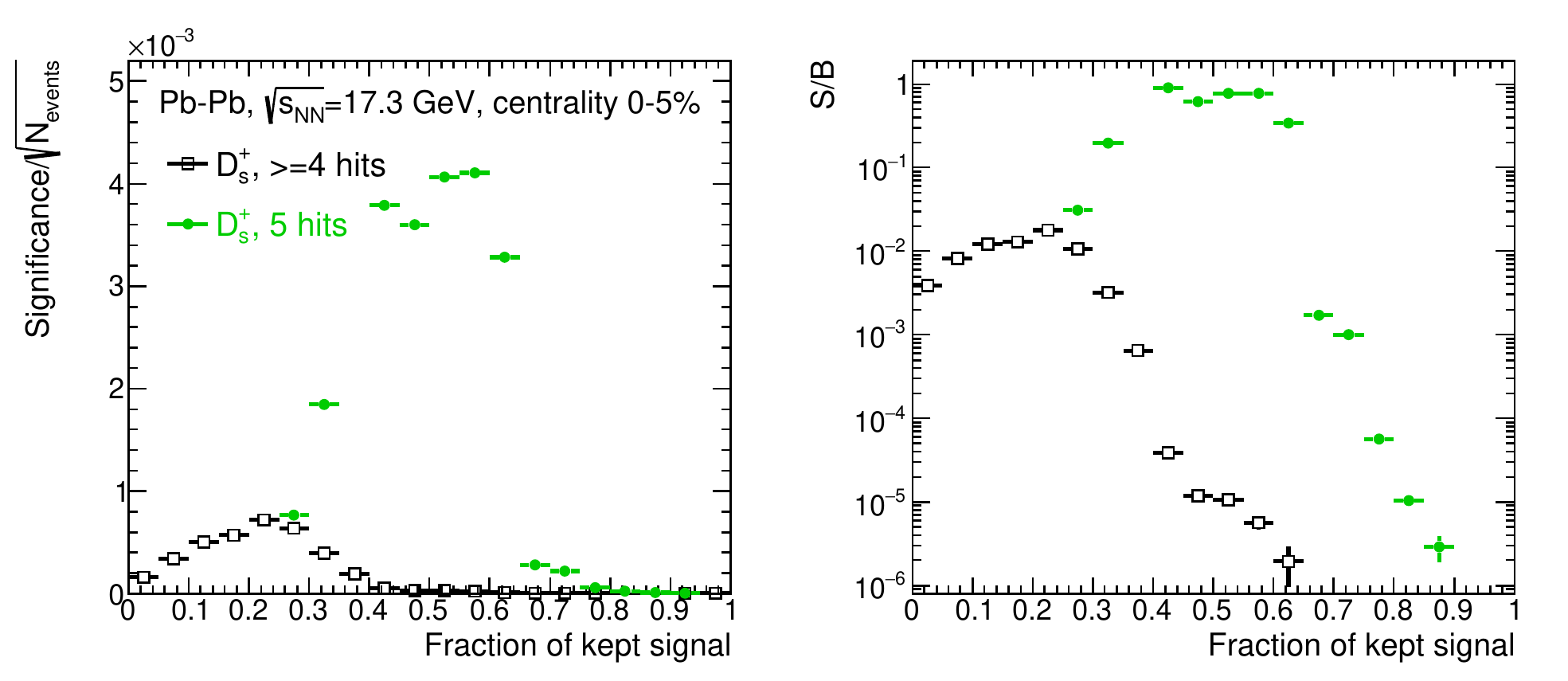}
\caption{\Ds-meson statistical significance (left, normalized to the square root of the number of events) and signal-over-background ratio (right) as a function of the selection efficiency for the 5\% most central \PbPb collisions at $\sqrtsNN = 17.3\GeV$ for two considered track selection criteria.
The signal and the background yields are evaluated in a $2\sigma$ invariant-mass region around the \Ds peak.}
\label{fig:Dssignif}
\end{center}
\end{figure}

\begin{figure}[tb!]
\begin{center}
\includegraphics[width=0.45\linewidth]{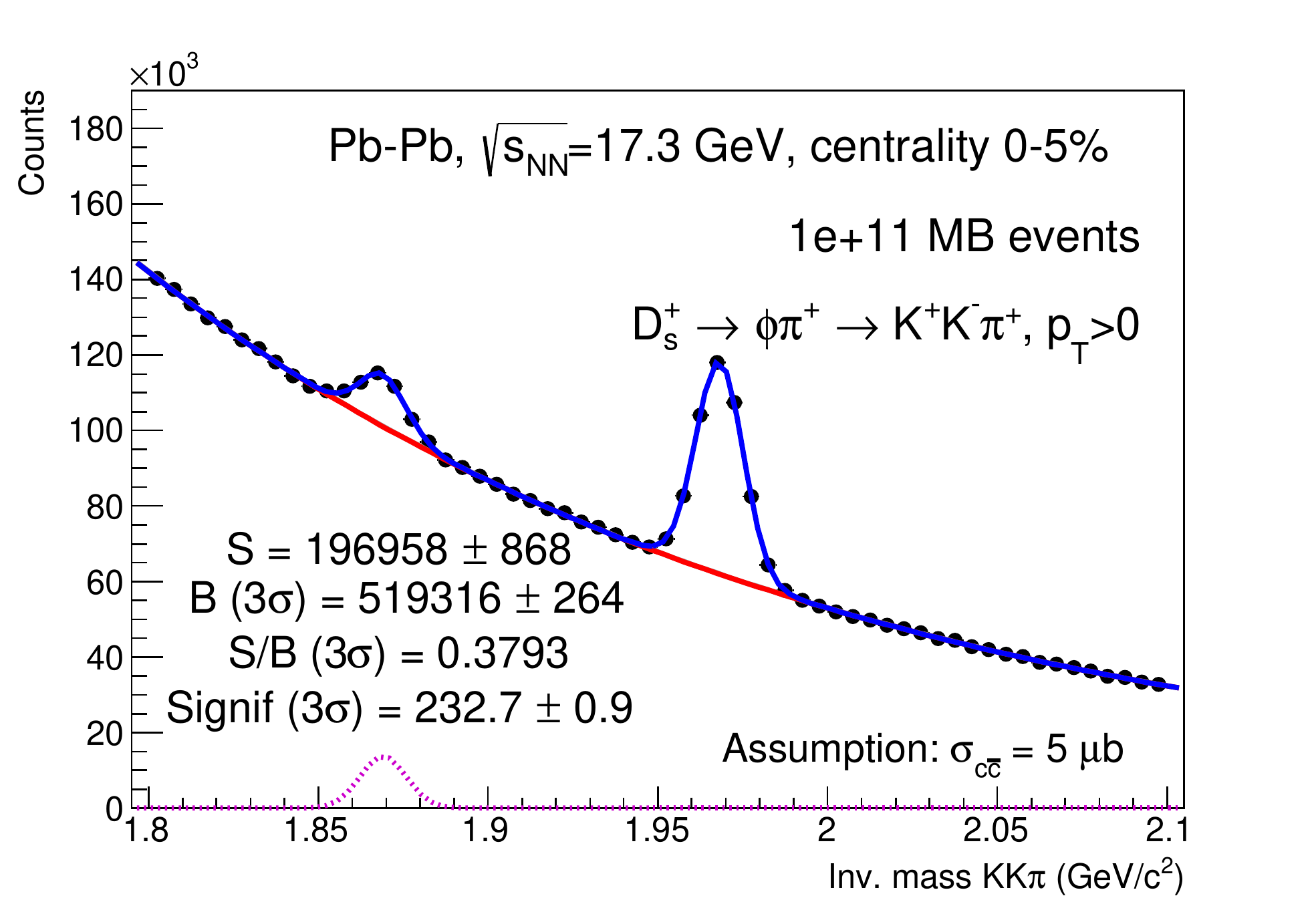}
\caption{Projection for the invariant-mass distribution of \Ds candidates in $5\times 10^9$ central \PbPb collisions at beam energies of 160 GeV/nucleon.}
\label{fig:Dsinvmass}
\end{center}
\end{figure}

As for the \Dzero meson studies, several sets of selection criteria were tested and for each of them the $S/B$ and the statistical significance were computed.
The results are reported in Fig.~\ref{fig:Dssignif}, where one can see that these selections provide an increase of the signal-to-background ratio in central \PbPb collisions from the initial expected value of $3 \times 10^{-10}$ to values larger than $10^{-1}$.
This corresponds to a significance of about 200 for $5 \times 10^9$ central \PbPb collisions at top SPS energy, which indicates that the \Ds yield can be measured with a statistical precision at the percent level.
An example of invariant-mass distribution of \Ds candidates in the 0--5\% centrality class for \PbPb collisions at beam energies of \SI{160}{\AGeV} is shown in Fig.~\ref{fig:Dsinvmass}. It can be noticed that also the signal of \Dplus $\rightarrow \phi \pi^+ \rightarrow {\rm K}^- {\rm K}^+ \pi^+$ decays is visible at the \Dplus mass of 1.870\GeVcc~\cite{PhysRevD.98.030001}.
This could also provide an alternative way of measuring the \Dplus meson production, even though the performance would be significantly worse than the one expected for the \Dplus reconstruction in the K$^-\pi^+\pi^+$ decay channel, which features a larger branching ratio.

\begin{figure}[tb!]
\begin{center}
\includegraphics[width=0.45\linewidth]{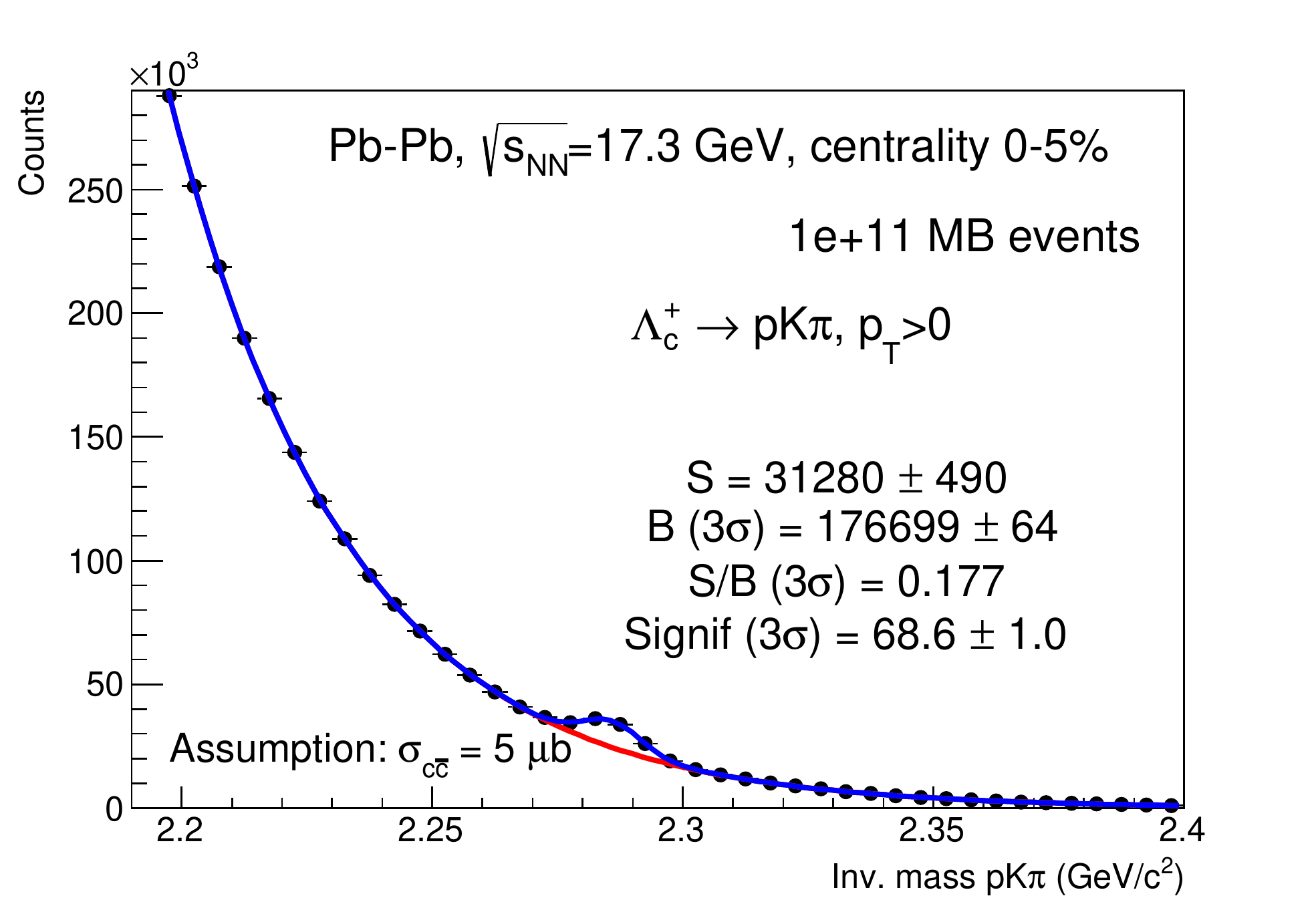}
\includegraphics[width=0.45\linewidth]{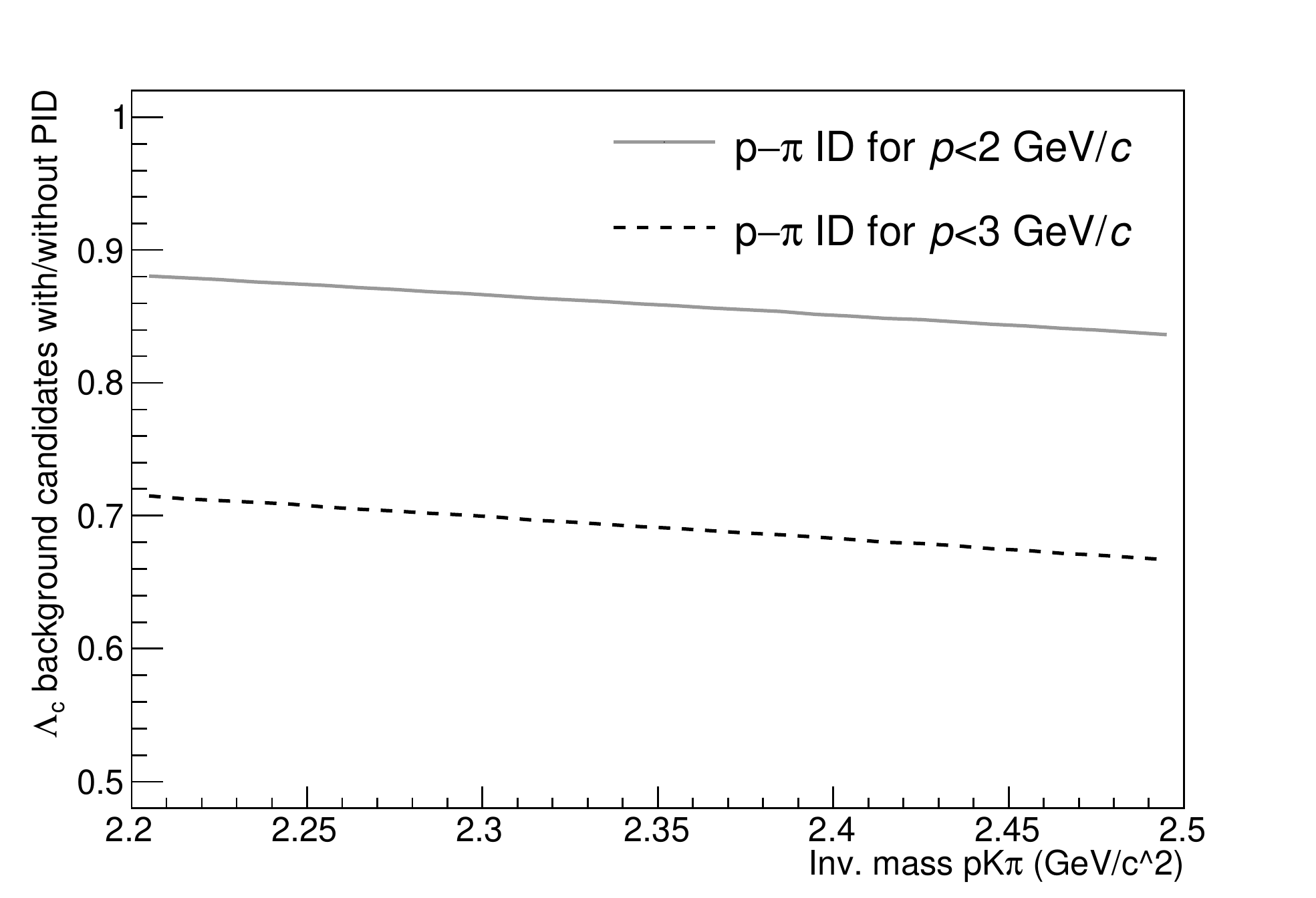}
\caption{Left: projection for invariant-mass distribution of \lambdac\ candidates in \num{5e9} central \PbPb collisions at beam energies of \SI{160}{\AGeV}. Right: ratio of the number of \lambdac\ background candidates with and without applying PID selections to separate protons from pions.}
\label{fig:Lcinvmass}
\end{center}
\end{figure}

Performance studies for charm baryon measurements were carried out for the reconstruction of the \mbox{$\lambdacplus \rightarrow \mathrm{pK^{-}}\pi^+$} decay (and its charge conjugate), which has a branching ratio of 6.28\%.
Given the shorter lifetime of \lambdacplus\ ($c\tau = 60~\mu$m) with respect to \Dzero, \Dplus, and \Ds\ mesons, the selection of the signal candidates poses more challenges.
Therefore, variables related to the decay kinematics were considered in the selection criteria in addition to those related to the displaced vertex topology. In particular, since in the \lambdac\ decay the proton is typically the particle emitted with the largest momentum, selections on the ratios of the momenta of the daughter particles ($p_{\pi}/p_{\rm p}$ and $p_{\rm K}/p_{\rm p}$) were applied. For the geometrical selections, the same variables considered for the \Ds (track impact parameter, distance between primary and decay vertices, cosine of the pointing angle, product of the impact parameters of the same-sign decay particles) were utilised.
An example of the invariant mass distribution of \lambdac\ candidates in the 0--5\% centrality class for \PbPb collisions at beam energies of \SI{160}{\AGeV} is shown in the left panel of Fig.~\ref{fig:Lcinvmass} for a set of selection criteria that provides good statistical significance of the signal and a selection efficiency of about 7.5\%. This result demonstrates that with the NA60+ setup it will be possible to reconstruct about 30,000 \lambdac\ baryons in central \PbPb collisions, reaching a precision of few percent in terms of statistical uncertainties.
Measurements of \lambdac\ production can also be carried out from two other decay channels with a long-lived neutral strange hadron in the final state, namely \mbox{$\lambdacplus \rightarrow \mathrm{p K^0_S}$} and \mbox{$\lambdacplus \rightarrow \Lambda \pi^+$} (and their charge conjugates). These decay modes have a smaller branching ratio (${\rm BR}=1.59$\% and 1.30\%, respectively) compared to the $\mathrm{pK^-}\pi^+$ channel and a worse precision on the determination of the \lambdac\ decay vertex, but they could benefit from the high purity that can be achieved for the reconstruction of $\mathrm{K^0_S}$ mesons and $\Lambda$ baryons from their V0 decay topologies, as discussed in the next Section.
This will enable a reconstruction of \lambdac\ independent from the one from the $\mathrm{pK^-}\pi^+$ decay, and will thus contribute to reduce the statistical uncertainties and to assess and control the systematic ones. Furthermore, the use of Boosted Decision Trees for the optimization of the significance of the signal is also planned.
Finally, an improved performance can be obtained utilising timing layers in the vertex telescope to provide particle identification capabilities. For instance, as it can be seen in the right panel of Fig.~\ref{fig:Lcinvmass}, the separation of proton and pion tracks up to momenta of 3(2)~GeV/$c$ allows us to reduce by about 30\%(15\%) the number of combinatorial background candidates in the \lambdac\ invariant mass region. This separating power could be obtained with two (one) layers, positioned at  40 cm from the target and providing a time resolution of 20 (35) ps. We are following the progress of a dedicated R\&D program aiming at reaching this performance that was started in the frame of the ALICE 3 project~\cite{ALICE:2022wwr}.

\subsubsection{Strangeness
}

\begin{figure}[h]
\begin{center}
\includegraphics[width=0.48\linewidth]{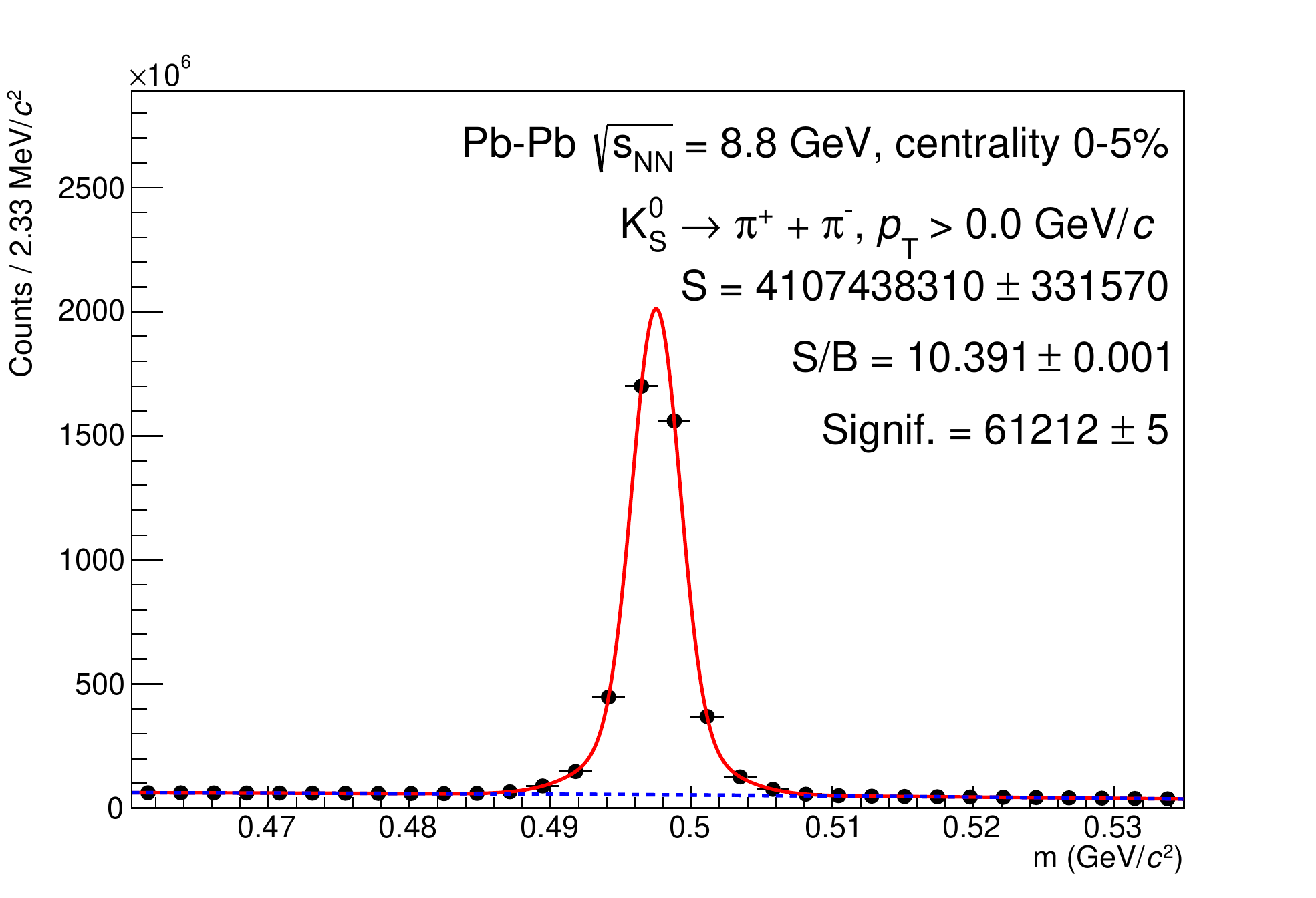}
\includegraphics[width=0.48\linewidth]{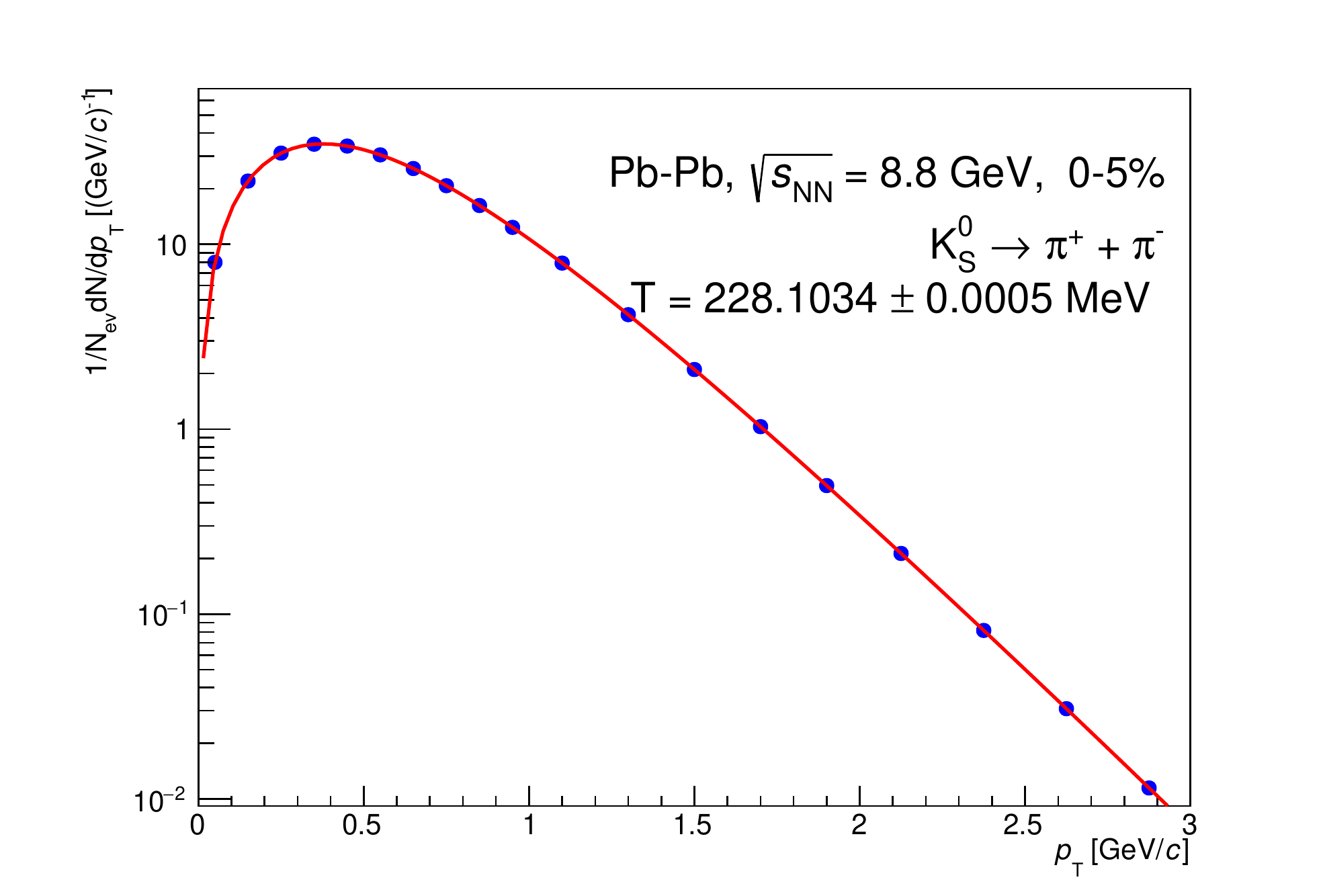}
\caption{Projection for the invariant-mass distribution of $K^0_S$ candidates and $p_T$ spectrum in $10^{10}$ central Pb-Pb collisions at beam energies of 40 GeV/nucleon.}
\label{fig:k0s}
\end{center}
\end{figure}

\begin{figure}[h]
\begin{center}
\includegraphics[width=0.48\linewidth]{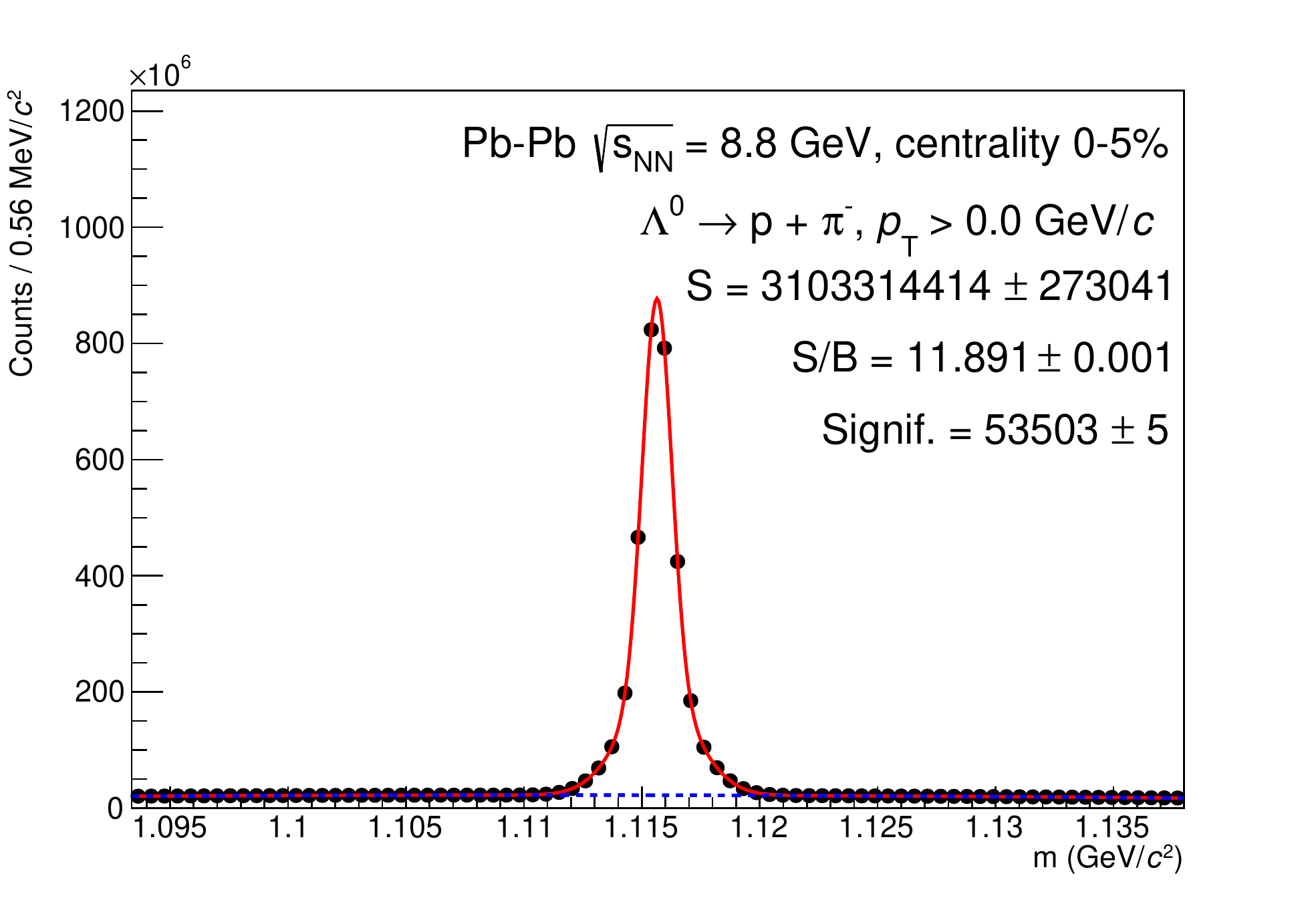}
\includegraphics[width=0.48\linewidth]{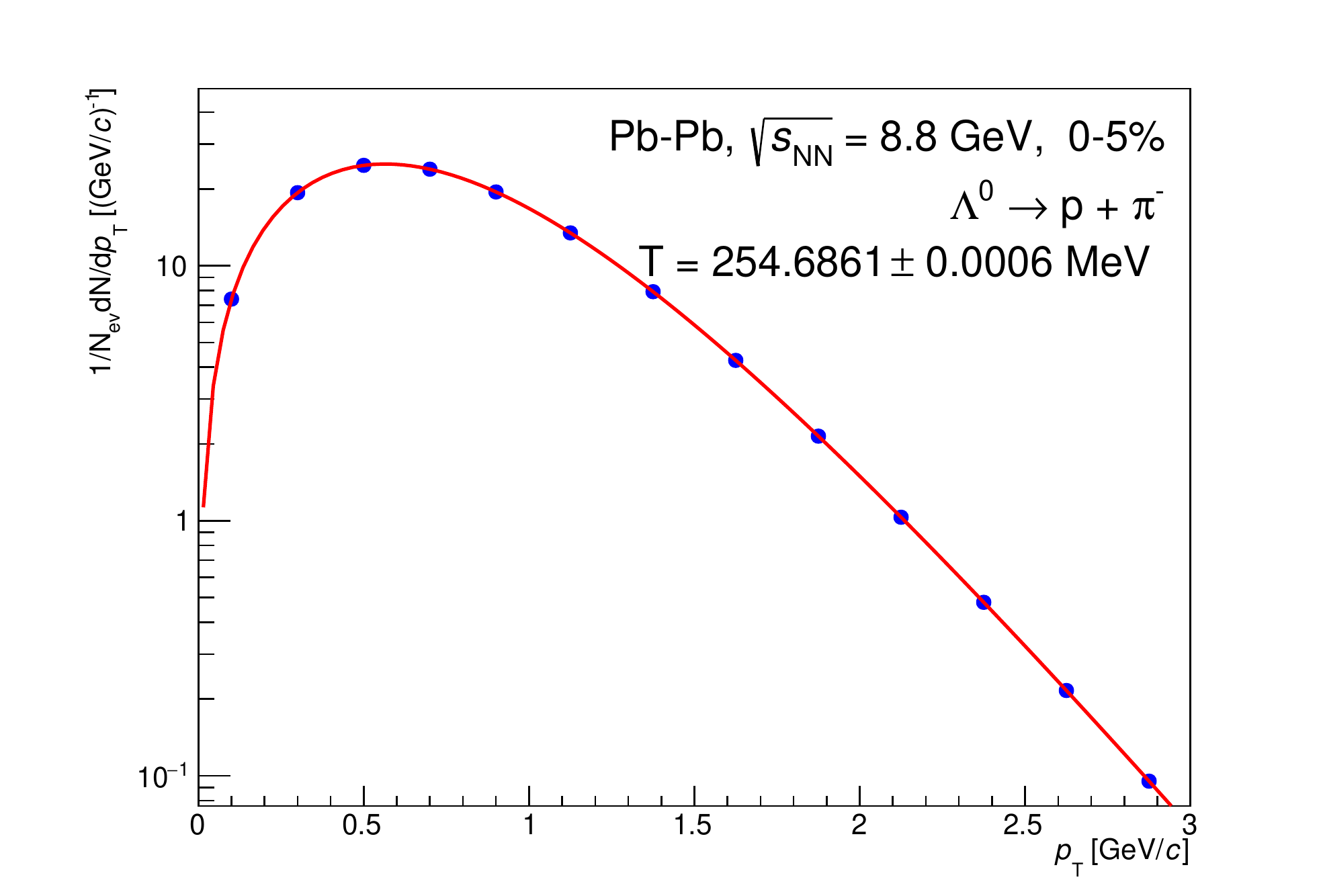}
\caption{Projection for the invariant-mass distribution of $\Lambda^0$ candidates and $p_T$ spectrum in $10^{10}$ central Pb-Pb collisions at beam energies of 40 GeV/nucleon.}
\label{fig:lambda}
\end{center}
\end{figure}

\begin{figure}[h!]
\begin{center}
\includegraphics[width=0.48\linewidth]{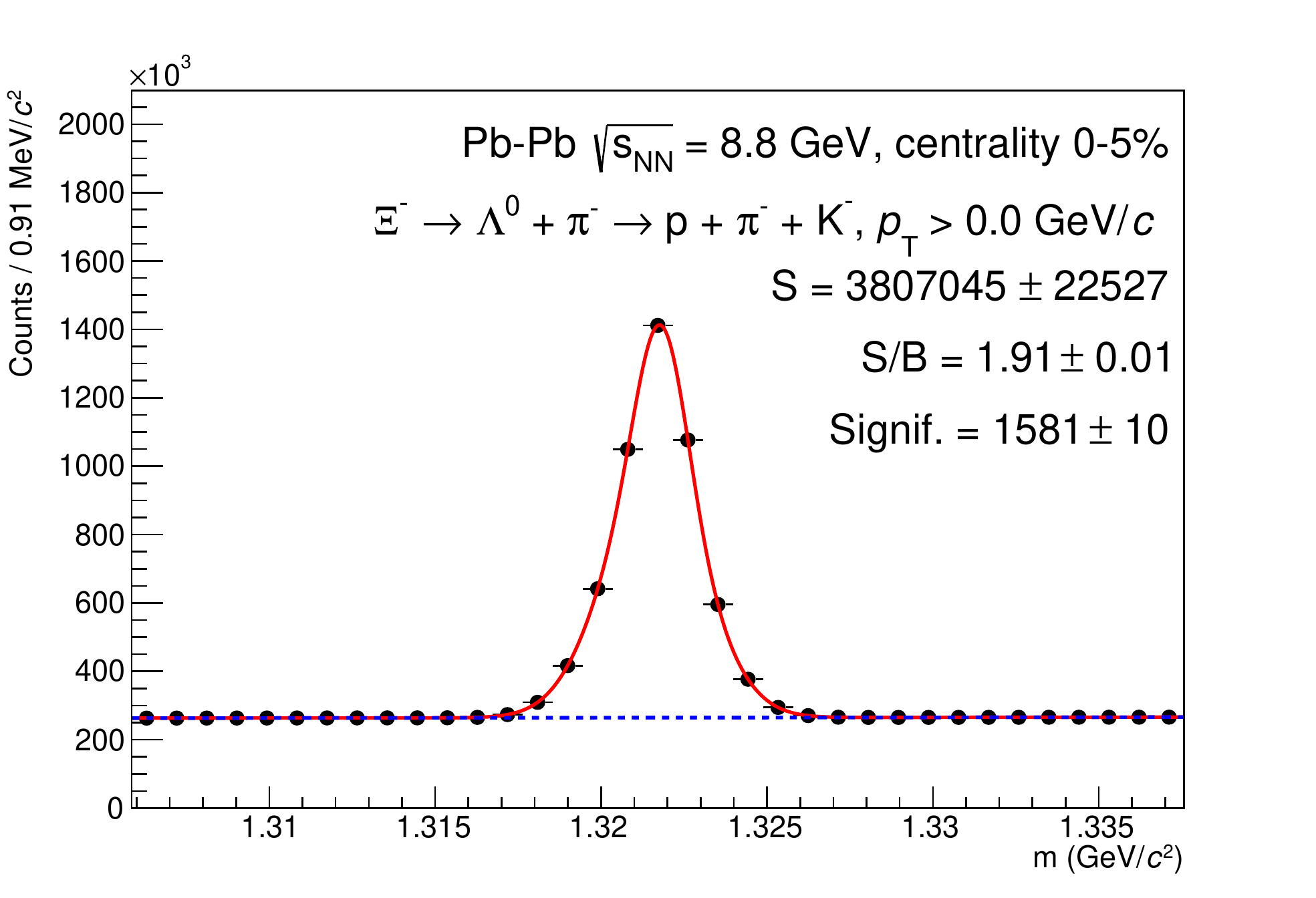}
\includegraphics[width=0.48\linewidth]{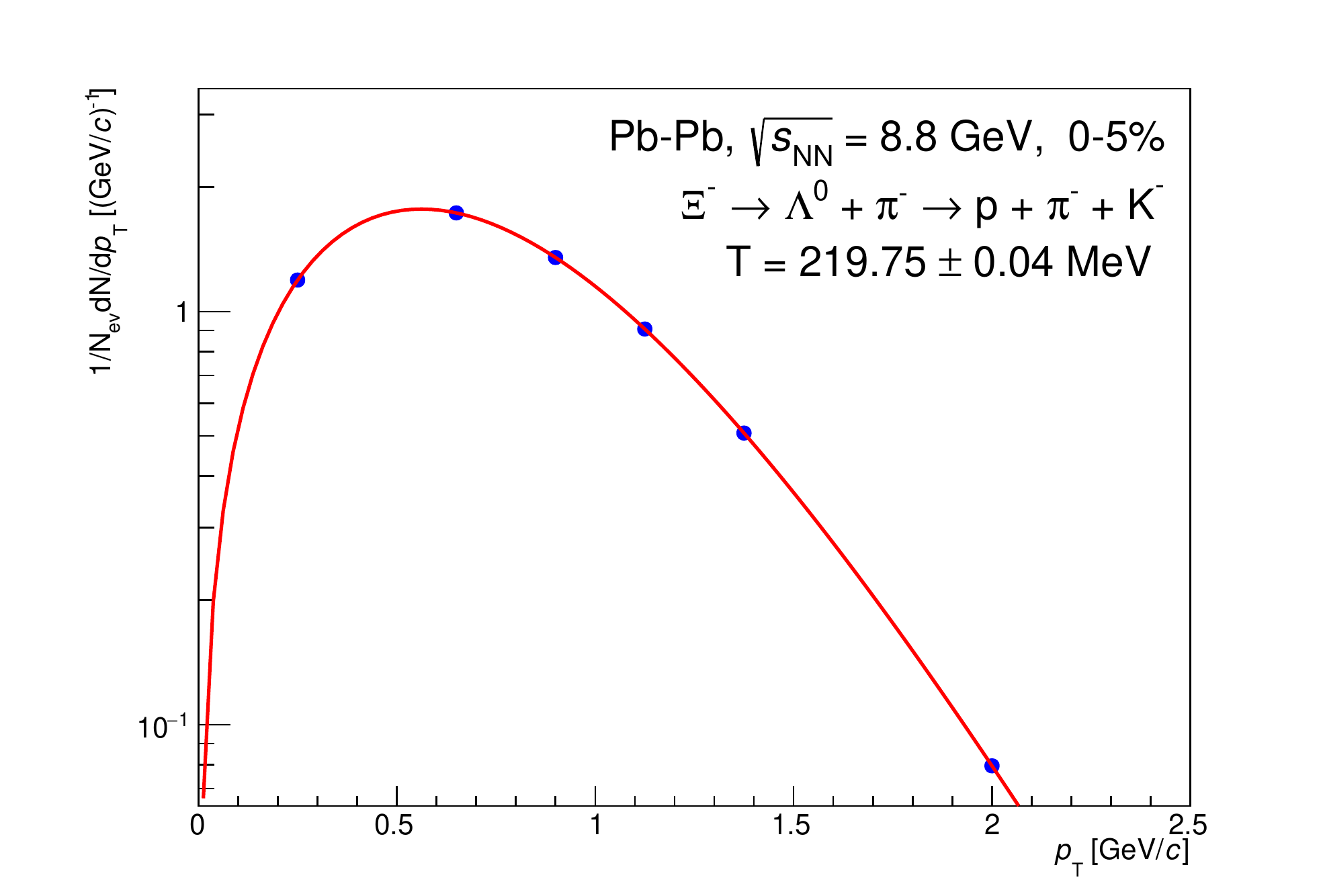}
\caption{Projection for the invariant-mass distribution of $\Xi^-$ candidates and $p_T$ spectrum in $10^{10}$ central Pb-Pb collisions at beam energies of 40 GeV/nucleon.}
\label{fig:ximinus}
\end{center}
\end{figure}

\begin{figure}[h!]
\begin{center}
\includegraphics[width=0.48\linewidth]{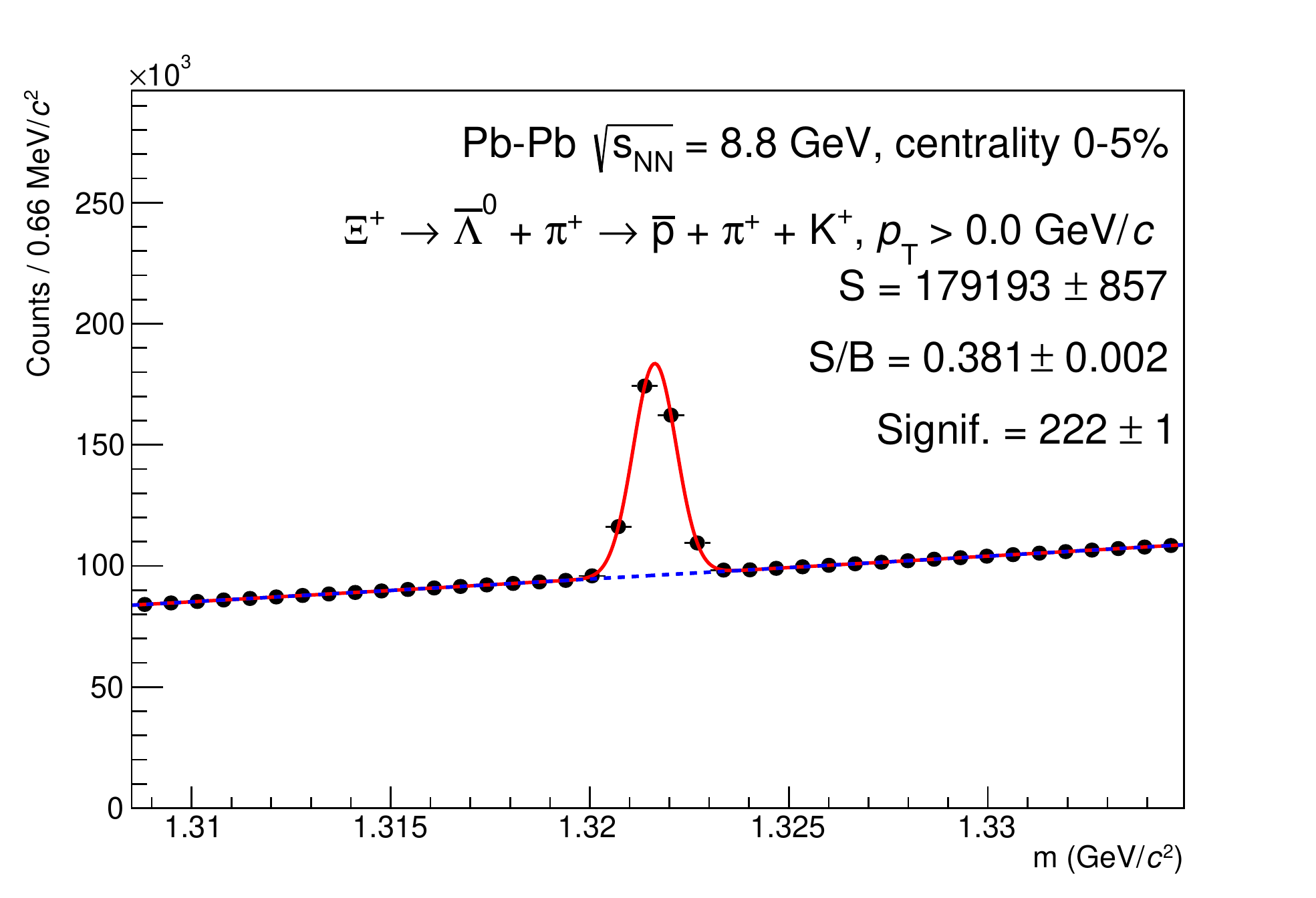}
\includegraphics[width=0.48\linewidth]{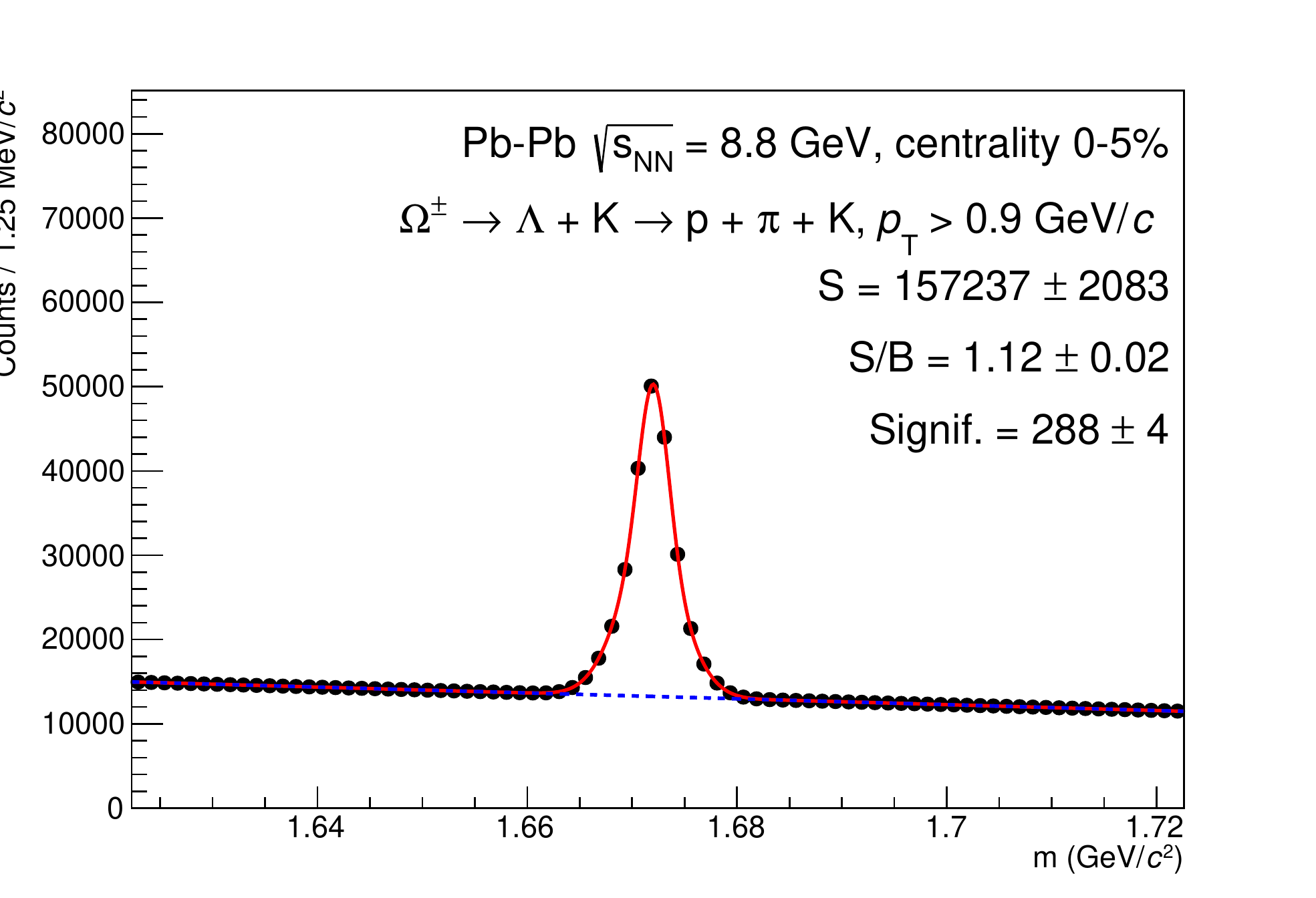}
\caption{Projection for the invariant-mass distribution of $\Xi^+$ (left) and $\Omega^-+\Omega^+$ (right) candidates in $10^{10}$ central Pb-Pb collisions at beam energies of 40 GeV/nucleon.}
\label{fig:antxiomega}
\end{center}
\end{figure}

The hadronic decays of strange hadrons can be studied with the VT of the NA60+ apparatus. The decay channel that were studied are $\Lambda^0 \rightarrow p + \pi^-,\bar{\Lambda}^0 \rightarrow \bar{p} + \pi^+, \Xi^- \rightarrow \Lambda^0 + \pi^-, \Xi^+ \rightarrow \bar{\Lambda}^{0} + \pi^+, \Omega \rightarrow \Lambda + K, K^0_S\rightarrow\pi^+\pi^- \text{and}\ \phi \rightarrow K^+K^-$, with the $\Lambda$ coming from the decay of hyperons then decaying into  $p+\pi$. The decays were simulated with the package EvtGen~\cite{Lange:2001uf} and the protons, kaons and pions were propagated through the VT utilizing the fast simulation of NA60+. The Pb-Pb collisions were simulated with a cocktail of prompt protons, kaons, pions superimposed to those coming from the hadronic decays of $\Lambda, \Xi, \Omega, K^0_S$ and $\phi$ under study. The yield, $p_T$ and rapidity distributions of the prompt $p, K$ and $\pi$ and from the decay of strange particles were generated according to the measurements performed by NA49~\cite{Afanasiev:2002mx,Alt:2006dk,NA49:2008goy,NA49:2008ysv,Mitrovski:2003ng}. The yields and kinematics of the $K^0_S$ were taken as the average values of those relative to the $K^+$ and $K^-$. The signal candidates were built combining pairs or triplets of tracks with the proper charge signs. The analysis was performed on $10^{7}$ simulated Pb-Pb collisions in the 0-5$\%$ centrality class at $\sqrt{s_{NN}} = 8.8$ GeV. The results obtained were scaled to represent the expected measurements after one month of data taking. A fit to the $p_T$ distribution was performed, using the same parametrization as for the NA49 results: $dN/dp_T \propto p_T e^{-({\sqrt{m^2+p_T^2}}/{T})}$.
\paragraph{Long-lived strange particles}
\label{subsubsection:longlivstrangeness}
$\Lambda$, $\Xi$, $\Omega$ and $K^0_S$ have a very long lifetime ($c\tau \sim \text{few}\ cm$) that allows topological selections to be applied to reduce the combinatorial background. The decay vertex position, and the kinematics of the candidate were computed by propagating the decay tracks to their point of closest approach and using the elements of the track covariance matrix as weights. For $\Omega$ and $\Xi$ this procedure was first done for the candidate decay $\Lambda$, which was then used to build the candidate $\Omega$ and $\Xi$. Boosted Decision Trees (BDT)~\cite{kotsiantis2013decision} were employed to enhance the significance of the signals. The BDT used in this analysis are the ones implemented in the python package XGBoost~\cite{xgboost}. The following variables were fed to the machine learning algorithm for the candidate selection: the rapidity of the candidate, the  product of the impact parameter of decay tracks, the distance of closest approach between the decay tracks, the decay length and the cosine of the pointing angle. In the case of the $\Omega$ and $\Xi$ the same topological variables were computed for the candidate daughter $\Lambda$ and fed to the machine learning algorithm. Selections on the Armenteros $\alpha$~\cite{podolanski1954iii} were applied to discriminate $\Lambda$ from $\bar{\Lambda}$. Further selections on invariant-mass of the $\Lambda$ candidates from $\Omega$ or $\Xi$ were applied. 
Out of the $10^7$ collisions that were generated, $10^6$ were used for the training of the BDT, while the remaining $9\cdot10^6$ were used for the analysis. The BDT applied to a candidate return a value called score which is linked to the probability of the candidate to be a signal. A selection was applied to the score that maximize the expected $S/\sqrt{S+B}$. Depending on the number of signal candidates, the analysis was carried out as a function of  $p_{T}$, training different BDT models for different $p_T$ intervals.
The projections for the measurement of $K^0_S \rightarrow \pi^+\pi^-$, $\Lambda^0\rightarrow p+\pi^-$, $\Xi^-\rightarrow \Lambda^0+\pi^-$ and $\Omega^-\rightarrow \Lambda^0+K^-$ (the latter including its charge conjugate) are shown in  Fig.~\ref{fig:k0s},~\ref{fig:lambda},
~\ref{fig:ximinus},~\ref{fig:antxiomega}.

\begin{figure}[h!]
\begin{center}
\includegraphics[width=0.48\linewidth]{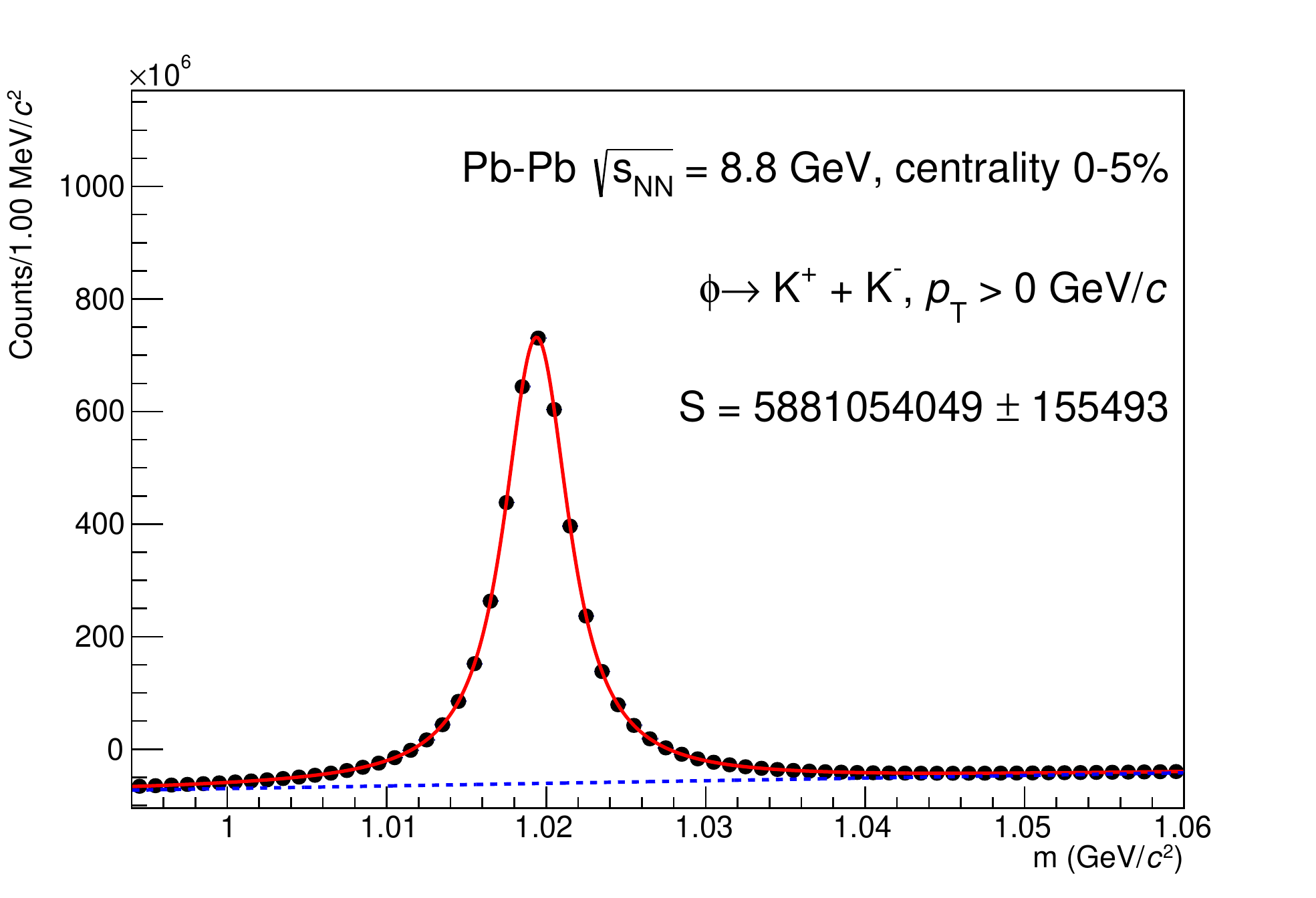}
\includegraphics[width=0.48\linewidth]{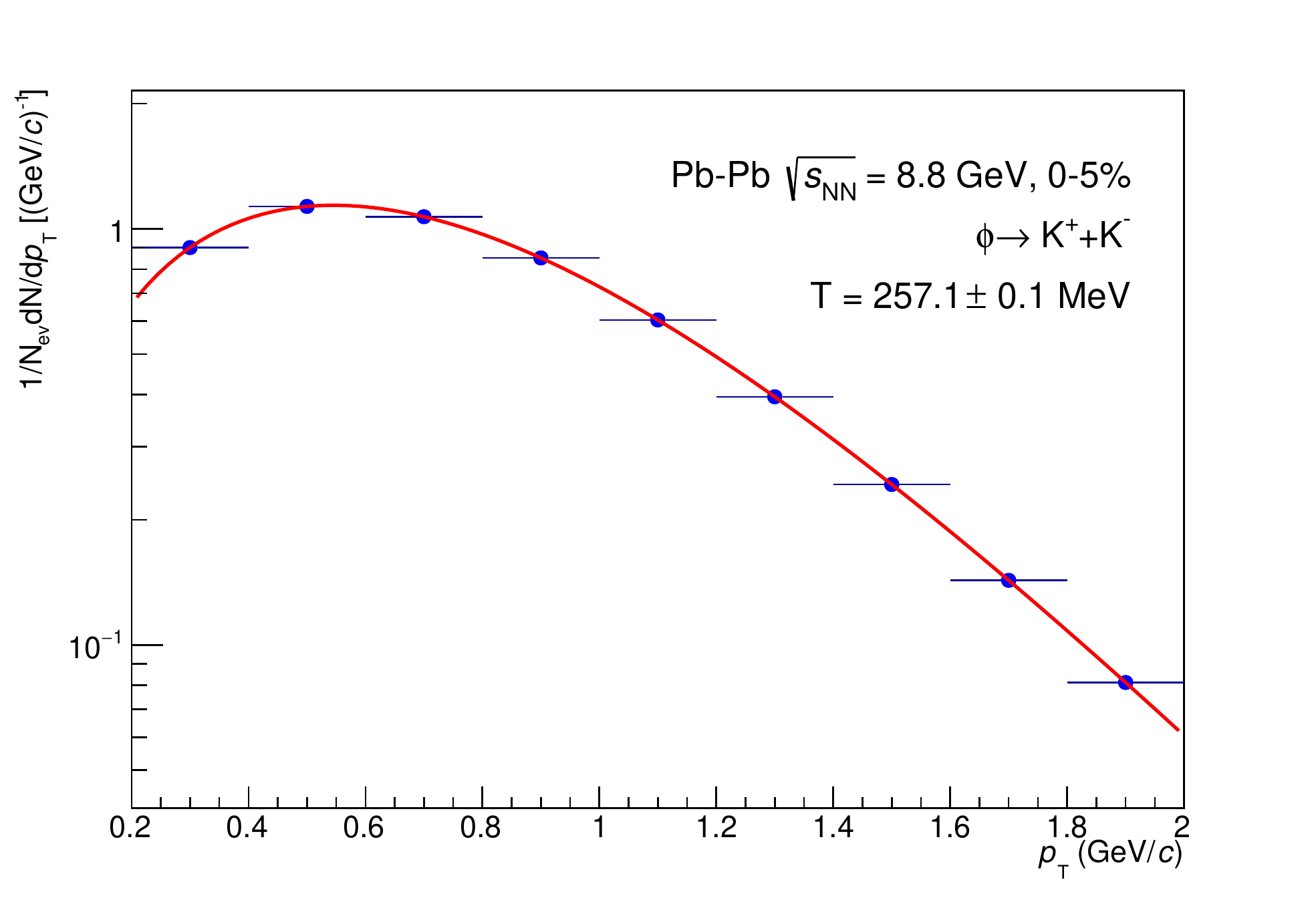}
\caption{Projection for the invariant-mass distribution of $\phi$ candidates in $10^{10}$ central Pb-Pb collisions at beam energies of 40 GeV/nucleon.}
\label{fig:phi}
\end{center}
\end{figure}

\paragraph{\texorpdfstring{$\bm{\phi\rightarrow}{\bf KK}$}{Strangemesons}}
Differently from the strange hadrons studied above, the $\phi$ meson has a very short lifetime $\tau = (1.55\pm0.01)\times 10^{-22}$s~\cite{PhysRevD.98.030001}. This does not allow to extract the signal by applying topological selections, therefore the background has been subtracted using the event mixing technique. The background has been reproduced building the candidate $\phi$ by pairing the tracks of an event with the tracks of the next four events. The event-mixed spectrum was normalized to the background counts outside the peak region ($0.98 < m < 0.99$ GeV$/c^2$ and $1.04 < m < 1.06$ GeV$/c^2$). The process is repeated for each $p_{\rm T}$ interval. The results are shown in Fig.~\ref{fig:phi}. The invariant mass resolution related to detector effects ranges between 1 and 2.5 MeV, a value smaller than the natural width of the resonance $\Gamma_{\phi} =$  4.26 MeV~\cite{PhysRevD.98.030001}. The high resolution and low statistical uncertainty could allow to observe modification of the mass that may be induced by the medium~\cite{lissauer1991k,Klingl:1997tm}. Moreover, it will be possible to extract the $\phi$ signal down to low $p_{\rm T}$, and perform a  comparison of the yield and spectra in the decay channels to kaons and muons. This will allow to finally solve the so called $\phi$-puzzle, a discrepancy in the inverse $T$ slopes and yields measured by NA49 and NA50 in the kaon~\cite{Friese:2002re,NA49:2004jzr} and muon~\cite{NA50:2003owh,Jouan:2008zz} channel, respectively.

\subsubsection{Hypernuclei}
\begin{figure}
    \centering
\includegraphics[width=0.8\textwidth]{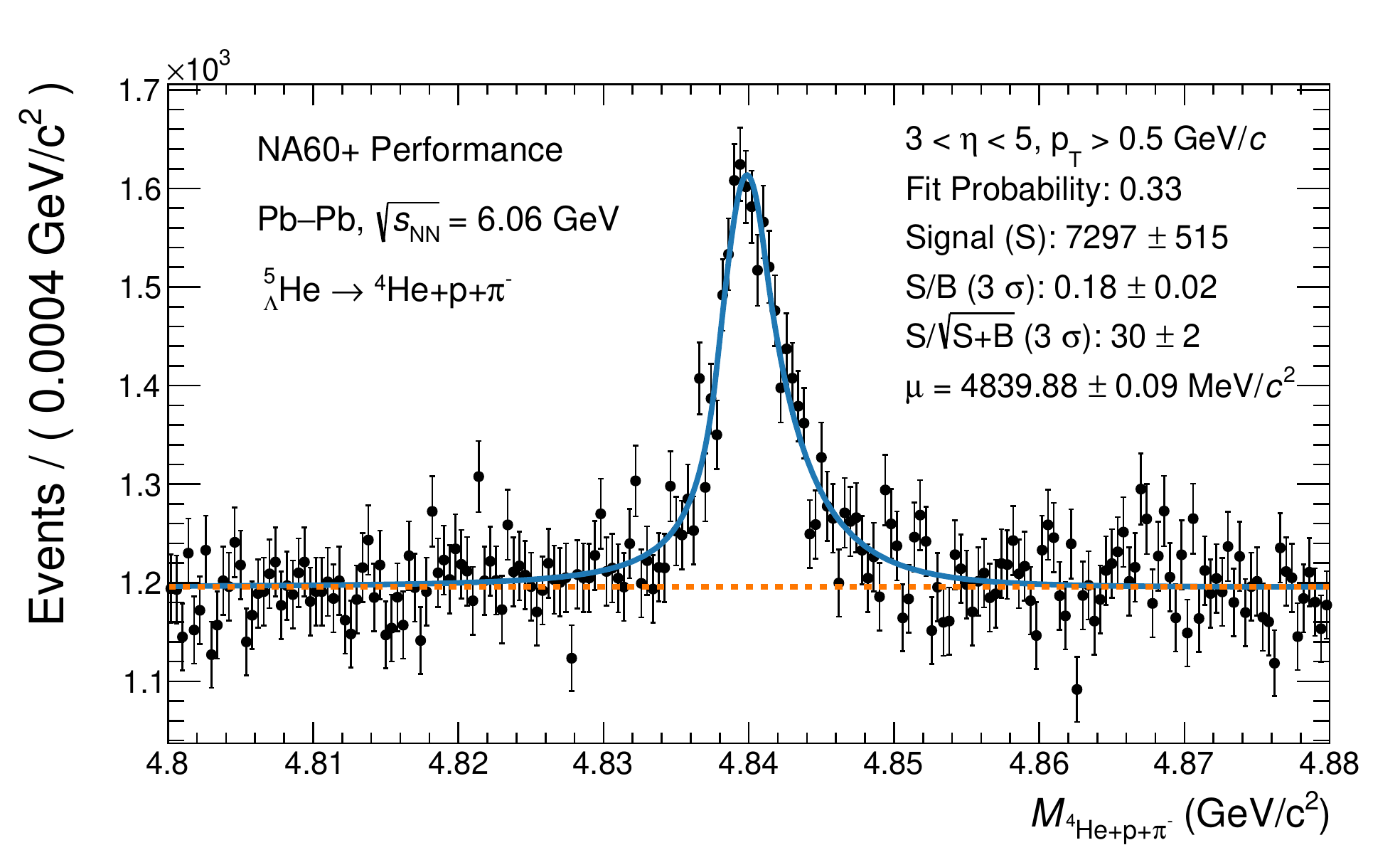}
    \caption{Projected invariant mass spectrum for $^{5}_{\Lambda}\mathrm{He}$ candidates in $10^{10}$ central Pb--Pb collisions at beam energies of 20 GeV/nucleon.}
    \label{fig:he5L}
\end{figure}

Hypernuclei represent some of the most elusive states that could be measured in heavy-ion collisions. The higher the hypernucleus mass, the rarer its production is. At the SPS energies, the hypernuclei production rate is less suppressed than at the top RHIC and LHC energies. However, even at the SPS energies, a sizeable integrated luminosity is required to detect hypernuclei with mass numbers larger than four.
In addition, a clean identification of nuclear fragments is instrumental to detecting the weak decays of hypernuclei.
The proposed VT of NA60+, with five tracking stations using MAPS detectors, allows the separation of heavily ionising particles from ordinary hadrons by looking at the size of the clusters associated with the tracks \cite{Abelevetal:2014dna}.
Assuming a perfect identification of the nuclear fragments and using the fast simulation of the VT, it is possible to establish the projected performance for identifying hypernuclei.
The $^{5}_\Lambda\mathrm{He}$ decay in its charged three-body decay ($^{5}_\Lambda\mathrm{He}\rightarrow^{4}\mathrm{He}+p+\pi^-$, B.R. of approximately 32\% \cite{FINUDA:2009aum}) is here proposed as a case study.
The expected yield and the momentum distribution for $^{5}_\Lambda\mathrm {He}$ are taken from the Thermal-FIST package event generator \cite{Vovchenko:2019pjl, Vovchenko:2021kxx}. The generator produces particles according to a Hadron Resonance Gas model abundances and with a momentum distribution dictated by the MUSIC hydrodynamical model.
The main background for the displaced ($^{5}_\Lambda\mathrm {He}$ $c\tau$ is around 7 cm) decays of the hypernuclei comes from the combinatorics of primary nuclei with hadrons from secondary vertices. The simulation of one million central Pb-Pb collisions using the Thermal-FIST package gave a representative background sample that could be used to devise a simple set of topological selections to reject it.
Figure \ref{fig:he5L} shows the projected performance for identifying $^{5}_\Lambda\mathrm {He}$ in 10$^{10}$ central Pb-Pb events at the lowest collision energy of SPS. With such a performance, a precision study of the production spectra, the measurement of the lifetime, and the measurement of the binding energy of hypernuclei with A=5 are within reach. As shown in Fig.~\ref{fig:he5L}, the expected precision on the mass of the $^{5}_\Lambda\mathrm {He}$ is of the order of 90 keV using a single heavy-ion data taking period.
This precision could be further refined by reconstructing the complete decay topology, including the charged hypernucleus track. This is especially important considering the recent discussions comparing heavy-ion and emulsion experiments measuring hypernuclei properties \cite{Gazda:2022jha} and the fact that there are currently no measurements of the properties of $^{5}_\Lambda\mathrm {He}$.

Similar analysis strategies can be adopted to study lighter and heavier hypernuclei and to look for evidence of the existence of light $\Xi$-nucleon bound states \cite{Hiyama:2019kpw, Le:2021gxa}.
The crucial aspect of all these studies will be the possibility of distinguishing Z=2 nuclear fragments from ordinary hadrons crossing the VT. This capability will need to be assessed during the testing of the detector.

\subsection{Dimuon measurements}
\label{dimuonmeasurements}
\vskip 0.2cm

\subsubsection{Reconstruction efficiencies and mass resolution}
\label{receffmassres}
\begin{figure}[hb]
\begin{center}
\includegraphics[width=1.0\linewidth]{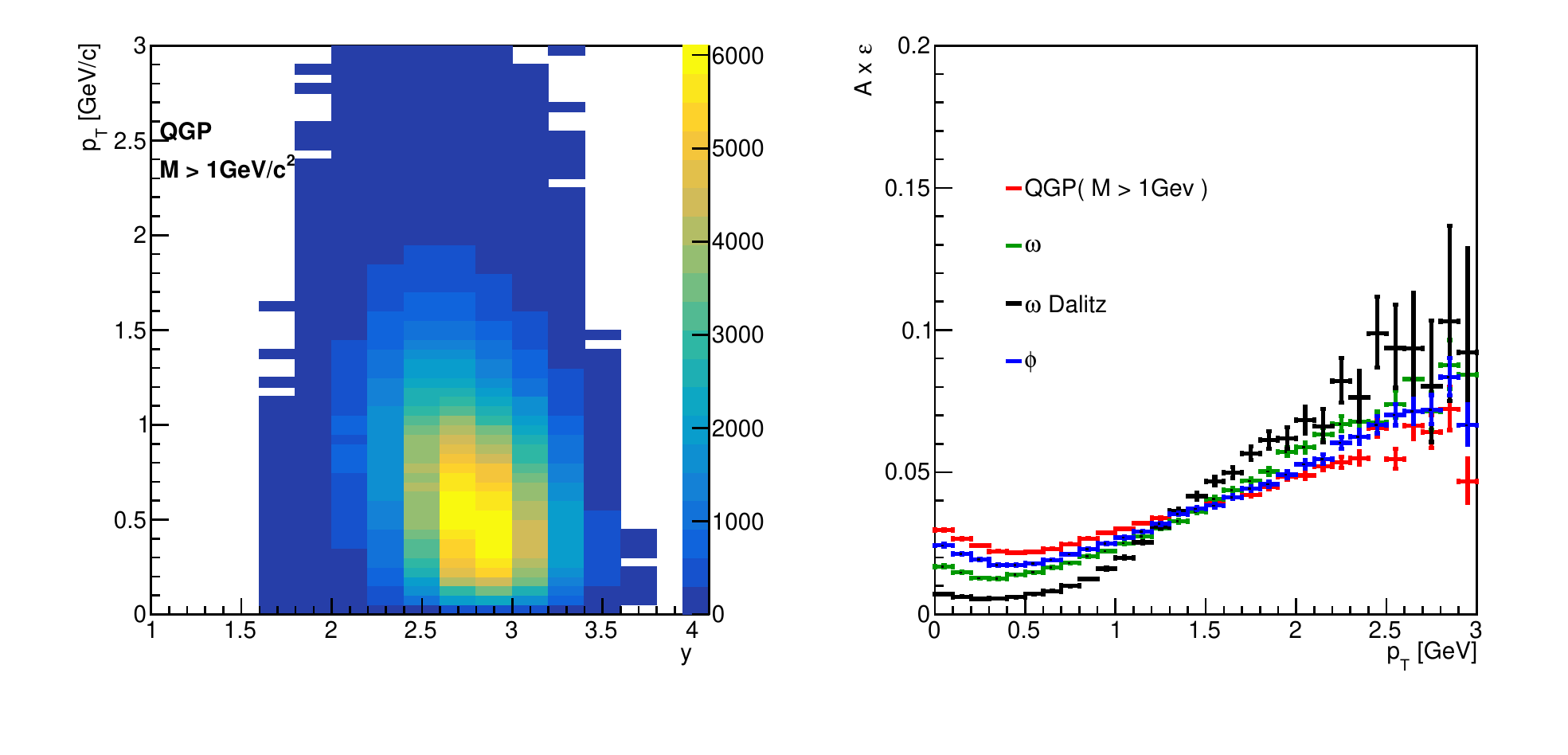}
\caption{(Left) Transverse momentum vs rapidity coverage for reconstructed dimuons with $M>1\GeVcc$ produced from the QGP phase in \PbPb central collisions at $\sqrtsNN=8.8$\GeV.
(Right) Acceptance times reconstruction efficiency as a function of \pt for processes in different mass ranges integrated over rapidity.}
\label{fig:DimuonRec}
\end{center}
\end{figure}

The transverse momentum--rapidity coverage is shown in the left panel of Fig.~\ref{fig:DimuonRec} for reconstructed dimuons with $M > 1\GeVcc$ produced from the QGP phase in central \PbPb collisions at $\sqrtsNN=8.8$\GeV.
In this mass region, the apparatus has a good coverage down to mid-rapidity ($y = 2.2$ in the lab system) and zero transverse momentum.
The right panel of Fig.~\ref{fig:DimuonRec} shows the pair acceptance times reconstruction efficiency as a function of \pt for processes in different mass ranges integrated over rapidity: this varies from 1\% at low masses and very small \pt to ${\sim}3-5\%$ for $M > 1\GeVcc$. 
The reconstruction efficiency is a complex interplay of several factors: low momentum muons stopped by the absorbers,  geometry of the toroidal magnet with a central dead region, matching efficiency of muon spectrometer to vertex tracks.

The dilepton spectrum is affected by a combinatorial background arising from muons produced by decays of primary or secondary hadrons.
Additionally, punch-through of primary or secondary hadrons produced in the absorber may occur.
In order to study this background, the \fluka package~\cite{Bohlen:2014buj,Ferrari:2005zk} has been used to simulate in detail the full hadronic shower development in the absorber.

Furthermore, it is possible in the signal reconstruction that hadronic hits in the silicon-pixel planes are associated to a muon track.
This potential contamination (fake matches) was taken into account at reconstruction level by including hadronic hits in the silicon stations according to the pion, kaon and proton multiplicities measured by the NA49 experiment at different collision energies.

\begin{figure}[ht]
\begin{center}
\includegraphics[width=0.9\linewidth]{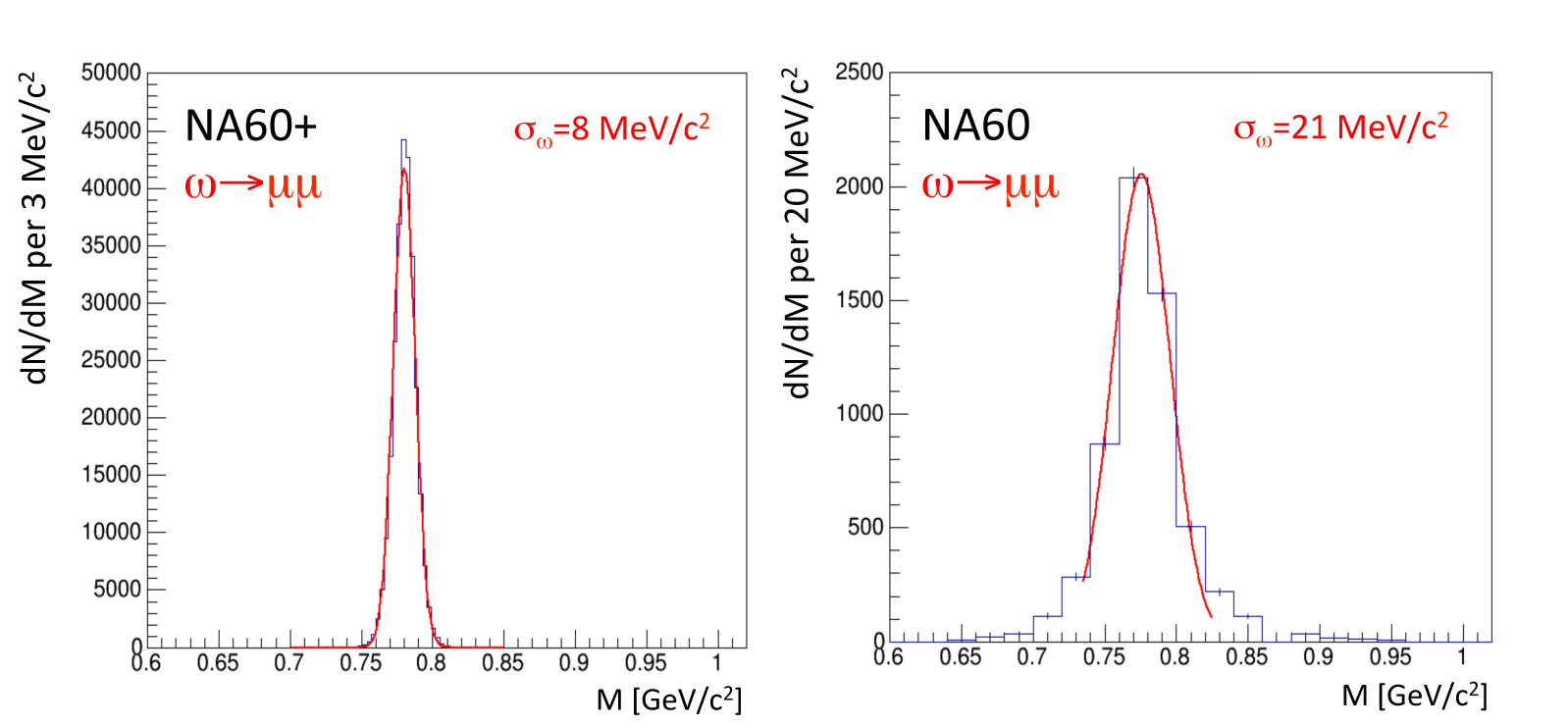}
\caption{(Left) Gaussian fit of the $\omega\rightarrow\mumu$ peak to extract the dimuon mass resolution at the $\omega$ mass.
(Right) Same as left for the original NA60 set-up for comparison.}
\label{fig:fig18}
\end{center}
\end{figure}

The mass resolution achievable with the NA60+ set-up for the $\omega$-meson measurement is shown in the left panel of Fig.~\ref{fig:fig18}, where a Gaussian fit to the dimuon mass spectrum gives a width of ${\sim}8\MeVcc$.
For comparison, the $\omega$-mass spectrum reconstructed by the original NA60 apparatus, shown in the right panel of Fig.~\ref{fig:fig18}, was ${\sim}21\MeVcc$.

\begin{figure}[!h]
\begin{center}
\includegraphics[width=0.4\linewidth]{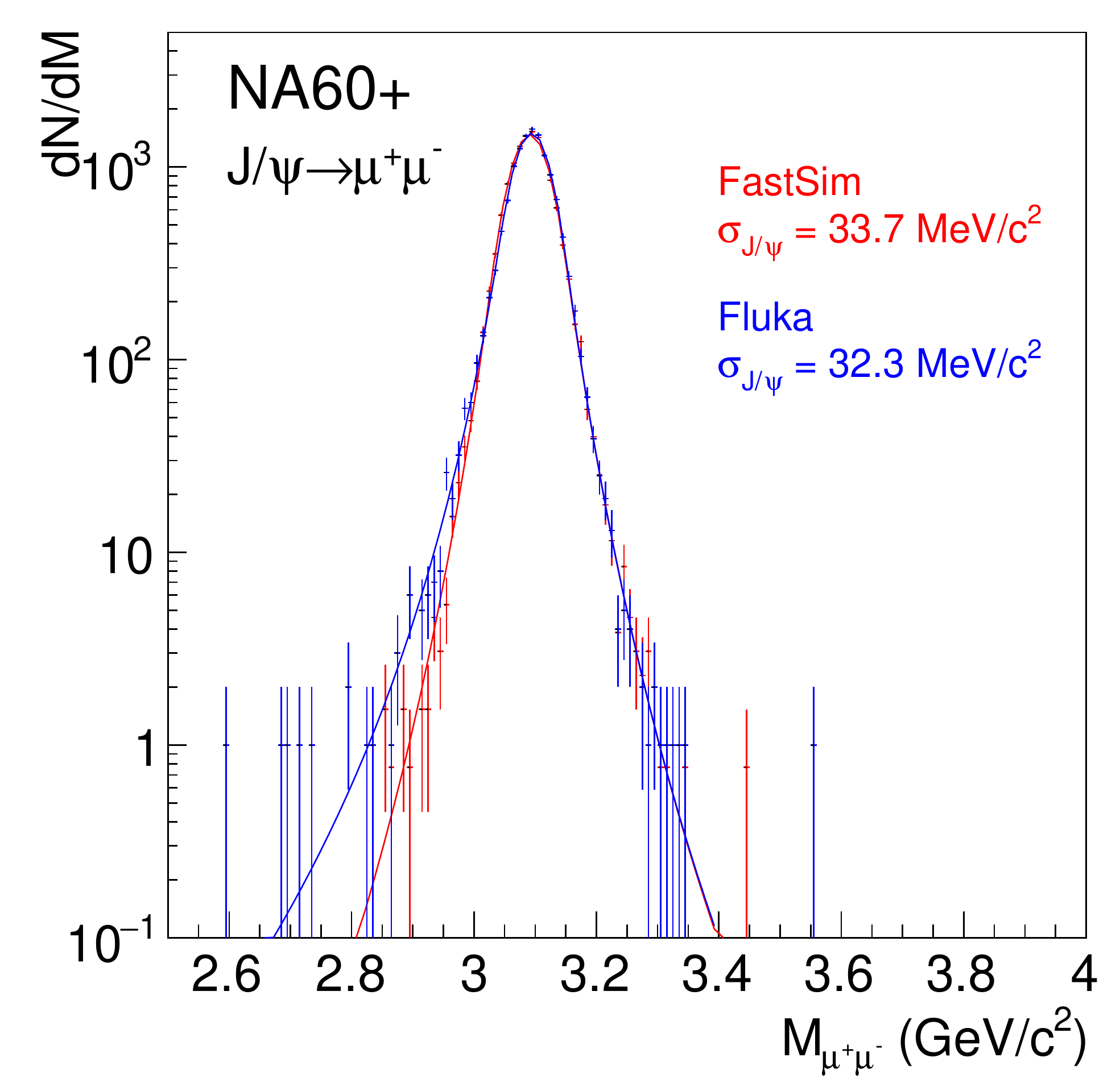}
\caption{Dimuon invariant-mass spectra for the decay $\jpsi\rightarrow\mumu$ based on  the fast-simulation framework (red) and \fluka (blue) for the muon tracking.
The spectra are fitted with a Crystal Ball function.}
\label{fig:jpsires}
\end{center}
\end{figure}

For the \jpsi, the foreseen mass resolution, as shown in Fig.~\ref{fig:jpsires}, is ${\sim}30\MeVcc$.
In order to test the possible limitations of the fast-simulation framework, in particular for what concerns the fluctuations of the muon energy loss in the apparatus (dominated by those in the hadron absorber), a simulation was performed using \fluka for the tracking of the decay muons in the set-up and the fast simulation for the reconstruction.
The resulting invariant-mass spectrum, also shown in Fig.~\ref{fig:jpsires}, exhibits the same mass resolution for the \jpsi, with only a slightly more extended tail on the left of the nominal mass of the meson.
As a further test, the fast-simulation framework was also used to simulate a set-up corresponding to that of the original NA60 experiment.
The obtained invariant-mass resolution for the \jpsi is ${\sim}70\MeVcc$, in remarkable agreement with the one measured in that experiment~\cite{Arnaldi:2010ky}.
We conclude that the fast-simulation tool is reliable enough to be used for the physics performance studies of NA60+ involving the propagation of muons in the set-up.

\subsubsection{Caloric curve, \texorpdfstring{$\rho$-${\rm a}1$}{rhoa1} chiral mixing and fireball lifetime
}
\label{thermal_dimu_performances}

Detailed performance studies  were carried out for the 5\% most central \PbPb collisions at $\sqrtsNN = \SIlist{6.3;8.8;17.3}{\GeV}$.
The differential spectra of thermal \mumu pairs, $\dd^3 N/(\dd M \dd \pt \dd y)$, are based on the in-medium $\rho$, $\omega$ and  4-pion spectral functions, QGP radiation and the expanding thermal fireball model of~\cite{Rapp:2014hha}.
The generator is based on the model calculation which assumes either no $\rho\text{--}\aone$ chiral mixing or full chiral mixing ($\epsilon=1/2$) in the mass region $1<M<1.5\GeVcc$.
For the performance of the temperature measurement, thermal dileptons were generated without chiral mixing.
The hadron cocktail generator for the 2-body decays of $\eta$, $\omega$,  and $\phi$ and the Dalitz decays  $\eta\to\gamma\mu^+\mu^-$ and $\omega\to\pi^0\mu^+\mu^-$ is based on the NA60 generator and on the statistical model of~\cite{Becattini:2005xt}.
The Drell--Yan process and open-charm production are simulated with the \pythia event generator.

We present results for data samples collected in one month data taking at $\sqrt s_{NN}=8.8$ and 17.3 GeV, and in two months of data taking at $\sqrt s_{NN}=6.3$ GeV. 
 The number of 
reconstructed thermal pairs in central collisions at each energy is summarized in Table~\ref{tab:ThermalRadiationStatistics}.

\begin{table}[ht]
\begin{center}
\caption{Number of reconstructed thermal pairs in central collisions at $\sqrt s_{NN}=6.3,8.8,17.3$ GeV and $T_{\rm slope}$ measurement. See text for details of the data taking conditions and the procedure for the $T_{\rm slope}$ measurement.}
\begin{tabular}{c c c}
\hline
Energy (GeV) & Thermal pairs & $T_{\rm slope}$ \\ \hline
6.3 & 3.52$\cdot10^6$ & $166\pm4.7\pm1$ \\
8.8 & 3.56$\cdot10^6$ & $169\pm4.4\pm1$ \\
17.3 & 9.70$\cdot10^6$ & $182\pm1.8\pm1$ \\
\hline
\end{tabular}
\label{tab:ThermalRadiationStatistics}
\end{center}
\end{table}


\begin{figure}[h]
\begin{center}
\includegraphics[width=0.45\textwidth]{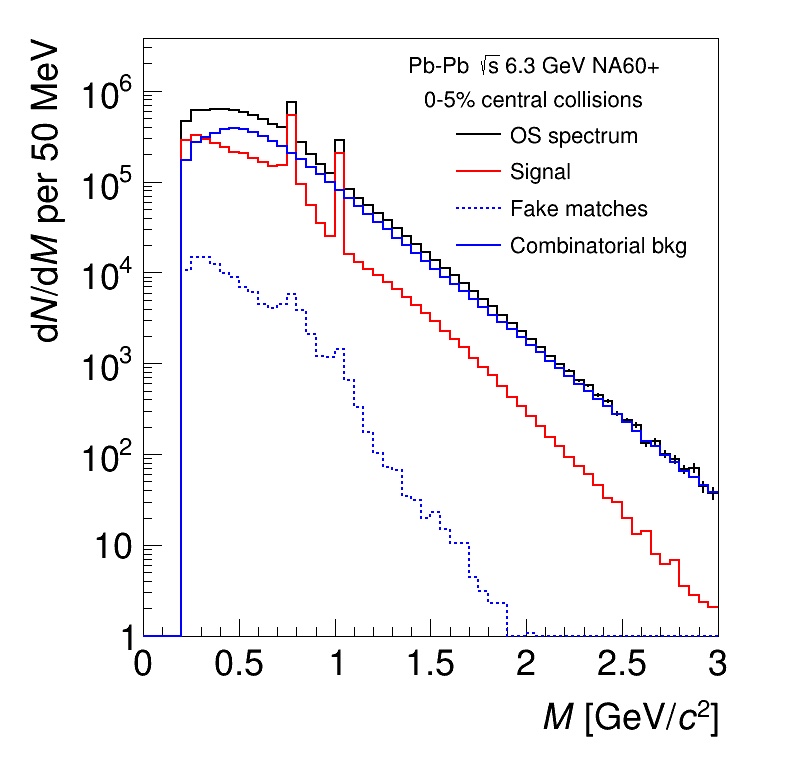}
\includegraphics[width=0.45\textwidth]{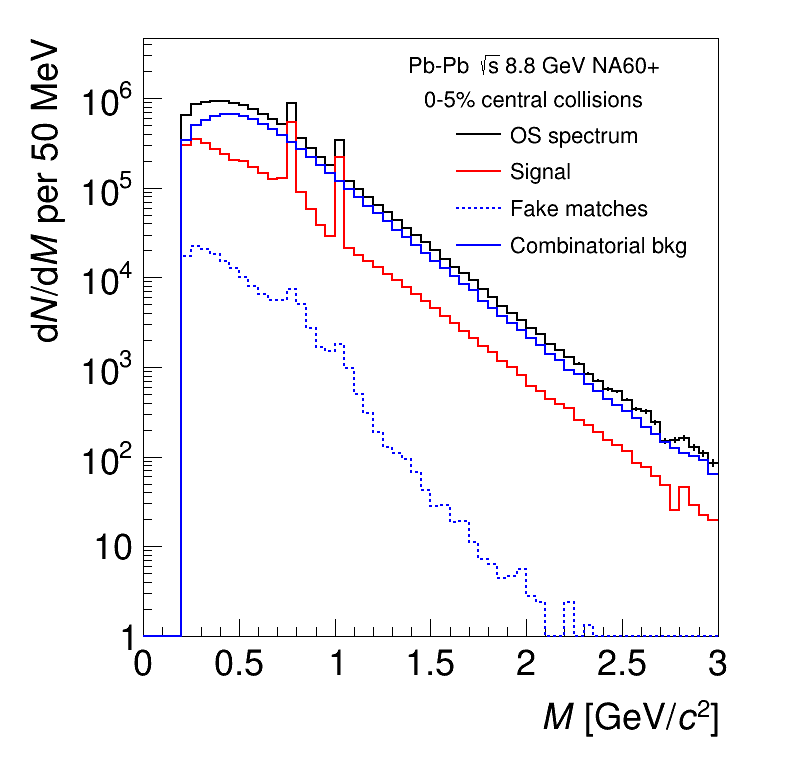}
\caption{(Left) Expected  sample in the 5\% most central \PbPb collision at $\sqrtsNN = 6.3\GeV$.
(Right) Same at $\sqrtsNN = 8.8\GeV$.}
\label{fig:RawMassSpectra}
\end{center}
\end{figure}

\begin{figure}[h]
\begin{center}
\includegraphics[width=0.45\textwidth]{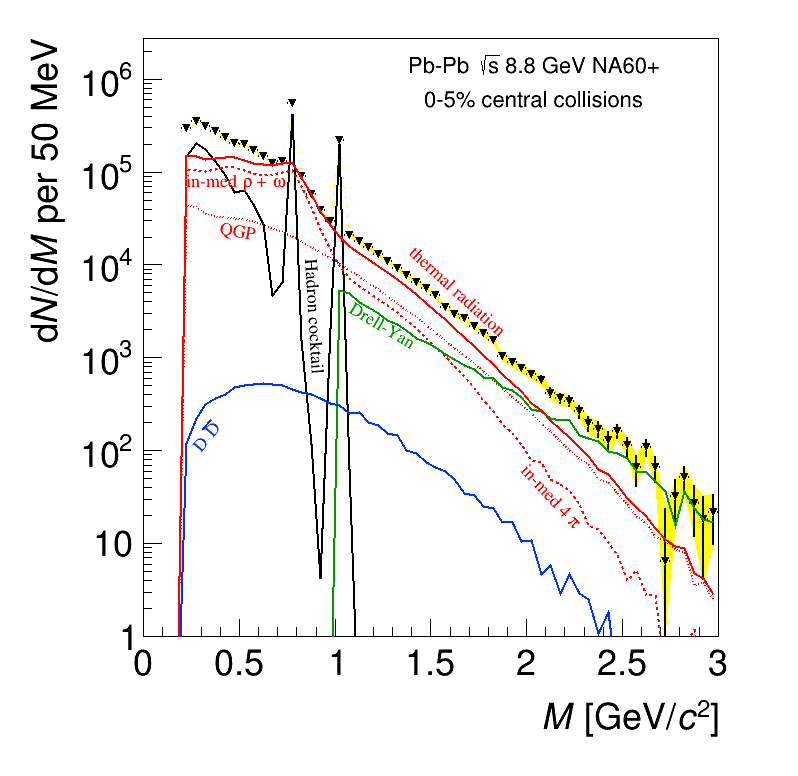}
\includegraphics[width=0.45\textwidth]{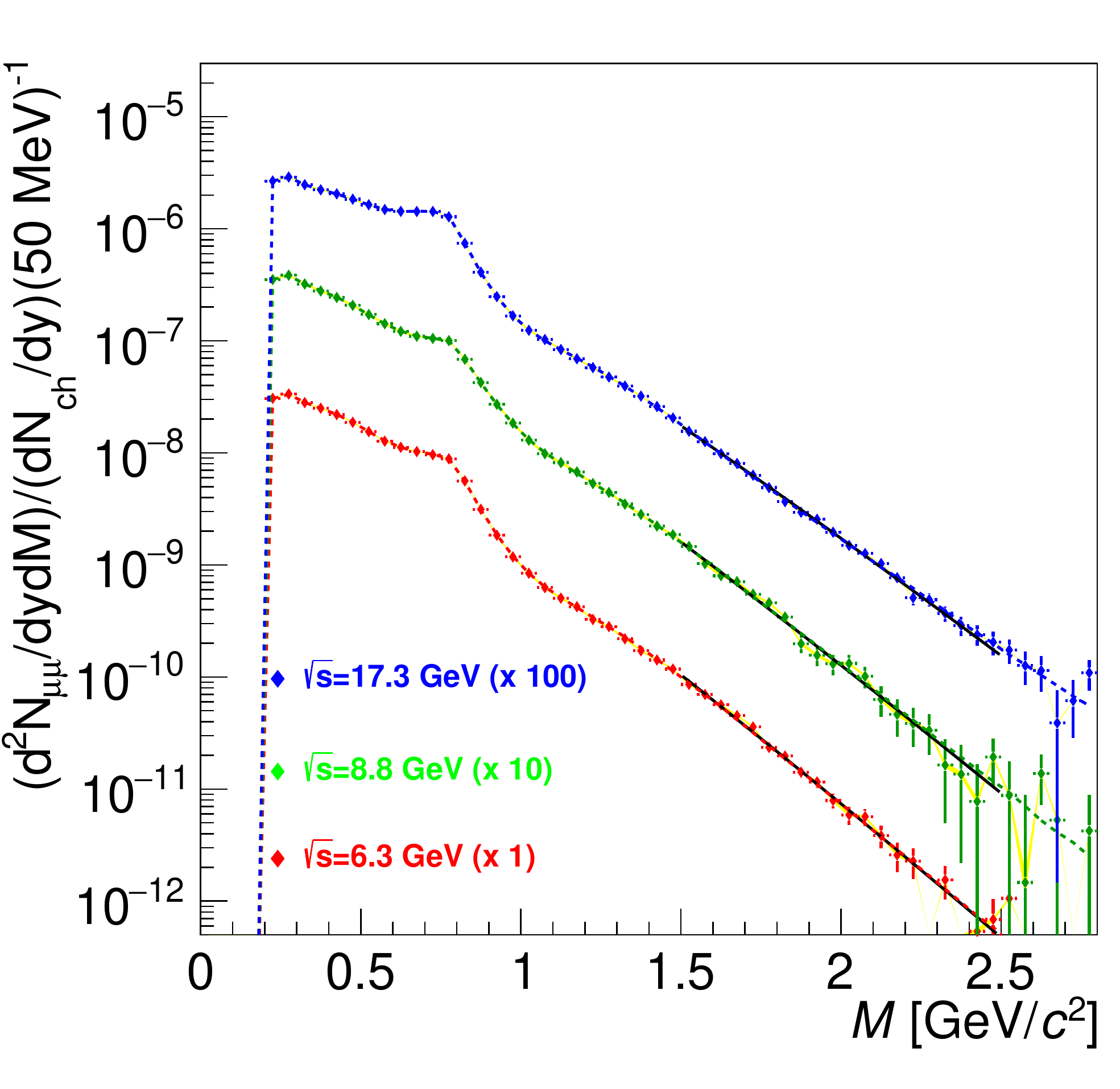}
\caption{(Left) Expected signal sample in the 5\% most central \PbPb collision at $\sqrtsNN = 6.3\GeV$ after subtraction of combinatorial and fake match background.
Various contributions are shown (see text for details). 
(Right) Acceptance corrected thermal spectra at $\sqrtsNN = 6.3, 8.8, 17.3\GeV$ obtained after subtraction of open charm, Drell--Yan and hadronic cocktail.
Model comparisons and exponential fits as discussed in the text are shown.
Systematic uncertainties are shown as yellow band.}
\label{fig:fig1-thermal-performance}
\end{center}
\end{figure}

 Fig.~\ref{fig:RawMassSpectra} shows the total reconstructed mass spectrum (black) in \PbPb collisions  at $\sqrtsNN = 6.3\GeV$ (left) and 
 $\sqrtsNN = 8.8\GeV$ (right).
The combinatorial background (continuous blue line) is estimated with \fluka simulations which take into account hadronic interactions in the absorber and the muon wall.
Primary pions, kaons and protons are generated and tracked through the spectrometer with kinematics as measured by NA49 at different energies~\cite{Afanasiev:2002mx,Alt:2006dk}.
Hits recorded in the detectors are injected in the fast simulation for track reconstruction.
Single track efficiencies (correct and fake matches) are evaluated and multiplied by the expected particle multiplicities in central collisions.
The background yield per event is then estimated by multiplying these quantities as obtained for  positively and negatively charged particle.
The background shape is obtained by sampling the momentum distributions of positively and negatively charged reconstructed particles and pairing the momenta to create a background particle pair.
The average signal-to-background ratio at $M=0.6\GeVcc$, in a region completely dominated by the continuum, is $\sim1/10$.
The combinatorial background is subtracted assuming a 0.5\% systematic uncertainty, based on the very conservative 1\% value estimated in the previous NA60 experiment.
The net signal after subtraction of the combinatorial background and fake matches is shown in red.
The contribution of the fake matches is very small in comparison to the combinatorial background, becoming completely negligible for $M>1\GeVcc$.
For what concerns minimum-bias collisions, the progress in statistics over the former NA60 experiment is a factor $\sim$20, with a significantly better mass resolution. 

The left panel of Fig.~\ref{fig:fig1-thermal-performance} 
shows the signal reconstructed mass spectra (black) for \PbPb collisions at $\sqrtsNN = 8.8\GeV$ after subtraction of the combinatorial background due to pion and kaon decays as well as fake matches.
The 0.5\% systematic uncertainty from the subtraction of combinatorial background is shown as a yellow band.
The figure shows all the expected signal components.
For $M<1\GeVcc$, the thermal radiation yield is dominated by the in-medium $\rho$.
The $\omega$ and $\phi$ peaks are well resolved with a resolution better than $10\MeVcc$ at the $\omega$ mass.
The  thermal spectrum is measurable up to \numrange{2.5}{3}\GeVcc.
The open-charm yield becomes totally negligible at low \sqrtsNN.
The Drell--Yan yield will be measured in dedicated \pA runs (see also Sec.~\ref{charmoniumperformance}).

The thermal spectra are obtained after (i) subtraction of the hadronic cocktail for $M<1\GeVcc$ of $\eta$, $\omega$ and $\phi$ decays into \mumu as well as the $\eta$ and $\omega$ Dalitz decays and (ii) subtraction of Drell--Yan as well as open-charm muon pairs for $M>1\GeVcc$.
The systematic uncertainty is larger at high energy due to the larger combinatorial background.
After acceptance correction, the spectra are fit with $\dd N/\dd M\propto M^{3/2}\exp(-M/T_{\rm slope})$ in the interval $M=\numrange{1.5}{2.5}\GeVcc$.
The resulting spectra at $\sqrtsNN = 6.3, 8.8, 17.3\GeV$  are shown
in the right panel of Fig.~\ref{fig:fig1-thermal-performance}. 
The theoretical spectra used as an input are shown as dashed lines, while
the exponential  fits are shown as black lines.

\begin{figure}[t]
\begin{center}
\includegraphics[width=0.85\textwidth]{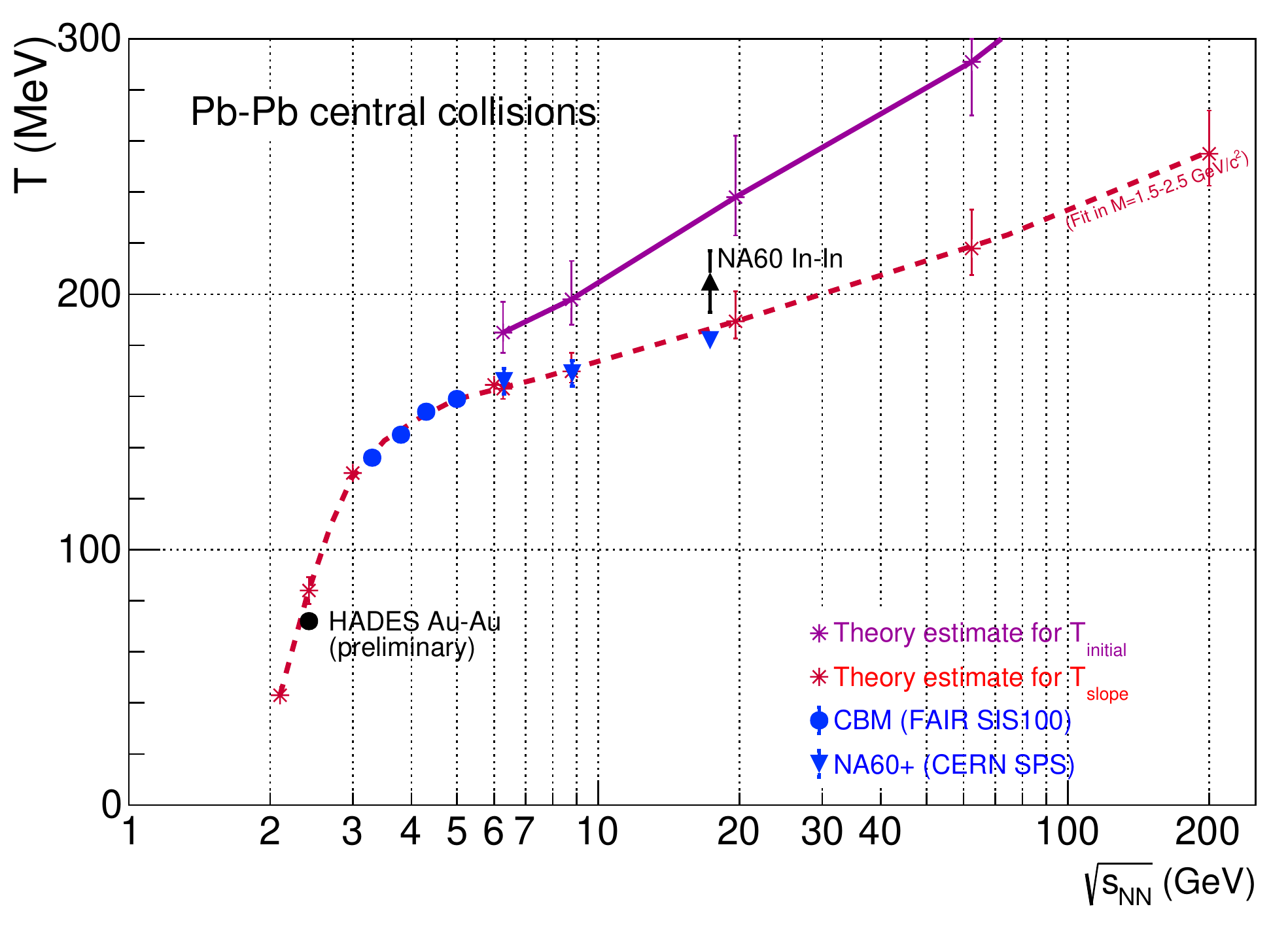}
\caption{
Caloric curve: medium temperature evolution vs \sqrtsNN in central \PbPb collisions.
\T{\rm initial} (magenta) and \T{\rm slope} (red) are theoretical estimates for the initial medium temperature and the temperature from dilepton spectra respectively, using Ref.~\cite{Rapp:2014hha} and a coarse graining approach in UrQMD~\cite{PhysRevC.92.014911}.
Blue triangles  are the expected performance from NA60+ (CBM performance is also shown~\cite{Galatyuk:2019lcf}).
The only existing measurements at present are from NA60 in \InIn~\cite{Arnaldi:2008er,Specht:2010xu} and from HADES in \AuAu collisions~\cite{Harabasz:2019lzg}.}
\label{fig:TvsSqrtS}
\end{center}
\end{figure}

The main result is the caloric curve of Fig.~\ref{fig:TvsSqrtS}, which displays the temperature evolution as a function of collision energy.
The dashed line is the $T_{\rm slope}$ from the theoretical model used as an input.
At low energies, the temperatures can be measured with a combined statistical and systematic uncertainty of just a few\MeV (see Tab.~\ref{tab:ThermalRadiationStatistics}), thus showing that the experiment has a strong sensitivity to a possible flattening of the caloric curve in a region complementary to the one which will be explored by CBM.  The expected temperature measurements for that experiment are shown as well.

\begin{figure}[h!]
\begin{center}
\includegraphics[width=0.65\textwidth]{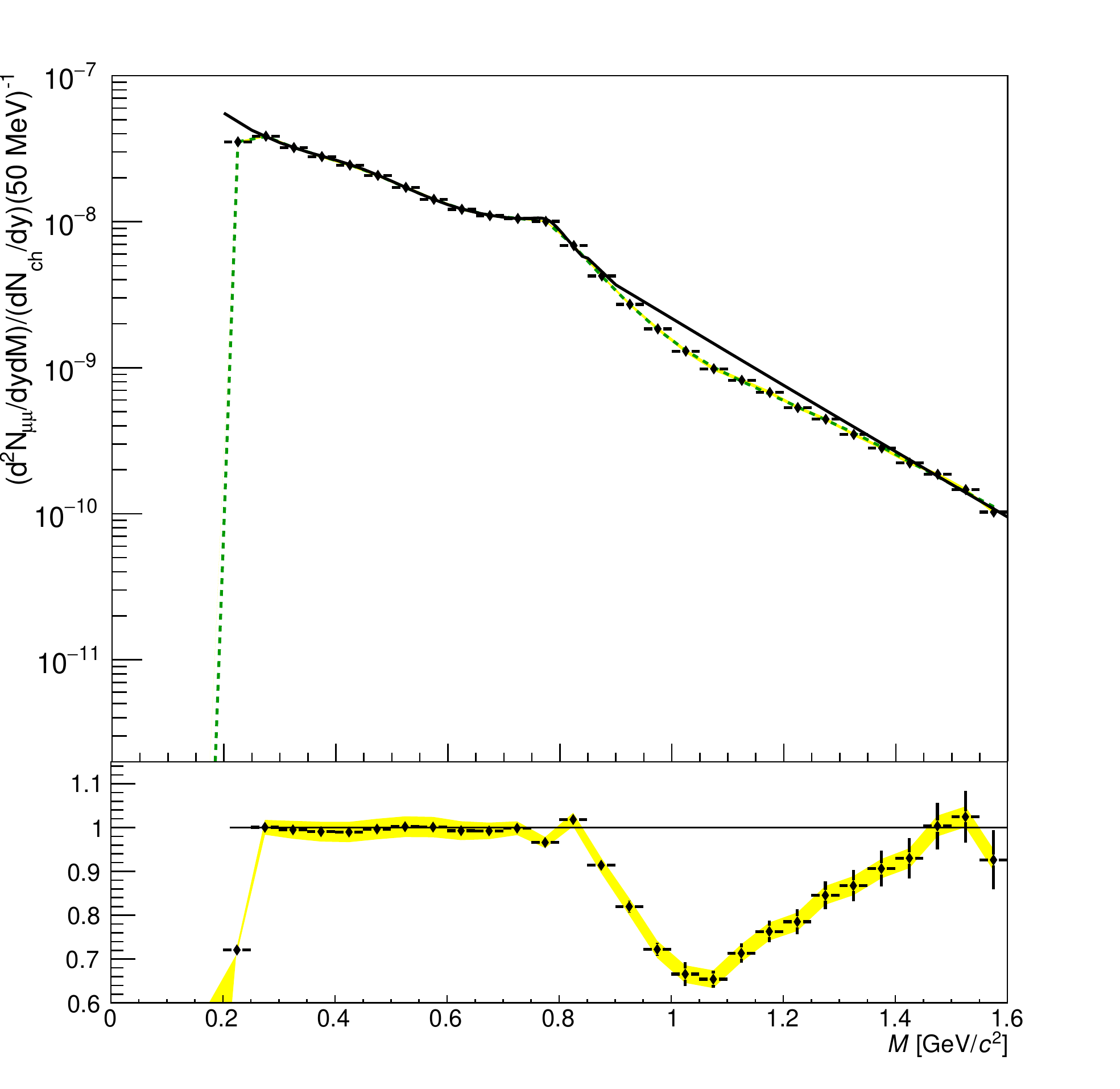}
\caption{
NA60+ projection for the acceptance corrected thermal dimuon mass spectrum at $\sqrtsNN = 8.8\GeV$ in case of no chiral mixing compared to the theoretical expectation (green line).
The black line above 1\GeVcc is the expectation from full chiral mixing~\cite{Rapp:2014hha}.}
\label{fig:fig3-thermal-performance}
\end{center}
\end{figure}

The acceptance corrected mass spectrum at $\sqrtsNN = 8.8\GeV$, based on the assumption of no chiral mixing, is compared to the expectation of full chiral mixing in  Fig.~\ref{fig:fig3-thermal-performance}.
As shown, the statistical and systematic uncertainty provide a very good sensitivity to an increase of the yield due to chiral mixing of ${\sim}20\text{--}30\%$. 

\begin{figure}[h!]
\begin{center}
\includegraphics[width=0.65\textwidth]{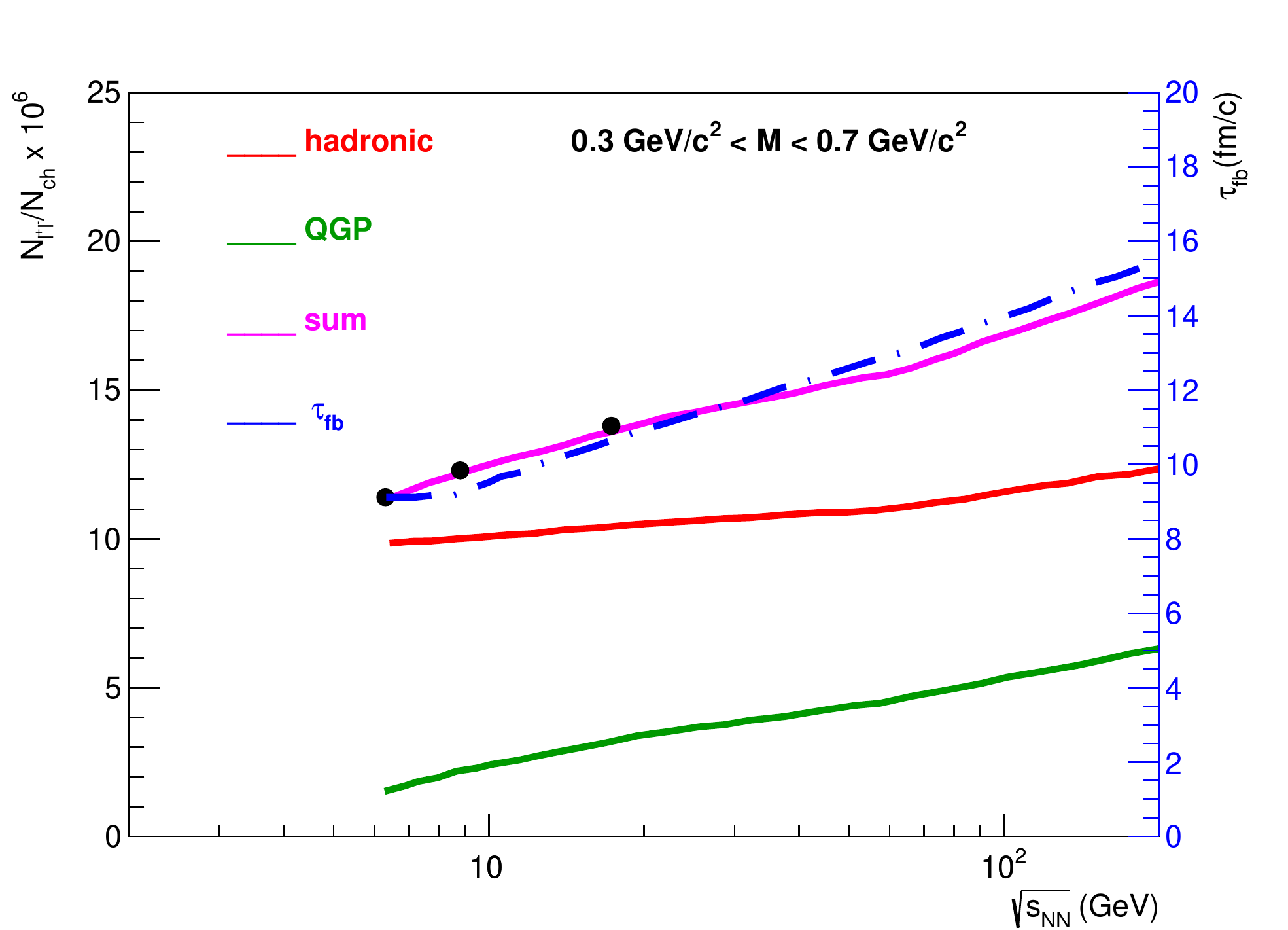}
\caption{
Performance for thermal dilepton yield measurement for the fireball lifetime measurement. The black circles correspond to measurements at $E_{\rm lab}=$ 40, 80 and 158 A\,GeV. The corresponding uncertainties are smaller than the size of the markers.}
\label{fig:fig4-thermal-performance}
\end{center}
\end{figure}
Finally the performance for the dilepton excitation function, \ie\ the total thermal yield measurement in the mass range \numrange{0.3}{0.7}\GeVcc, is compared to the fireball lifetime estimate of Ref.~\cite{Rapp:2014hha} in  Fig.~\ref{fig:fig4-thermal-performance}. 
The plot shows the correlation between the fireball lifetime (dashed blue line, values are read from right vertical scale) and  the thermal yield from the calculation (magenta line, values are read from the left scale). The black points show the accuracy of the corresponding NA60+ yield measurement  at different energies.
The  uncertainty is dominated by the systematic error from the background subtraction.
The estimated precision of the measurement at low energies provides very good sensitivity to possible anomalies in the fireball lifetime, as explained in Sec.~\ref{sec:fireball_lifetime}.

\subsubsection{Elliptic flow of thermal dimuons
}
\label{v2_thermal_performances}

The performances on the measurement of the elliptic flow of thermal dimuons were studied at $\sqrt{s_{\rm NN}}=8$ and 17.3~GeV.  
With respect to the simulations described  in~\ref{thermal_dimu_performances}, a mass-dependent  elliptic flow parameter is introduced for thermal dimuons coming from the hadronic gas and the QGP. The calculation is performed in the centrality range 20-30\%.

To describe the dimuon mass spectrum in such semi-central collisions, the same kinematic distributions as in central collisions are taken for all the processes, while the normalization of each process is scaled from the one in central collisions.  
The normalization of the resonances is assumed to scale as the number of charged particles $N_{\rm ch}$, which is calculated as a function of centrality using the measurements in Ref.~\cite{NA57:2005jac}. The thermal dimuon yield is expected to scale with centrality as $N_{\rm ch}^{1.4}$~\cite{Rapp:2013nxa}. The open charm and Drell-Yan contributions are scaled as the number of binary collisions.  

At present, no theoretical calculation of the elliptic flow of thermal dimuons is available at the SPS energies. For this reason, two scenarios were considered: an elliptic flow $v_2$ comparable to the expectations at RHIC energies and a vanishing $v_2$. 
In the first scenario, the mass dependence of $v_2$ for a hadron gas and a QGP is parameterized according to the calculations in~\cite{Xu:2014ada}. 

\begin{figure}[ht]
    \begin{center}

    \includegraphics[width=0.49\linewidth]{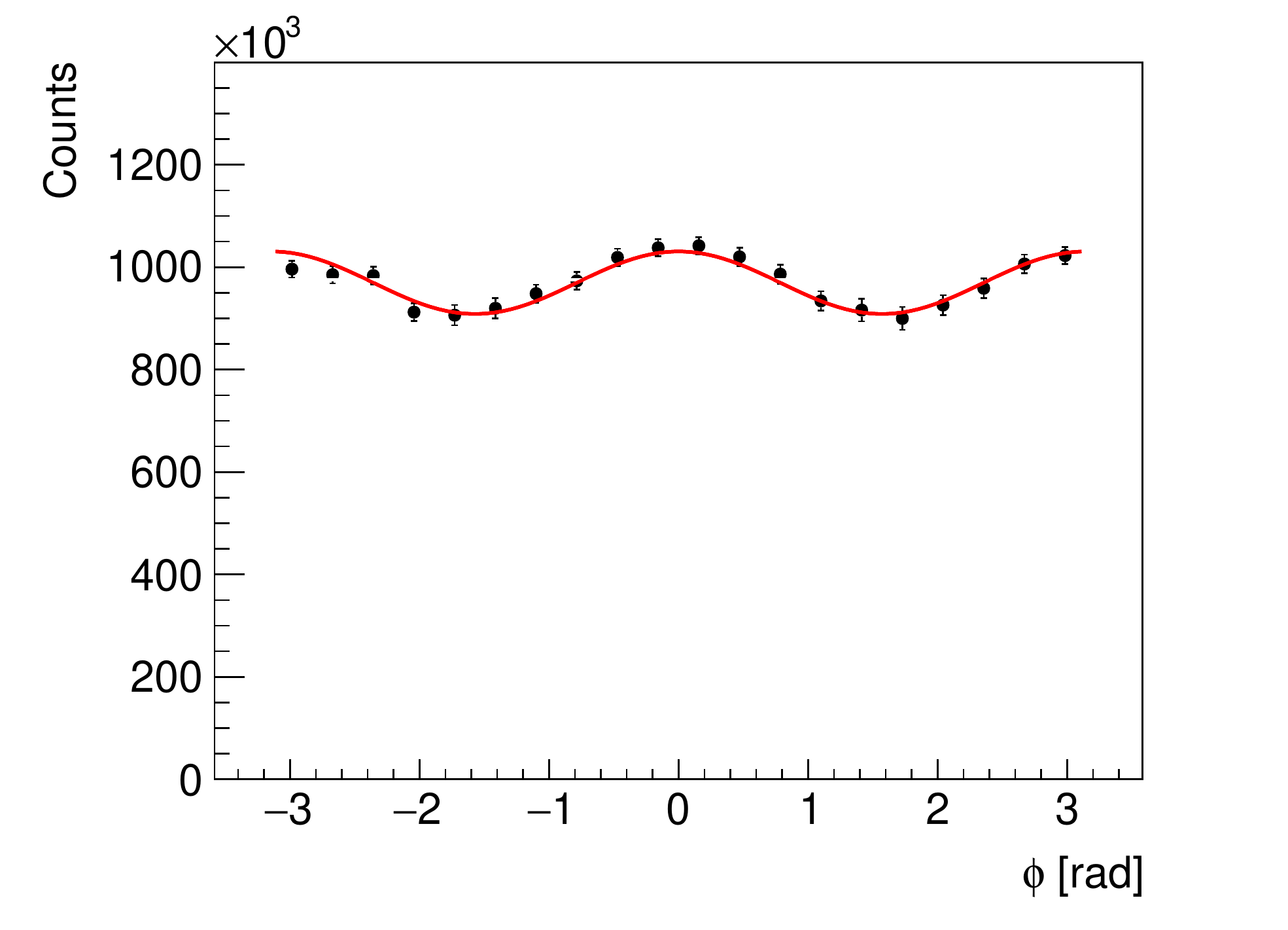}
    \includegraphics[width=0.49\linewidth]{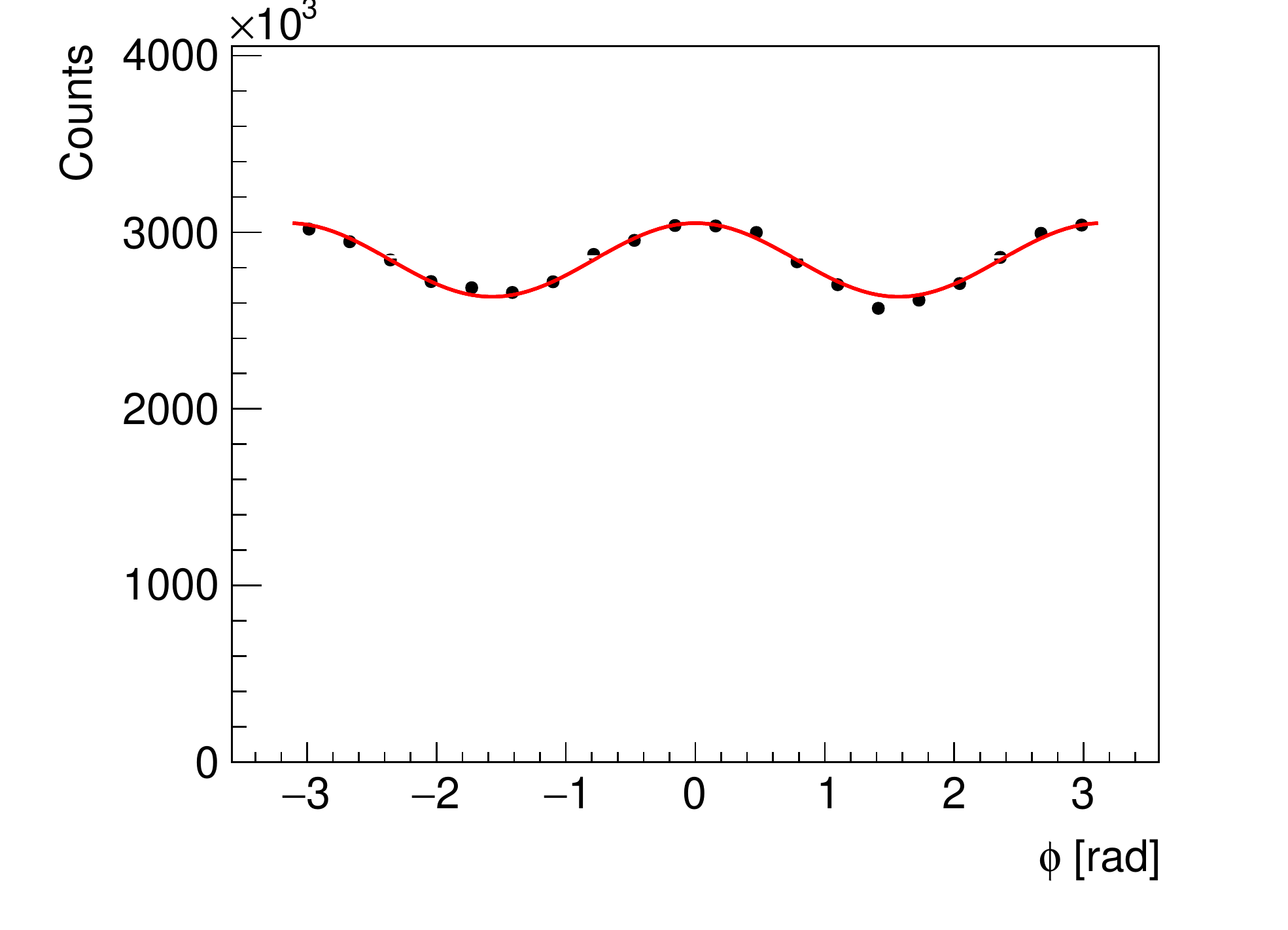}
    \caption{Azimuthal angle distribution of thermal dimuons for $0.6< M < 1.2$~GeV/$c^2$ at $\sqrt{s_{NN}}=8.8$~GeV (left) and 17.3~GeV (right), assuming $v_2\sim 0.025$.}

\label{fig:phi_thermal}
\end{center}
\end{figure}

Fig.~\ref{fig:phi_thermal} shows the expected acceptance-corrected azimuthal angle distribution for $0.6< M < 1.2$~GeV/$c^2$ at $\sqrt{s_{NN}}=8.8$~GeV (left) and 17.3~GeV (right), in the hypothesis of non-null $v_2$. In this mass range, the elliptic flow parameter can be obtained with a statistical uncertainty of $\sim 10\%$ and $\sim 4\%$, respectively. 

\begin{figure}[ht]
    \begin{center}
    \includegraphics[width=0.49\linewidth]{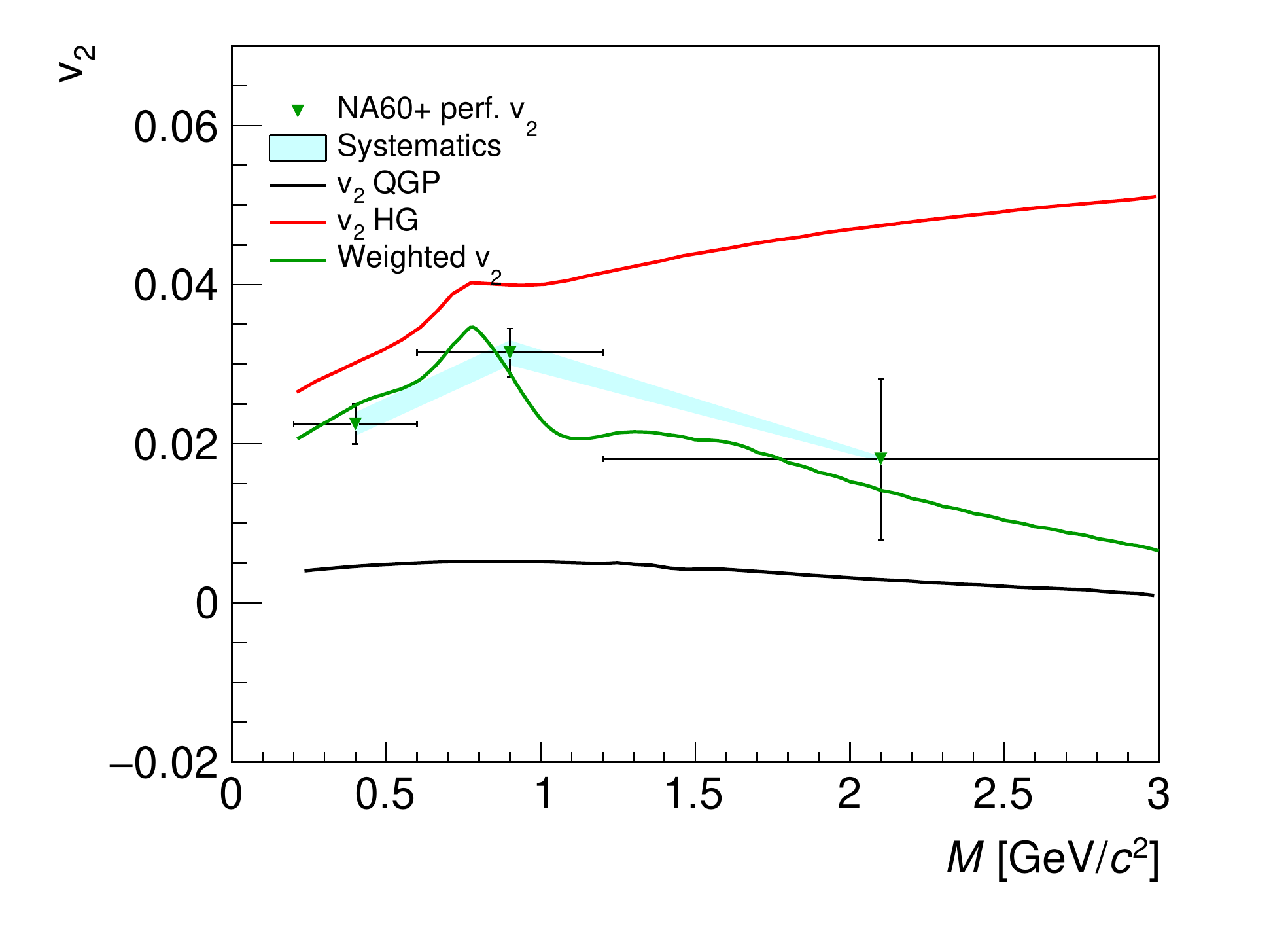}
    \includegraphics[width=0.49\linewidth]{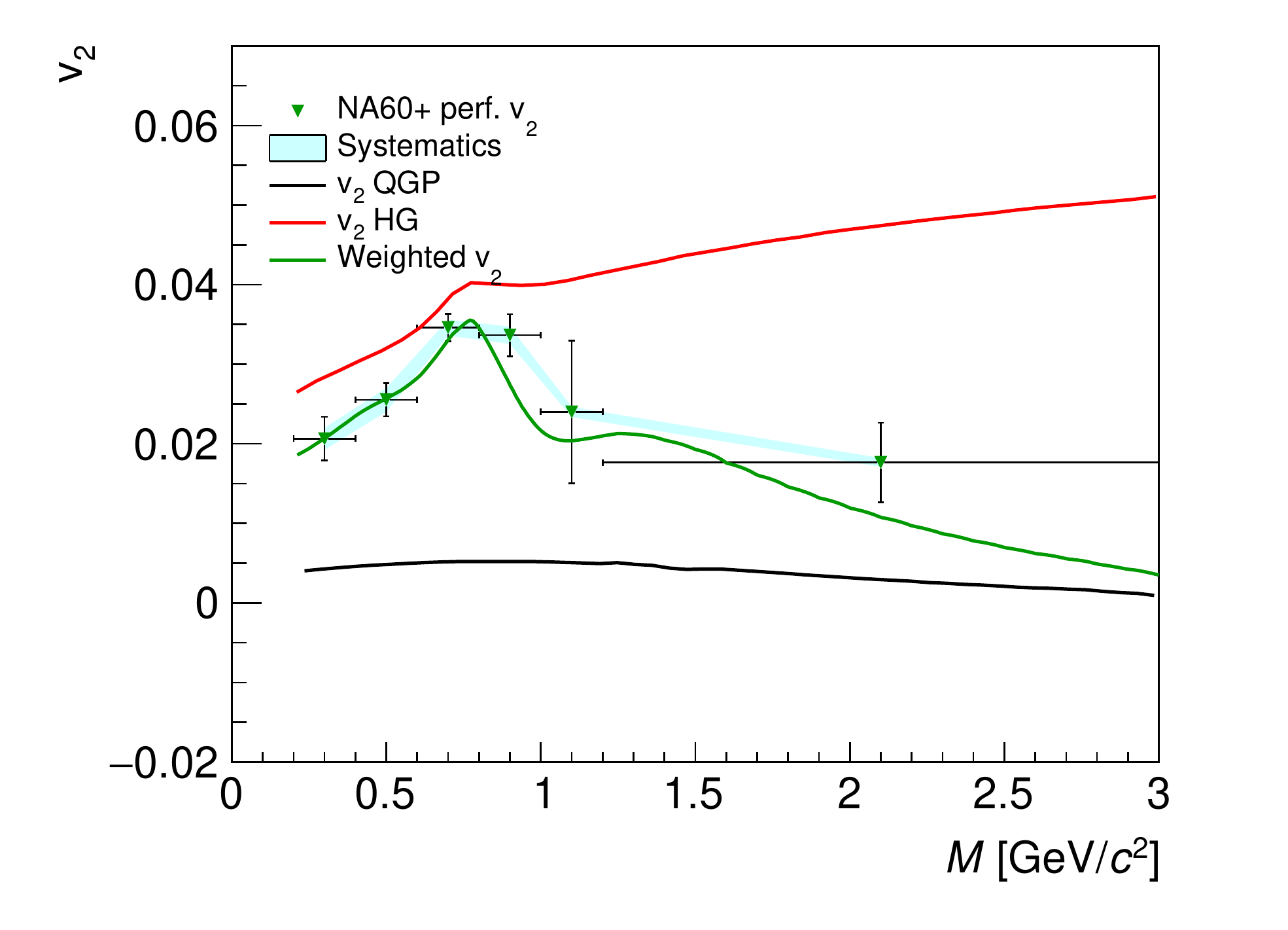}
    \includegraphics[width=0.49\linewidth]{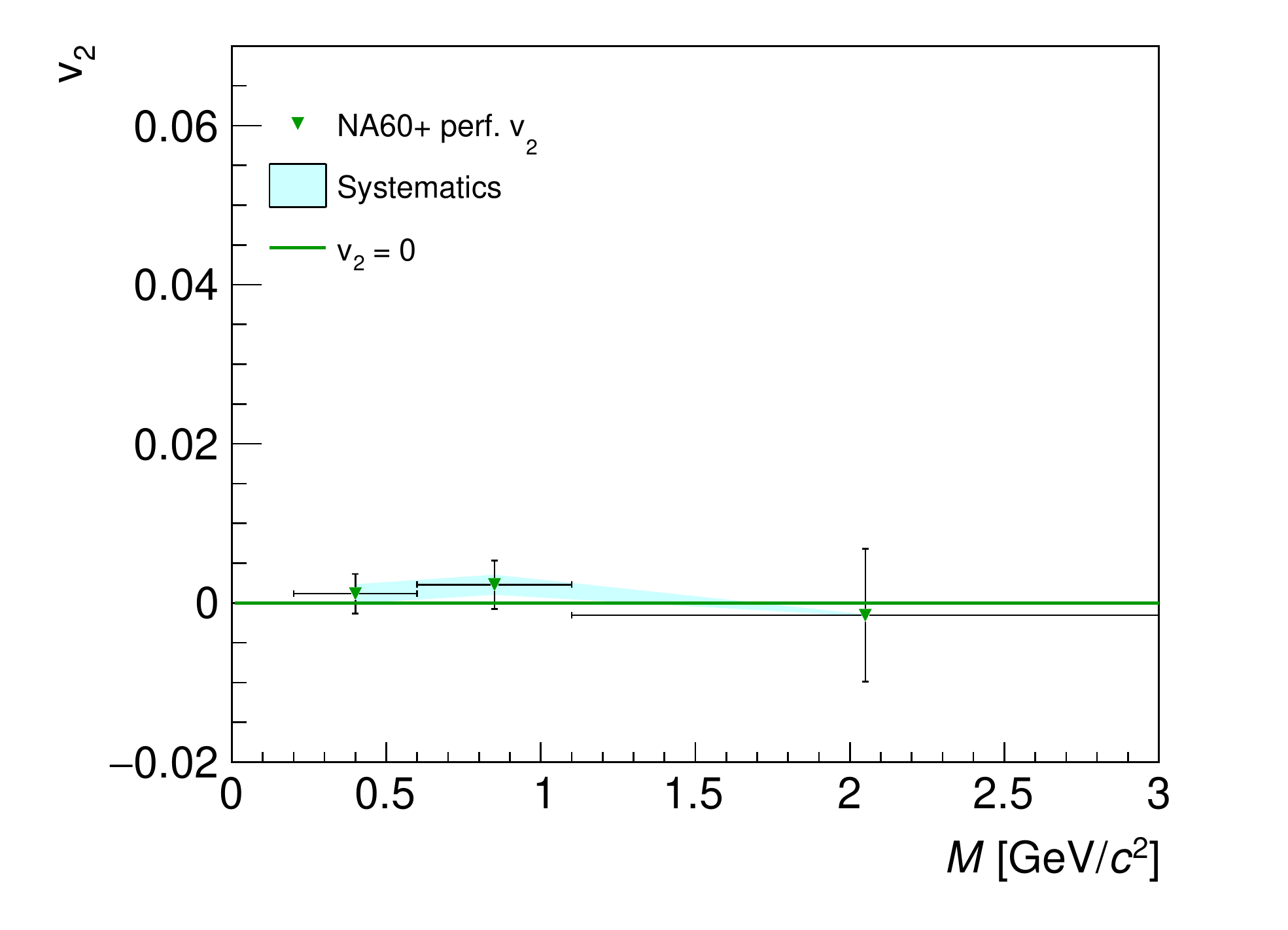}
    \includegraphics[width=0.49\linewidth]{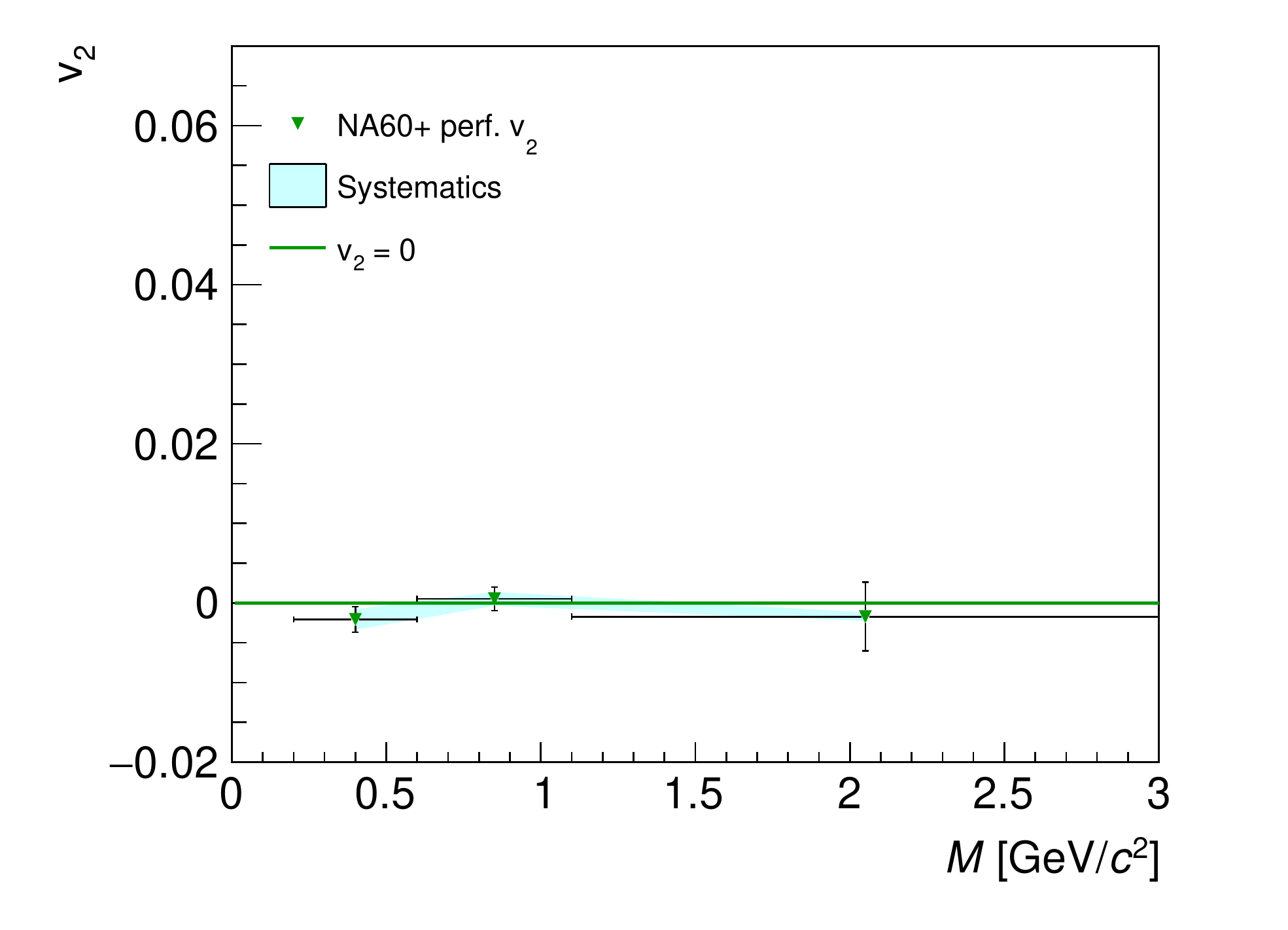}
    \caption{Elliptic flow of thermal dimuons versus mass at $\sqrt{s_{NN}}=8.8$~GeV (left) and 17.3~GeV (right), under the hypothesis of $v_2$ expected at RHIC (top) or vanishing (bottom).}

\label{fig:v2_thermal}
\end{center}
\end{figure}

The elliptic flow parameter can be measured in several mass intervals at the two considered energies. In the top panel of Fig.~\ref{fig:v2_thermal}, the expected performances are shown under the hypothesis of the $v_2$ expected at RHIC. The mass bins are optimized for the expected statistics in one month of data taking. The statistics is sufficient to observe an increase of $v_2$ as a function of mass in the region dominated by the hadronic gas, and a drop where the radiation from QGP dominates.   
If no elliptic flow is assumed (bottom panel), a measurement can be performed with an uncertainty between 0.002 and 0.008, depending on the mass and the energy.  

\subsection{Charmonia}
\label{charmoniumperformance}
\vskip 0.2cm

Quarkonium production studies require a relatively large integrated luminosity, due to the small production cross section at low incident beam energy.
On the other hand, past SPS experiments (NA50/NA60) showed that the background levels in the dimuon invariant mass spectrum in the \jpsi region are very small (${<}5\%$)~\cite{Alessandro:2004ap,Arnaldi:2007zz}. The NA60 experiment was able to perform a very accurate measurement of the centrality dependence of the suppression by collecting \num{\sim3e4} \jpsi in \InIn collisions at \SI{158}{\AGeV} incident energy, \ie\ $\sqrtsNN = 17.3\GeV$~\cite{Arnaldi:2007zz}.

The statistics that can be obtained in a \jpsi measurement in Pb--Pb collisions for various energies in the SPS domain can be calculated with an evaluation of the production cross section and of the detection efficiency in the NA60+ set-up. For the \jpsi cross section the parameterization $\sigma(x_{\rm F}>0)=\sigma_0(1-m/\sqrt{s})^n$, commonly adopted at fixed-target energies~\cite{Vogt:1999cu} was used, with $n=12.0\pm0.9$ and $\sigma_0=638\pm104$ nb. The fraction of the total cross section falling in the fiducial region $0<y_{\rm CM}<1$ was estimated according to the empirical formula ${\rm d}N/{\rm d}x_{\rm F}=(1-|x_{\rm F}|)^c$, with $c=a/(1+b/\sqrt{s})$, $a=13.5\pm4.5$ and $b=44.9\pm21.9$ GeV~\cite{Vogt:1999cu}. The detection efficiency was estimated by means of the fast simulation and reconstruction tool, obtaining the value 0.14 in the range $0<y_{\rm CM}<1$. In addition, a suppression of the \jpsi yield by a factor 3 with respect to binary collision scaling was assumed at all energies. This factor includes both cold nuclear matter and possible QGP-related effects. It was evaluated from results at top SPS energy and approximately assumed to be energy independent, since for decreasing energy the two effects are expected to vary in opposite directions. 

\begin{figure}[ht]
\begin{center}
\includegraphics[width=0.7\linewidth]{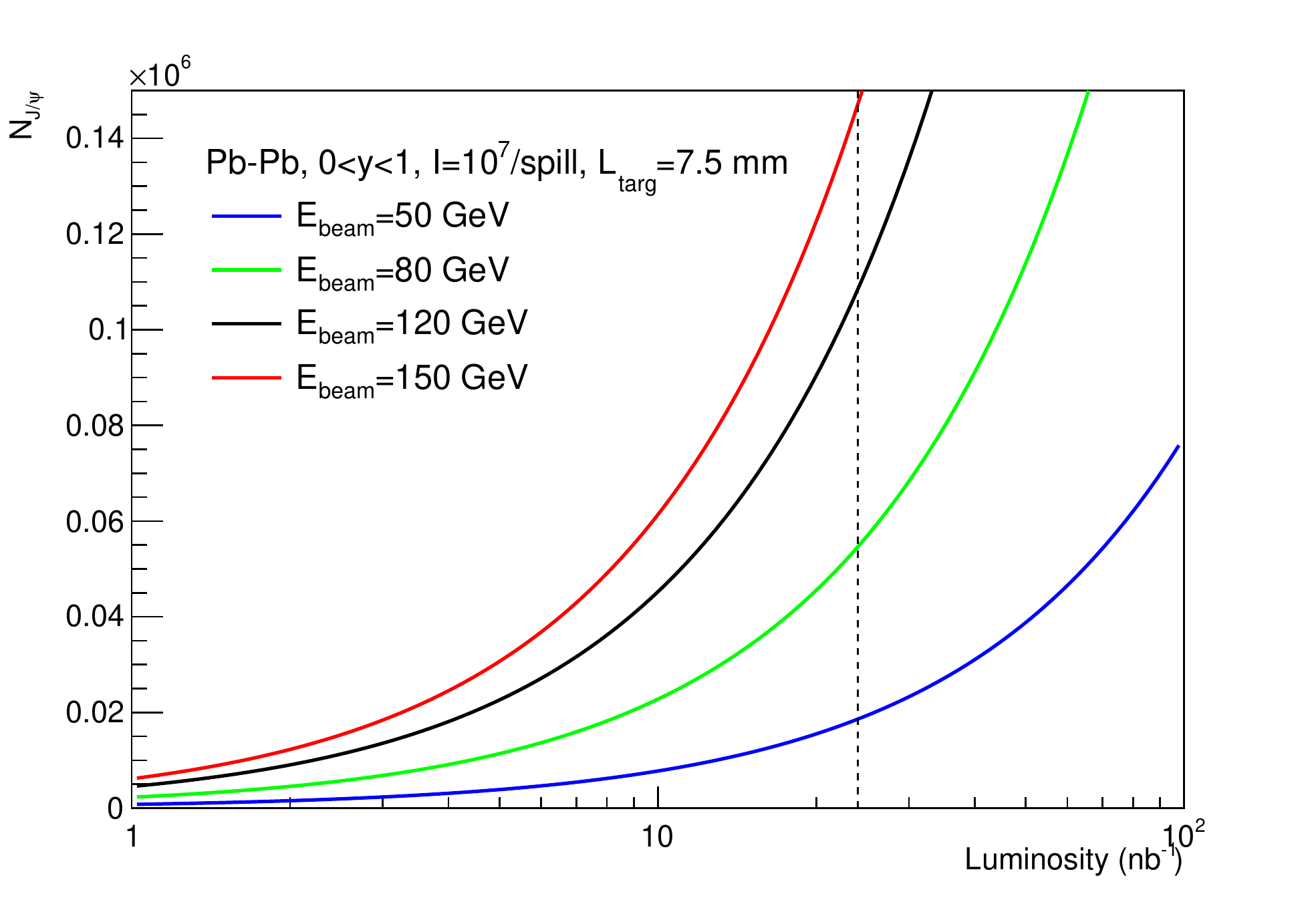}
\caption{The number of \jpsi that can be detected in NA60+ (Pb--Pb collisions) for various collision energies as a function of the integrated luminosity. The dashed line represents the expected luminosity value corresponding to one month of data taking. }
\label{fig:Jpsiperf}
\end{center}
\end{figure}

In Fig.~\ref{fig:Jpsiperf}, the number of \jpsi that can be collected at various energies, as a function of the integrated luminosity, is shown. The dashed vertical line corresponds to the expected luminosity for one month of Pb--Pb data taking, $L_{\rm int}\sim 24$ nb$^{-1}$. The expected statistics varies from $\sim1.8\cdot10^4$ at $E_{\rm beam}=50$ GeV to $1.5\cdot10^5$ at top SPS energy. 

At SPS energies, it is well known that the break-up of \jpsi mesons in cold nuclear matter plays an important role in determining the final observed yields in nucleus--nucleus collisions.
Therefore, data taking with \pA collisions are mandatory to calibrate such an effect.
We assumed to have 7 nuclear targets, each one 1 mm thick, simultaneously exposed to the incident proton beam and a total number of $5\cdot 10^{13}$ protons on target at each energy, with a beam intensity of $\sim 8\cdot 10^8$ protons/spill (see Sec.~\ref{Beamconditions}).

\begin{figure}[ht]
\begin{center}
\includegraphics[width=0.45\linewidth]{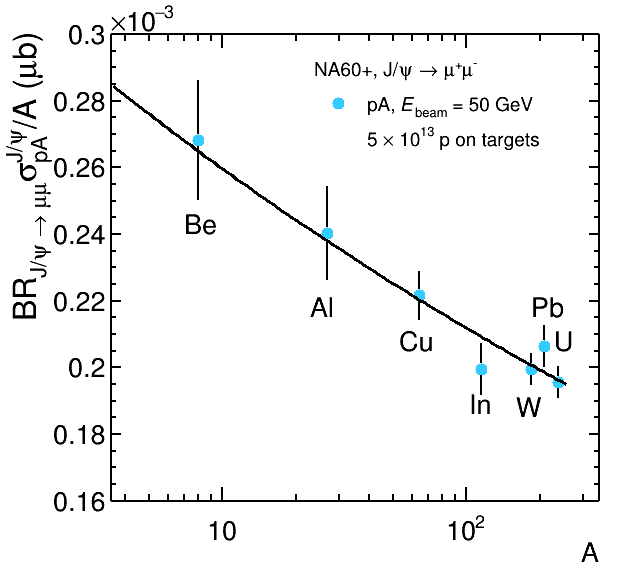}
\includegraphics[width=0.45\linewidth]{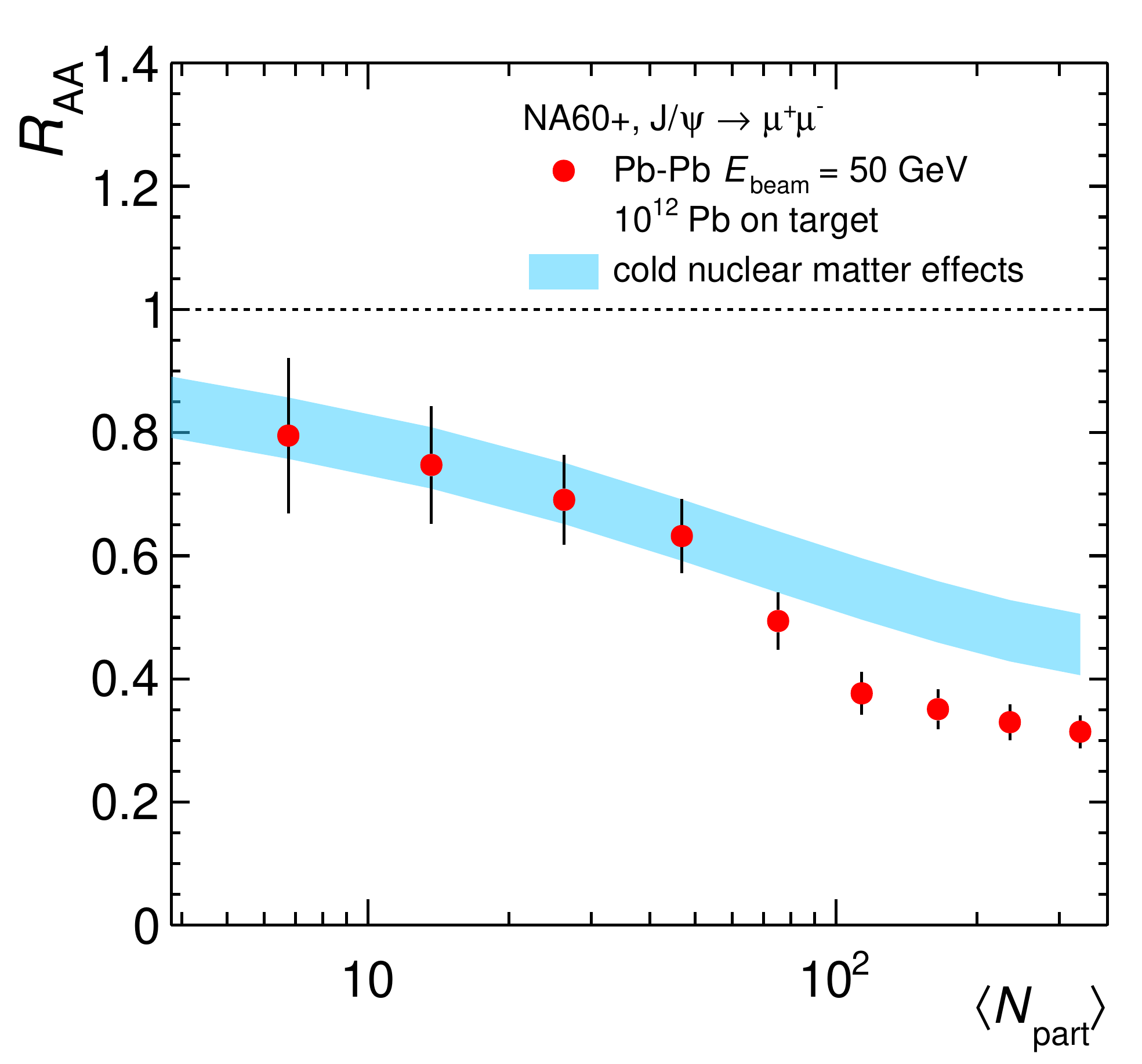}
\caption{(Left) \jpsi cross section normalized to the mass number A for \pA collisions at $E_{\rm beam} = 50\GeV$.
The results of a fit using the parameterisation $\sigma^{\jpsi}_{\pA}=\sigma^{\jpsi}_{\pp}\cdot A^{\alpha}$ are shown.
(Right) The nuclear modification factor for \jpsi production in \PbPb collisions at $E_{\rm beam} = \SI{50}{\AGeV}$ as a function of \Npart, compared with expectations from cold nuclear matter effects, obtained from the \pA results and shown as a blue band.}
\label{fig:Jpsiresult}
\end{center}
\end{figure}

In the left panel of Fig.~\ref{fig:Jpsiresult}, the expected cross sections as a function of the mass number A are shown, for an incident beam energy $E_{\rm beam} = 50\GeV$, or $\sqrtsNN = 9.8\GeV$, and assuming a dissociation cross section of $\sigma_{\jpsi\text{--}\rm N}=4.3\mb$, as measured at top SPS energy~\cite{Alessandro:2003pi}.
Such measurements are necessary in order to evaluate the \jpsi production cross section in pp collisions $\sigma_{\pp\rightarrow \jpsi \rm X}$, that can be obtained by extrapolating pA results to A=1. This quantity is needed for the calculation of the nuclear modification factor in Pb--Pb collisions. The pA results are also used to extrapolate the J/$\psi$ break-up effects in cold nuclear matter to the conditions relative to \PbPb results. These procedures were well tested in the past NA50/NA60 SPS experiments~\cite{Alessandro:2003pi}.

Finally, the right panel of Fig.~\ref{fig:Jpsiresult} shows the results of a simulation of the \jpsi\ \raa for \PbPb collisions as a function of the number of participant nucleons, \Npart, assuming that for $\Npart \lesssim 50$ the suppression is entirely due to cold nuclear matter effects, while for more central events an extra-suppression reaching 20\% sets in.
The simulation assumes 10$^{12}$ Pb ions incident on a 15\% interaction probability Pb target
at $E_{\rm beam} = \SI{50}{\AGeV}$. 
The uncertainties shown for \raa in the right panel of Fig.~\ref{fig:Jpsiresult} include, in addition to the statistical uncertainty on the \jpsi yield, conservatively evaluated assuming a 20\% background level, those on the evaluation of the centrality variables (Glauber model) and on the \pp cross section, calculated from the \pA results displayed in the left panel of Fig.~\ref{fig:Jpsiresult}.
One can clearly see that a precise evaluation of a relatively small anomalous \jpsi suppression is within reach down to low $E_{\rm beam}$.

\begin{figure}[h]
\begin{center}
\includegraphics[width=0.55\linewidth]
{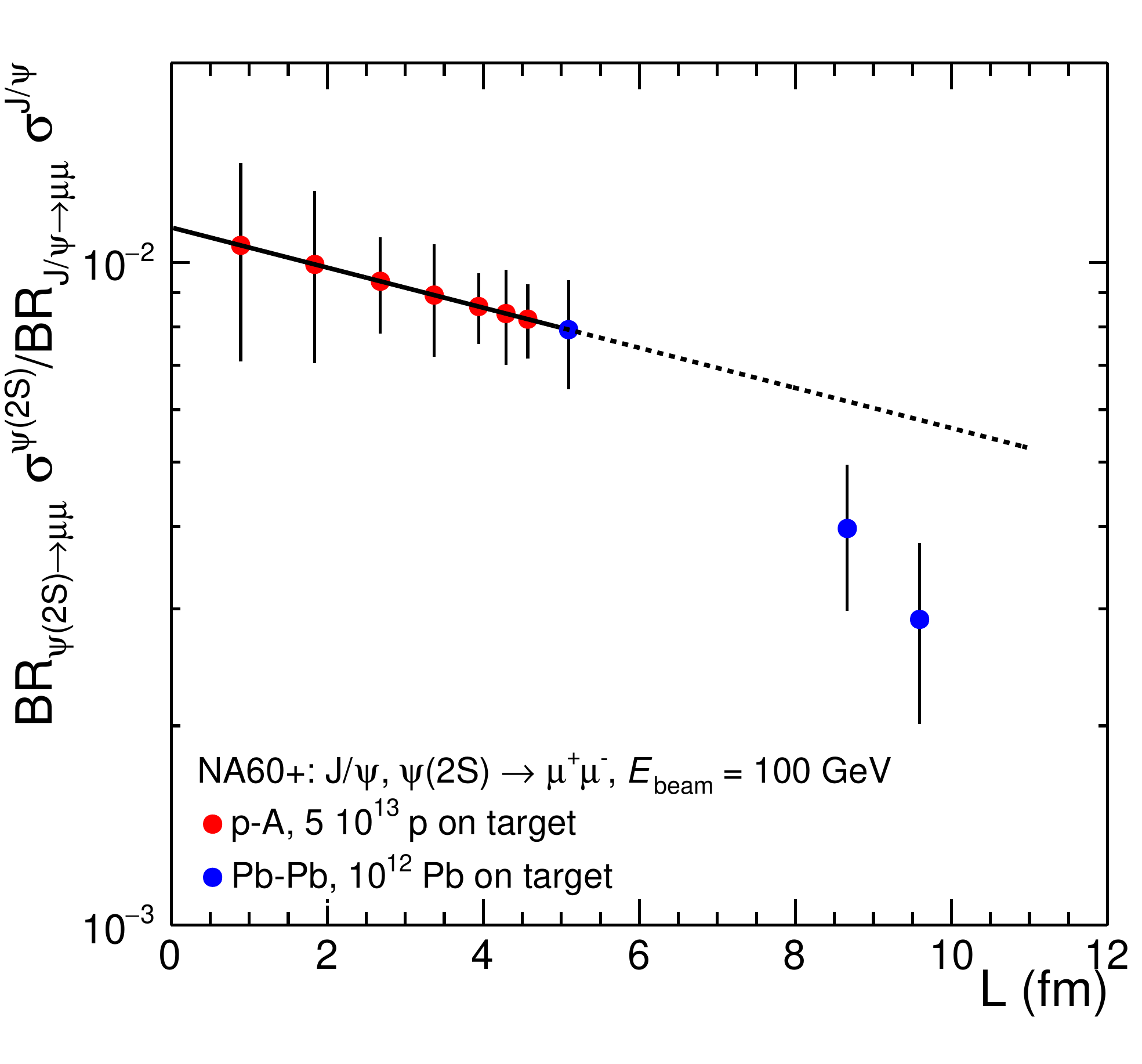}
\caption{The ratio between the \psiP and \jpsi production cross sections, not corrected for the branching ratios to dimuons, for \pA and \PbPb collisions at $\Elab = 100\GeV$, as a function of $L$ (see text for details).}
\label{fig:psi2s}
\end{center}
\end{figure}

The study of the \psiP meson can also be performed analysing the invariant-mass spectrum of muon pairs.
A performance study was carried out, calculating the ratio between the \psiP and \jpsi yields in \pA and \PbPb collisions. A systematically stronger suppression of the \psiP with respect to the \jpsi, increasing from \pA to \PbPb , is assumed in this study. This behaviour was discovered by NA50 at the top SPS energy~\cite{Alessandro:2006ju} and is expected to hold also at lower energies. However, no quantitative model calculations exist in this regime.
The results can be presented as a function of the variable $L$, corresponding to the mean thickness of nuclear matter crossed by the mesons or by their pre-resonant states, and can be calculated in the frame of the Glauber model.
In Fig.~\ref{fig:psi2s}, assuming the same beam intensity and running time as for the \jpsi studies, and considering $\Elab = \SI{100}{\AGeV}$, the decrease of the ratio \psiP/\jpsi between \pA and \PbPb can be detected. Clearly, the quantitative amount of such a decrease has to be considered for the moment as an educated guess, so that the result shown in Fig.~\ref{fig:psi2s} is essentially meant to demonstrate the expected statistical accuracy of this measurement.
A measurement at even lower energies would require significantly larger beam intensities and/or running times.

Corresponding studies for the detection of  the $\chic\rightarrow\mumu\gamma$ decays are ongoing and are expected to complement the physics programme of NA60+ in the charmonium sector.

\newpage

\section{Detectors and systems}
\label{Detectors}
\vskip 0.4cm

In this section the current choices for the detector technologies to be used for the main elements of the experimental set-up are discussed, together with the status of the corresponding R\&D studies. In Sec.~\ref{FLUKArate} we describe the results of simulations, carried out with the FLUKA event generator, to determine the particle fluence in the various detector elements, a crucial information for the definition of their main features. In Sec.~\ref{targetsystem} we propose a technical solution for the target system, while in Sec.~\ref{dipolemagnet} the characteristics of the MEP48 dipole magnet are briefly described. Section~\ref{sec:vertex_telescope} describes in detail the specifications, the technology choice, the R\&D plans, the cooling studies and the foreseen mechanics for the vertex spectrometer. Then, in Sec.~\ref{sec:muonspectrometer} the main characteristics of the muon spectrometer are detailed, including two possible technology choices for the tracking detectors and the description of corresponding facilities that could manage their construction. In Sec.~\ref{Toroid} the foreseen features of the toroidal magnet, as well as the results of studies carried out on a first working prototype, are discussed. Finally, in Sec.~\ref{daq}, preliminary considerations and estimates on the data acquisition and processing are shown.

\subsection{FLUKA rate calculations}
\label{FLUKArate}
\vskip 0.2cm

For the choice of the technology to be used for the tracking detectors and for the estimate of the required granularity it is mandatory to perform an accurate evaluation of the expected charged-particle rates. For the vertex spectrometer, in addition to the secondary hadrons produced in inelastic Pb--Pb interactions, a non-negligible contribution comes from $\delta$ rays. The muon spectrometer detectors are ``protected'' by the hadron absorber, but the contribution of punch-through hadrons and background muons needs to be evaluated as well. In particular, one has to make sure that the central high-density plug of the absorber can efficiently stop the hadronic cascade generated by those Pb ions ($\sim$ 85\%) that do not interact in the targets.

\begin{figure}[ht]
\begin{center}
\includegraphics[width=1.\linewidth]{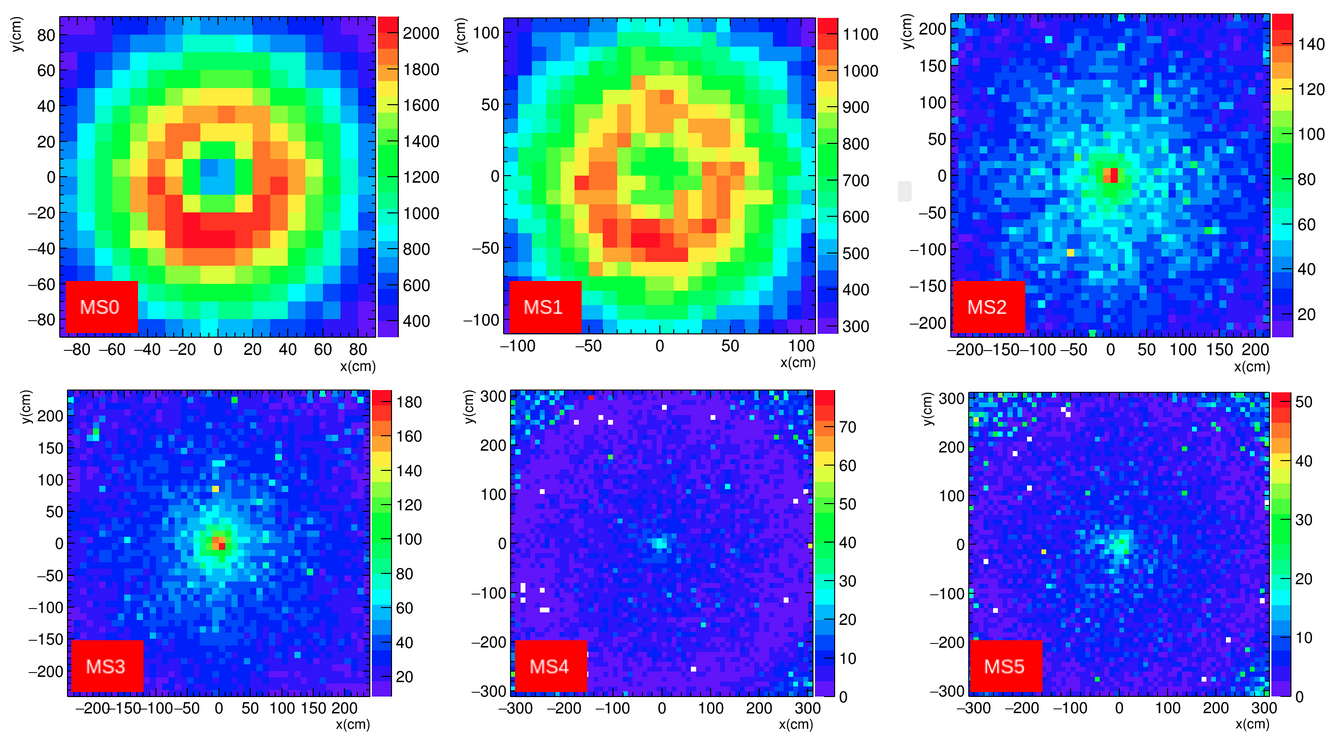}
\caption{Number of charged particles per cm$^2$s crossing the muon tracking chambers, assuming a 10$^6$ s$^{-1}$ 40 AGeV Pb beam. The shaded central area represents the region outside the acceptance of the detectors. The  MS0 and MS1 stations are positioned upstream of the toroidal magnet, while MS2 and MS3 are located downstream of the magnet. Finally MS4 and MS5 are installed downstream of the graphite muon wall.}
\label{fig:flukams}
\end{center}
\end{figure}

Making use of the scoring options provided by \fluka (see Sec.~\ref{flukastudies}), the charged particle fluence was evaluated at the z-axis position of the various detector elements, by shooting Pb ions of 40 AGeV and 160 AGeV inside the low- and high-energy set-ups of the experiment, respectively. As an example of the results, we show in Fig.~\ref{fig:flukams} the number of charged particles per cm$^2$s crossing the set-up at the z-position of the muon tracking chambers, assuming a 10$^6$ s$^{-1}$ Pb-beam intensity. One can see that the maximum rate does not exceed 2 kHz/cm$^2$ at the position of the muon tracking station closer to the end of the hadron absorber. This value becomes lower for the downstream stations, and in particular for those positioned after the toroidal magnet, which sweeps away low-momentum hadrons exiting the absorber. A slight asymmetry between the upper and lower section of MS0 and MS1 can be remarked, generated by the presence of the dipole field in the target region. 

\begin{figure}[ht]
\begin{center}
\includegraphics[width=1.\linewidth]{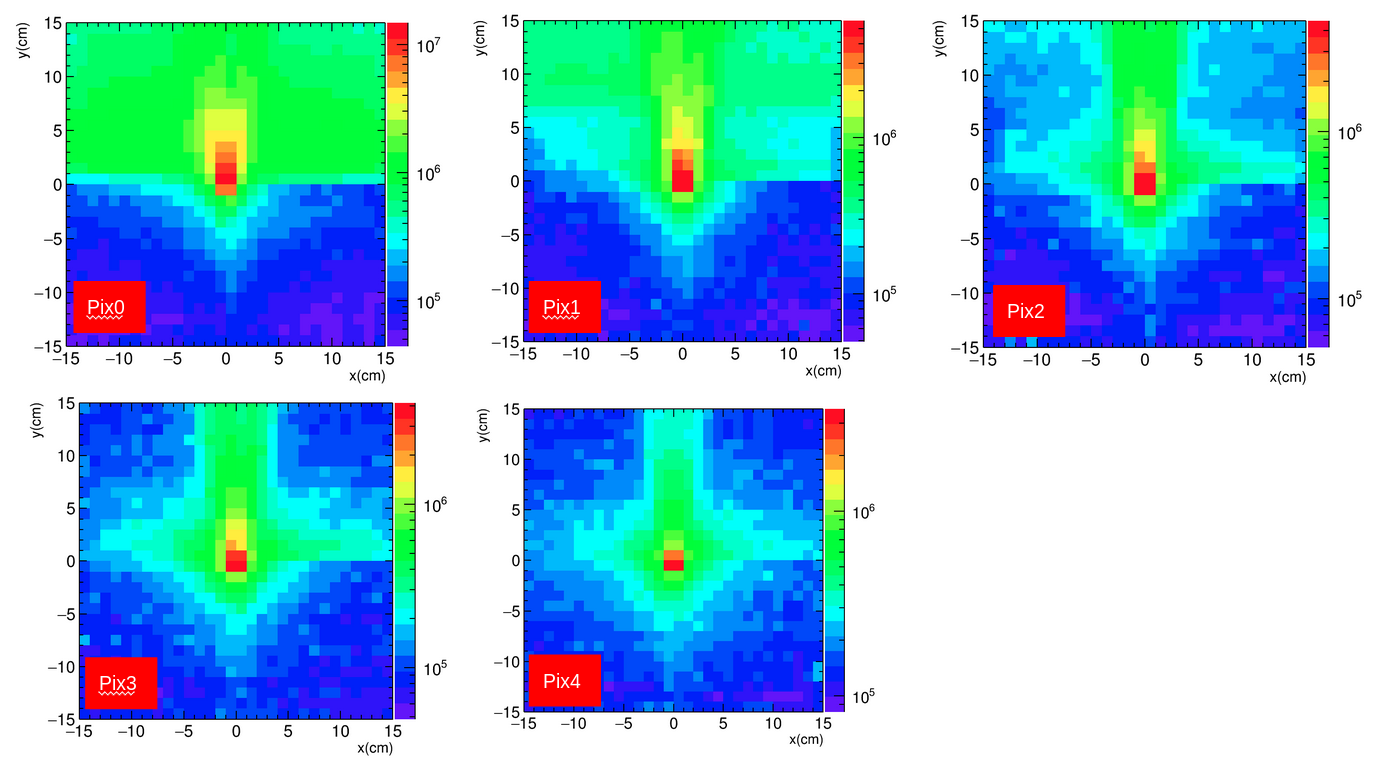}
\caption{Number of charged particles per cm$^2$s crossing the tracking planes of the vertex spectrometer, assuming a 10$^6$ s$^{-1}$ 40 AGeV Pb beam. }
\label{fig:flukavt}
\end{center}
\end{figure}

A similar evaluation was performed for the vertex spectrometer region, where the charged-particle rates are obviously much larger. In Fig.~\ref{fig:flukavt} we show, at the foreseen positions of the five tracking stations, the number of charged particle per unit surface and time, again assuming a 10$^6$ s$^{-1}$ Pb-beam intensity. A clear up-down asymmetry can be seen, with the larger fluence at positive y-values related to $\delta$-rays that are swept by the dipole field. Values of $\sim$10$^6$ cm$^{-2}$s$^{-1}$ are reached over a significant fraction of the detector surface.

\subsection{Target system}
\label{targetsystem}
\vskip 0.2cm
As introduced in Sec.~\ref{TargetSystem}, for the heavy-ion runs, the target system will be composed of five 1.5 mm thick Pb disks spaced by 12 mm. The first one has a radius of 3 mm, 
and has reference holes, while the other targets have 1 mm diameter. Due to their small transverse size, they should be aligned with respect to the beam axis with a precision of $\sim100$ $\mu$m.

For proton--nucleus runs, the system will be composed of a number of sub-targets of different nuclear species like Be, Cu, In, W and Pb, simultaneously exposed to an incident proton beam. The sub-targets will have a diameter of \SI{\sim1}{\mm} with a spacing of 12 mm.

\begin{figure}[ht]
\begin{center}
\includegraphics[width=0.45\linewidth]{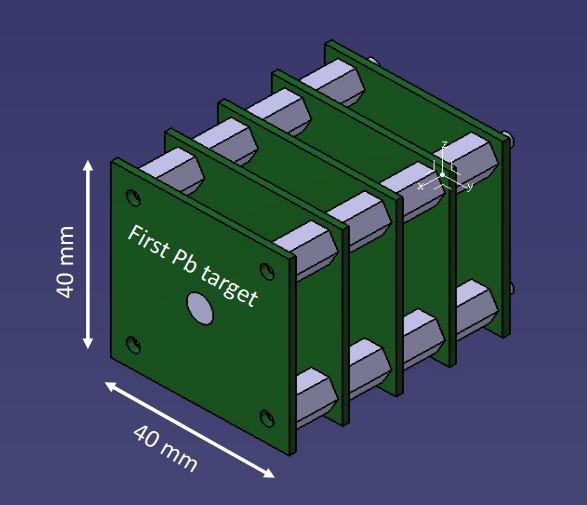}
\caption{Sketch of the NA60+ target system (Pb--Pb collisions set-up).}
\label{fig:Targetsystem}
\end{center}
\end{figure}

The mechanical solutions foreseen are the same for both collision systems. More in detail,  each Pb target will be fixed at the center of an FR4 panel, with the various panels being connected by aluminum bolts, as shown in Fig.~\ref{fig:Targetsystem}. The target ensemble will then be fixed to an FR4 panel with a large central hole, placed in front of the vertex telescope frame. As an alternative solution, the target disks can be held in the center of larger diameter holes in the FR4 panels by means of 100 $\mu$m diameter tungsten wires, copper-soldered on the surface of frames.
The ensemble composed by the target system and the vertex telescope frame will be then mounted on non-magnetic linear guides, allowing the system to slide away if interventions will be necessary. The guides will be mounted on the dipole magnet structure, with the mechanical fixation inside the magnet gap still to be worked out.

\subsection{Dipole magnet}
\label{dipolemagnet}
\vskip 0.2cm

\begin{figure}[ht]
\begin{center}
\includegraphics[width=0.55\linewidth]{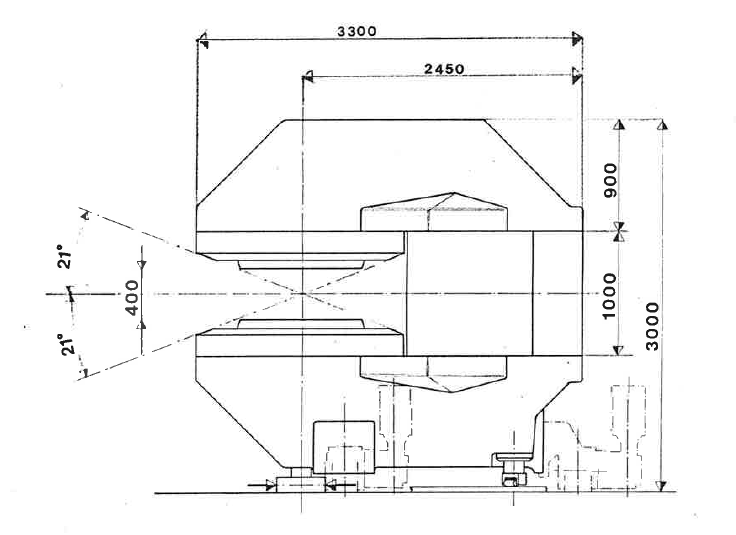}
\includegraphics[width=0.4\linewidth]{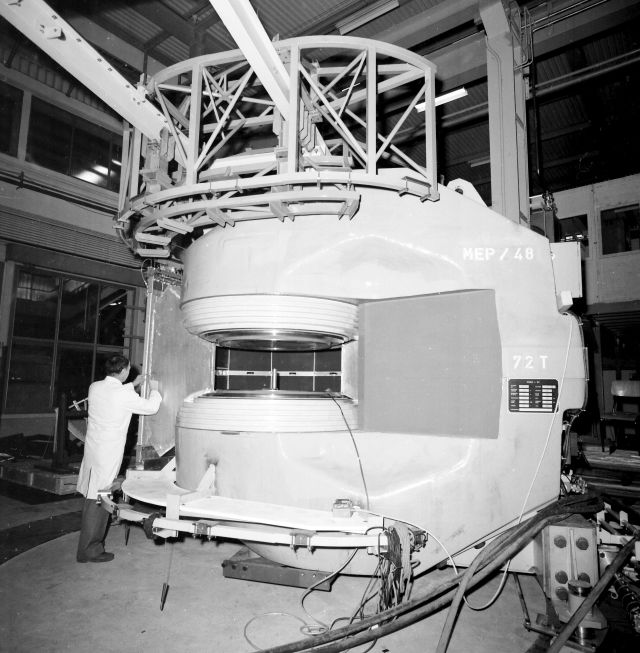}
\caption{The MEP48 dipole magnet.}
\label{fig:MEP48}
\end{center}
\end{figure}

The magnetic field for the momentum measurement in the vertex spectrometer will be provided by a dipole magnet. The PT7 magnet~\cite{Banicz:2001}, used by the former NA60 experiment, provided a rather intense field (2.5 T) but had a rather narrow gap (106 mm), which would make it unsuitable for the wide angular coverage needed at low energy by NA60+. The current choice is the MEP48 magnet, originally built for the PS170 experiment, which can deliver a smaller field (1.5 T) over a much wider gap (400 mm). A sketch and a photograph of the magnet can be seen in Fig.~\ref{fig:MEP48}. A preliminary survey of the magnet, presently stored in building 190 at CERN, has shown that a refurbishment, due to a short-circuit in the lower expansion, will be necessary. 

\subsection{Vertex telescope
}
\label{sec:vertex_telescope}
\begin{figure}[ht]
    \centering
    \includegraphics[width=\textwidth]{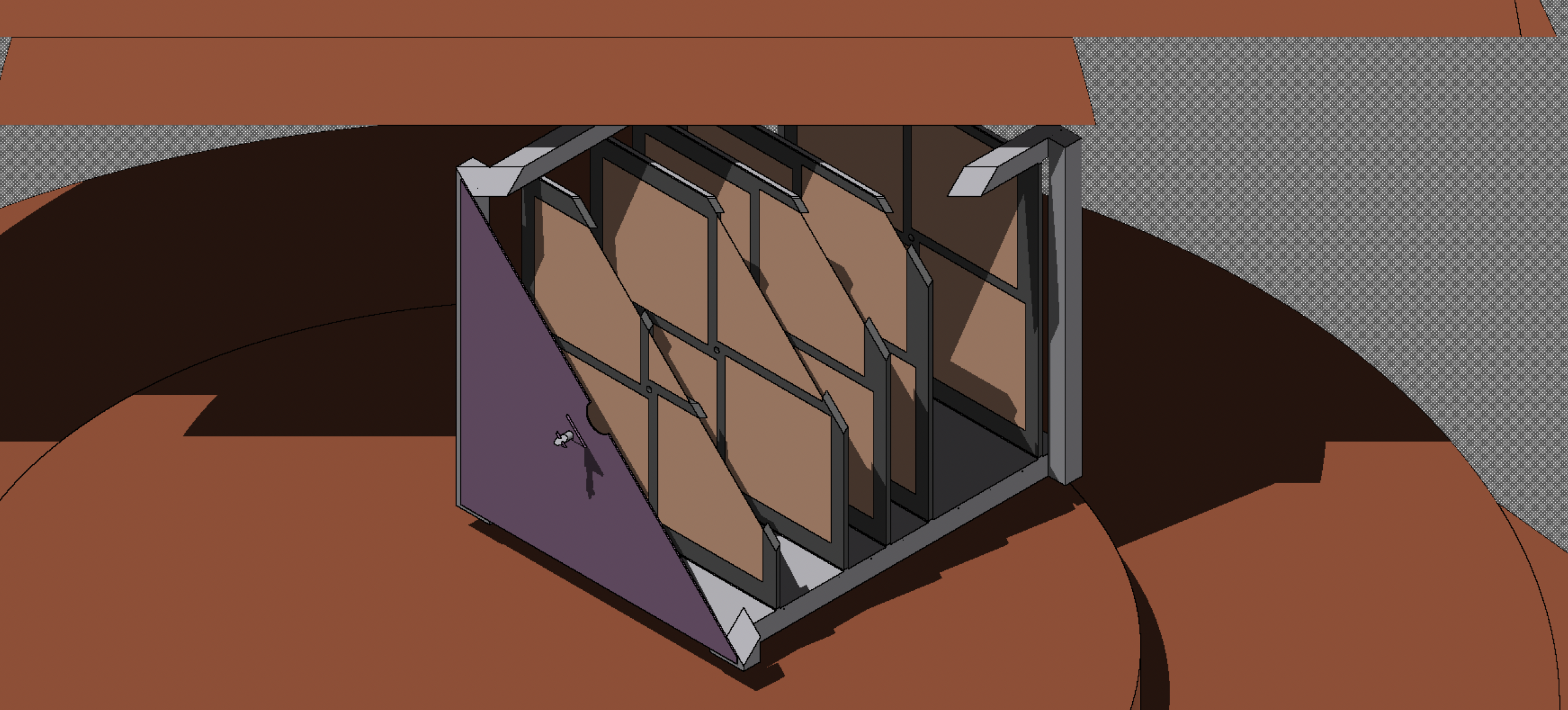}
    \caption{Exploded prospective view of the silicon pixel planes inside the vertex telescope. Each plane is formed by 4 large area MAPS and is housed in a aluminum crate. The target system sits in front of the crate.}
    \label{fig:detectors:vertex-telescope-fig2}
\end{figure}

\vskip 0.2cm
The performance studies detailed in this document are based on a vertex telescope consisting of 5 identical silicon pixel  planes positioned at $7<z<38~\si{\cm}$ starting from the most downstream target.
The absorber  starts at \SI{\sim45}{cm} from the last target, providing a good rejection of background muons from pion and kaon decays.
The planes are immersed in the 1.5 T dipole field of MEP48, providing a field integral of about 1.2 Tm.
Each plane, featuring a material budget of 0.1\% X$_0$ and intrinsic space resolution of $\sim$5 $\mu$m,  is formed by 4 large area monolithic pixel sensors of 15x15 cm$^2$ each.
 The total active area is $\sim$ 0.5 m$^2$.
An exploded view of the silicon telescope together with the target system integrated inside MEP48 is shown in 
Fig.~\ref{fig:detectors:vertex-telescope-fig2}. 

\subsubsection{Specifications}

{\bf Material budget} The vertex telescope is conceived as silicon sensors without support frame in most of the active area, targeting a 0.1\% radiation length per plane.

{\bf Intrinsic spatial resolution} The planes of the vertex telescope will provide intrinsic position resolutions of 5 $\mu$m. Assuming a cluster size of one pixel, this translates to pixel pitches around 20 $\mu$m.

{\bf Hit time resolution} Considering the beam intensity of $10^6$ ions/s, the sensors must provide a (rms) time resolution of $\sim200$ ns.

{\bf Rate capability} The sensors in the most exposed region of the vertex detector must be able to read a maximum rate, in the most exposed region of sensors around the beam hole, of 5-10 MHz cm$^{-2}$ in order to record all interactions in triggered mode.

{\bf Data throughput} Assuming an encoding with 25 bytes/hit and a fake hit rate of $\sim10^{-8}$ per pixel per event, the total expected data rate is $\sim$ 8-17 Gbits/s in case of a triggered readout mode (50 and 80\% centrality selection, respectively). With a triggerless readout mode the expected data throughput is $\sim$ 26 Gbits/s.

{\bf Power consumption and powering scheme} In order to keep the material thickness within the specification, the power consumption of the sensors should not exceed $\sim70$ mW cm$^{-2}$.

{\bf Radiation hardness} The maximum radiation load per operational year (1 month of data taking) will be $\sim 10^{13}$ 1 MeV n$_{\rm eq}/$cm$^2$ on the first plane, starting at a radial distance of 3 mm from the beam axis. Considering a lifetime of a decade the maximum radiation load is $\sim 10^{14}$ 
n$_{\rm eq}/$cm$^2$ .

\subsubsection{Pixel sensor}

CMOS monolithic active pixel sensors (MAPS) are the default choice for the vertex telescope.
In MAPS, the sensing volume and readout electronics are in the same silicon wafer. The detection volume is typically a very thin (20 $\mu$m or so) epitaxial layer grown over a
standard substrate silicon wafer, processed in commercial CMOS technologies. Chips are routinely
thinned down to 50 $\mu$m with pixel sizes down to 20$\times$20 $\mu$m$^2$ or even less, providing very good
performance in terms of material budget and spatial resolution.

The state-of-the-art MAPS is the ALPIDE chip developed for  the Inner Tracking System upgrade of the ALICE experiment at CERN~\cite{Abelevetal:2014dna,AGLIERIRINELLA2017583}.
It is a matrix of approximately \num{500000} pixels, each measuring 27$\times$29 $\mu$m$^2$ and arranged in 512 rows and 1024 columns, for a total active area of 30$\times$15 mm$^2$.
The sensor is based on the TowerJazz 180 nm CMOS imaging process, which offers full flexibility for CMOS design of complex in-pixel circuits.

CMOS imaging processes are offering the possibility to design large area sensors with widespread applications by means of the {\it stitching} technology.
In the next years, this same technology will allow very large area sensors for particle physics experiments to be produced.
This will pave the way to the construction of cheaper and high performance large area silicon trackers, simplifying the system integration aspects and reducing the material budget considerably.
For these reasons, we propose to base the NA60+ silicon tracker on MAPS exploiting the stitching.

\begin{figure}[ht]
    \centering
    \includegraphics[width=\textwidth]{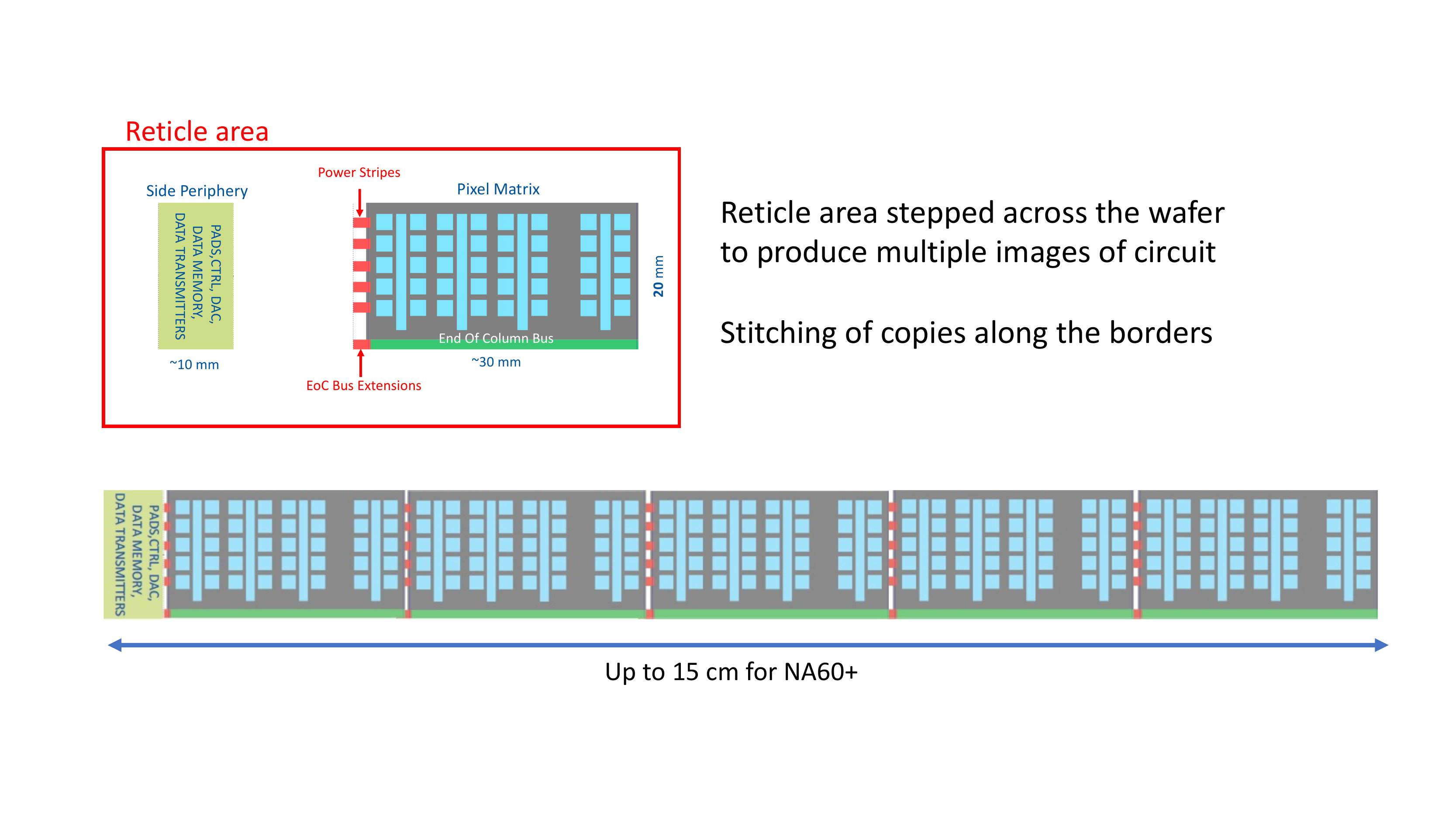}
    \caption{Stitching principle and stitched sensor proposed for NA60+}
    \label{fig:detectors:vertex-telescope-fig3}
\end{figure}

Stitching  allows the fabrication of an image sensor that is larger than the field of view of the lithographic equipment.
In this way, sensors of arbitrary size can be manufactured, the only limit being the wafer size.
The basic unit is a sensor of size 1.5$\times$15 cm$^2$  as shown in Fig.~\ref{fig:detectors:vertex-telescope-fig3}~\cite{ALICE-PUBLIC-2018-013}, obtained replicating several times a sub-sensor unit of size similar to ALPIDE.

\begin{figure}[ht]
    \centering
    \includegraphics[width=\textwidth]{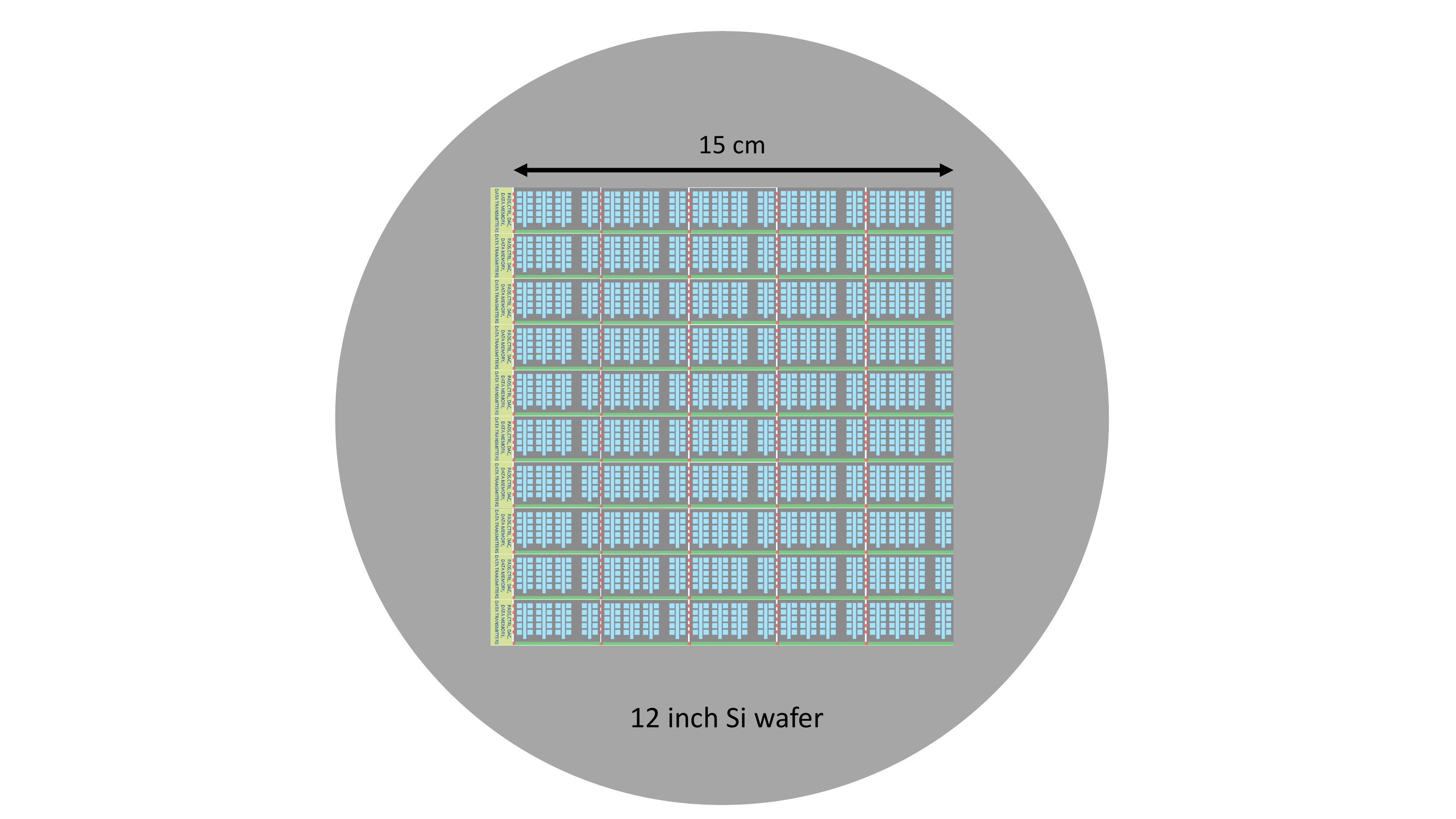}
    \caption{15x15 cm$^2$ sensor obtained replicating the stitched sensor vertically 10 times in a 12" silicon wafer.}
    \label{fig:detectors:vertex-telescope-fig4}
\end{figure}

The periphery, shown in the left part of the sensor in Fig.~\ref{fig:detectors:vertex-telescope-fig3}, contains the control logic to steer the priority encoders, the interfaces for the configuration of the chip and serial data transmitters.
The partitioning of the sensor in different regions, memory buffers and serial transmitters is determined on the basis of the  data throughput.
Since the periphery would be outside the acceptance, there is no particular constraint on its size, which can be adjusted on the basis of the requirements.

The 15$\times$15 cm$^2$ sensor is obtained replicating the stitched sensor chip several times on a silicon wafer as shown in Fig.~\ref{fig:detectors:vertex-telescope-fig4}. Such a matrix contains $\sim40\times 10^6$~pixels.

\subsubsection{Required R\&D for the pixel sensor}
Most sensor requirements can be met with present technologies.
The required R\&D is presently performed in synergy with the ALICE experiment for the development of a stitched sensor for the ALICE ITS3 project.
The 65 nm TowerJazz technology is under test on small scale pixel matrices in the lab and in test beams. The results are encouraging. The first stitched sensor prototype, with the aim to study the yield of the process, is scheduled at the end of 2022.

\subsubsection{Mechanics and interconnections}

The 30$\times$30 cm$^2$ stations will be realized by 4 wafer-scale sensors. For the stations positioned closest to the targets, a fraction of the outer region of the sensor area will be outside of the acceptance of the experiment. However, this drawback is compensated by the relatively simple and modular design, allowing the replacement of non-working stations to be performed easily.

\begin{figure}[h!]
    \centering
    \includegraphics[width=\textwidth]{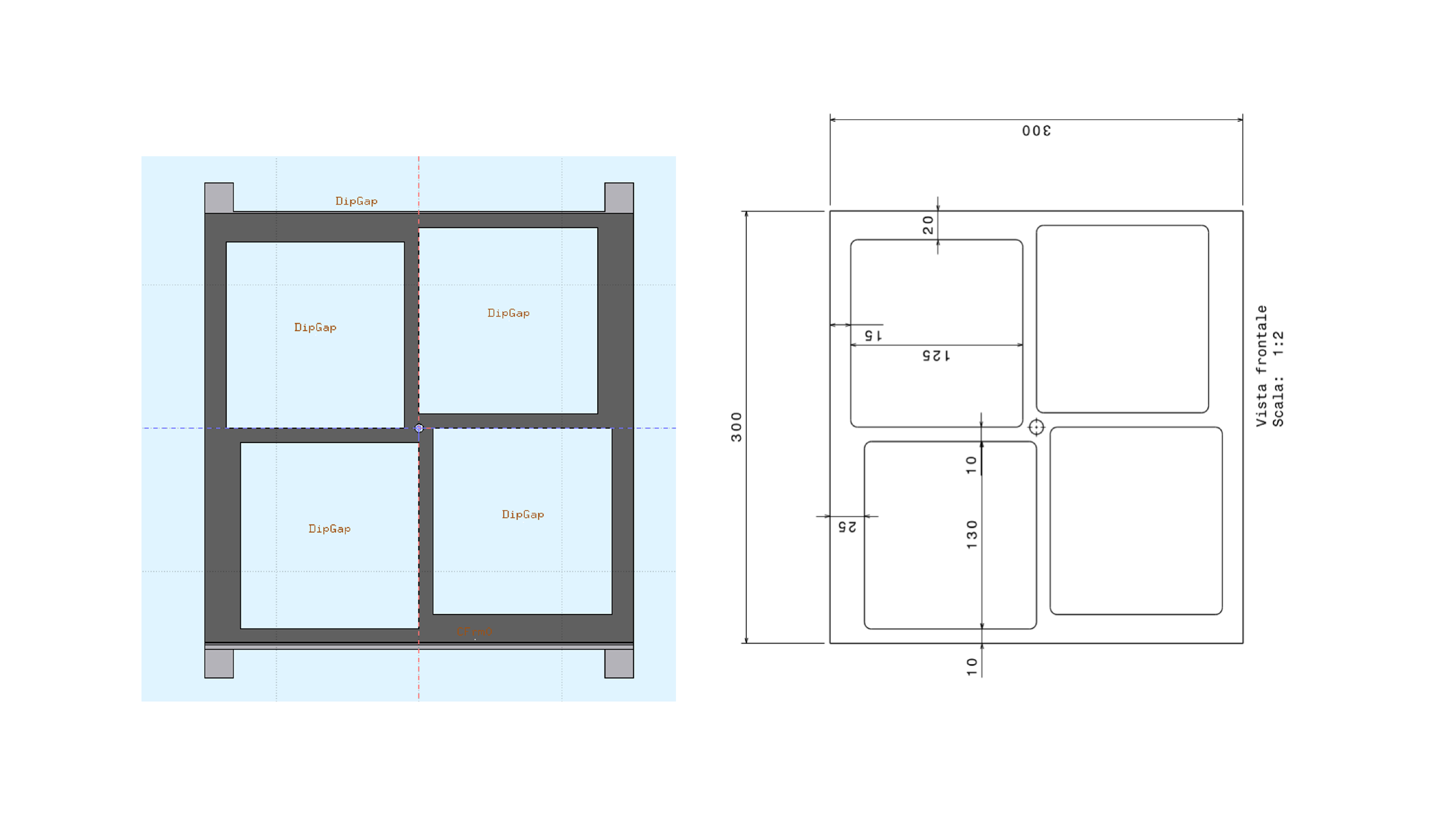}
    \caption{Detail of the carbon foam/fiber support frame on which 4 MAPS are glued. In the left panel it is shown inserted inside the aluminum crate.}
    \label{fig:detectors:vertex-telescope-fig5}
\end{figure}

\begin{figure}[h!]
    \centering
    \includegraphics[width=\textwidth]{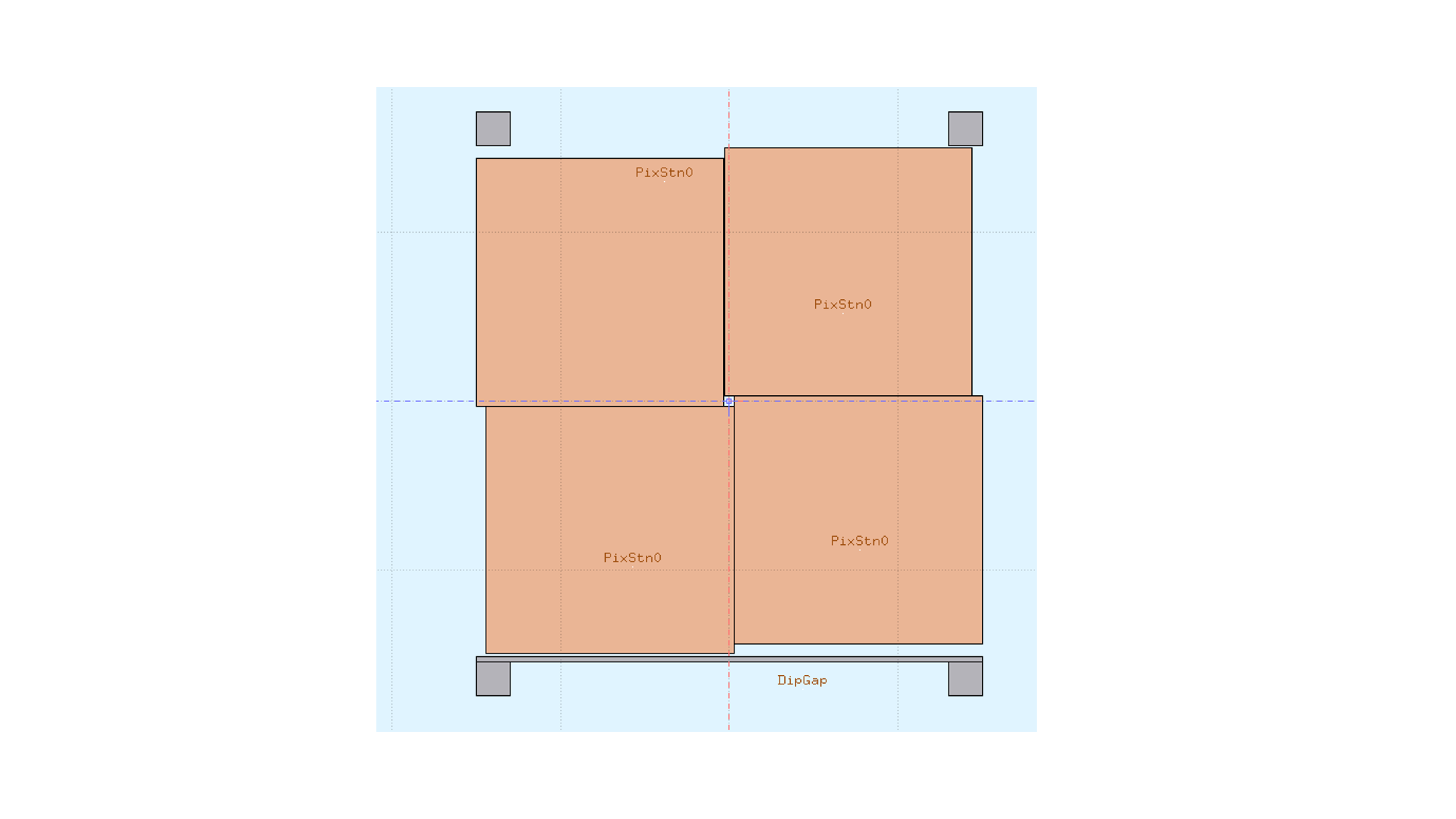}
    \caption{Arrangement of the 4 sensors, to form a detection plane.}
    \label{fig:detectors:vertex-telescope-fig6}
\end{figure}

An outer frame of carbon foam, reinforced with carbon fibre if needed, runs along the borders and is glued to the area of the digital periphery of the sensors.
Vertical and/or horizontal strips of carbon foam (the cross structure in Fig.~\ref{fig:detectors:vertex-telescope-fig5}) are glued to  the sensors surface close to their borders to provide rigidity to the central zone of the sensors. In order to completely eliminate dead zones, the sensors might be mounted  in a slightly overlapping arrangement, by gluing one pair of sensors on one side of the frame, the other pair on the other side. 
The arrangement of 4 sensors glued on the support frame is shown in
Fig.~\ref{fig:detectors:vertex-telescope-fig6}. They are positioned in such a way to leave a central 6 mm square hole for the beam passage.

The sensor periphery and interface pads will be all located on one edge.
On this edge, the chip is glued over a length of about a few mm to a flexible printed circuit (FPC) to which it is electrically interconnected using aluminium wedge wire bonding.
The FPC extends laterally from the chip edge to  a patch panel, where interconnections to the electrical data cables and power cables are realized.

\subsubsection{Cooling }
Ansys, Fluent \cite{Ansys} and COMSOL 
Multiphysics \cite{Comsol} simulation tools have been used to develop numerical simulations to study the  heat conditions and the required cooling for a vertex detector plane. 
\begin{figure}[h]
    \centering
    \includegraphics[width=\textwidth]{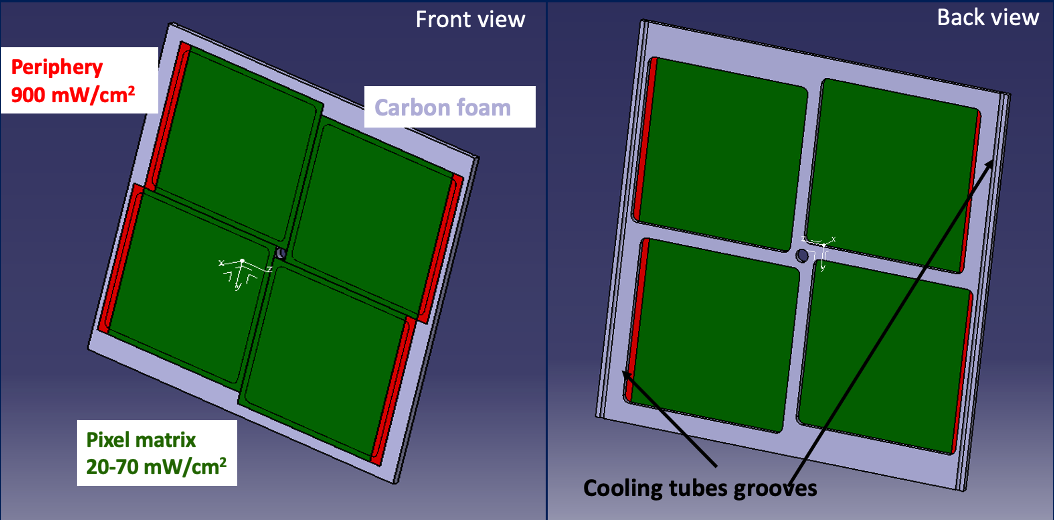}
    \caption{Vertex telescope plane scheme used for the simulations composed by a carbon foam frame on which 4 silicon sensors are placed (left). The sensor periphery is shown in red, the pixel matrix in green. The cooling tubes grooves are placed on the back of the frame near to the sensors periphery (right).}
    \label{fig:TelescopeScheme}
\end{figure}
The structure modelled for the simulations is shown in Figure \ref{fig:TelescopeScheme}. It consists of a frame of carbon foam on which  4 silicon sensors are placed. The carbon foam frame is 300$\times$300 mm$^2$, while the silicon sensors are composed by two parts: one is the pixel matrix having a surface of 130$\times$135 mm$^2$, the other one is the periphery with a surface of 10$\times$135 mm$^2$. For the pixel matrix, several values of power dissipation were considered, from \SI{20}{\milli\watt\per\cm\squared} to \SI{70}{\milli\watt\per\cm\squared}, while the periphery has a power consumption of \SI{900}{\milli\watt\per\cm\squared}. The back-side of the carbon frame hosts the space for the cooling tubes. The tube grooves are placed on the frame side as close as possible to the periphery of sensors, that 
has the highest power consumption.
\begin{figure}[h]
    \centering
    \includegraphics[width=\textwidth]{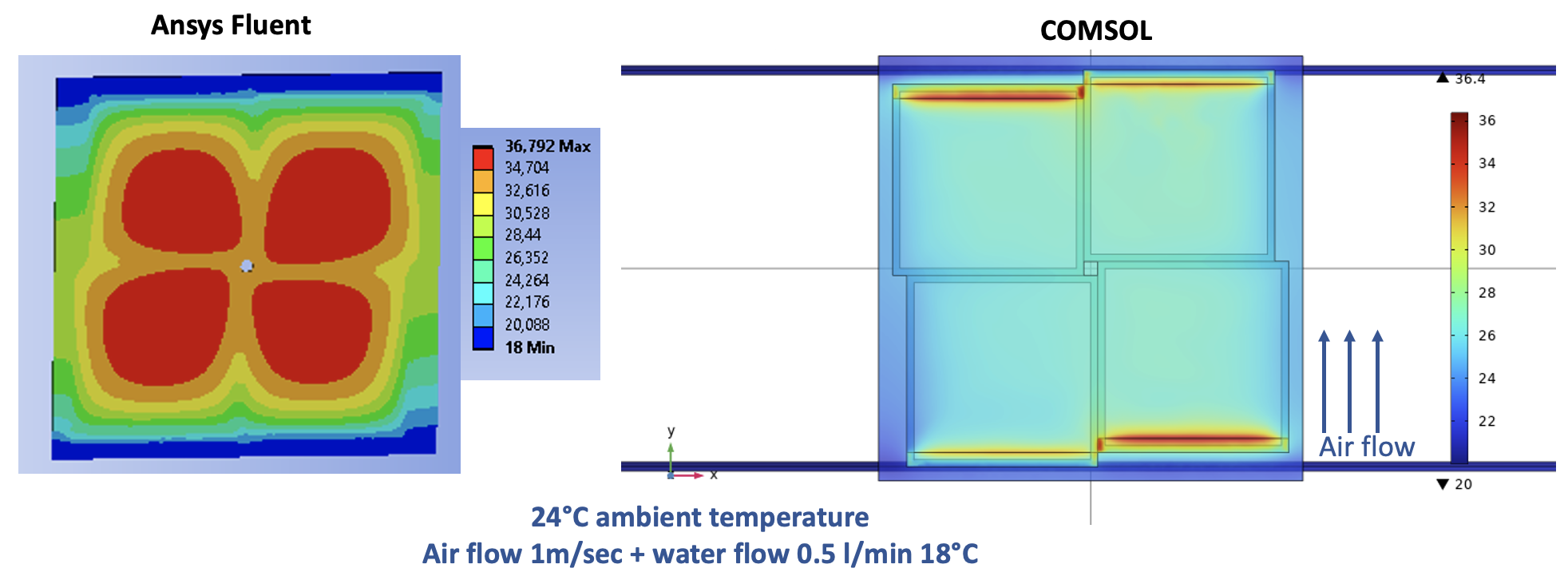}
    \caption{Stabilized temperature profiles with an air flow of 1 m/sec (24°C) and a water (18°C) flow rate of 0.5 l/min. Results obtained with Ansys Fluent (left) and COMSOL Multiphysics (right). The temperature profile assumes a \SI{20}{\milli\watt\per\cm\squared} power dissipation in the pixel matrix.}
    \label{fig:SimulationResult}
\end{figure}
With an air flow of 0.1 m/s (natural air flow) and this power consumption, a maximum temperature of $\sim$ 500 ºC is reached in the periphery of the sensor. Several cooling possibilities were considered to keep the silicon planes temperature below 40 ºC. Simulations show that the best solution is to use an air flow together with a water cooling. Water flows inside copper tubes (with an outer diameter of 3 mm and an inner diameter of 2.6 mm) at a temperature of 18°C and a rate of 0.5 l/min. The temperature and speed of the air flow depends on the power dissipated in the pixel matrix. For the maximum (\SI{70}{\milli\watt\per\cm\squared}) and minimum (\SI{20}{\milli\watt\per\cm\squared}) values of the considered dissipated power, an air flow at 3 m/s and 17 °C and an air flow at 1 m/s and 24°C (ambient temperature) respectively are required to maintain the temperature of the entire sensor below 40 °C. An example of temperature profiles obtained from the simulations with Ansys Fluent and COMSOL are shown in Figure \ref{fig:SimulationResult}. The two simulation tools are in agreement for the maximum temperature reached which is $\sim$ 37°C. On the other hand, the  temperature profiles are different because the two simulation tools handle the geometry and divide it into finite volumes to solve the heat transfer problem in different ways. 

\subsubsection{Readout electronics}
At present  we assume to adopt the ALICE data acquisition chain developed for the LHC Run 3~\cite{Antonioli:1603472}, which is shortly summarized in the following. The interface between sensors and the ALICE readout chain is the ALICE ITS Readout Unit (RU), which is an FPGA-based board for control, trigger and readout.  Each stiched sensor could be connected to one RU, which can control up to 9 data serial links. The RUs receive commands and deliver data directly from/to the Common Readout Unit (CRU), via optical links. Each RU has it own power unit for power distribution. Eight RUs are connected to each CRU. The CRU is a PCIe board which acts as an interface to the ALICE Run 3 Data Acquisition chain. The CRU is based on high performance
FPGA processors equipped with multi-gigabit optical inputs and outputs. It also acts as the interface to the Detector Control System (DCS) and the trigger system. The CRU is hosted in commercial servers called the First Level Processor (FLP) - each can host up to 2 CRUs. The accumulated data from different FLPs are then shipped to another commercial server called the Event Processing Node (EPN) which processes the data online to make them analysis-ready. 

\subsubsection{Cost estimate}

The development and R\&D of the pixel sensor will profit of the synergy with the ITS3 project of ALICE, which has a similar timescale. The cost estimate is based on the 65 nm option.  For the final sensor we plan an  engineering run for a fully functional prototype and foresee the possibility of a second run containing possible optimizations (2$\times$600 kCHF). 
The wafer post-processing, in particular the thinning of the wafer to about 30 $\mu$m, will require a total investment of the order of 300 kCHF.
The interconnection of the sensor to
the FPC can be based on  conventional aluminum wedge wire bonding. This activity is estimated to cost 200 kCHF.
The detector mechanical support structures, the tooling for
the assembly and installation of the detector and the cooling system will require a total investment of about 200 kCHF.
Data copper cables, power-cable patch panels in front of the detector,  needed to match the mechanical layout, have an estimated cost of the order of 300 kCHF.
The cost of the readout and power distribution system based on the ALICE ITS is estimated to be of the order of 900 kCHF. The summary of the costs for the vertex telescope is reported in Table~\ref{VT_cost_table}. 

\begin{table}[h]
\caption {The cost breakdown structure for the vertex telescope.}
\centering
\begin{tabular}{|l|l|}
\hline
                   & kCHF       \\ \hline
Engineering runs        & 600-1200   \\ \hline
Wafer post-processing   & 300       \\ \hline
FPC and wire bonding            & 200        \\ \hline
Mechanical support      & 200        \\ \hline
Cables, patch panels                  & 300        \\ \hline
Readout and power distribution  & 900 \\ \hline
\textbf{TOTAL}          & 2500-3100       \\ \hline
\end{tabular}

\label{VT_cost_table}
\end{table}

\subsubsection{Lab facilities and experience}

INFN Cagliari, Padova and Torino were involved since the very beginning in the design and testing of the ALICE ALPIDE sensor, and the design and assembly of mechanical structures for the ALICE ITS.
At present the teams are involved directly also in the R\&D of the stitched MAPS within the ALICE ITS3 project.
INFN has vast amount of lab facilities that allow performing sensor testing as well as assembly of detectors. Additionally, there is availability of clean rooms with an area of $\sim$50 m$^2$ for the assembly of pixel stations.

\subsection{Muon spectrometer
}
\label{sec:muonspectrometer}
\vskip 0.2cm

\newcommand*{\sqs}{\ensuremath{\sqrt{s}}\xspace}
\newcommand*{\sqn}{\ensuremath{\sqrt{s_{_{\mathrm{NN}}}}}\xspace}

The primary function of the muon spectrometer is to identify muons among other particles, measure their momentum, and match them to the vertex telescope tracks. This is done by reconstructing charged particle tracks after the absorber and measuring their curvature in the magnetic field of the toroidal magnet.

\begin{figure}[h]
    \centering
    \includegraphics[width=\textwidth]{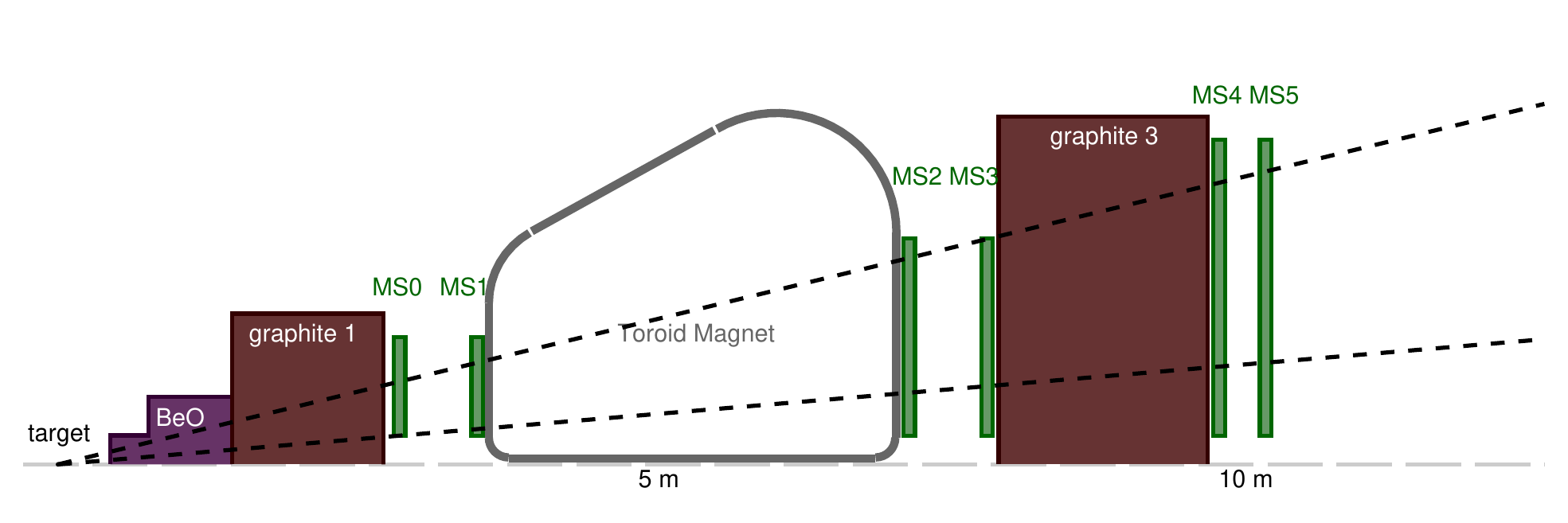}
    \includegraphics[width=\textwidth]{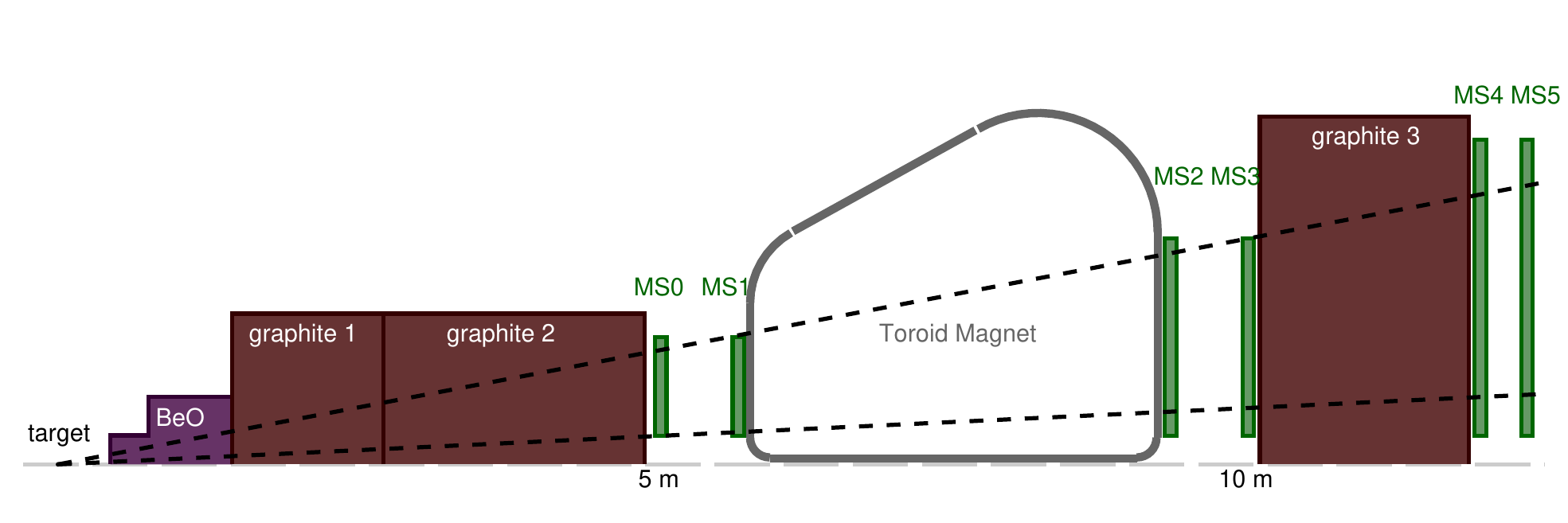}
    \caption{Schematic layout of the muon spectrometer. The top setup is adapted for a beam energy of 20 AGeV or $\sqrt{s_{_{\mathrm{NN}}}}=6.3$ GeV. The setup shown in the bottom includes an extended graphite absorber for high beam energies, here 158 AGeV or $\sqrt{s_{_{\mathrm{NN}}}}=17.3$ GeV. }
    \label{fig:mutracker}
\end{figure}

The schematic setup of the NA60+ muon spectrometer is shown in Fig.~\ref{fig:mutracker}. It closely follows the design of its successful predecessor NA60, but will deploy new and significantly better-performing detector technologies and can be adapted to different beam energies. 

The muon spectrometer is mounted on a rail system that allows for an adjustment of the aperture to the beam energies, by displacing its elements and changing the thickness of the hadron absorber. For smaller beam energies the muon spectrometer is moved closer toward the target such that the aperture in rapidity remains similar and covers at least one unit near mid-rapidity. Details of the $\sqrt{s_{_{\mathrm{NN}}}}$ dependent aperture are given in Tab.~\ref{tab:muon_spectrometer_dimensions}. 

\begin{table}[htb]  
\caption{
Acceptance of the low and high beam energy configurations of the muon spectrometer. Rapidities are given in the laboratory system.\label{tab:muon_spectrometer_dimensions}
} 
\begin{center} 
\begin{tabular}{cccccccccc} 
 configuration    & beam energy & $\sqrt{s_{_{\mathrm{NN}}}}$ & $y_{cm}$ & $\theta_{min}$ & $\theta_{max}$ & $y_{min}$ & $y_{max}$ \\ 
            &  [GeV] &  [GeV] & & [mrad] & [mrad] \\ 
\hline
low energy   & 20   & 6.3  &  1.90   & 84 & 245   & 2.1   & 3.16 \\
high energy  & 158  & 17.3 &  2.91   & 47 & 191   & 2.35  & 3.74 \\
\end{tabular} 
\end{center} 
\end{table} 

The hadron absorber starts immediately after the vertex spectrometer. Its purpose is to stop hadrons from reaching the muon spectrometer, and also to prevent pions and kaons from decaying into muons before they are stopped.
In the current design, the absorber starts 7 cm downstream of the last station of the vertex spectrometer or 45 cm downstream of the last target. The materials for the absorber are selected to have high density and low Z, to minimize multiple scattering of muons. The absorber includes an upstream section composed of BeO with a fixed length of 105 cm. It is followed by a graphite absorber with a length between 130 cm and 355 cm, depending on the setup. Details of the hadron absorber, for the low and high energy configurations, are given in Tab.~\ref{tab:hadron_absorber}. Intermediate set-ups in terms of absorber thickness and position of detector elements can be envisaged.

\begin{table}[h!]
\caption{
Properties of the hadron absorber made from BeO ($X_{\rm 0}=13.7$ cm, $\rho=3.01$ g cm$^{-3}$) and graphite ($X_{\rm 0}=19.3$ cm, $\rho\sim 2.01$ g cm$^{-3}$).  \label{tab:hadron_absorber}
} 
\begin{center} 
\begin{tabular}{cccccccccc} 
 Setting    & \multicolumn{3}{c}{BeO absorber}& \multicolumn{3}{c}{Graphite absorber} \\
  & \multicolumn{2}{c} {thickness} & $E_{loss}(E=10{\rm\ GeV})$ & \multicolumn{2}{c} {thickness} & $E_{loss}(E=10{\rm\ GeV})$ \\
  & [cm]      & [$X_{\rm 0}$] & [GeV] & [cm]  & [$X_{\rm 0}$]   & [GeV] \\
\hline
low energy  & 105  & 7.7 & 0.66   & 130 & 6.7 & 0.63  \\
high energy & 105 & 7.7 & 0.66 & 355 & 18.4 & 1.73 \\
\end{tabular}
\end{center}
\end{table}

The variable thickness of the absorber is a key feature of the NA60+ setup that allows to optimize the setup for different beam energies. The total thickness of the absorber increases with beam energy and the muon spectrometer is moved further away from the target. Therefore the rapidity coverage is shifted to larger rapidity such that the spectrometer covers more than one unit near center-of-mass rapidity at all beam energies, while simultaneously 
providing more absorption length for the more energetic pions and kaons.   

At very large rapidities $\eta >4.2$ ($\theta<30$ mrad), outside of the spectrometer acceptance, a high-density absorber material is used to stop the non-interacting beam ions as well as the spectator nucleons produced in beam-target interactions. This central ``plug'' will be made of tungsten, possibly in a sintered ($\rho> 15$ g cm$^{-3}$) form to avoid machining issues.

The muon spectrometer is azimuthally symmetric around the beam axis. In the muon spectrometer tracks are measured by six tracking layers that are grouped in three sets of two quasi-planar tracking layers. The first and second layers MS0 and MS1 form the first pair, are separated by 67 cm and located in front of the toroidal magnet. The third and fourth layers MS2 and MS3, also spaced by 67 cm, form the second pair that follows the magnet. The last two layers spaced by 40 cm are located downstream of an additional absorber. The last pair is primarily used to identify muons. Locations and dimensions of the tracking chambers are summarized in Tab.~\ref{tab:muon_spectrometer_trackers}. \begin{table}[htb]  
\caption{
Summary of location and dimensions of the muon tracking stations chambers for the high beam energy and low beam energy configurations. Values are approximate and are subject to further optimization.
\label{tab:muon_spectrometer_trackers}
} 
\begin{center} 
\begin{tabular}{ccccc} 
Station & r$_{in}$ [cm] & r$_{out}$ [cm] & z$_{\text{low energy}}$ [cm] & z$_{\text{high energy}}$ [cm]\\ 
\hline
MS0     & 24    & 110   &  294  & 519  \\
MS1     & 24    & 110   &  361  & 586  \\
MS2     & 24    & 195   &  733  & 958  \\  
MS3     & 24    & 195   &  800  & 1025 \\
MS4     & 24    & 280   &  1000 & 1225 \\
MS5     & 24    & 280   &  1040 & 1265 \\ 
\end{tabular} 
\end{center} 
\end{table} 

The muon tracker provides the track's momentum by measuring the difference in the polar angle between the first and the second pairs of layers. This difference is determined by the magnetic field of the toroidal magnet 
$B_{\phi} = 0.75\text{ [T m]}/r$.

The intrinsic resolution with which the momentum of the muon is given is defined by the magnetic field and the radial single-hit resolution of the muon tracker stations. However, the actual resolution of the muon momentum measurement at the target is physically limited by the energy loss straggling in the hadron absorber. The fluctuations of the energy loss exceed the intrinsic momentum resolution of the muon tracker for a single-hit resolution of about 200 $\mu$m, so further improvement of the single-hit resolution does not improve the muon momentum resolution. Thus the muon tracker is  designed to measure space points with a resolution of 200 $\mu$m in the radial direction, which can be easily achieved with Multi-Wire Proportional Chambers (MWPC) or Micro-Pattern Gas Detectors (MPGD). 


The effect of multiple scattering in the absorber is overcome by 
measuring the muon tracks before the hadron absorber with the vertex telescope, which is the key element for the precision measurement of muons. Muon tracks are matched to the tracks measured in the vertex telescope in coordinate and momentum space. Establishing a precise measurement of the muons before the absorber improves the opening angle measurement of muon pairs and thus the pair mass resolution.
With the combination of the muon tracker and vertex telescope, the optimum momentum and mass resolution are achieved, about 8 MeV/c$^2$ for $\omega\rightarrow\mu\mu$ and 34 MeV/c$^2$ for $J/\psi\rightarrow\mu\mu$ (see Sec.~\ref{receffmassres} for more details on the physics performance). In addition, matching the muon tracks to the vertex telescope significantly reduces residual muon background from hadron decays. 

The following subsections first introduce the general layout of the  tracking stations (Sec.~\ref{sec:MSdesign}), two possible technology choices, MPGD (Sec.~\ref{sec:MS-MPGD}) and MWPC (Sec.~\ref{sec:MS-MWPC}) for the tracking stations, and an estimate of the cost (Sec.~\ref{sec:MScost}). Details of the toroidal magnet design are discussed in Sec.~\ref{Toroid}. 
\begin{figure}[h]
    \centering
    \includegraphics[width=0.9\textwidth]{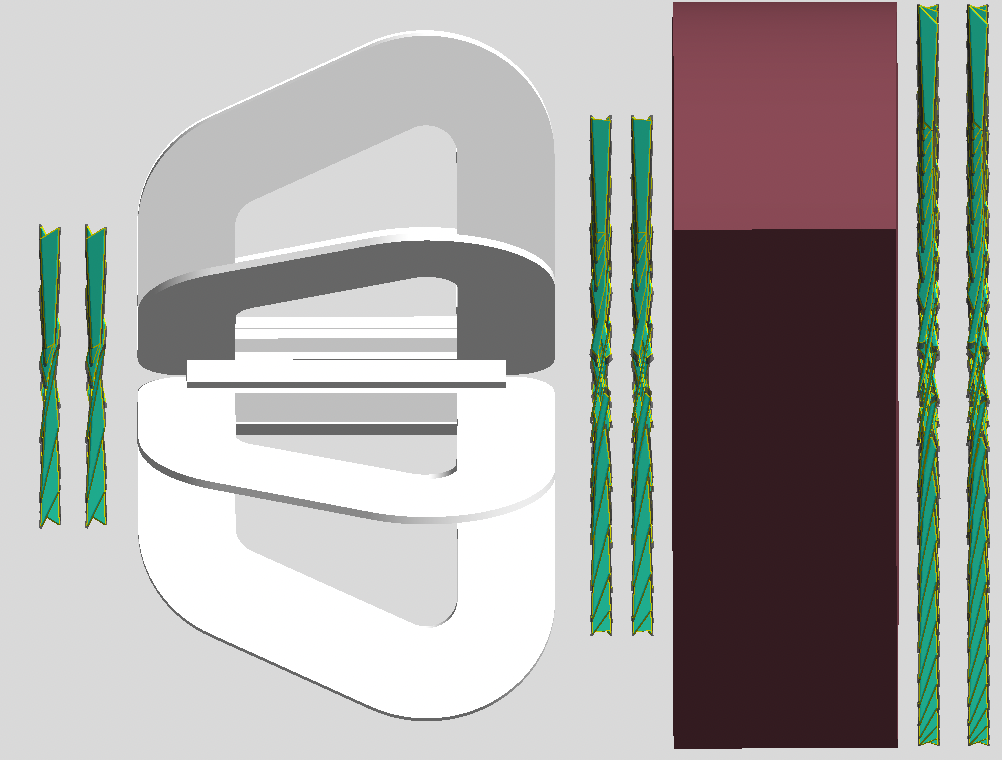}
    \caption{The layout of NA60+ as modelled in the GEANT4 simulation for the low energy set-up with zoom on the muon spectrometer.}
    \label{fig:g4setup}
\end{figure}

\subsubsection{Muon spectrometer tracking stations}
\label{sec:MSdesign}

The main requirements for the muon spectrometer tracking stations are: (i) measuring space points with a resolution in the radial direction of about 200 $\mu$m and (ii) handling a maximum flux of particles of the order of a kHz/cm$^2$ (see Sec.~\ref{FLUKArate}). 
Both requirements can be achieved with MPGDs or MWPCs technologies and a final choice has not yet been made. The 6 muon tracking stations (layers) combined together must cover an area of $\sim 100$~m$^2$. To minimize the production and maintenance costs, the design of the muon tracking stations is based on one standard tracking module that could be built using either MPGD or MWPC technologies. Figure~\ref{fig:g4setup} shows the GEANT4 implementation of the muon spectrometer for the low-energy setup. Currently, the GEANT4 setup includes a detailed implementation of the tracking stations as sensitive volumes and the essential elements of the module design. 

\begin{figure}[h]
    \centering
    \includegraphics[width=0.45\textwidth]{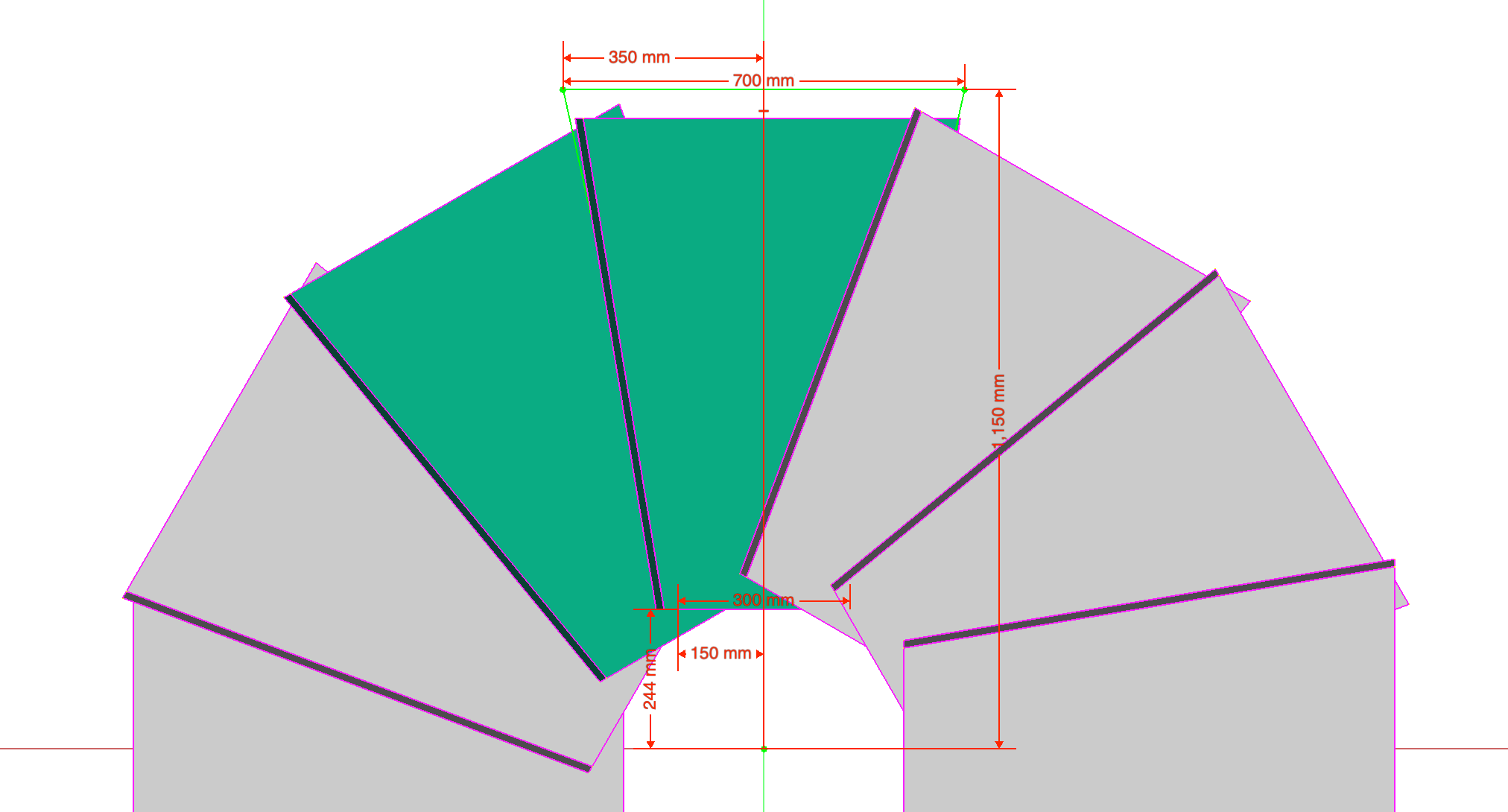}
    \includegraphics[width=0.45\textwidth]{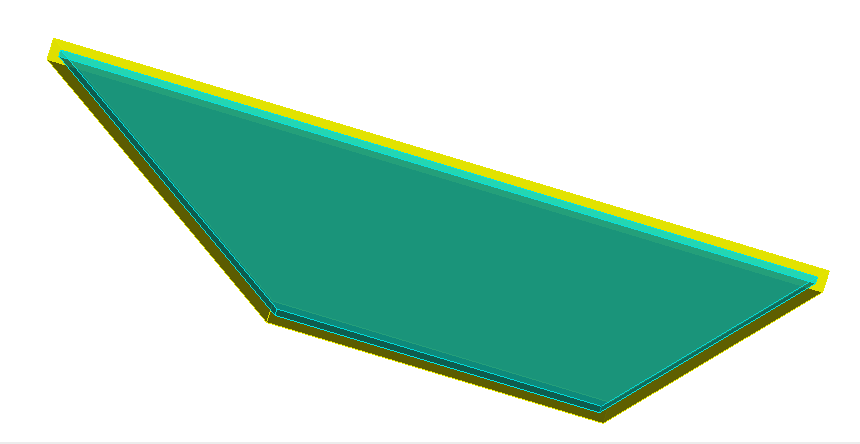}    
    \caption{The front and top view of the detector module. }
    \label{fig:petal}
\end{figure}

All tracking stations are built from standard detector modules, referred to as petals. Each petal has a symmetric trapezoidal shape with an overall height of 90 cm, a base width of 30 cm, and 70 cm at the top (see Fig.~\ref{fig:petal}, left). The petal thickness is 2 cm, with the sensitive layer of 0.6 cm being filled with ArCO$_{2}$ gas (green insert in Fig.~\ref{fig:petal}, right) and with honeycomb panels shown in yellow. 

\begin{figure}[h]
    \centering
    \includegraphics[width=0.21\textwidth]{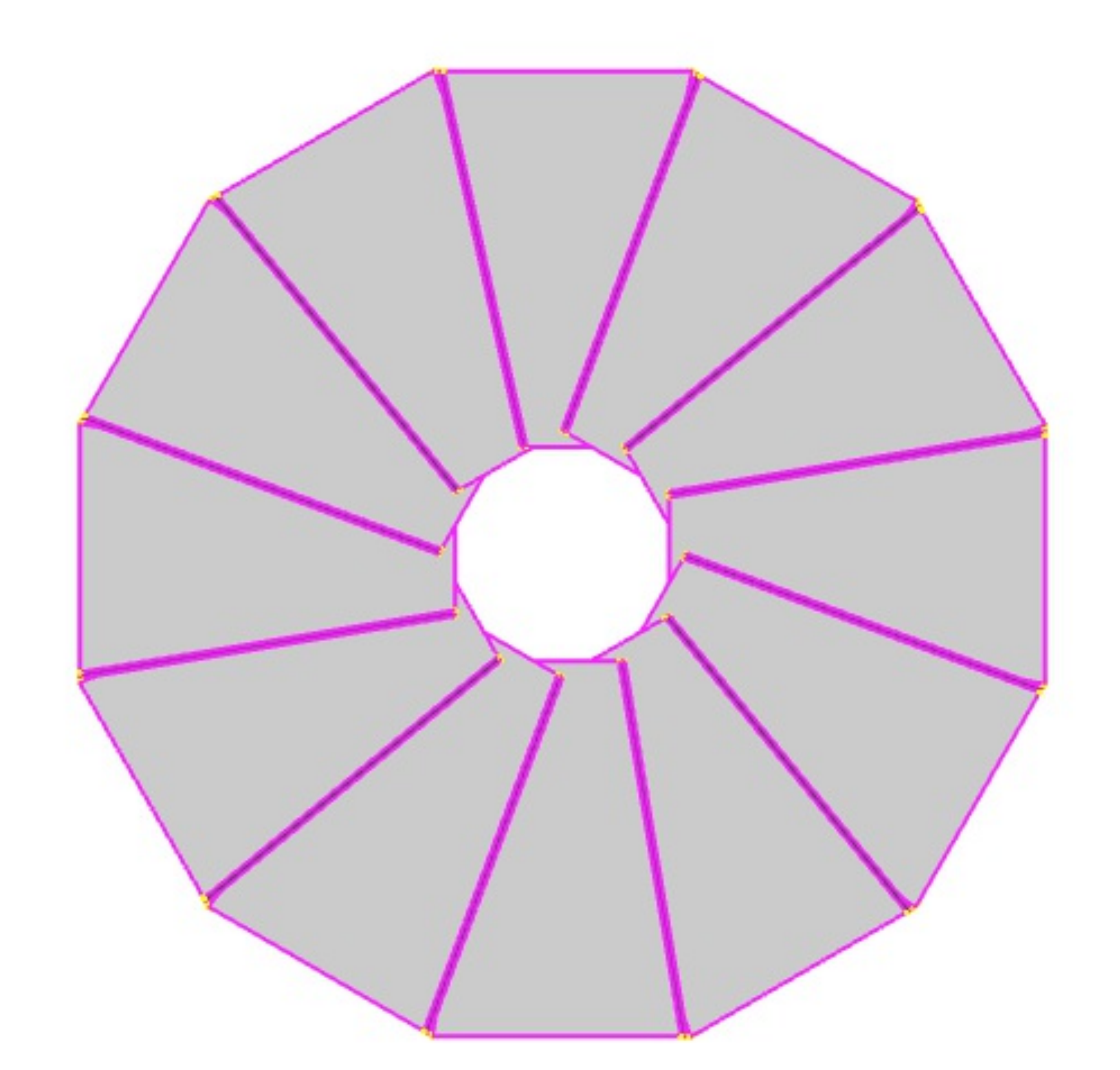}
    \includegraphics[width=0.32\textwidth]{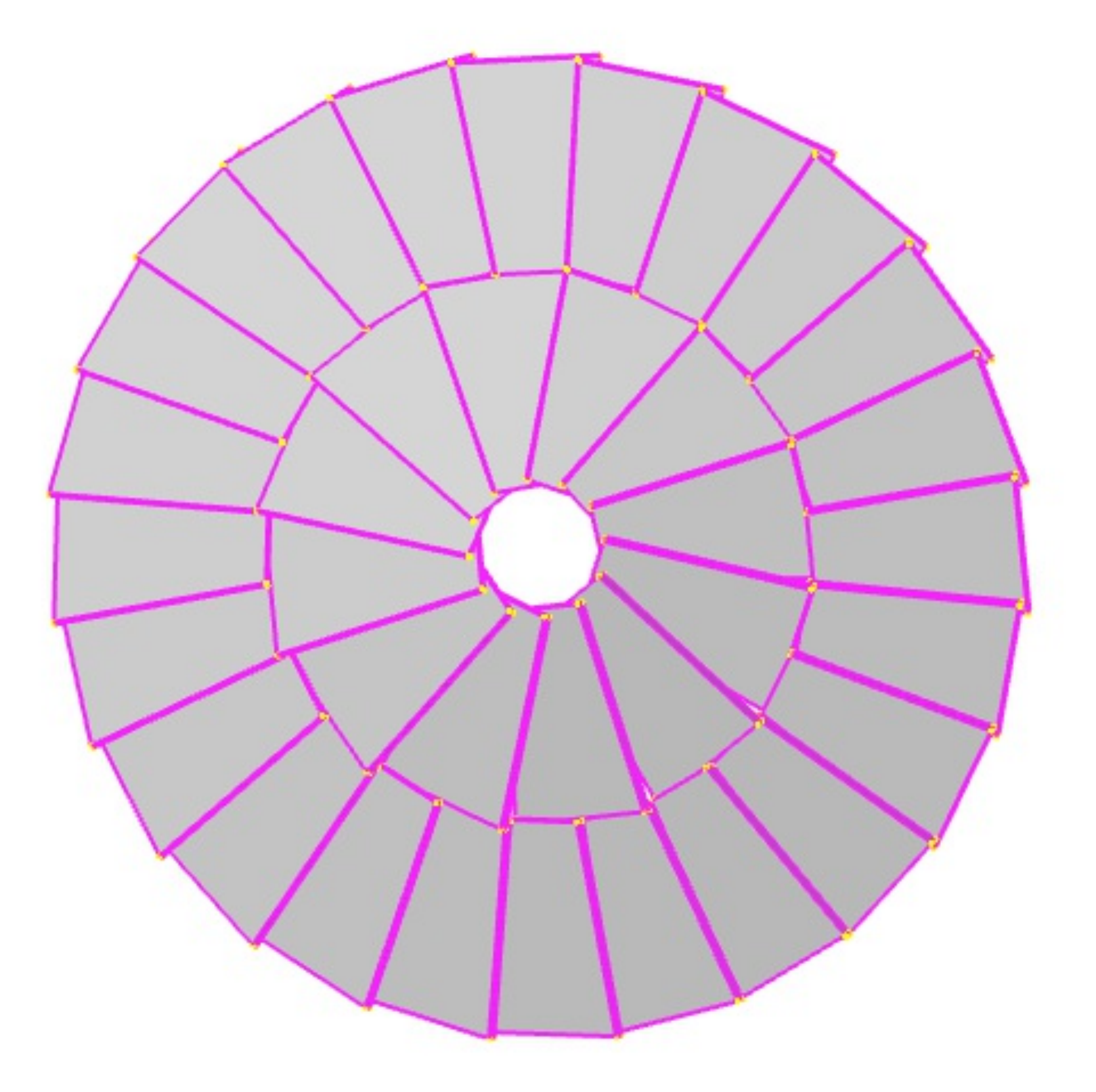}    
      \includegraphics[width=.43\textwidth]{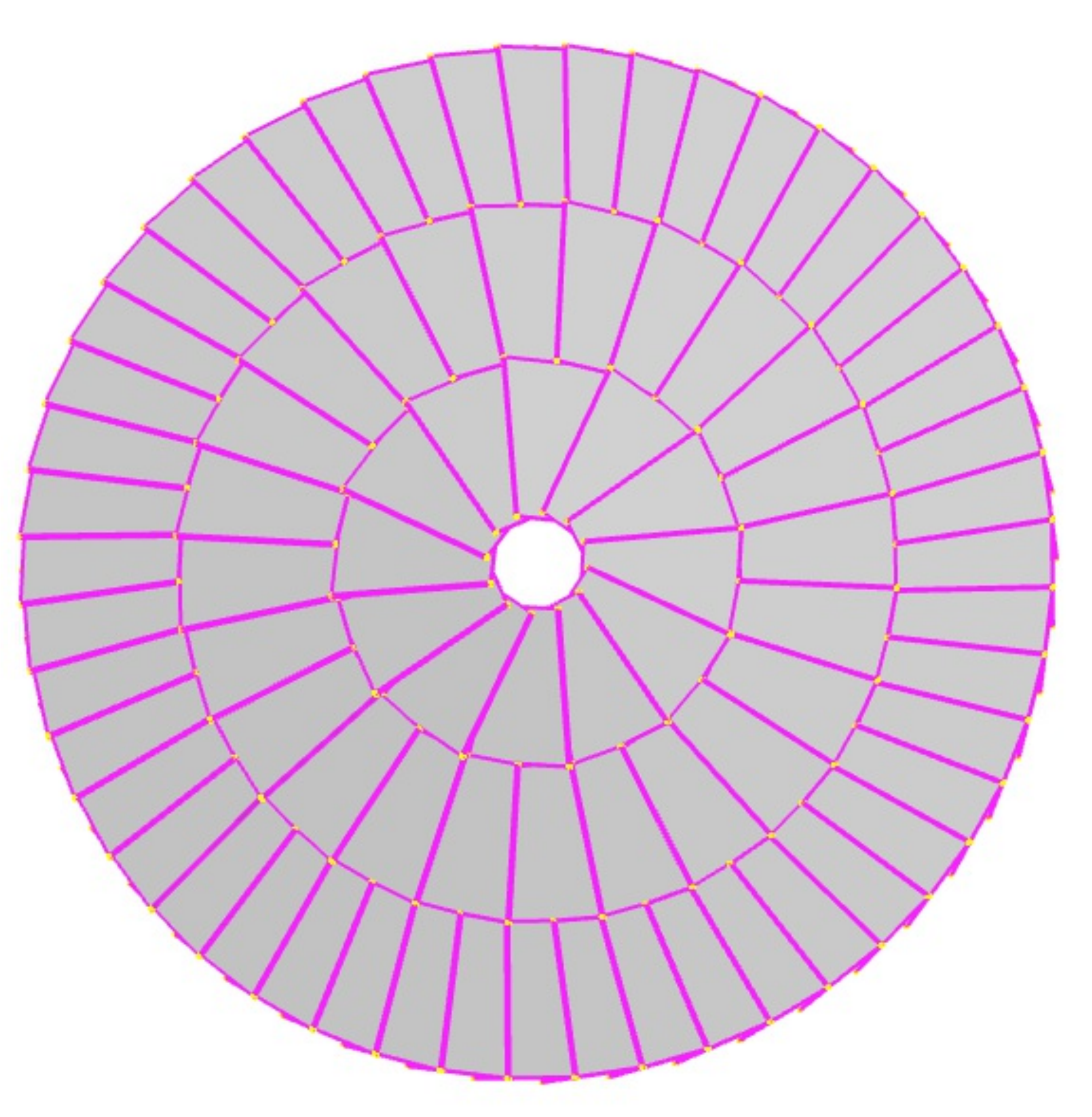}
    \caption{The rendering of the MS0, MS2 and MS4 muon tracking stations (not to scale). }
    \label{fig:wheels}
\end{figure}

The first two stations (MS0, MS1) are composed of 12 overlapping petals, each rotated by 11.0\,$^{\circ}$ perpendicular to the beam axis. A view of MS0 is shown in Fig.~\ref{fig:wheels}, left. The next two stations (MS2, MS3) have two rows of petals, the inner one with 12 and the outer one with 24 petals (see Fig.~\ref{fig:wheels}, middle). For the final stations (MS4, MS5) a third row with 48 petals is added. The set-up of the last stations is shown in Fig.~\ref{fig:wheels}, right. 
There is an approximately 20-30\% overlap in the area between the petals of the same layer, predominantly in the MS4 and MS5 stations. 
This drawback is compensated by the advantage of having a universal design of all elements, leading to a reduction of production costs and simplifying the replacement of the petals if this should become necessary.

\subsubsection{Micro Pattern Gas Detectors}
\label{sec:MS-MPGD}
\begin{figure}[ht!b]
\begin{center}
\includegraphics[width=0.6\textwidth]{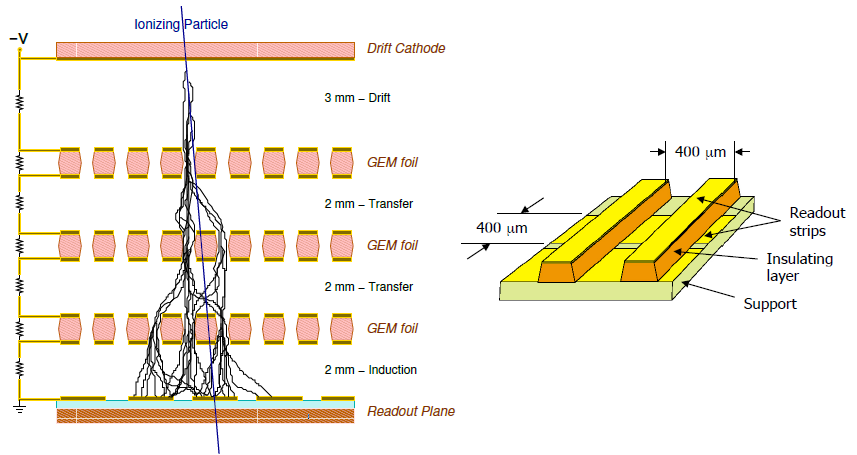}
\end{center}
\caption{Left: principle of a triple-GEM operation. Right: 2D Cartesian readout structure as being used in COMPASS.}
\label{fig:GEMprincipleAndRO}
\end{figure}

\begin{figure}[ht!b]
\begin{center}
\includegraphics[width=0.65\textwidth]{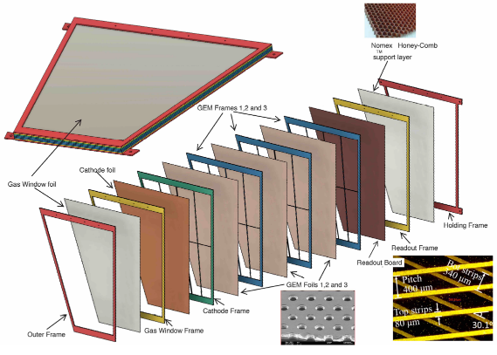}
\end{center}
\caption{Exploded view of the MOLLER GEM tracker. The bottom-right insert shows a microscopic picture of a 2-dimensional readout board similar to what would be used for MOLLER GEMs. The application in the MOLLER setup requires only pitches of 800 $\mu$m as opposed to the pitch of 400 $\mu$m as shown here.}
\label{fig:MOLLER_GEM_explView}
\end{figure}

\begin{figure}[h]
    \centering
      \includegraphics[width=0.45\textwidth]{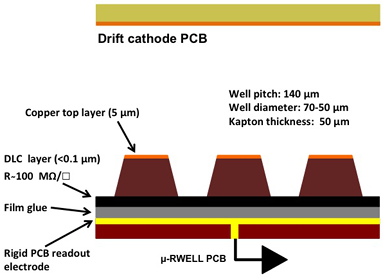}
    \caption{Principle of the uRWell amplification. }
    \label{fig:urwell_proto}
\end{figure}

\begin{figure}[h]
    \centering
      \includegraphics[width=\textwidth]{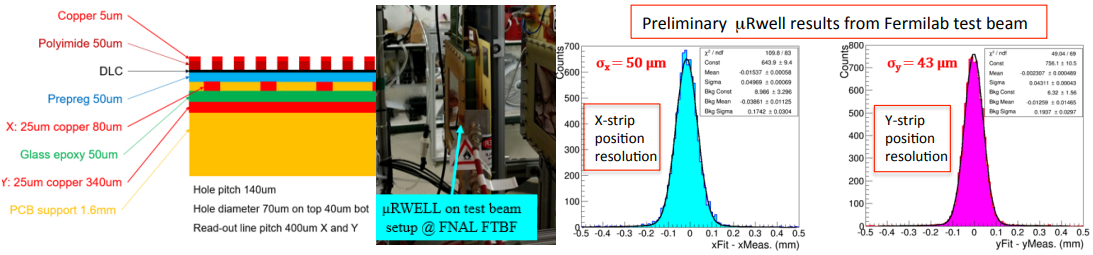}
    \caption{uRWELL prototype with 2D readout: (left) Cross section of the prototype; (center)
Prototype installed in test beam area at Fermilab (June-July 2018); (right) Preliminary results of
spatial resolution performances of the µRWELL prototype with 2D X-Y strip readout layer.}
    \label{fig:urwell}
\end{figure}

Micro Pattern Gas Detectors (MPGD) have been successfully deployed as trackers, particle identification devices, and in other applications. Applications concerning trackers have found their way in all experiments at the LHC, at many experiments at the SPS, as well at RHIC and JLab, and other large accelerator facilities. The Gas Electron Multiplier (GEM) and the Micro Mesh Gaseous Device (MicroMegas) are the main MPGDs which have been in use for decades at above mentioned facilities. 
\paragraph{Gas Electron Multiplier detectors}
A cost effective solution for relatively large area trackers is provided by the Gas Electron Multiplier (GEM) technology invented by F. Sauli \cite{Sauli:1997qp}, in 1997. The GEM is based on gas avalanche multiplication within small holes (on a scale of 100 $\mu$m), etched in a Kapton foil with a thin layer of copper on both sides. The avalanche is confined in the holes, resulting in fast (about 10 ns rise time) signals. Several GEM foils (amplification stages) can be cascaded to achieve high gain and stability in operation. The relatively small transparency of GEM foils reduces the occurrence of secondary avalanches in cascaded GEM chambers. All these properties result in very high rate capabilities of over 100 MHz per cm$^2$ and an excellent position resolution of approximately 70 $\mu$m. Fig.~\ref{fig:GEMprincipleAndRO} left, illustrates the principle of operation of a triple (three foils) GEM chamber, while Fig.~\ref{fig:GEMprincipleAndRO} right, shows a zoom-in view of an X-Y type (90\textdegree~angle between the two readout strip layers) of a 2-D readout layer used to capture the amplified electron cloud and register its position with high accuracy in both dimensions. The geometry of the readout plane can be chosen in a very flexible manner and provides the opportunity of designing it to the need of the experiment.\newline
GEM chambers have been pioneered by the COMPASS experiment at CERN \cite{Ketzer:2004jk}, and are now routinely used in a variety of high-rate experiments at various laboratories; the KLOE-2, PRad and the SBS experiment at Jefferson Lab; the STAR and PHENIX/sPHENIX experiment at RHIC; the LHCb, ALICE, and CMS experiments at CERN; and many other experimental setups around the world.
\paragraph{GEM tracker candidate}
The GEM trackers for the MOLLER experiment might serve as a template for part of the muon tracker in the NA60+ experiment because of its similarity in size and shape.
\begin{table}[hbt]
\caption{Requirements for the MOLLER GEM-tracker.}
\begin{center}
\begin{tabular}{|l|r|}
    \hline
    Parameter & Value \\
    \hline
    \hline
     Radial coverage & $\sim$ 50 cm to $\sim$ 120 cm\\
     Area/sector & 0.5 m$^2$ \\
     Max rate & $<$ 200 kHz/cm$^2$ \\
     Detection efficiency per plane & $>$ 90\% \\
     Detector thickness & $<$ 2\%X$_0$ \\
     Position resolution & $<$ 1 mm\\
     Angular resolution & $<$ 1 mrad \\
    \hline     
    \end{tabular}
    \label{tab:GEMspecs}
\end{center}
\end{table}
The MOLLER GEM tracker consists of three standard GEM foils, i.e., highly perforated Kapton-Cu-sandwiched foils which are attached to a framing structure and complemented by a drift and readout plane, see Fig.~\ref{fig:MOLLER_GEM_explView}.
Specifications for this detector are listed to Table~\ref{tab:GEMspecs}.
Detectors like these have been extensively tested (\cite{Gnanvo:2015xda,Zhang:2015pqa}) and are being used in experiments at JLab (PRad) and at the LHC (CMS).
\paragraph{uRWell}
Another MPGD technology has been well investigated and is now being considered as upgrades and for applications in future facilities. The micro-resistive well (uRWell) detector basically consists of a single GEM-foil attached to a PCB (Fig.~\ref{fig:urwell_proto}). This assembly provides a very simplified and robust mechanism to perform, for instance, as a tracking device. Furthermore, it will reduce the complexity of a tracking detector and this results in a reduction in costs. A uRWell device can take on any shape, for instance as shown in Fig.~\ref{fig:wheels}. The technology has been studied (Fig.~\ref{fig:urwell}) and is considered for upgrades and installation at facilities like JLab \cite{PRAD-IIprop} and EIC.
\paragraph{Stony Brook University - Lab Facilities and Experience}
SBU has vast amount of facilities that allow performing R\&D on as well as mass production of detectors. The SBU group has just finished the production of the GEM based readout chambers of the sPHENIX TPC~\cite{Klest:2020sdb} and is currently characterizing and performing QA procedures before installation into the final detector. In parallel, the group is building up a detector production facility aiming to mass produce triple-GEM detectors for the MOLLER experiment at JLab~\cite{MOLLER:2014iki}. These detectors coincidentally have similar properties that would be needed for the NA60+ experiment and might serve as a good candidate for the muon tracker at NA60+.\newline
The main facility encompasses about 1000 m$^2$ for detector R\&D and production and has a large class 1000 (ISO class 6) clean room with an area of $\sim$200 m$^2$. In addition, the group is establishing the production facility for the MOLLER GEM tracker with an additional class 1000 (ISO class 6) clean room with an area of $\sim$20 m$^2$. An irradiation facility with a high intensity X-ray tube as well as a cosmic ray telescope add onto the detector capabilities.\newline
The SBU group has also vast amount of experience in performing R\&D on detectors and has been vital in the production of several large scale detector applications. Members of the group were involved in the very first application of GEMs within the HERA-B experiment (DESY), the first large scale application of GEM detectors in COMPASS (SPS), the forward muon tracker for CMS (LHC), the HBD project for PHENIX (RHIC) (together with the Weizmann Institute), and the GEM modules for the sPHENIX TPC (RHIC) (together with the Weizmann Institute). Presently, the group is responsible for the production of 12 large scale triple-GEM detectors for the MOLLER experiment at CEBAF/JLab as described in Fig.~\ref{fig:MOLLER_GEM_explView}.

\subsubsection{Multi-Wire Proportional Chambers (MWPC)
}
\label{sec:MS-MWPC}


{\bf Advantages of MWPC technology.} Position-sensitive gaseous detectors, based on the multi-wire proportional chamber (MWPC), are widely used in particle physics experiments, including those currently running at the LHC, like ATLAS, CMS, and LHCb Muon Spectrometers. Sub-millimeter localization accuracies in detecting ionizing radiation are routinely achieved in these apparatuses. Still, a significant effort is continuously made to further improve the detector parameters by using suitable geometries and operational conditions.

The creation of the electron-ion pairs is at the origin of avalanche multiplication, and it is exploited as a signal amplification mechanism in gaseous detectors. The photon-mediated diffusion contributes to the spread of the growing avalanche. The whole process, which begins a few wire radii from the anode, is over after a short period (fraction of a nanosecond), leaving the cloud of positive ions receding from the anode plane at decreasing speed as a consequence of the decrease of the electric field strength. This motion is responsible for the largest fraction of charge, induced by the avalanche, detected on the anode and all surrounding electrodes, which are the cathode planes. The measurement of the charge profile, induced on the cathode plane suitably segmented, allows for two-dimensional localization of the ionizing cloud to be achieved. And here the so-called center-of-gravity (COG) method, which allows the attainment of the highest localization accuracy, plays a very important role.

There are many advantages of using MWPC technology for NA60+ Muon Spectrometer tracking chambers, several of them are listed below:

\begin{itemize}
\setlength\itemsep{1em}
  
\item mature technology mastered for more than 50 years;
    
\item high flexibility in choosing the detector parameters (gain, resolution, readout pattern) and the detector geometry;

\item excellent detector longevity, the authors of the LoI have experience in operating analogue detectors that lasted for nearly 20 years;

\item required parameters (200 $\mu$m resolution and $<10^4$ cm$^{-2}$s$^{-1}$ rate) are readily  achievable;
    

 \item low cost per unit of sensitive area;
 

  \item simplicity of the design, availability of world-class production facilities, author's hands-on expertise; 
  
\end{itemize}

Considering all the above, one can conclude that the NA60+ 
Muon Spectrometer tracking chambers can be built using wire chambers technology.

\begin{figure}[ht]
\begin{center}
\includegraphics[width=0.9\linewidth]{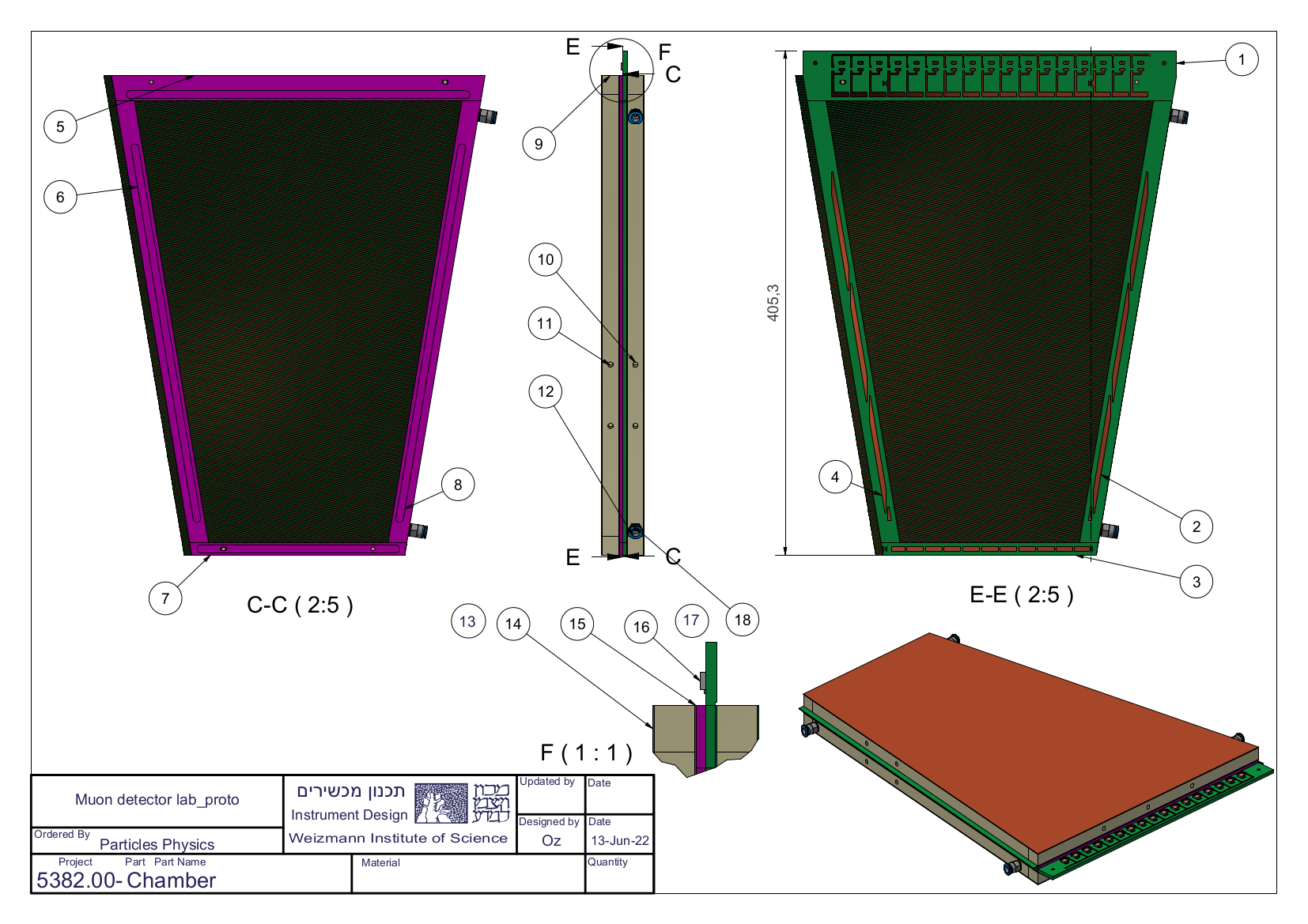}
\caption{The design of the first MWPC prototype.}
\label{fig:detectors:prototype_design}
\end{center}
\end{figure}

{\bf MWPC prototype design and expected lab and beam tests.} The preliminary design of the first MWPC prototype is shown Fig.~\ref{fig:detectors:prototype_design}. It consists of two half detectors, later glued together, each half is a sandwich of 12.7 mm Nomex honeycomb covered from one (outer) side with a 0.5 mm FR4 sheet with 35 microns thick copper looking outside. The other (inner) side is an actual cathode with also 0.5 mm FR4 sheet with strips of 1 mm pitch and 17 microns thick. The strips on each cathode are running on a different direction, providing a small angle stereo readout. The wires in this detector are running on a vertical direction between two bases of the trapezoid, their pitch for this MWPC prototype is 3 mm with a distance to each cathode of 3 mm, so the total gas volume gap is 6 mm thick.

The glued detector are equipped with the multi-layer strip and wire adapter boards for the signal readout. The strip adapter boards for both cathodes are soldered on both sides of the trapezoid, whereas the wire adapter board is soldered on its big base. Readout electronics cards with VMM3a ASIC are installed on these adapter boards. Each cathode adapter board has three such electronics cards (384 channels), whereas the wire adapter board is equipped with only one (128 channels).

The detailed laboratory tests at Weizmann Institute test bench will include HV stability studies, the choice for the optimal and sufficient gas gain, the detector intrinsic noise studies, and then the analysis of wire and strip signals from triggered cosmic muons. After validating all the above steps, the detector will be installed inside the tracking telescope, equipped with three standard GEM detector stations, and the resolution of this MWPC prototype will be measured.

After all these detailed and time consuming laboratory tests will be done, this tracking telescope setup together with the MWPC prototype will be shipped to CERN for muon and pion beam tests.


{\bf Weizmann Institute detectors construction facility.} The detector construction facility at the Weizmann Institute of Science (Mexico laboratory, hereafter) has been established in 1997 for further development and mass production of the Thin Gap Chambers (TGC) for the ATLAS muon endcap trigger. The TGC, developed at WIS many years ago for the OPAL experiment at the CERN LEP, are multi-wire proportional chambers (MWPC) working in semi-proportional regime. The small gap of 1.4 mm between the wires and the cathodes, relative to the 1.8 mm wire pitch, make them unique MWPC capable of operating as the trigger chambers at the high collision rate (40 MHz) provided by the LHC. The main TGC production campaign was carried out between 1998-2005. About 2500 large area chambers were produced by a team of about 15 technicians. Between 2005-2008, the Mexico laboratory team has been responsible for the assembly of the chambers on the ATLAS Big Wheels sectors at CERN. Since then, a routine production and installation of the spare TGC chambers is ongoing.

The small-strip Thin Gap Chambers (sTGC) for the ATLAS New Small Wheel (NSW) Phase-I Upgrade project have been developed at the Mexico laboratory as well. The challenging mass production of these chambers, which has been finished a year ago, has been splitted between five countries, with Mexico laboratory serving as a leading construction site, producing more than 200 large area single gaps, and a know-how center.

In addition to a routine construction of 10 TGC spare chambers for ATLAS each year, a new type of TGC detector is being developed these days for the Phase-II Upgrade of ATLAS Muon Spectrometer. The current R$\&$D phase is expected to end by 2022 and it will be followed by a mass production campaign ($\sim$150 large area single gaps) scheduled in 2023-2024, to be ready for installation in the ATLAS cavern in May 2025.

\begin{figure}[ht]
\begin{center}
\includegraphics[width=0.9\linewidth]{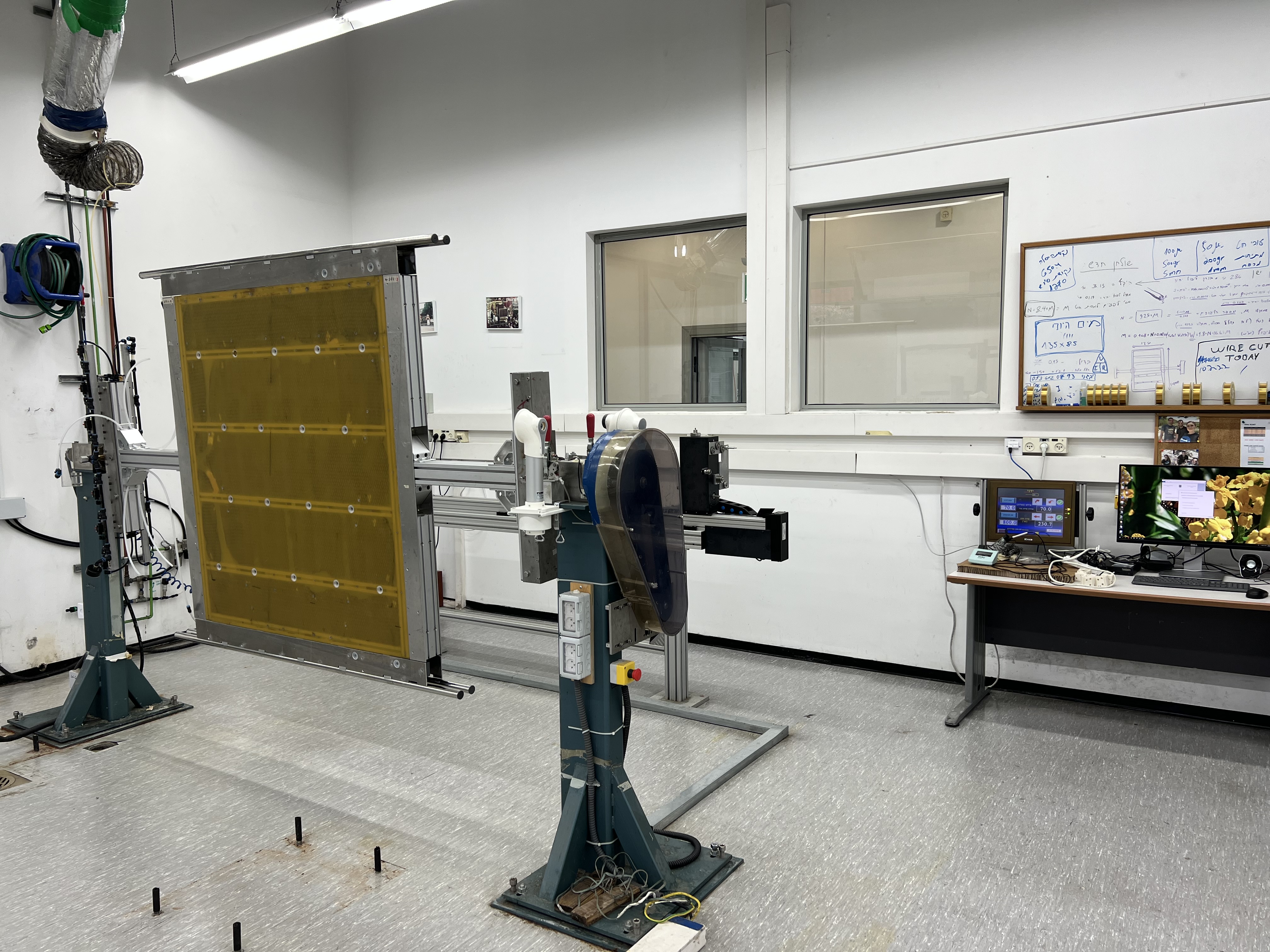}
\caption{The winding machine in Mexico laboratory.}
\label{fig:detectors:winding_machine}
\end{center}
\end{figure}

\begin{figure}[ht]
\begin{center}
\includegraphics[width=0.7\linewidth]{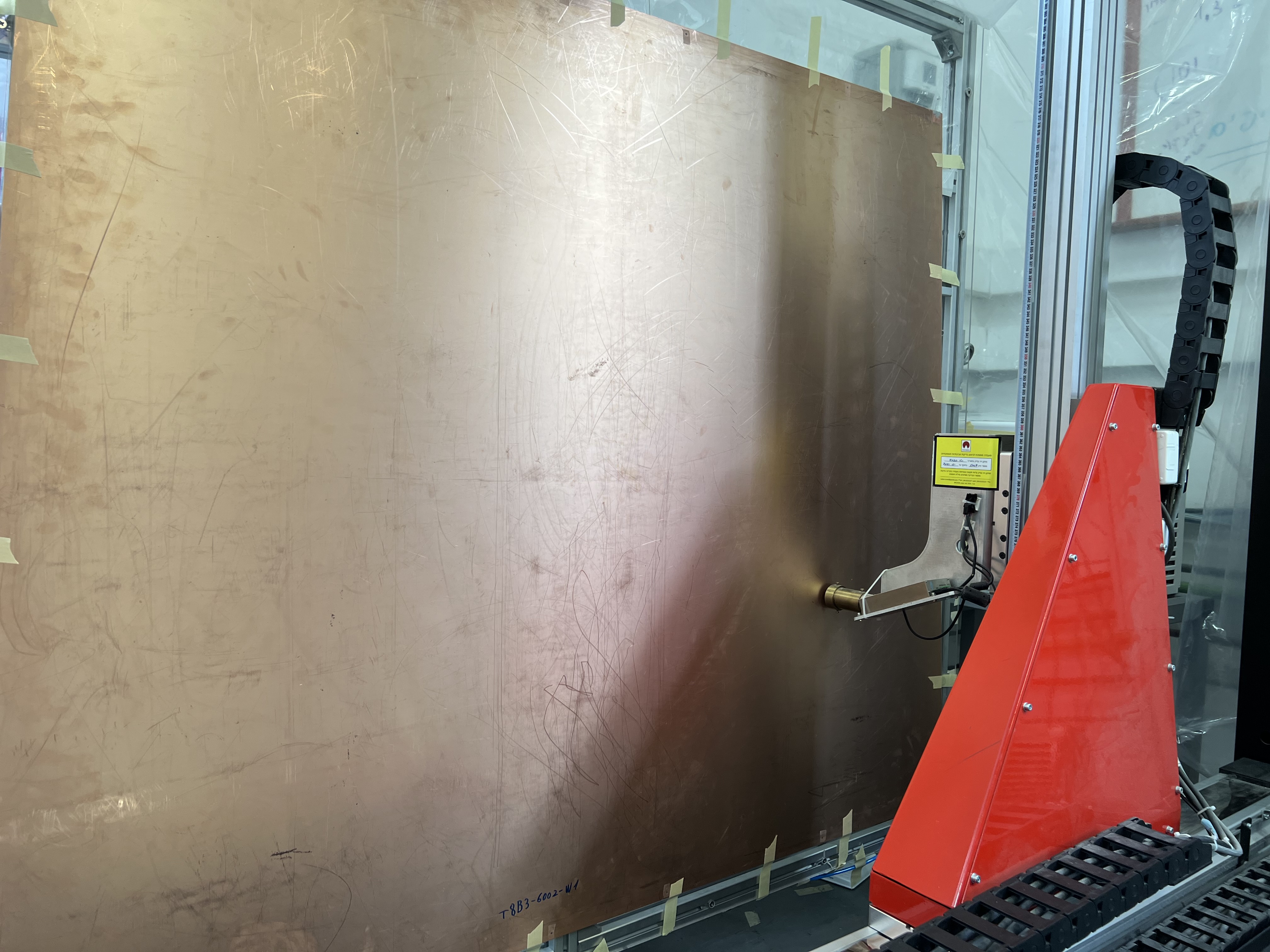}
\caption{The X-ray scanner in Mexico laboratory.}
\label{fig:detectors:scanner}
\end{center}
\end{figure}

Nowadays, the Mexico laboratory is the part of the Physics Faculty Core Facility unit. This world-leading infrastructure for detector construction as well as the position of 5 permanent team members (3 engineers and 2 technicians with more than 20 years of experience for detector construction) are supported by the Institute. Additional person manpower is hired in periods of the mass productions. 
The Mexico laboratory has a total working area of $\sim$500 m$^2$ and additional storage area of $\sim$120 m$^2$ for raw material, components, produced gaps etc. It includes:

\begin{itemize}
  \setlength\itemsep{0.5em}
  
  \item a large area clean room, equipped with all needed for mass production systems like gas, crane, vacuum, high and low voltage, filtered air, air conditioning and humidity control, allowing up to 2 large area gas gaps closed per day;
  
  \item 10 precise and large surface area granite tables equipped with a vacuum system, including pumps, sensors, HV and gas supplier;
  
  \item a fully robotic and computer control winding machine shown in Fig.~\ref{fig:detectors:winding_machine}. It allows to wind the wires with a given precise pitch and with a required tension control for  large detectors up to several square meters of active area;
  
  \item a fully robotic and computer controlled X-ray scanner, shown in Fig.~\ref{fig:detectors:scanner}. It has been designed for the discovery of the various production defects prior to the readout electronics installation, which usually happens at the last stage of detector assembly. Thus, the X-ray scanning allows checking the quality of the produced chambers, identifying defects and then possibly fixing them already at early stage. It provides a precise and detailed scan on a large active area gas detectors up to several square meters;
  
  \item an electronics laboratory for detector R$\&$D and final testing of produced gaps, fully equipped with the final readout electronics. Some details of this lab will be given in the next subsection.
  
\end{itemize}


\begin{figure}[ht]
\begin{center}
\includegraphics[width=0.9\linewidth, angle=270]{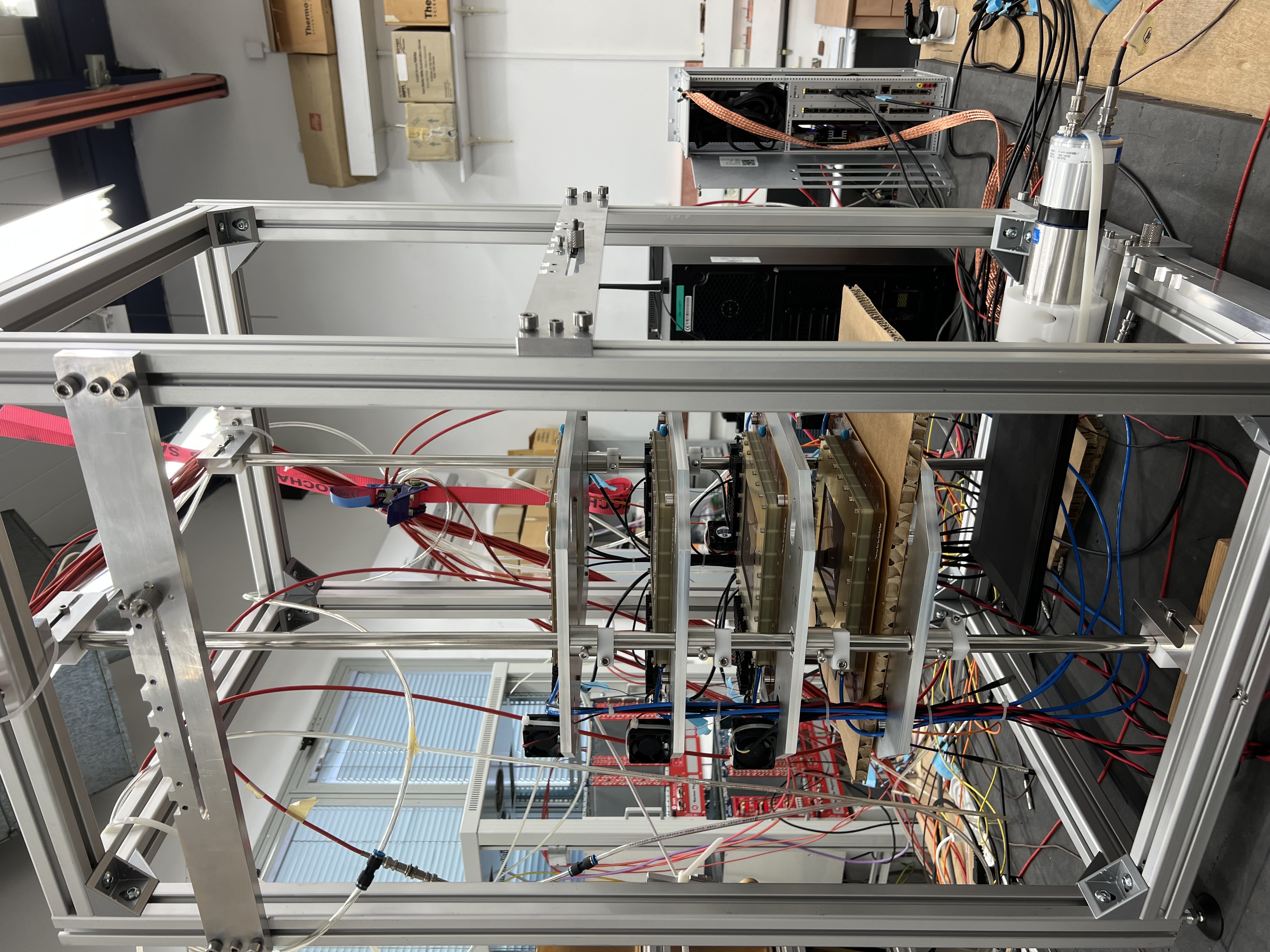}
\caption{The tracking telescope in Mexico laboratory.}
\label{fig:detectors:tracking_telescope}
\end{center}
\end{figure}

\begin{figure}[ht]
\begin{center}
\includegraphics[width=0.6\linewidth]{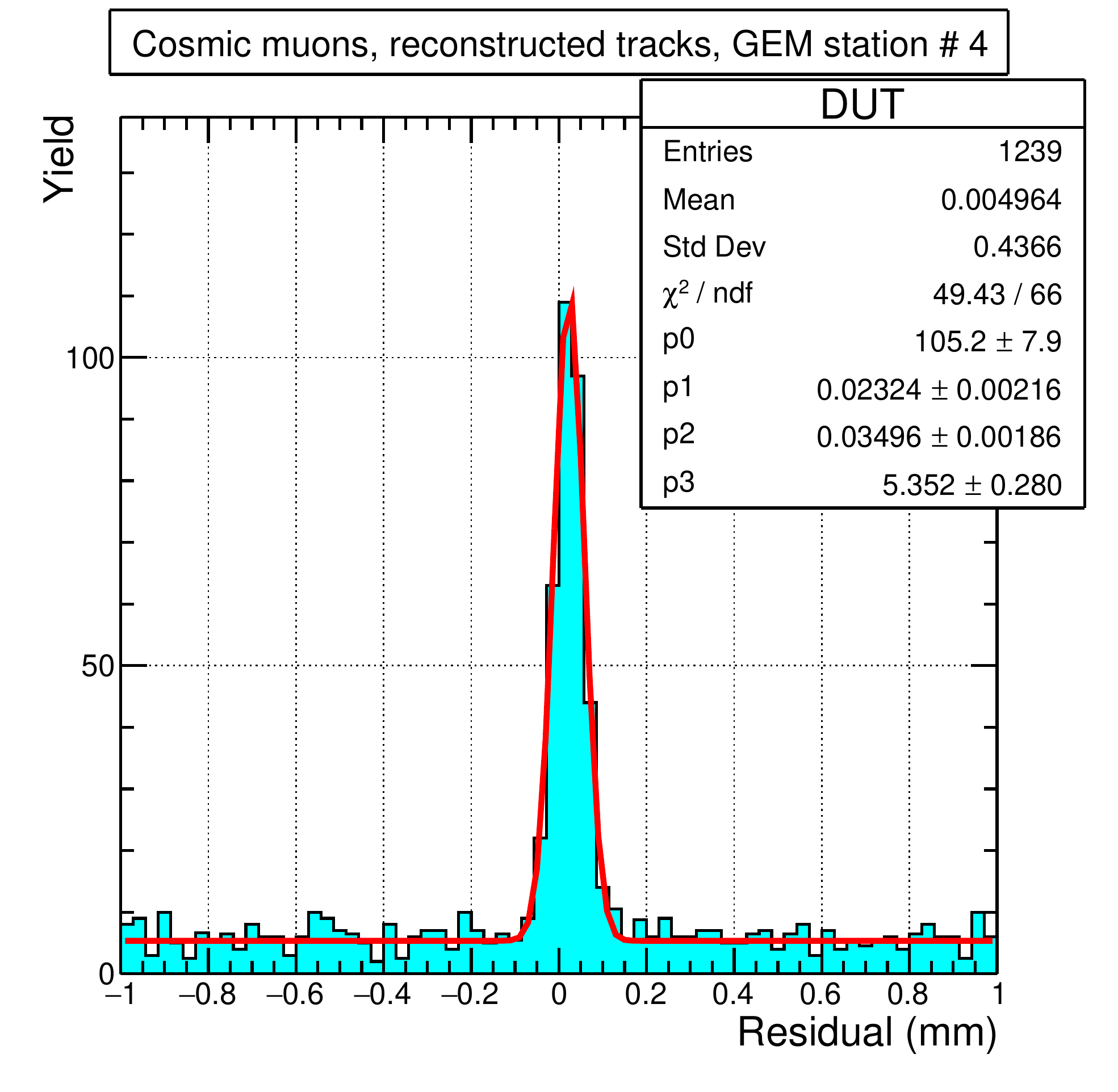}
\caption{The obtained residual distribution for the standard triple GEM detector.}
\label{fig:detectors:residual}
\end{center}
\end{figure}


{\bf Test bench.} The cosmic test bench in Mexico laboratory has been recently upgraded with a tracking telescope (shown in Fig.~\ref{fig:detectors:tracking_telescope}) based on triple GEM stations with two-dimensional readout. The active area of these stations is dictated by the size of GEM foils of 100$\times$100 mm$^2$, used in cascades. The electrons transferred from the avalanche in GEM foils are then collected by the pick-up electrode. This consists in one set of parallel metal strips sitting on thin kapton ridges, and a second set of perpendicular metal strips on the bottom of the ridge, 256 strips in each direction with a pitch of 400 microns. The collected charge is shared between strips in the two layers, and a center-of-gravity calculation provides the central avalanche coordinate in two projections. The new version of the Scalable Readout System (SRS), which is a powerful and versatile electronic readout system dedicated to MPGD, is used in order to readout the signals from the GEM stations. The original reference implementation with the APV25 as front-end ASIC is currently replaced with a new version using the VMM3a ASIC. The VMM3a, originally developed for the ATLAS New Small Wheel Upgrade, is a readout ASIC, specifically developed for the readout of gaseous detectors, offering with its integration into the SRS new capabilities for the readout of MPGD. Out of the plurality of VMM3a/SRS features, two prominent ones are the high rate capability and a continuous self-triggered readout. The cosmic muon trigger for this tracking telescope is based on a coincidence of three scintillators with PMT readout.




{\bf First results with tracking telescope.} The three triple GEM detectors in the tracking telescope, mentioned in the previous subsection, have been used as the tracker GEM stations for pointing to the detector-under-test (DUT). In this particular case, devoted to the testing of the telescope abilities and its performances, the DUT was a fourth GEM station. All GEM stations have been operated in an Ar/CO$_2$ (70:30) gas mixture at a gas gain close to 10$^4$. The obtained result for the residual distribution in DUT is shown in Fig.~\ref{fig:detectors:residual}. The measured resolution of better than 40 $\mu$m demonstrates the excellent performance of the detectors and readout electronics used for the tracking telescope, which can then be used to test any gaseous tracking detector considered as a possible candidate for NA60+ muon spectrometer tracking chambers.


\subsubsection{Cost estimates for muon spectrometer
}
\label{sec:MScost}


{\bf Cost estimate for MWPC option.} As it has been mentioned in the previous subsection, the wire based gas detectors are relatively cheap to produce assuming the facility for the mass production is in place. The cost estimate per detector element in units of one square meter is very preliminary and it is based on a similar estimation done in 2020 for the production of spare TGC chambers with an active area close to 2 m$^2$. What will be the prices in the coming years is hard to estimate considering the current uncertain situation on the market, but this will be true for any technology chosen for the NA60+ muon spectrometer tracking chambers. 

  
  
  


For the current estimate it is assumed that for the first two muon stations the required resolution is $\sim$200 $\mu$m, for the following two it is $\sim$ 400 $\mu$m and for the last two it is $\sim$ 800 $\mu$m, bringing the total number of channels to $\sim$100K. The breakdown of the estimated project cost, which includes 10$\%$ spares and a 30$\%$ safety factor, is presented in Table \ref{mwpc_cost_table:1}.


\begin{table}[h]
\caption {The cost breakdown structure for MWPC option. The estimates do not include the local expenditures and manpower costs.}
\centering
\begin{tabular}{|l|l|}
\hline
                   & kCHF       \\ \hline
Detectors          & 500        \\ \hline
FEE                & 1000       \\ \hline
HV system          & 150        \\ \hline
Mechanical support & 750        \\ \hline
Gas system         & 300        \\ \hline
\textbf{TOTAL}     & 2700       \\ \hline
\end{tabular}

\label{mwpc_cost_table:1}
\end{table}

{\bf Cost estimate for MPGD option.}
For the uRWell we received a cost estimate based on a preliminary quote from the CERN-PCB workshop. The price for one petal, i.e., one detector module of the wheel as shown in Fig.~\ref{fig:wheels} is estimated to be 4,800 CHF. However, the dimensions for the MPGD detector have to be adapted because of a limited width for the detector production. This would result in 16 detector modules around the azimuth for the inner part of the stations.\\
Each module would accommodate about 2250 readout stripes, respectively (in r- and $\phi$-direction). Consequently, for reading out MS0 and MS1 a total of about 72k channels of electronics is needed. The readout electronics for MPGD detectors is identical to the readout electronics for the MWPC, i.e., the VMM3a ASIC or their upgrades can be used. Each detector module based on the uRWell and VMM3/SRS technology would contribute with about 30 kCHF.\\
For equipping the tracking stations with GEM detectors one can scale the costs for the module with a factor of three. Consequently, the costs for one triple-GEM module will be about 15 kCHF. The electronic readout part for this technology is not affected and remains the same as for the uRWell. The effective cost for a single module based on the GEM and VMM3/SRS technology would be about 41 kCHF. In summary, Table \ref{table:uRWell_gem_cost_table} shows the breakdown of the costs for equipping MS0 and MS1 with uRWell or GEM detectors, based on estimates at the time of this write-up.\\
\begin{table}[h]
\caption {The cost breakdown structure for the MPGD (uRWell and GEM)  option. The cost estimate is based on equipping MS0 and MS1 with GEM. The cost factors include 10\% additional yield for the detectors and readout electronics.}
\centering
\begin{tabular}{|l|r|r|}
\hline
                    & uRWell: kCHF  & GEM: kCHF\\ \hline
Detectors           & 170           & 530        \\ \hline
Readout electronics & 790           & 790       \\ \hline
HV system           & 20            & 20        \\ \hline
Mechanical support  & 50            & 50       \\ \hline
Gas system          & 50            & 50       \\ \hline
\textbf{TOTAL}      & 1,080         & 1,440     \\ \hline
\end{tabular}

\label{table:uRWell_gem_cost_table}
\end{table}
Eventually, if the uRWell or the GEM technology would be used to equip all muon tracker stations the cost scheme as in Table ~\ref{table:mpgd_cost_table} can be obtained, by simply scaling the costs with the number of detector modules needed.
\begin{table}[h]
\caption {The cost breakdown structure for the MPGD options for all muon tracker stations. The costs are based on scaled cost estimates from Table~\ref{table:uRWell_gem_cost_table}.}
\centering
\begin{tabular}{|l|r|r|}
\hline
                & uRWell: kCHF  &   GEM: kCHF     \\ \hline
MS0/1           & 1,080         &   1,440\\ \hline
MS2/3           & 3,090         &   4,058\\ \hline
MS4/5           & 6,112         &   7,960\\ \hline
\textbf{TOTAL}  & 10,282        &   13,450\\ \hline
\end{tabular}

\label{table:mpgd_cost_table}
\end{table}

\subsection{Toroidal magnet}
\label{Toroid}
\vskip 0.2cm
We consider a normal-conducting toroidal magnet producing 0.5 T of magnetic field magnitude over a volume of 120 m$^3$.
One of the main challenges is the heat generated by the current and the cooling of the conductor. In principle, it would be possible to increase the cross-section of the conductors sufficiently to mitigate the problem. However, the dimensions are limited by the constraints from the aperture and the physics acceptance. The optimisation of the design consists of finding the right balance between these conflicting requirements. 
To test the technology needed to wind the large normal-conducting coil, a small-size demonstrator (scale 1:5) was constructed and tested. This demonstrator allows to cross-check various aspects of the design.

\subsubsection{Toroid design }
\begin{figure}[h]
    \centering
    \includegraphics[scale=1.0]{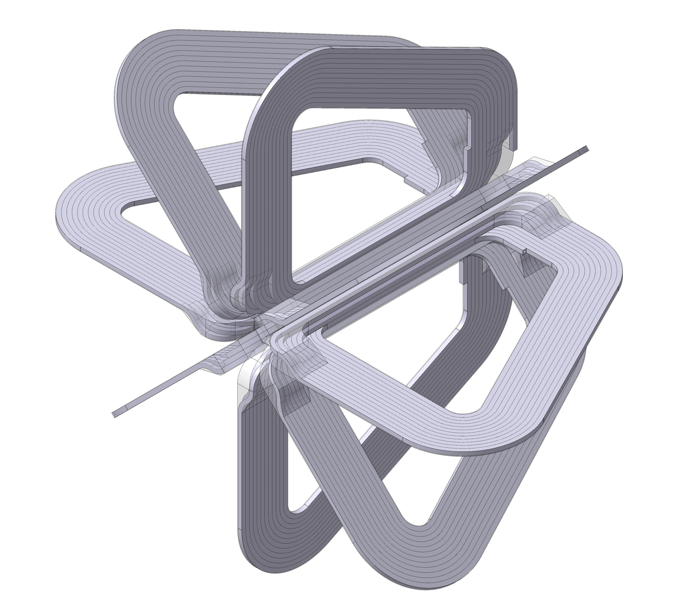}
    \caption{Perspective view of the arrangement of the 8 coils.}
    \label{fig:magnets:magnet1}
\end{figure}

\begin{figure}[h]
\begin{subfigure} {}
\includegraphics[scale=0.8]{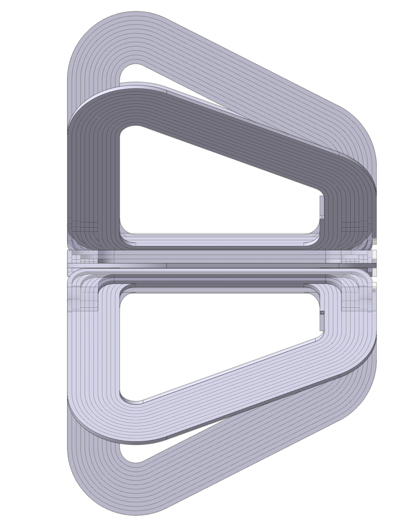} 
\label{fig:magnet2}
\end{subfigure}
\begin{subfigure}{}
\includegraphics[scale=0.25]{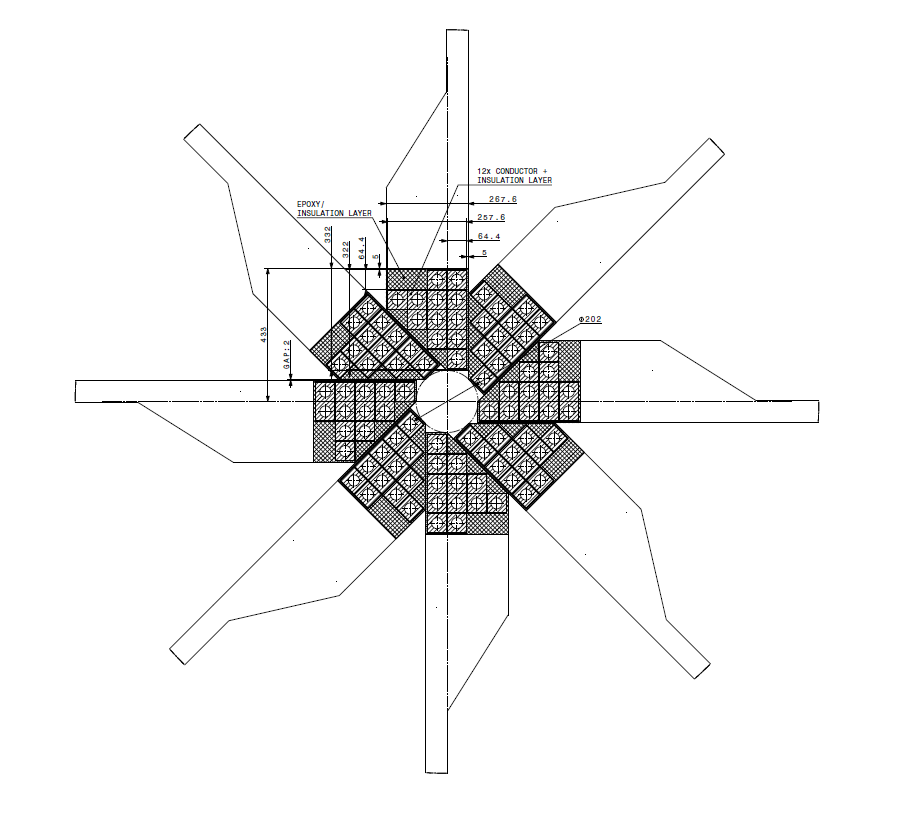}
\label{fig:magnet3}
\end{subfigure}
    \caption{Arrangement of the coils. The inner diameter of the centre bore is ~20 cm.}
    \label{fig:magnets:magnets}
\end{figure}

The toroid is made from eight sectors~\cite{ToroidalPrototypeNa60plus}. 
A 3D view of the magnet layout and the cross-section at the inner radius is shown in Fig.~\ref{fig:magnets:magnet1} and~\ref{fig:magnets:magnets}. The sectors are in contact at the inner radius and the dimensions are chosen to give little obstruction for incoming particles. Each coil has 12 turns and the conductor has a square copper section with 50 mm sides and a circular cooling channel in the centre.
The cooling and electrical requirements are somewhat similar to those of the LHCb magnet~\cite{Amato:424338}.  
The current is 195 kA (16.25 kA per turn) and 
all coils are connected in series, giving a total power of 3 MW. Concerning the cooling, a limit of    30 \textdegree C temperature difference between inlet and outlet of the cooling water was used. The main parameters of the toroid are listed in Table~\ref{table:1}.

\begin{table}[h!]
\caption {The toroid magnet characteristics.}
\centering
\begin{tabular}{|l|l|}
\hline
\textbf{Magnetic Parameters}                   &                                               \\ \hline
Bending power                                  & $\sim$0.65 Tm               \\ \hline
Central conductor region                 & $\sim$10 cm \textless r \textless $\sim$45 cm \\ \hline
Electric power dissipation                     & 3.07 MW                                       \\ \hline
Total conductor length                         & 1000 m                                        \\ \hline
\textbf{Coil and Current}                      &                                               \\ \hline
Length of each coil conductor                       & 125 m                                         \\ \hline
Number of coils                                & 8                                             \\ \hline
Number of turns/coil                           & 12                                            \\ \hline
Conductor material                             & Copper                                        \\ \hline
Current in conductor                           & 16250 A                                       \\ \hline
Tension per coil                               & 23.6 V                                        \\ \hline
Current density                                & 10.96 A/mm$^2$                                   \\ \hline
\textbf{Cooling}                               &                                               \\ \hline
Requirements                                   & All coils in parallel                         \\ \hline
Water channel diameter                         & 36 mm                                         \\ \hline
Power per coil                                 & 384 kW                                        \\ \hline
$\Delta$ T                                             & 30 K                                          \\ \hline
Fanning friction factor,  f                    & 0.0054                                        \\ \hline
Reyenolds number, Re                           & \(1.0^{5}\)                                           \\ \hline
Flow                                           & 3.1 l/s                                       \\ \hline
Water speed                                    & 3.0 m/s                                       \\ \hline
$\Delta$ P                                             & 3.4 bar                                       \\ \hline
\end{tabular}
\label{table:1}
\end{table}

\subsubsection {Cooling considerations and calculations }

The calculation of the cooling requirements starts from the power dissipation in the coils and the acceptable temperature gradient of the magnet. Following other similar working magnets, a limit of 30 \textdegree C was chosen. The heat dissipation per coil can easily be calculated, then giving the required water flow. The pressure drop ($\Delta P$) in the cooling channel and the Fanning friction factor, $f$, can be calculated from the formulas below: 

\begin{equation}
\Delta P=2\times\frac{f\times L\times\rho\times v^2}{D}
\end{equation}
where
\begin{equation}
\frac{1}{\sqrt{f}}=-4\times\log_{10}\left[ \frac{\epsilon/D}{3.7}+\frac{1.26}{Re \times\sqrt{f}}\right] 
\end{equation}
assuming an estimated surface roughness $\epsilon$ of 0.03 mm (a copper pipe was used).
The list of parameters and the results of the  calculations are also  summarized in Table~\ref{table:1}.

\subsubsection{Magnet demonstrator (scale 1:5)}
\begin{figure}[h]
    \centering
    \includegraphics[scale=1.0]{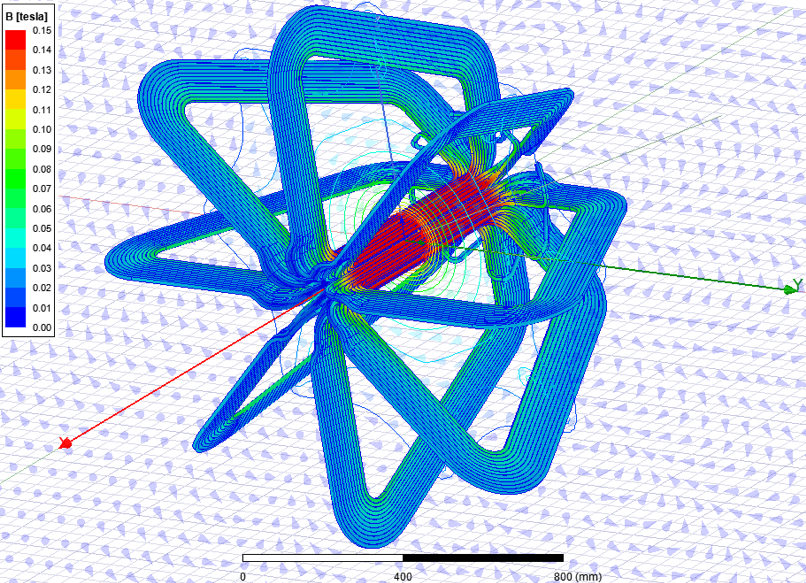}
    \caption{Magnetic field on the surface of the coils and in the surrounding volume. At a current of 500 A the peak magnetic field is 0.15 T.}
    \label{fig:magnets:field1}
\end{figure}

\begin{figure}[h]
    \centering
    \includegraphics[scale=1.0]{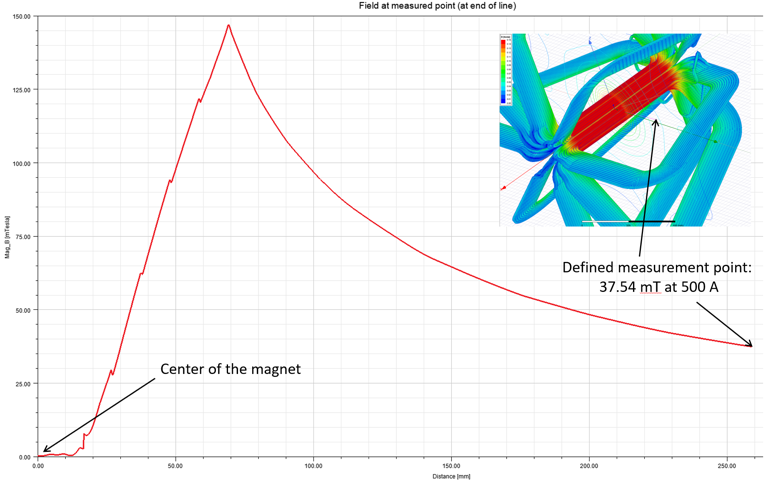}
    \caption{Magnetic field along a line, between the center of the magnet and a defined measurement point, later-on measured as part of the experimental campaign.}
    \label{fig:magnets:field2}
\end{figure}

A small-size demonstrator (scale 1:5) was constructed and tested,  allowing a cross-check of various aspects of the design.
Simulations were performed with an applied current of 500 A. After the simulation model had successfully converged to a solution, various magnetic field plots were generated ( Fig.\ref{fig:magnets:field1} and \ref{fig:magnets:field2}).

\begin{figure}[h]
    \centering
    \includegraphics[scale=1.2]{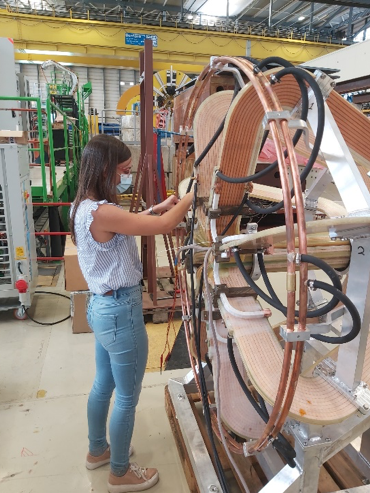}
    \caption{Step-by-step measurement of coil, busbar, and joint resistances.}
    \label{fig:magnets:measure1}
\end{figure}

Initially, the coils were powered at low current, up to 20 A, and an  overall magnet resistance of 49.3 m$\Omega$ was measured. To ascertain the individual coil, busbar and joint resistances, the magnet was powered at a constant current of 10 A. The voltage drop over each element was recorded (Fig.~\ref{fig:magnets:measure1}). At first, large deviations in the contact resistance were observed, but after application of indium foils in the joints, a resistance in the range (0.39 $\pm$ 0.37) $\mu\Omega$ was measured.  This was a factor of 190 lower compared to the situation without indium foils. Finally, the joint with highest resistance constitutes 0.0015 \% of the overall resistance, which is considered a very good result.

\subsubsection{Magnetic field measurement}
\begin{figure}[h]
\begin{subfigure} {}
\includegraphics[scale=0.9]{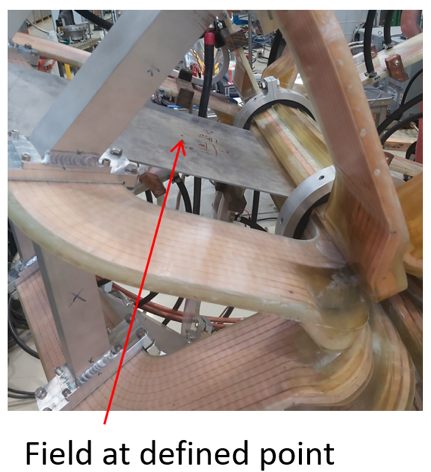}
\label{fig:measure2}
\end{subfigure}
\begin{subfigure}  {}
\includegraphics[scale=1.0]{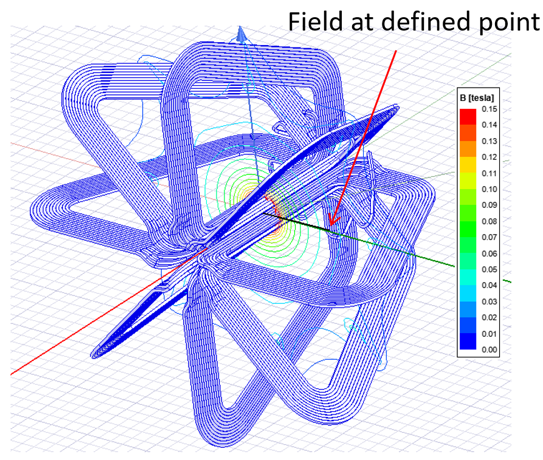}
\label{fig:measure3}
\end{subfigure}
    \caption{Magnetic field measured and simulated at a defined point.}
    \label{fig:magnets:measure}
\end{figure}

A specific point was defined inside the coil volume, to allow a comparison of magnetic field measurement and simulation (Fig.~\ref{fig:magnets:measure}). The magnetic field was measured with a gauss probe at various applied currents. The trend was found to be linear as expected. At 408 A, the measured magnetic field at the defined point is 29.9 mT, whereas the simulation gives 30.6 mT, a 3\% difference. This minor difference can be understood in terms of the different busbar layout in the simulated geometry versus reality. Nevertheless, the consistency between the two numbers is encouraging.

\subsubsection{Next step and future}
 The main tasks for the next step are the following:
\begin{itemize}
  \setlength\itemsep{0.5em}
 \item Defining the coil production procedures together with the required tooling and infrastructure;
 \item Designing and calculating the mechanical structure support of the toroid, needed to hold the coils in place.
 \item Defining the assembly tooling and procedure needed to align and fix the coils with the required position.
 \item Designing the power and control systems. 
\end{itemize}

Given the recent discussion concerning power consumption at CERN and in particular the use of electricity, innovative methods to power the magnet may need to be studied. One obvious possibility to reduce the power consumption is a pulsed operation of the toroid (the magnet is only ON when there is beam). Experimental requirements  like field stability and requirements on the power system need to be studied and evaluated in detail. The main cost items of the toroid magnet are listed in Table \ref{table:2}.

\begin{table}[h]
\caption {The estimated cost of the magnet.}
\centering
\begin{tabular}{|l|l|}
\hline
\textbf{Estimated cost (MCHF)}        &         \\ \hline
Copper Conductor                      & 0.6     \\ \hline
Manufacturing of coils              & 1.7     \\ \hline
Power converter (confirmation $\sim$1/8)                     & 0.8      \\ \hline
Mechanical structure                & 0.4       \\ \hline
Cooling system                     & 0.3        \\ \hline
\textbf {TOTAL}             & \textbf {3.8} \\ \hline
\end{tabular}
\label{table:2}
\end{table}

\subsection{Data acquisition, processing and computing
}
\label{daq}
\vskip 0.2cm

The data rate as well as the data storage and processing requirements of  NA60+ are fully dominated by the contribution from the vertex telescope.
The estimates presented in this Section are based on the FLUKA simulations (see Sec.~\ref{FLUKArate}), on the experience gained in operating the Internal Tracking System (ITS) of the ALICE experiment and in the proposal for a new central detector (ITS3) for the LHC run 4 ~\cite{ALICE-PUBLIC-2018-013}.

In the following we assume that the vertex telescope readout will pack the data to the level of 25 bits/hit.
FLUKA simulations have shown that an incoming Pb ion of 40 AGeV energy leaves in the vertex telescope on average $\sim 2.1\cdot 10^3$ hits in the case of a hadronic interaction taking place in one of the targets (with $15\%$ probability) and $\sim 8.3\cdot 10^2$ hits otherwise (due to the $\delta$-rays and conversions from electromagnetic interactions). This would lead on average to $\sim 1.03\cdot 10^3$ hits per incoming ion if we would use a trigger-less continuous readout. For the assumed $10^6$ ions/s beam intensity the corresponding data rate is $\sim 3.3$ GB/s, leading to $\sim 3.3$ PB of data collected per year.

We are currently considering to read out only events triggered by a segmented scintillator multiplicity detector placed close to beam line after the last pixel station. As one can see on Fig.~\ref{fig:flukavt} in Sec.~\ref{FLUKArate}, due to the bending of the soft $\delta$-rays in the dipole field, the occupancy is highly asymmetric in the bending plane.
Preliminary estimates of triggering capabilities are based on a cut on the multiplicity seen by the scintillator detector in the acceptance region least affected by $\delta$-rays. For simplicity it is assumed that the geometric coverage of such detector corresponds to the region covered by one of the two lower sensors ($y<0$, referring to Fig.~\ref{fig:flukavt}, where the bending direction of the dipole was assumed to be vertical).
More elaborate schemes accounting for the observed multiplicity asymmetry may allow better rejection of ions suffering only electromagnetic interactions and are currently being studied.

\begin{table}[h]
\caption{
Performance of triggering on the multiplicity observed in the scintillator detector. \label{tab:trig_scenario}
}
\begin{center}
\begin{tabular}{cccccc}
selection,\%  & trigger   &  purity, \% & hits readout     & hits readout & readout rate, GB/s\\
              & rate, kHz &             & per incoming ion & per trigger  &                   \\
\hline
50   & 100  & 80 &  300 & 2960 & 0.94   \\
\hline
80   & 365  & 35 &  675 & 1541 & 2.1 \\
\hline
100 & 1000 & 16 & 1030 & 1030 & 3.3 \\
\hline
\end{tabular}
\end{center}
\end{table} 

\begin{figure}[h]
    \centering
    \includegraphics[scale=0.5]{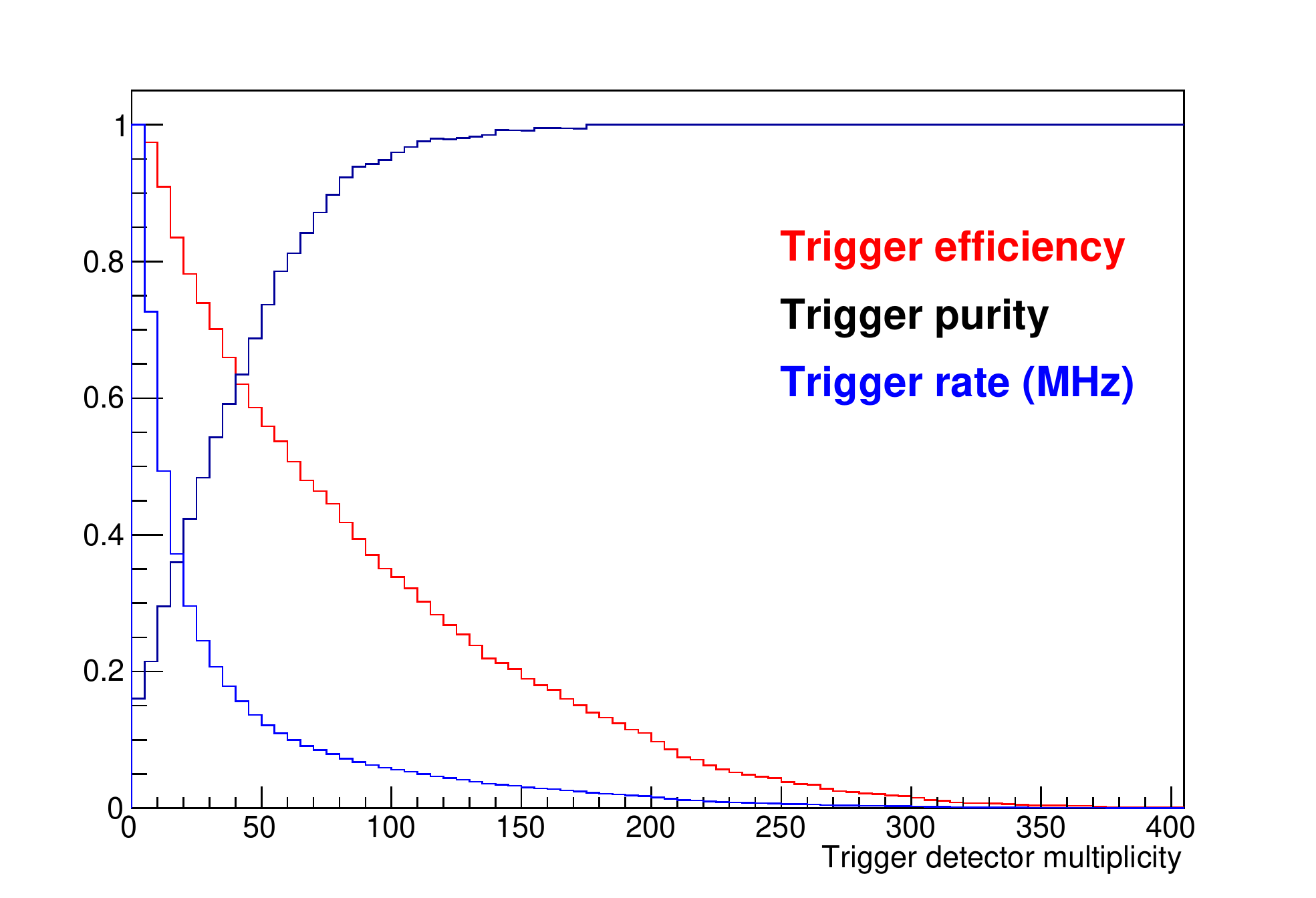}
    \caption{Trigger efficiency, purity and trigger rate as a function of the multiplicity cut in the trigger detector.}
    \label{fig:trigeff}
\end{figure}

Fig.~\ref{fig:trigeff} shows, as a function of the multiplicity cut, the fraction of selected interacting ions (i.e. the efficiency of the trigger) as well as the ratio between the interacting and total incident Pb ions  (i.e. the purity of the selection). Table \ref{tab:trig_scenario} gives numerical values for two scenarios corresponding to multiplicity thresholds allowing to read out $50\%$ and $80\%$ most central PbPb interactions. Values corresponding to a trigger-less continuous readout are also shown (100\% selection).

Processing of the collected data will be performed off-line. The needed computing resources are also dominated by the requirements of data reconstruction in the vertex telescope. We plan to use a modified version of the {\it Cellular Automaton} track finder developed for the ALICE ITS~\cite{Puccio:2016biw}, assuming three passes per trigger to find long primary, long secondary and short tracks.
Using the benchmarks prepared for the reconstruction of the ALICE ITS data we estimate the data decoding and cluster-finding to require $\sim 240$ ($\sim 450$) CPU seconds for 50\% (80\%) efficiency triggering scenarios, for $10^6$ incoming ions.  The corresponding track finding time should be $\sim 4200$ CPU seconds, from   a conservative extrapolation of track finding with the 7 layers of the ALICE ITS. The numbers are quoted for a single core of the {\it Intel i7-8700K @ 3.7 GHz} processor. Both clustering and track finding are parallelizable (in ALICE the tracking is also ported on GPUs). As a rough estimate, the data collected per period can be fully processed in  2--3 months by a farm of $\sim 100$ modern multicore processors or equivalent GRID jobs.

\newpage
\section{Experimental site}
\label{ExperimentalSite}
\graphicspath{ {./Figures Chapter: Experimental Site/} }
\vskip 0.4cm
\subsection{EHN1 hall and current layout of zone PPE138}
\begin{figure}[h]
\begin{center}
\includegraphics[width=0.8\textwidth]{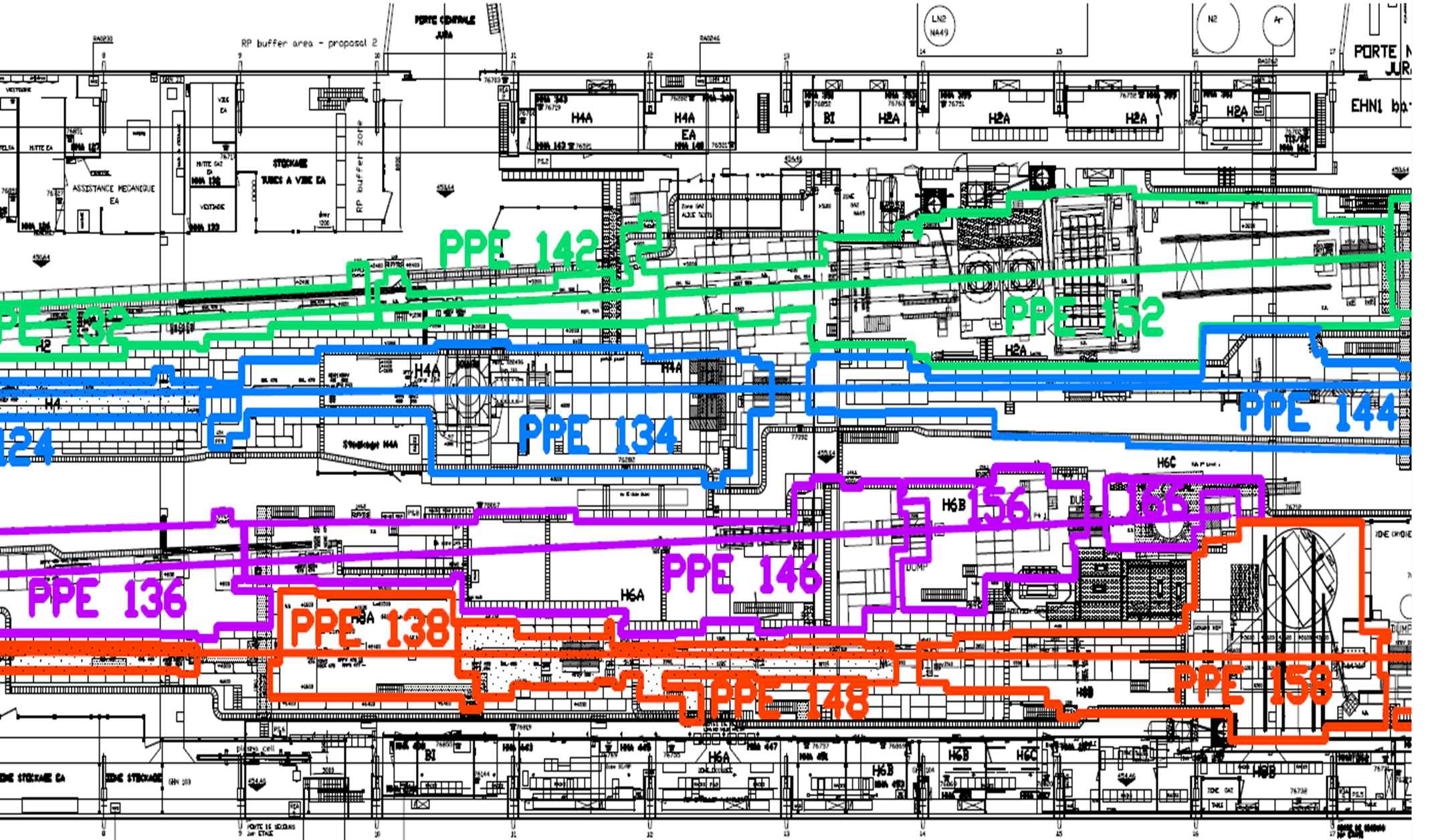}
\end{center}
\caption{Layout of EHN1 hall with its user zones.}
\label{fig:EHN1_Layout}
\end{figure}

\vskip 0.2cm
Several locations have been considered for the potential accommodation of the NA60+ detector in the EHN1 hall of the CERN SPS North Area. 
Figure~\ref{fig:EHN1_Layout} shows the layout of the EHN1 hall and indicates the user zones within the hall. The four beamlines H2, H4, H6 and H8 run across the hall from the left towards the right side of the diagram. Their user zones are marked by green, blue, violet and red colours, respectively.  
The PPE138 zone (in the bottom left quarter of Figure \ref{fig:EHN1_Layout}) was identified to be the most promising candidate, considering the absence of another major fixed target experiment on this beamline (competing for space and beam time) and the full spectrum of requests within the Physics Beyond Colliders programme for the other zones.

\begin{figure}[h]
\begin{center}
\includegraphics[width=0.8\textwidth]{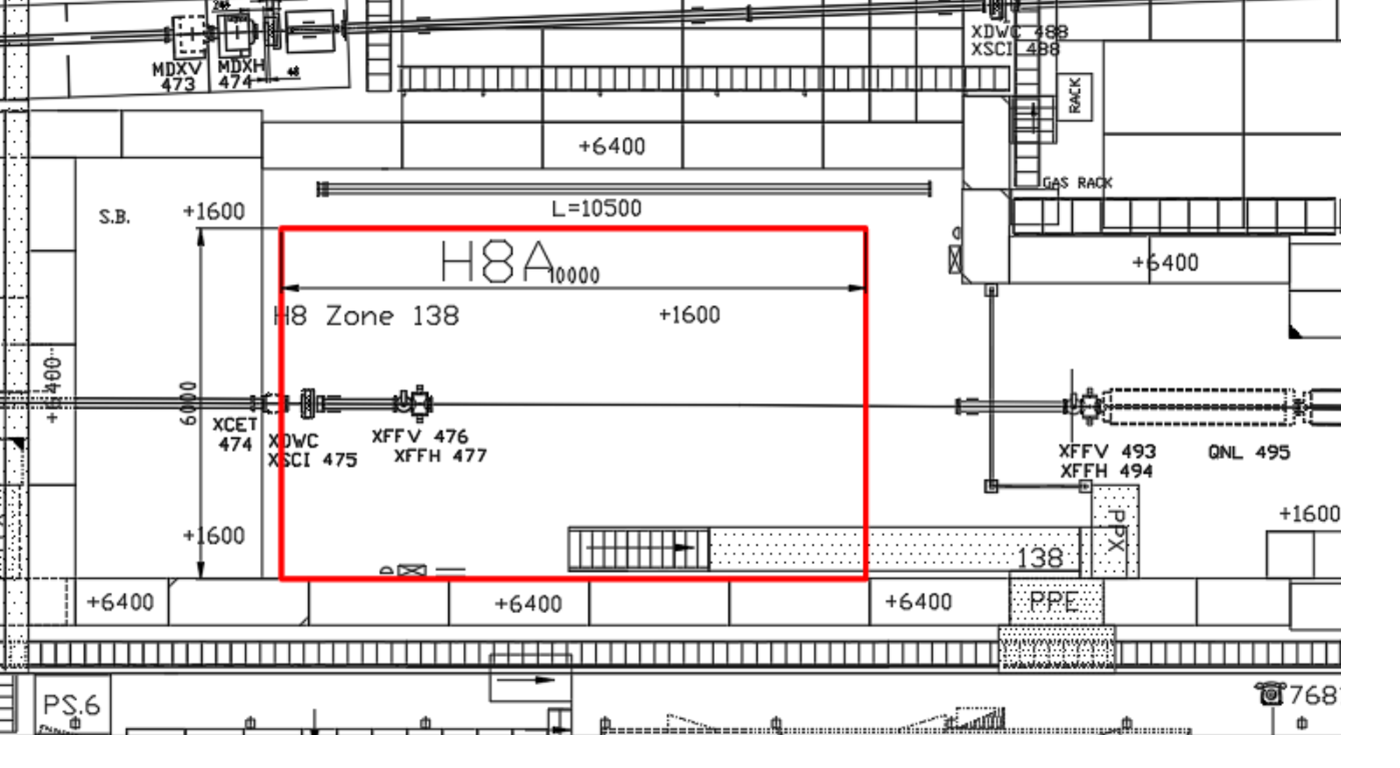}
\end{center}
\caption{Current layout of zone PPE138 of EHN1 hall.}
\label{fig:PPE138_Layout}
\end{figure}

The current layout of the zone is shown in Figure \ref{fig:PPE138_Layout}. The beam enters the zone from the left side and travels towards the right. The entrance door of the zone is indicated in the bottom right corner as PPE. The potential location of the NA60+ detector is marked with a red rectangle. In order to accommodate the experiment, the zone would need to be substantially modified in regard to its shielding, access and layout. The proposed beam setup, zone layout and integration studies, as well as radioprotection studies, are described in the following sections and summarized in Ref.~\cite{Gerbershagen:2022zbq}.

\subsection{Proposed beam setup}
\label{beam_setup}
\vskip 0.2cm
The slow extraction of the ions from the SPS into the CERN North Area is performed in debunched spills of about 10 seconds duration. The maximal duty cycle is 50\% and consists of two spills within a supercycle interval of 40 seconds. This value has been taken as a baseline for the investigation of Radiation Protection (RP) related issues, beam optics calculation and integration design. 

\begin{figure}[h]
\includegraphics[width=\textwidth]{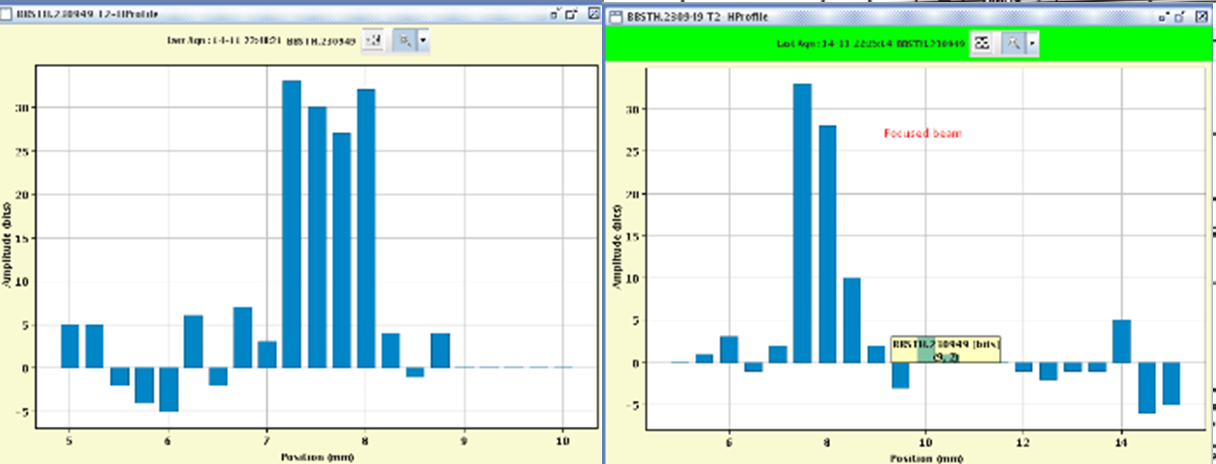}
\caption{Histogram profile of 150A GeV/c lead ion beam measured at the T2 target location.}
\label{fig:Beam_at_T2}
\end{figure}

The beam intensity required to fulfil the NA60+ physics programme for the setup that can be installed in EHN1 is $10^7$ primary lead ions per spill.
The beam spot size at the experiment location needs to be as small as possible, with the complete beam intensity fitting within a square 6 mm side hole in the central part of the vertex detector stations.
The requested beam intensity can routinely be delivered by the accelerator chain and a strong collimation will even be needed to reduce the intensity delivered by the SPS to  $10^7$ ions per spill at the experiment, where the intensity is limited by the RP considerations. The beam parameters at the beginning of the H8 transfer line (location of T4 target, which would be moved out of the beam path) are not well known due to the lack of precise beam instrumentation. At the T2 target, where the beam conditions are not identical, but comparable, a measurement of the beam size has been performed in 2017. The 150A GeV/c lead ion beam had the profile displayed in Figure \ref{fig:Beam_at_T2} and an overall beam size of approx. 1 mm.

Based on this measurement and for the purpose of this study, a conservative assumption has been made that the 150A GeV/c lead ion beam at the T4 location would have an RMS beam size of 0.5 mm (full size of ~2 mm) and that the beam divergence has a comparably large RMS value of 0.5 mrad (see Table \ref{tab:Beam_at_T2}). In order to estimate the values for a low energy beam of 30A GeV/c, only the geometrical change of divergence (proportional to $1/\sqrt{p}$) has been considered. In reality, additional changes of beam size and divergence can be expected due to the limited precision of SPS rectifiers and the reduced response from the beam instrumentation at lower momenta; however, the exact amount of their contributions is difficult to estimate. It should also be noted that the RMS size of the initial beam divergence is a less important parameter for the estimation of the beam size, since the maximal divergence is anyway limited by the H8 beamline acceptance, which is dependent on the apertures and optics settings of the H8 beamline. However, the initial divergence is relevant for the estimation of the relative transmission through the beamline for the different beam optics options. 

\begin{table}[h!]
\caption{Assumptions of initial lead ion beam parameters at the location of the T4 target (starting point of the H8 beamline).}
\centering
\begin{tabular}{||c c c ||} 
 \hline
 Parameter & 160A GeV/c & 30A GeV/c  \\ [0.5ex] 
 \hline\hline
 $\sigma _x$ (mm)	 & 0.5 & 1.15  \\ 
 \hline
 $\sigma _y$ (mm) & 0.5  & 1.15  \\
 \hline
 $\sigma p_x$ (mrad) & 0.5 & 0.5  \\
 \hline
 $\sigma p_y$ (mrad) & 0.5 & 0.5  \\
 \hline
 $\sigma _{p}/p$ (\%) & 0.1 & 0.1  \\
 \hline
\end{tabular} \\
\label{tab:Beam_at_T2}
\end{table}

\begin{figure}[h]
\includegraphics[width=0.5\textwidth]{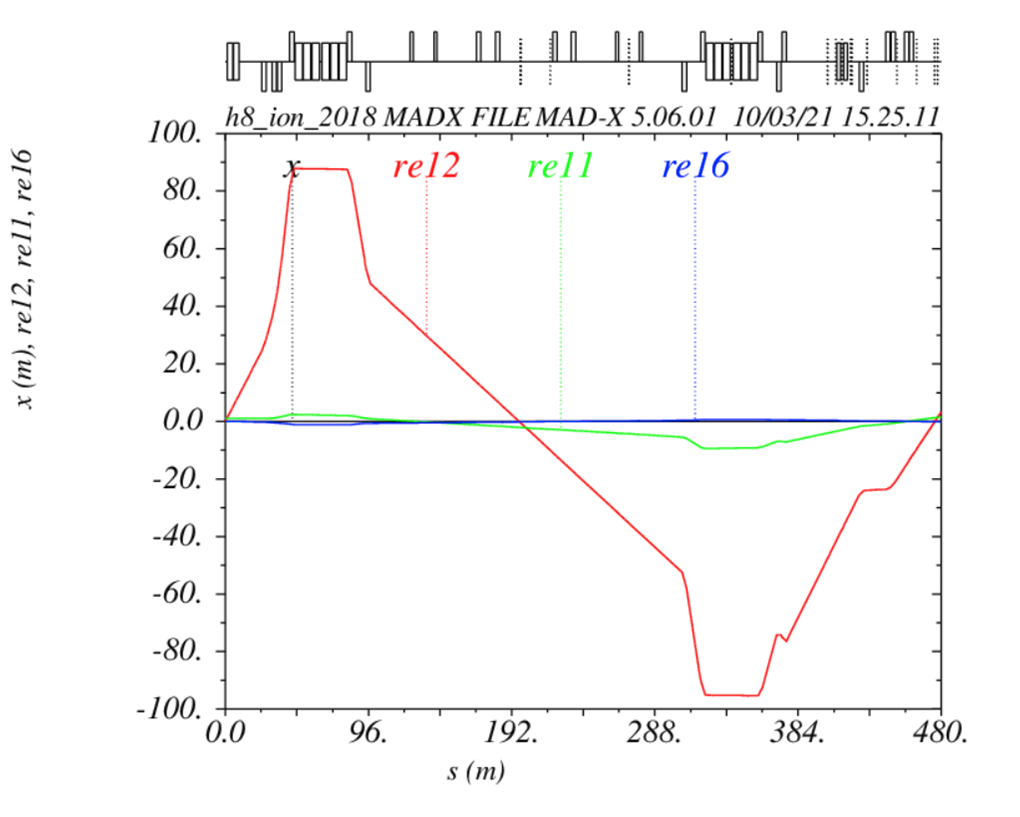}
\includegraphics[width=0.5\textwidth]{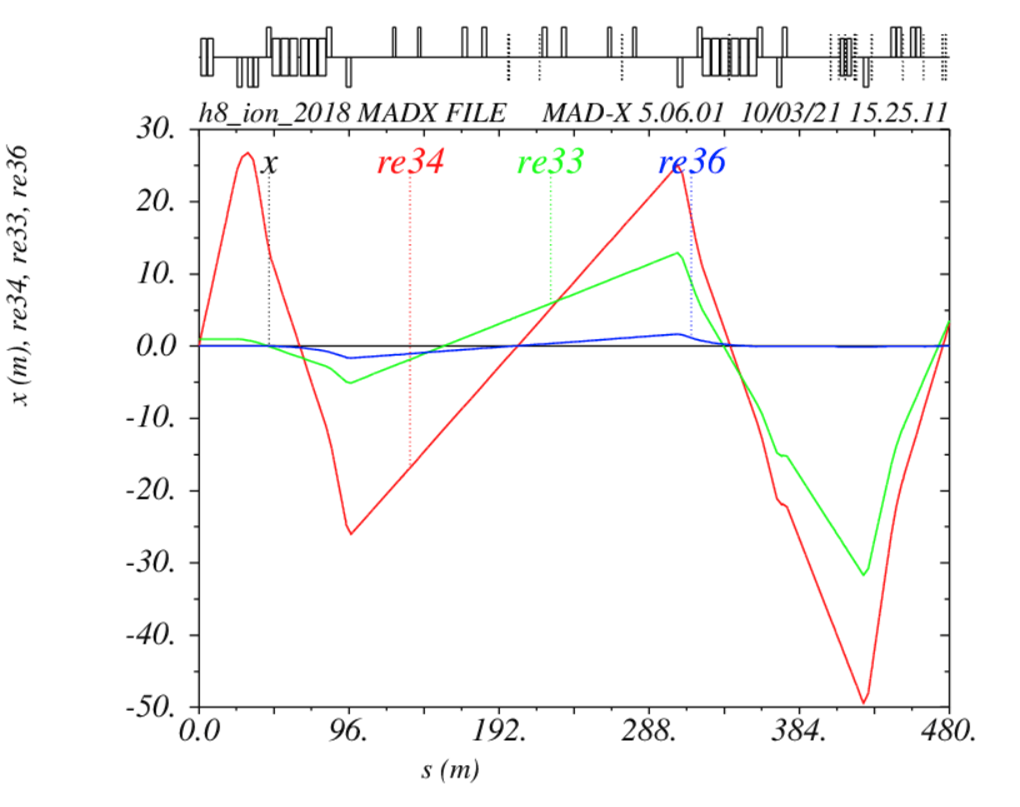}
\caption{Optical transfer matrix functions for the horizontal (left) and vertical (right) plane for the new Microcollimator ion beam optics for NA60+. The 
horizontal axis is the position in meters along the H8 beamline. The green, red and blue curve represent the contribution of the initial beam size, beam divergence and momentum spread, respectively, to the beam size at a given location.}
\label{fig: Microcollimator_optics}
\end{figure}

The currently used beam optics settings would deliver a beam of 0.8 mm RMS transverse size at the location of the experiment, which is larger than required by NA60+. Hence, two new optics settings have been developed, aiming at a reduction of the beam size. Both optics versions are achromatic in first order. One is based on the use of the so-called Microcollimator, a very small and precisely aligned collimator used for the primary proton beam operation in H8 (see Figure~\ref{fig: Microcollimator_optics}). It provides a comparably low transmission (which might be good in case a strong reduction of the primary beam intensity is required) as well as high beam stability, since the beam is imaged from the well-defined physical gap of the Microcollimator to the location of the experiment. The setting of the Microcollimator gap can be also directly used to modify the beam size at NA60+.

\begin{figure}[h]
\includegraphics[width=0.5\textwidth ]{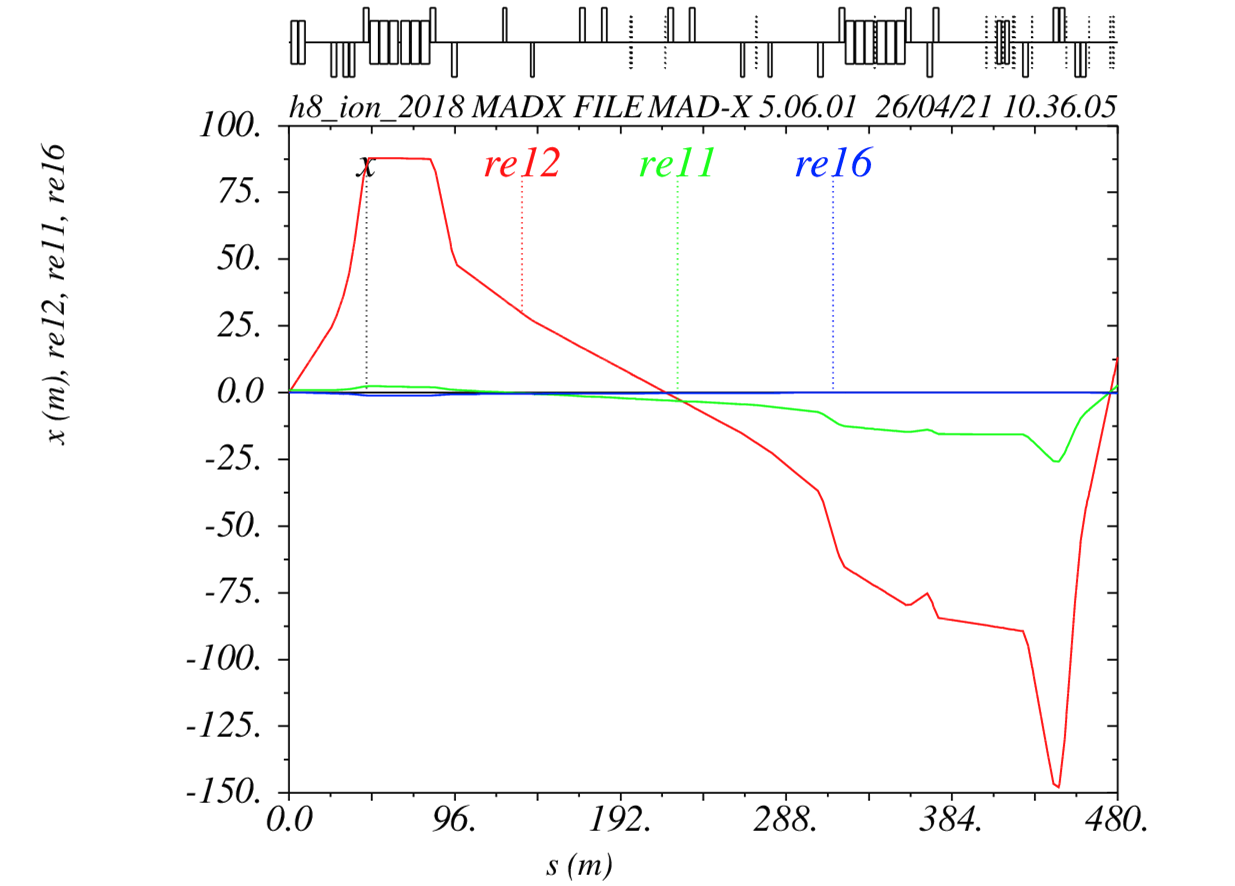}
\includegraphics[width=0.5\textwidth ]{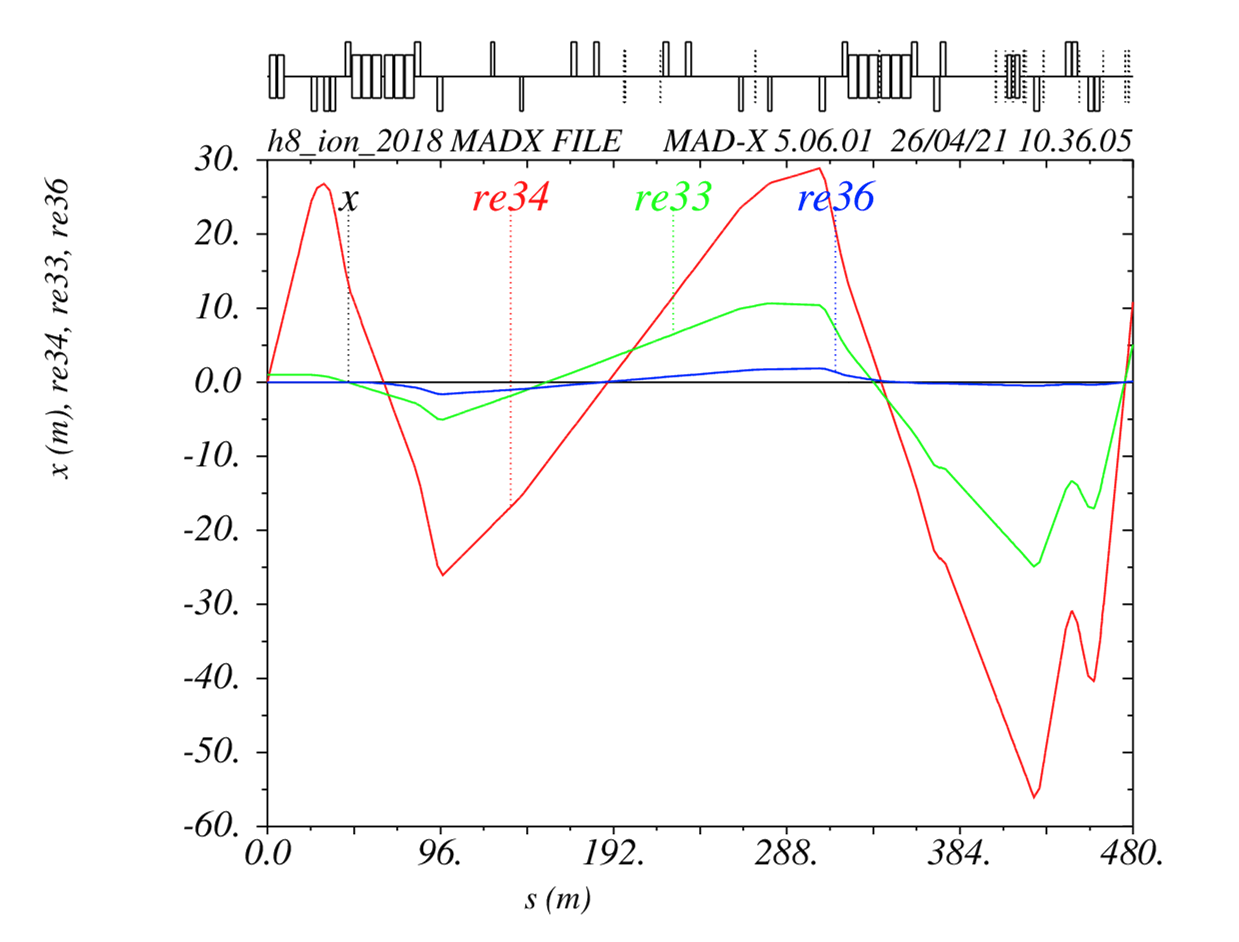}
\caption{Optical transfer matrix functions for the horizontal (left) and vertical (right) plane for the new strongly focussed ion beam optics for NA60+. The 
horizontal axis is the position in meters along the H8 beamline. The green, red and blue curve represent the contribution of the initial beam size, beam divergence and momentum spread, respectively, to the beam size at a given location.}
\label{fig: Focussing_optics}
\end{figure}

The second optics does not use the Microcollimator, but instead relies on stronger focussing of the beam at the experiment location (see Figure \ref{fig: Focussing_optics}). The resulting beam sizes at the experiment location and transmissions through the H8 beamline are summarized in Tables \ref{tab:Beam_with_Microcollimator_optics} and \ref{tab:Beam_with_new_optics}.

\begin{table}[h!]
\caption{Summary of the beam parameters at the potential location of NA60+ for the Microcollimator design.}
\centering
\begin{tabular}{||c c c ||} 
 \hline
 Parameter & 160A GeV/c & 30A GeV/c  \\ [0.5ex] 
 \hline\hline
 $\sigma _x$ (mm)	 & 0.33 & 0.35  \\ 
 \hline
 $\sigma _y$ (mm) & 0.34  & 0.36  \\
 \hline
Transmission from T4 (\%) & 12.22 & 2.91  \\
 \hline
\end{tabular} \\
\label{tab:Beam_with_Microcollimator_optics}
\end{table}

\begin{table}[h!]
\caption{Summary of the beam parameters at the potential location of NA60+ for the focusing optics design.}
\centering
\begin{tabular}{||c c c ||} 
 \hline
 Parameter & 160A GeV/c & 30A GeV/c  \\ [0.5ex] 
 \hline\hline
 $\sigma _x$ (mm)	 & 0.19 & 0.33  \\ 
 \hline
 $\sigma _y$ (mm) & 0.19  & 0.36 \\
 \hline
Transmission from T4 (\%) & 32.43 & 23.5  \\
 \hline
\end{tabular} \\
\label{tab:Beam_with_new_optics}
\end{table}

\subsection{Proposed zone layout and integration studies}
\label{Proposed_zone_layout}
\vskip 0.2cm

\begin{figure}[h!]
\begin{center}
\includegraphics[width=0.8\textwidth]{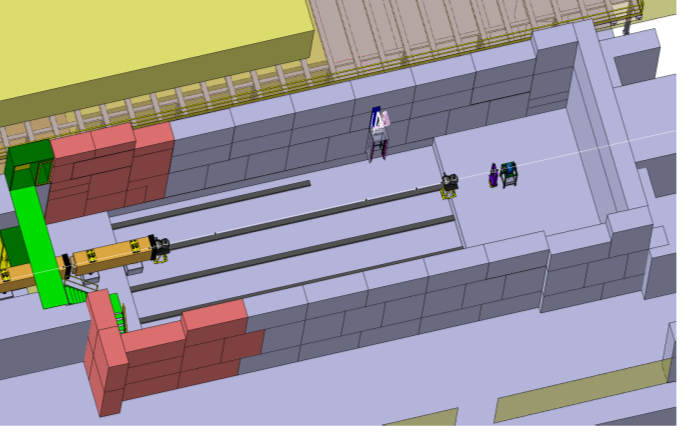}
\end{center}
\caption{3D drawing of the PPE138 zone after the proposed modification for accommodating the NA60+ detector.}
\label{fig:PPE138_with_NA60}
\end{figure}

Currently, the PPE138 zone is utilized for providing test beams to several users, such as LHCb, R2E, UA9, and others, and hence is not optimally laid out for the installation of a major detector. An integration study has been conducted, revealing that the placement of the NA60+ detector (radius of 3.1 m and maximal length of 13.7 m) is feasible, provided the zone layout is modified substantially. The 3D-drawing of the modified design is displayed in 
Fig.~ \ref{fig:PPE138_with_NA60}.

\begin{figure}[h]
\begin{center}
\includegraphics[width=0.8\textwidth]{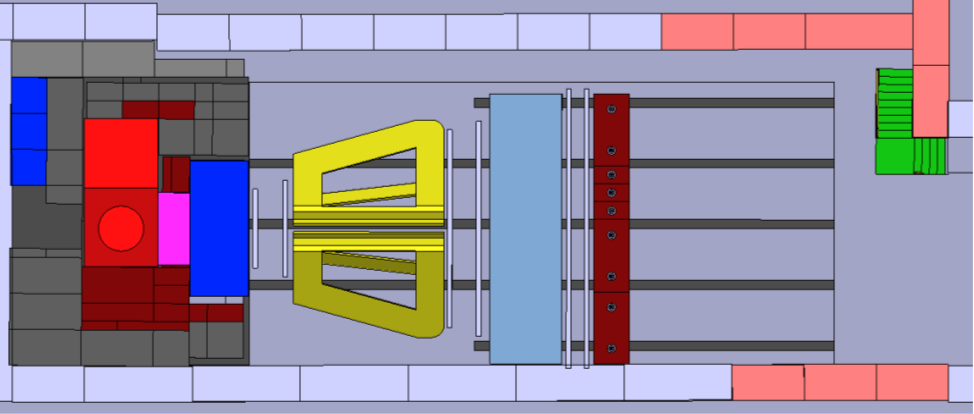}
\end{center}
\caption{Drawing of the short (10.4 m) setup of NA60+ installed in the modified PPE1138 zone, top view.}
\label{fig:NA60_short}
\end{figure}

\begin{figure}[h]
\begin{center} 
\includegraphics[width=0.8\textwidth]{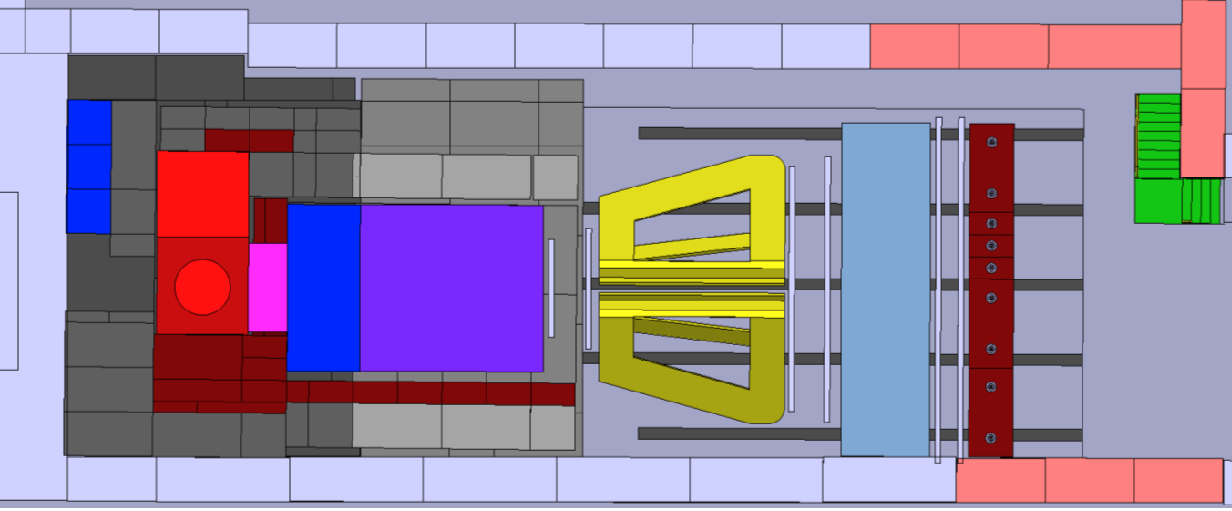}
\end{center}
\caption{Drawing of the long (13.7 m) setup of NA60+ installed in the modified PPE1138 zone, top view.}
\label{fig:NA60_long}
\end{figure}

The shielding wall on the Jura side of the zone, separating it from the user zones of the H6 beam line, would have to be moved towards the H6 beam line by 80 cm, which in itself does not present a major problem for the H6 beam line zones, apart from the displacement of the cable trays. A new bridge and stairs to access the detector will need to be installed (depicted in green in Fig.~\ref{fig:PPE138_with_NA60}). Since the H8 beam height above the hall ground is 2.88 m, an excavation needs to be performed for accommodating a detector with 3.1 m radius and its mounting structure. The depth of the excavation has been set to 1 m and the transverse dimension to 6 m. The longitudinal extent of the excavation needs to cover the two setup lengths, the short one (10.4 m total length, see Figure \ref{fig:NA60_short}) for the low energy run and the full length one (13.7 m total length, see Figure \ref{fig:NA60_long}) for the high energy run. Rails must be installed on the bottom of the excavated area to enable the longitudinal movement of the toroidal magnet and of the muon wall for the modification of the setup between the short and the long versions.
    
\begin{figure}[h]
\includegraphics[width=\textwidth]{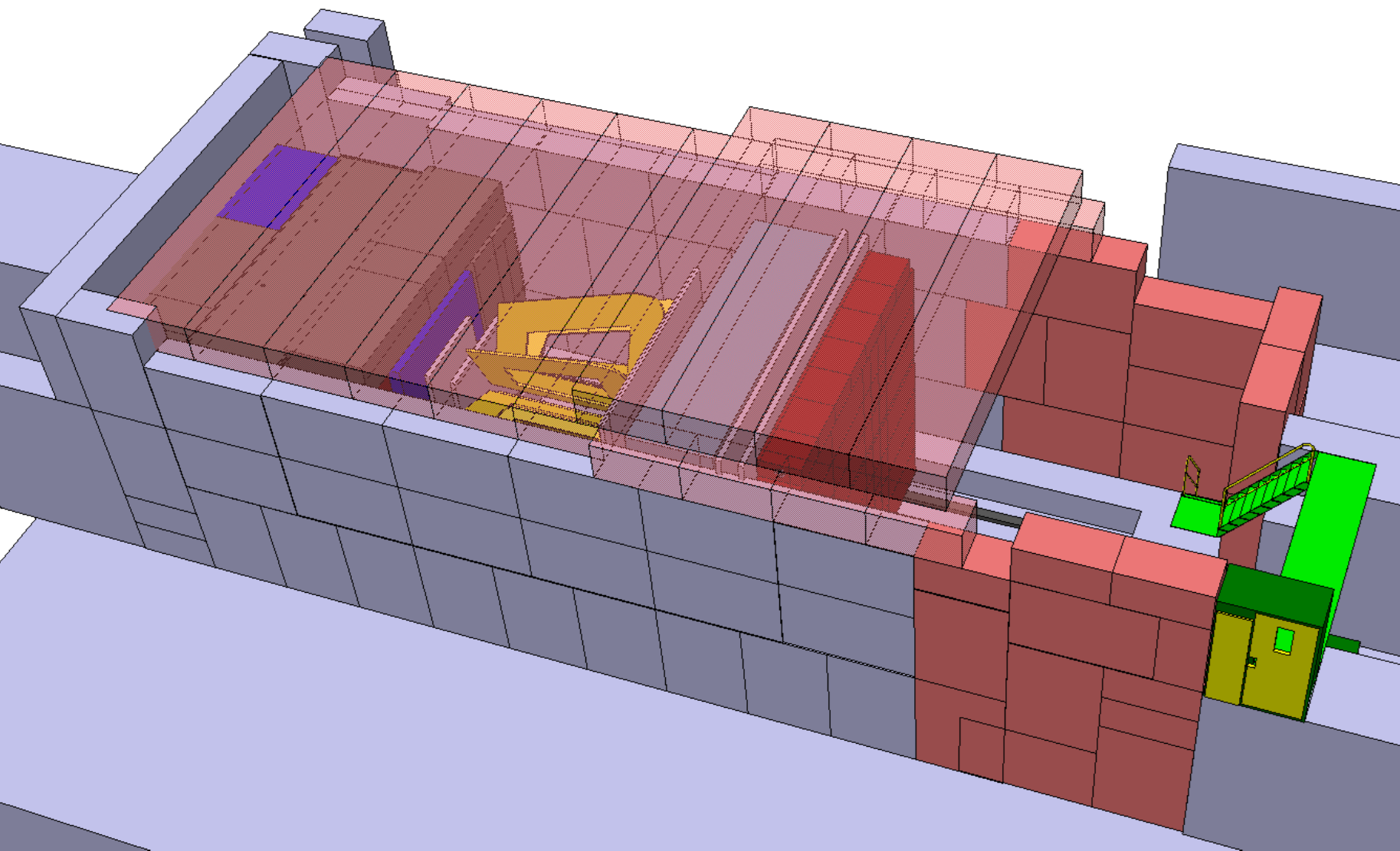}
\caption{3D drawing of the PPE138 zone with the NA60+ detector and the additional shielding.}
\label{fig:NA60_with_roof}
\end{figure}

The dipole magnet around the NA60+ target is shown in Fig.~\ref{fig:NA60_short}  and Fig.~\ref{fig:NA60_long} in light red colour. The integration includes the installation of the additional shielding, required due to the radiation protection considerations described in Sec.~\ref{sec:RP}. It includes the concrete and iron shielding blocks in the region around the target and behind the muon wall, marked as grey and dark-red blocks in Fig.~\ref{fig:NA60_short} and Fig.~\ref{fig:NA60_long}, as well as the installation of a roof shielding, shown in Fig.~\ref{fig:NA60_with_roof}.
 
\subsection{Radiation protection studies}
\label{sec:RP}
\vskip 0.2cm
A detailed radiation protection assessment was performed since NA60+ aims at pushing the beam intensity by at least one order of magnitude with respect to what is currently delivered to EHN1. The area surrounding the proposed location of NA60+ is classified according to CERN’s radiological classification \cite{RPclassification} as a Supervised Radiation Area with a low occupancy zone on one side (15~$\mu$Sv/h limit) and permanent workplaces on the other (3~$\mu$Sv/h limit). The shielding structure therefore has to be designed in such a way as to sufficiently reduce the prompt radiation to be compatible with the ambient dose equivalent rate limits linked to the area classification. The residual dose rates and air activation were also analysed, along with accidental beam losses in the beamline upstream of the experiment. The assessment was based on the \fluka Monte Carlo particle transport code~\cite{Ahdida:2022gjl, Battistoni:2015epi}. The \fluka simulations were performed using the latest released version (FLUKA 4-1.0), while the geometry was created using Flair~\cite{Vlachoudis:2009qga}.

\subsubsection{Shielding layout}

\begin{figure}[h]
\begin{center}
\includegraphics[width=0.7\textwidth]{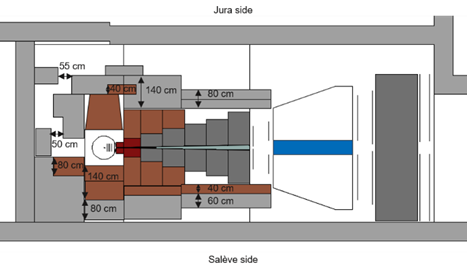}
\includegraphics[width=0.7\textwidth]{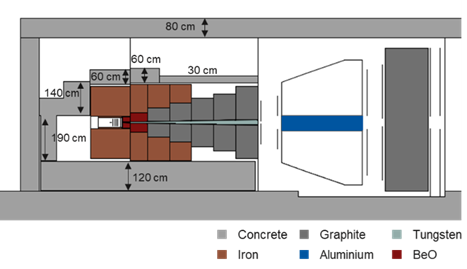}
\end{center}
\caption{Shielding layout of NA60+ for a 160A GeV/c Pb beam, as implemented in \fluka, with view from the top (left) and from the side (right).}
\label{fig:RP_setup}
\end{figure}

The final shielding layout for NA60+ for the case of a 160 A GeV/c Pb beam is depicted in Figure~\ref{fig:RP_setup}. In the most critical region around the target and the absorber a ﬁrst layer of iron shielding, providing a higher attenuation of the radiation than concrete shielding, was implemented, which is then followed by additional concrete shielding. A chicane upstream of the target was added to allow access to the target region under certain conditions. A concrete shielding roof spanning the whole detector setup was added to reduce skyshine radiation.

\subsubsection{Prompt dose rates}

\begin{figure}[h]
\includegraphics[width=0.5\textwidth]{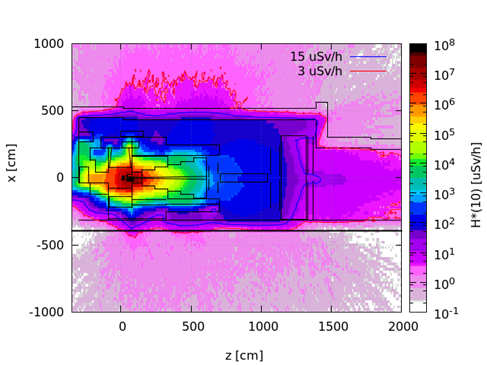}
\includegraphics[width=0.5\textwidth]{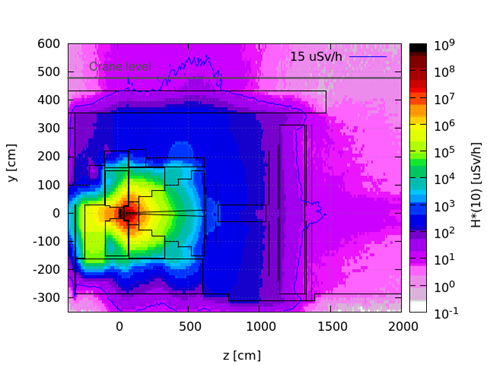}
\caption{View from above (left) and side (right) of the prompt ambient dose equivalent rate in $\mu$Sv/h for $1 \times 10^7$ Pb ions/spill of 160A GeV/c with 2 spills of 10 s every 40 s. The horizontal cut in the left figure is vertically averaged over $\pm$50 cm around the beam axis. The vertical cut in the right figure is horizontally averaged over $\pm$40 cm around the beam axis. The red and blue lines illustrate the 3~$\mu$Sv/h and 15~$\mu$Sv/h dose rate limits for a Supervised Radiation Area with permanent and low occupancy workplaces, respectively.}
\label{fig:RP_prompt}
\end{figure}

\begin{figure}[h!]
\centering
\includegraphics[width=0.8\textwidth]{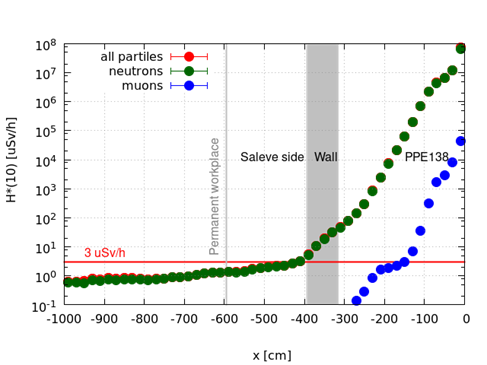}
\caption{Vertical transverse cut of  the prompt ambient dose equivalent rate in $\mu$Sv/h for $1 \times 10^7$ Pb ions/spill of 160A GeV/c with 2 spills of 10 s every 40 s. The values are obtained by averaging vertically over $\pm$50 cm around the beam axis and longitudinally over 50-100 cm behind the front of the target. The red line illustrates the 3~$\mu$Sv/h dose rate limit for a Supervised Radiation Area with permanent workplaces.}
\label{fig:RP_promptx}
\end{figure}

Figure~\ref{fig:RP_prompt} and~\ref{fig:RP_promptx} depict the prompt ambient dose equivalent rate distributions H*(10) (used as a standard quantity by CERN RP) for the experimental zone and its surroundings. A safety factor of 3 was taken into account for uncertainties related to material densities, geometry, beam parameters, simulations, etc., meaning that the displayed dose rates are 3 times greater than the simulation output for nominal intensity. The results show that the shielding allows sufficient reduction of the ambient dose rates to comply with the 3~$\mu$Sv/h and 15~$\mu$Sv/h dose rate limits. However, towards the top at the level of the crane driver cabin, which is located at a height of 7.65 m from the ﬂoor, the dose rate slightly exceeds the 15~$\mu$Sv/h dose rate limit. During beam operation with 160 A GeV/c lead ions a crane exclusion zone above the experiment will therefore need to be put in place.

\subsubsection{Residual dose rates}

\begin{figure}[h]
\centering
\includegraphics[width=0.45\textwidth]{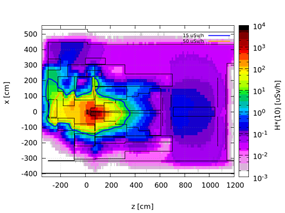}
\includegraphics[width=0.45\textwidth]{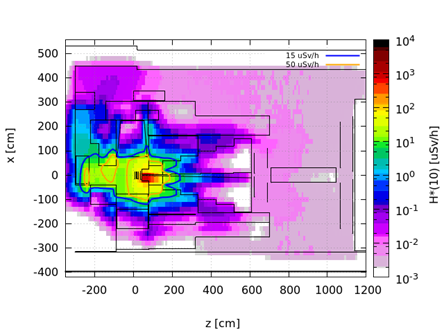}
\includegraphics[width=0.45\textwidth]{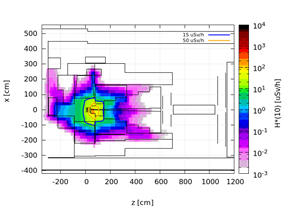}
\caption{View from above of the residual ambient dose equivalent rate H*(10) (in µSv/h) for 1 minute (top left), 1 day (top right) and 1 week (bottom) of decay after 4 weeks of operation $1.2\times 10^{12}$ ions on target. The horizontal cut is vertically averaged over $\pm$50 cm around the beam axis. The blue and yellow lines illustrate the 15~$\mu$Sv/h and 50~$\mu$Sv/h dose rate limits for a Supervised and Simple Controlled Radiation Area, respectively.}
\label{fig:RP_res}
\end{figure}

The residual dose rates after 4 weeks of beam operation with 160 A GeV/c lead ions are presented in Figure~\ref{fig:RP_res} for different decay times. Also here a safety factor 3 was taken into account. The results show that the area outside of the shielding is compatible with a Supervised Radiation Area even for short decay times. However, close to the target the dose rates largely exceed the given 15 µSv/h limit of a Supervised Radiation Area. Directly upstream of the MEP48 dipole magnet, the dose rates reach approximately 440~$\mu$Sv/h and 12~$\mu$Sv/h after 1 minute and 1 week of cooling, respectively.  That implies that one week of cooldown should be foreseen before general controlled access to the area is given.
In view of the high residual dose rates in the target area, any access to the chicane leading to the target area is foreseen to be regulated by a specialized procedure. Access will be granted only with the required training for work in such high radiation areas and under supervision by a representative from the CERN Radiation Protection Group (a condition not applicable for the majority of the user zones in EHN1). For shorter cooling times, where the ambient dose rates exceed the limit of a Simple Controlled Radiation Area (50~$\mu$Sv/h limit), an operational dosimeter (DMC) is required next to the passive dosimeter (DIS). Furthermore, any work in the highly activated area must be optimized. Next to that, measures to prevent uncontrolled access to the area are to be foreseen.

\subsubsection{Air activation}

To evaluate air activation, the particle ﬂuences were scored in the air regions of the experimental zone and then combined with the energy-dependent radionuclide production cross-sections using the ActiWiz Creator tool~\cite{Vincke:2014vwm}. The dose due to inhalation of activated air was calculated by using the guidance value for airborne activity CA  and the inhalation dose coefﬁcients $e_{\rm inh}$ from the Swiss Radiological Protection Ordinance~\cite{OFSP}. The dose was estimated conservatively with 4 weeks of beam operation at maximum intensity and with no air exchange. The intensity per year has been calculated based on the assumptions of $1\times10^7$ Pb ions/spill of 160A GeV/c with 2 spills each 40 s, 4 weeks (28 days) of operation per year, thus yielding $1.2\times10^{12}$ ions per year on the NA60+ target.
When assuming full mixing between the air regions and no cooling, the speciﬁc airborne radioactivity amounts to 0.02 CA and therefore lies below the given limit of 0.1 CA for a Supervised Radiation Area at CERN~\cite{RPclassification}. Here, the largest contribution comes from the short-lived radionuclides $^{41}$Ar, $^{13}$N, $^{15}$O and $^{11}$C. The dose from inhalation during 1 hour of stay was estimated to be of 0.006~$\mu$Sv, with the main contribution coming from $^{14}$C, $^{32}$P, $^{7}$Be, $^{33}$P and $^{35}$S.

\begin{figure}[h!]
\centering
\includegraphics[width=0.8\textwidth]{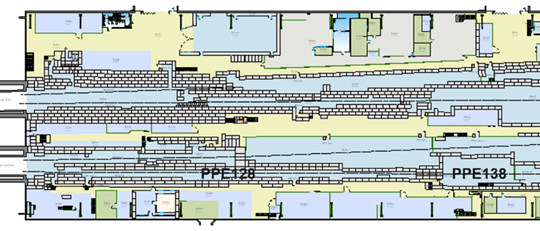}
\caption{Layout of the shielding of the beamline zone PPE128. The shielding towards the top amounts to 80 cm of concrete.}
\label{fig:loss1}
\end{figure}

\subsubsection{Accidental beam loss}

Protective measures need to be put in place to mitigate the impact of accidental beam loss upstream of the NA60+ experiment.The shielding layout of the upstream region (PPE128) is depicted in Figure~\ref{fig:loss1}. At several locations only 80 cm of concrete shielding are present and there are unshielded access doors to the zone.

To study the radiation levels for an accidental beam loss upstream of the experiment, it was assumed that the beam is lost on a massive object, such as a magnet (iron cylinder of R = 30 cm, L = 200 cm), in a part of the zone where only 80 cm of concrete shielding is present. The results show that the accidental loss of one single spill can cause a dose exceeding 15 µSv.
At least 160 cm of concrete shielding in the upstream region is therefore required. Furthermore, shielding chicanes for the access doors need to be implemented, and a crane exclusion zone has to be established to prevent the crane to travel above the zone during the NA60+ beam operation.


While there is an extensive radiation protection monitoring system covering the most critical areas of EHN1~\cite{monitoring1,Boukabache:2017msn}, which raises an alarm in case the radiation levels exceed the set limits, there are currently no dedicated monitors to detect beam losses upstream of the experiment. Additional  monitors would therefore need to be installed in this zone.

\clearpage

\newpage
\section{Timeline and preliminary cost estimate}
\label{TimelineCost}
\vskip 0.4cm
\subsection{Foreseen timeline: R\&D, construction, installation and data taking 
}
\vskip 0.2cm
At the submission time of this Letter of Intent, the R\&D phase is already started for all the sub-detectors involved in the project. Considering the level of detail already reached in this document, we expect to be able to submit an experiment proposal by the end of 2024. Should this proposal be approved our aim is to be able to build and install the experiment for a first data taking to be held after the Long Shutdown 3 of the LHC. Current estimates place this date at around 2029. Hereafter we briefly summarize the present status and the next steps for the realization of the experiment, as they can be foreseen today (end of 2022).

\begin{itemize}
    \item For the vertex spectrometer, studies carried out in synergy between ALICE and NA60+ have converged up to the production of a first stitched MAPS prototype that will become available at the beginning of 2023.  The main steps of the project are planned as follows: finalization of the pixel sensor and finalization of R\&D on mechanics, cooling and interconnections within 2025. The assembly of the final system could start from 2026. We stress that the timelines for the construction of the vertex spectrometer of NA60+ and of the corresponding ALICE detector making use of the same technology (ITS3) are identical, helping in this way the parallel development of the two objects.
\vskip 0.3cm
\item For the muon spectrometer, the MWPC option is currently more advanced and the prototype testing phase has already started in the lab, to establish the working parameters and measure the detector resolution. Beam studies will then be carried out at the CERN SPS. After the design and test of the final detector modules, the mass production of detector elements may start. The total surface of the tracking detectors is $\sim 100$ m$^2$, requiring the participation of at least two major facilities dedicated to gas detector construction. The other main option currently considered, three-foil GEMS, was already validated for the MOLLER experiment at JLab and represents an alternative or complementary solution to MWPCs. The final design of the muon tracker may include both types of detectors, to cover for example angular regions with different expected particle rates.
\vskip 0.3cm
\item The toroidal magnet represents another key element of the set-up. The construction and test of the 1:5 scale prototype has allowed the validation of the mechanical feasibility of the design. Comparisons of various parameters (resistance, inductance, magnetic field) with simulation results have shown that a good control over these quantities was reached. The next goal is the design of the full-scale object, which requires a detailed study of several engineering issues, due to the large magnetic forces acting on the structure. The construction of the magnet will likely have to be outsourced to a specialized firm.  
\vskip 0.3cm
\item The study of beam-related issues, including optics adapted to the experimental requirements and the co-existence with other fixed-target and collider heavy-ion projects, has started. A joint document by the experimental collaborations and the CERN accelerator groups is being prepared, as an input for SPSC/LHCC and the CERN management. The aim is to define the ion beam requirements after the Long Shutdown 3, when NA60+ should take its first data.
\vskip 0.3cm
\item The integration studies carried out until now have shown that the experimental set-up can be installed in the PPE138 experimental area, on the H8 beam line. Keeping into account the height of the beam line (265 cm) and the presently foreseen dimension of the larger tracking stations ($r\sim$ 315 cm), an excavation of the floor of PPE138 is required, in the region to be occupied by the muon spectrometer. A thick iron/concrete shielding, surrounding the experiment was also designed, in order to have dose levels outside the experimental area compliant with the CERN regulations. 
\vskip 0.3cm
\end{itemize}

Once the data taking will start, we foresee an experimental run with a 10$^6$ s$^{-1}$ primary Pb beam each year, assuming the usual $\sim 1$ month availability of ion beams at CERN. We would like to take data for at least six different beam energies, with one collision energy explored each year, exploring the interval between $\sim 20$ and 158 GeV/nucleon. Should the corresponding expected integrated luminosity not be reachable, as it might be the case for the lower energy points, some of the measurements may span more than one year of data taking. 
Corresponding periods with a proton beam, at the same energy per nucleon of the ion beam, with an integrated luminosity roughly equivalent to the Pb one, are also a fundamental ingredient of the NA60+ physics program.   

\subsection{Preliminary cost estimate
}
\vskip 0.2cm
    A precise cost estimate for the NA60+ experiment is presently not feasible, due to the fact that a final definition of the set-up details is still not possible. In any case estimates performed today may evolve in the next future due, for example, to oscillations in the cost of raw materials, electronics, and so on. With this in mind, and based on the considerations reported in Chapter~\ref{Detectors}, we have estimated a total cost between 10 and 13  MCHF. The evaluation of the costs related to data acquistion, storage and computing is still in progress. The quoted range is related to: (i) the potential need of a second engineering run for the vertex spectrometer MAPS; (ii) using MWPCs for all the tracking stations of the muon spectrometer, or GEMs for the two upstream stations and MWPCs for the others.  A 1:1 rate exchange with Euro and USD was assumed. In Table~\ref{tab:Costs} a breakup of the cost is presented.

\begin{table}[t]
    \centering
    \begin{tabular}{|c|c|}
    \hline
      Sub-system   &  Estimated cost (MCHF) \\
    \hline \hline
     Vertex spectrometer & 2.5 -- 3.1\\
     \hline
     Muon spectrometer & 2.7 -- 4.0 \\
     \hline
     Toroidal magnet & 3.8 \\
     \hline
     RP monitors, Shielding & 1.5 \\
     \hline
     \hline
     {\bf Total} & 10.5 -- 12.4 \\
     \hline
   \end{tabular}
    \caption{Estimated costs of the various NA60+ subsystems. }
    \label{tab:Costs}
\end{table}

\newpage
%

\bibliographystyle{utphys}
\bibliography{main}

\newpage
\section*{Appendix: NA60+ Collaboration}
\label{app:collab}
\bigskip
\begingroup
\begin{flushleft}
C.~Ahdida\Irefn{cern}\And
G.~Alocco\Irefnn{ucagliari}{cagliari}\And
F.~Antinori\Irefn{padova}\And
M.~Arba\Irefn{cagliari}\And
M.~Aresti\Irefnn{ucagliari}{cagliari}\And
R.~Arnaldi\Irefn{torino}\And
A.~Baratto Roldan\Irefn{cern}\And
S.~Beol\`e\Irefnn{utorino}{torino}\And
A.~Beraudo\Irefn{torino}\And
J.~Bernhard\Irefn{cern}\And
L.~Bianchi\Irefnn{utorino}{torino}\And
M.~Borysova\Irefnn{weizmann}{kyiv}\And
S.~Bressler\Irefn{weizmann}\And
S.~Bufalino\Irefnn{disat}{torino}\And
E.~Casula\Irefnn{ucagliari}{cagliari}\And
C.~Cical\`o\Irefn{cagliari}\And
S.~Coli\Irefn{torino}\And
P.~Cortese\Irefnn{piemonte}{torino}\And
A.~Dainese\Irefn{padova}\And
H.~Danielsson\Irefn{cern}\And
A.~De Falco\Irefnn{ucagliari}{cagliari}\And
K.~Dehmelt\Irefn{stonybrook}\And
A.~Drees\Irefn{stonybrook}\And
A.~Ferretti\Irefnn{utorino}{torino}\And
F.~Fionda\Irefnn{ucagliari}{cagliari}\And
M.~Gagliardi\Irefnn{utorino}{torino}\And
A.~Gerbershagen\Irefn{groningen}\And
F.~Geurts\Irefn{rice}\And
V.~Greco\Irefnn{ucatania}{catania}\And
W.~Li\Irefn{rice}\And
M.P.~Lombardo\Irefn{lfn}\And
D.~Marras\Irefn{cagliari}\And
M.~Masera\Irefnn{utorino}{torino}\And
A.~Masoni\Irefn{cagliari}\And
L.~Micheletti\Irefn{cern}\And
L.~Mirasola\Irefnn{ucagliari}{cagliari}\And
F.~Mazzaschi\Irefnn{cern}{utorino}\And
M.~Mentink\Irefn{cern}\And
P.~Mereu\Irefn{torino}\And
A.~Milov\Irefn{weizmann}\And
A.~Mulliri\Irefnn{ucagliari}{cagliari}\And
L.~Musa\Irefn{cern}\And 
C.~Oppedisano\Irefn{torino}\And
B.~Paul\Irefnn{ucagliari}{cagliari}\And
M.~Pennisi\Irefnn{utorino}{torino}\And
S.~Plumari\Irefn{ucatania}\And
F.~Prino\Irefn{torino}\And
M.~Puccio\Irefn{cern}\And
C.~Puggioni\Irefn{cagliari}\And
R.~Rapp\Irefn{tamu}\And
I.~Ravinovich\Irefn{weizmann}\And
A.~Rossi\Irefn{padova}\And
V.~Sarritzu\Irefnn{ucagliari}{cagliari}\And
B.~Schmidt\Irefn{cern}\And
E.~Scomparin\Irefn{torino}\And
S.~Siddhanta\Irefn{cagliari}\And
R.~Shahoyan\Irefn{cern}\And
M.~Tuveri\Irefn{cagliari}\And
A.~Uras\Irefn{lyon}\And
G.~Usai\Irefnn{ucagliari}{cagliari}\And
H.~Vincke\Irefn{cern}\And
I.~Vorobyev\Irefn{cern}
\renewcommand\labelenumi{\textsuperscript{\theenumi}~}
\end{flushleft}

\bigskip
\renewcommand\theenumi{\arabic{enumi}~}
\begin{Authlist}
\item \Idef{cern}European Organization for Nuclear Research (CERN), Geneva, Switzerland
\item \Idef{ucagliari}Dipartimento di Fisica dell'Universit\`{a} di Cagliari, Cagliari, Italy
\item \Idef{cagliari}INFN, Sezione di Cagliari, Cagliari, Italy
\item \Idef{padova}INFN, Sezione di Padova, Padova, Italy
\item \Idef{torino}INFN, Sezione di Torino, Turin, Italy
\item \Idef{utorino}Dipartimento di Fisica dell Universit\`{a} di Torino, Turin, Italy
\item \Idef{weizmann}{Department of Particle Physics and Astrophysics, Weizmann Insitute of Science, Rehovot, Israel}
\item \Idef{kyiv}{Kyiv Institute for Nuclear Research (KINR), Natl. Acad. of Sci. of Ukraine (NASU)}
\item \Idef{disat}Dipartimento DISAT del Politecnico di Torino, Turin, Italy
\item \Idef{piemonte}Dipartimento di Scienze e Innovazione Tecnologica dell'Universit\`{a} del Piemonte Orientale, \\Alessandria, Italy
\item \Idef{stonybrook}Department of Physics and Astronomy, Stony Brook University, SUNY, Stony Brook, New York, USA
\item \Idef{groningen}{Department of Radiation Oncology, University of Groningen, Groningen, The Netherlands}
\item \Idef{rice}Department of Physics and Astronomy, Rice University, Houston, Texas, USA
\item \Idef{ucatania}Dipartimento di Fisica e Astronomia dell'Universit\`{a} di Catania, Catania, Italy
\item \Idef{catania}INFN, Laboratori Nazionali del Sud, Catania, Italy
\item \Idef{lfn}INFN, Laboratori Nazionali di Frascati, Frascati, Italy
\item \Idef{tamu}Cyclotron Institute and Department of Physics and Astronomy, Texas A\&M University, College Station, Texas, USA
\item \Idef{lyon}Institut de Physique des 2 Infinis de Lyon, Université de Lyon, CNRS/IN2P3, Lyon, France

\end{Authlist}
\endgroup

\end{document}